\documentclass[3p,times]{elsarticle}
\pdfoutput=1
\usepackage{hypernat,mathtools}
\usepackage[pdftex]{hyperref}
\usepackage[applemac]{inputenc}

\graphicspath{{figures/}}

\usepackage[title,titletoc]{appendix}

\def\vec#1{{\bf #1}}

\newcommand{\CCCl}{CsCuCl$_3$}
\newcommand{\CCC}{Cs$_2$CuCl$_4$}
\newcommand{\CCB}{Cs$_2$CuBr$_4$}
\newcommand{\CCCB}{Cs$_2$CuCl$_{4-x}$Br$_x$}
\newcommand{\Cupz}{Cu(pz)$_{\text{2}}$(ClO$_{\text{4}})_{\text{2}}$}
\newcommand{\LiX}{Li$_{\text 2}$VOXO$_{\text 4}$}
\newcommand{\LiS}{Li$_{\text 2}$VOSiO$_{\text 4}$}

\newcommand{\AAVO}{AA'VO(PO$_{\text 4}$)$_{\text 2}$}
\newcommand{\BCVO}{BaCdVO(PO$_{\text 4}$)$_{\text 2}$}

\begin{document}


\begin{frontmatter}
\journal{Physics Reports}
\title{Frustrated two dimensional quantum magnets}
\author{Burkhard Schmidt}
\ead{bs@cpfs.mpg.de}
\author{Peter Thalmeier}
\address{Max-Planck-Institut f\"ur chemische Physik fester Stoffe, 01187 Dresden, Germany}
\begin{abstract}
We overview physical effects of exchange frustration and quantum spin fluctuations in (quasi-) two dimensional (2D) quantum magnets ($S=1/2$) with square, rectangular and triangular structure. Our discussion is based on the $J_1$-$J_2$ type frustrated exchange model and its generalizations. These models are closely related and allow to tune between different phases, magnetically ordered as well as more exotic nonmagnetic quantum phases by changing only one or two control parameters. We survey ground state properties like magnetization, saturation fields, ordered moment and structure factor in the full phase diagram as obtained from numerical exact diagonalization computations and analytical linear spin wave theory. We also review finite temperature properties like susceptibility, specific heat and magnetocaloric effect using the finite temperature Lanczos method. This method is powerful to determine the exchange parameters and g-factors from experimental results. We focus mostly on the observable physical frustration effects in magnetic phases where plenty of quasi-2D material examples exist to identify the influence of quantum fluctuations on magnetism.
\end{abstract}

\begin{keyword}
frustrated magnetism \sep magnetocaloric properties \sep linear spin wave theory \sep numerical exact diagonalization \sep finite temperature Lanczos method \sep layered vanadium compounds.
\end{keyword}
\end{frontmatter}

\tableofcontents


\section{Introduction}
\label{sect:introduction}

At sufficiently low energies the magnetic degrees of freedom in compounds with transition elements can be mapped to effective spin exchange models. In the insulating state the range of these interactions $J_1$, $J_2$ etc. does not extend beyond a few neighbors. Then due to the directional character of classical spins (either single component Ising or multiple component vector type)  commonly there is no unique ground state  which simultaneously minimizes the energy of all exchange bonds if at least some of them are of the antiferromagnetic type. Such magnets are called `frustrated'~\cite{diep:94,diep:13}. As a consequence there are macroscopic number of many body states with low energies making it difficult for the magnet to develop long range order at low temperatures. In quasi-two-dimensional (2D) or quasi-one-dimensional (1D) compounds the 3D ordering eventually may appear due to interlayer or interchain coupling.

Commonly two characteristic temperatures are accessible experimentally: The paramagnetic Curie-Weiss temperature $\Theta_\text{CW}$ and the magnetic ordering temperature $T_\text N$. Within mean-field approximation to a simple Néel-type Hamiltonian, the ratio $f:=\Theta_\text{CW}/T_\text N$ is of the order of 1. Therefore, in case of predominantly antiferromagnetic exchange, a simple empirical signature of frustrated magnets might be that $f$ is considerably larger~\cite{obradors:88,ramirez:94,kaul:05} because the moments do not have a unique ordered ground state to select leading to a small $T_\text N$. However, $f$ is difficult to quantify generally because the influence of frustration on the 3D bulk $T_\text N$ depends on the details of the model. 

Naturally frustration and quantum effects are particularly pronounced in low dimension and in this review we restrict ourselves to $S=\frac{1}{2}$ Heisenberg quantum spin models in the simple and most important 2D (Bravais) lattices: square (rectangular) or triangular structures. Spin models on 2D lattices with basis like kagome~\cite{norman:16,chernyshev:15,nishimoto:13}, honeycomb (Heisenberg and Kitaev)~\cite{kalz:12,zhang:13,li:14,rau:16} Shastry-Sutherland~\cite{koga:00,matsuda:13,corboz:14} or checkerboard~\cite{bishop:12} will not be discussed here. They belong to the larger class of 2D `Archimedean lattices'~\cite{richter:04,farnell:14}.
However to obtain a proper perspective we also survey some other types of frustrated magnets in the introduction. 

Generally one distinguishes two types of spin exchange frustration, firstly  `geometric frustration' when only AF nearest neighbor (n.n.) bonds are present and the exchange bonds cannot be simultaneously minimized because of local geometric constraints enforced by the lattice coordination as e.g. in the triangular or kagome lattice. Secondly `interaction frustration' or competition when the n.n. interactions themselves are unfrustrated, as in the square lattice, but the exchange bonds to the next nearest (n.n.n.) or further neighbors introduce a conflict of spin orientation. In both cases the physical consequences are similar. In fact, as discussed later geometrically frustrated triangular and interaction frustrated square lattice models are related to each other.

Physically it is more important whether spin degrees of freedom have only a discrete symmetry as in Ising-type models or continuous rotation symmetry as in the Heisenberg case. In the former case (if transverse magnetic fields are excluded) the problem is one of classical statistical mechanics determined by the interplay of thermal fluctuations and magnetic frustration. This leads to typical effects like nonzero residual entropy in macroscopically degenerate ground states,  complicated modulated or incommensurate (IC) magnetic structures in the strong frustration case and appearance of magnetization steps and plateaux. The latter are common in Ising systems~\cite{kudasov:06} because spins cannot continuously rotate but must flip directions. If order appears at low temperature the ordered moments correspond to the classical value due to the absence of quantum fluctuations.

On the other hand in the non-classical Heisenberg spin systems one has the unavoidable simultaneous influence of frustration and quantum fluctuations (zero point motion of spin waves) at zero temperature complemented by thermal fluctuations (thermally excited spin wave modes) at finite temperature. For quantum spin systems ($S=1/2$) even the pure unfrustrated N\'eel antiferromagnet (AF) on the square lattice has an ordered antiferromagnetic (AF) moment $m_{\vec Q}\approx0.606S$ in units of $g\mu_\text B$ much reduced from the classical value $S$ due to zero point fluctuations. The influence of  frustration may dramatically enhance the zero point fluctuations, physically through the appearance of low energy spin wave branches that are flat along lines in momentum space. The magnetization is mostly smooth as function of field due to the possibility of continuous canting of moments but nonlinear due to field dependent quantum effects. However, at certain rational values of the magnetization (in units of $g\mu_\text BS$) like $1/2$, $1/3$, $\ldots$ etc. rather narrow plateaux in the magnetization may appear. In contrast to the Ising case they have a quantum origin and may be understood by a stabilization of spin-wave bound states in narrow field intervals. However as discussed in Sec.~\ref{sect:extended} when the spin-space anisotropy of the model  is tuned continuously  from Heisenberg to Ising case the narrow plateau widens into an Ising type magnetization step.

\begin{figure}
    \centering
     \includegraphics[width=.5\columnwidth]{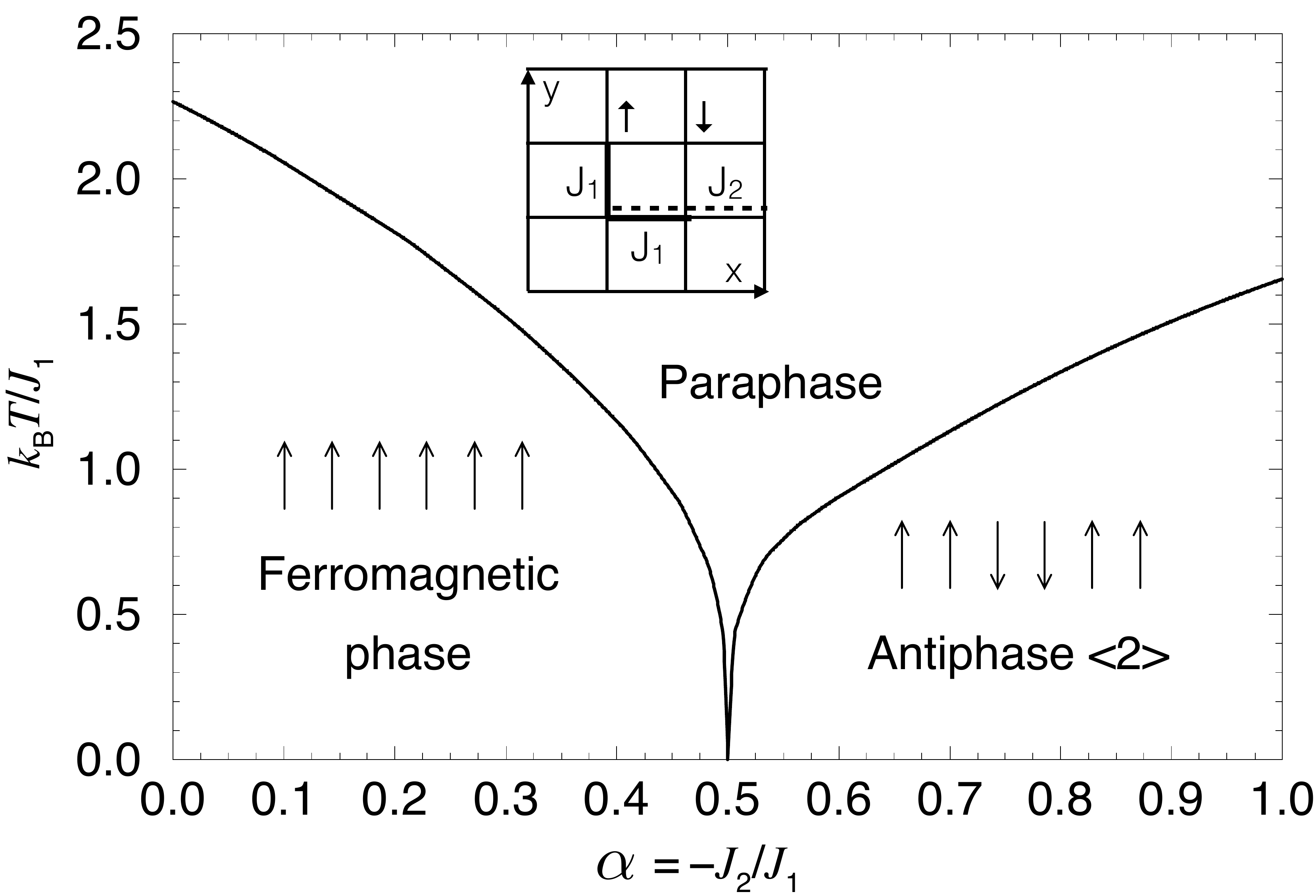}
     \caption{Phase diagram of the 2D ANNNI model $(J_1<0, J_2>0$) on the square lattice (inset). FM antiphase consist of FM chains ($\uparrow,\downarrow$) along y stacked along x as indicated. Here $\alpha_\text c=1/2$ is the critical point. Above the antiphase line an infinitely small region of a `floating phase' with algebraically decaying spin correlations may exists~\cite{derian:06}. In the 3D ANNNI model the FM chains become FM planes and then infinitely many phase lines for more complicated ( with larger unit cell than $\langle 2\rangle$) structures emanate from the critical point~\cite{bak:80,selke:84} (adapted from~\cite{derian:06}).}
    \label{fig:ANNNI2D}
\end{figure}

Because in Ising systems the effect of frustration is not entangled with those of quantum fluctuations we devote some space to mostly 2D frustrated Ising models in this introduction. It is also the reason why historically they have been considered first, namely the $J_1$-$J_2$ ANNNI (anisotropic (or axial) next nearest neighbor Ising) model on the square (rectangular) lattice~\cite{bak:82,selke:88}  and the $J_1$-$J_2$ (anisotropic) triangular lattice Ising antiferromagnet (TLIA)~\cite{wannier:50}. Furthermore, due to their classical nature they are accessible for standard Monte-Carlo (MC) numerical simulations~\cite{shirahata:01,sato:13}. The 2D ANNNI model is described by the Hamiltonian 
\begin{equation}
{\cal H}_\text{ANNNI}=J_1\sum_{\langle ij \rangle_{xy}\in\square}S_iS_j + J_2\sum_{\langle\langle ij\rangle\rangle_{x}\in\square}S_iS_j
\end{equation}
where $S_i$ are Ising spins and the FM n.n. couplings $\langle...\rangle$ $J_1<0$ extend along x,y directions while the AFM coupling $\langle\langle...\rangle\rangle$ J$_2 >0$ is present only for  n.n.n along x direction.
The phase diagram of this generic frustration model  is shown in Fig.~\ref{fig:ANNNI2D}. For all control parameters $\alpha=-J_2/J_1$ the ground state is magnetically ordered except at the critical point $\alpha_\text c=0.5$ where infinitely many phases with larger magnetic unit cell become degenerate.  The frustration or competition between $J_1, J_2$ coupling (inset) leads to the possibility of the antiphase denoted $\langle 2\rangle$. The temperature region immediately above the antiphase transition line was much discussed in the past~\cite{bak:82,selke:88}. It was proposed to host a `floating phase' without true long range order but algebraic decay of spin correlations. Over time this region has shrunk and is now believed to be exceedingly small~\cite{shirahata:01,derian:06}.
The model may be generalized to 3D, then it corresponds to FM (yz) planes coupled FM and AFM along $x$ (as in the inset of Fig.~\ref{fig:ANNNI2D}). It demonstrates the appearance of complex magnetic structures in order to compromise between frustrated exchange interactions. Mean field calculations~\cite{bak:80,selke:84} and MC simulations~\cite{gendiar:05} have shown that besides FM and antiphase $\langle 2\rangle$ many more phases with larger unit cells are stabilized and their phase lines emanate all from the critical point $\alpha_\text c$ and show further branching at higher temperature. Finally they terminate in a boundary line to the paraphase. Close to this line the order parameter behaves essentially as a sinusoidal incommensurate (IC) structure. At a fixed $\alpha >\alpha_\text c$ the $T$-dependence of the modulation wave vector $\vec q^*$ is first continuous and then shows step like `devil's stair case' behavior when locking in at various commensurate values, finally at $\vec q^*=\vec a^*/4=\pi/2$ for the antiphase $\langle 2\rangle$.
 
The interaction frustrated ANNNI model is a generic classical statistical model applicable not only to magnetism but also to charge order~\cite{ohwada:01} and structural order~\cite{janssen:87}, whenever sensible Ising degrees of freedom may be present. In magnetism the Ising behavior is frequent among rare earth compounds which have an incomplete $4f^n$ ($n=1-13$) shell with total angular momentum $J$.  The crystalline electric field (CEF) effect splits the $(2J+1)$-fold degenerate multiplet. When the ensuing ground-state, separated by a large CEF splitting from other states,  is a (Kramers or non-Kramers) doublet, an Ising pseudo-spin corresponding to the doublet degeneracy may be introduced which describes the low energy physics. A number of $4f$ compounds of this kind like monopnictides CeSb, CeP~\cite{kohgi:00}, triarsenide EuAs$_3$~\cite{chattopadhyay:86} and recently CeSbSe~\cite{chen:17} have been found where the magnetic phase diagram shows a similarity to the frustration driven devil's stair case phenomenon and commensurate lock-in transition of the ANNNI model.

\begin{figure}
    \centering
     \includegraphics[width=.45\columnwidth]{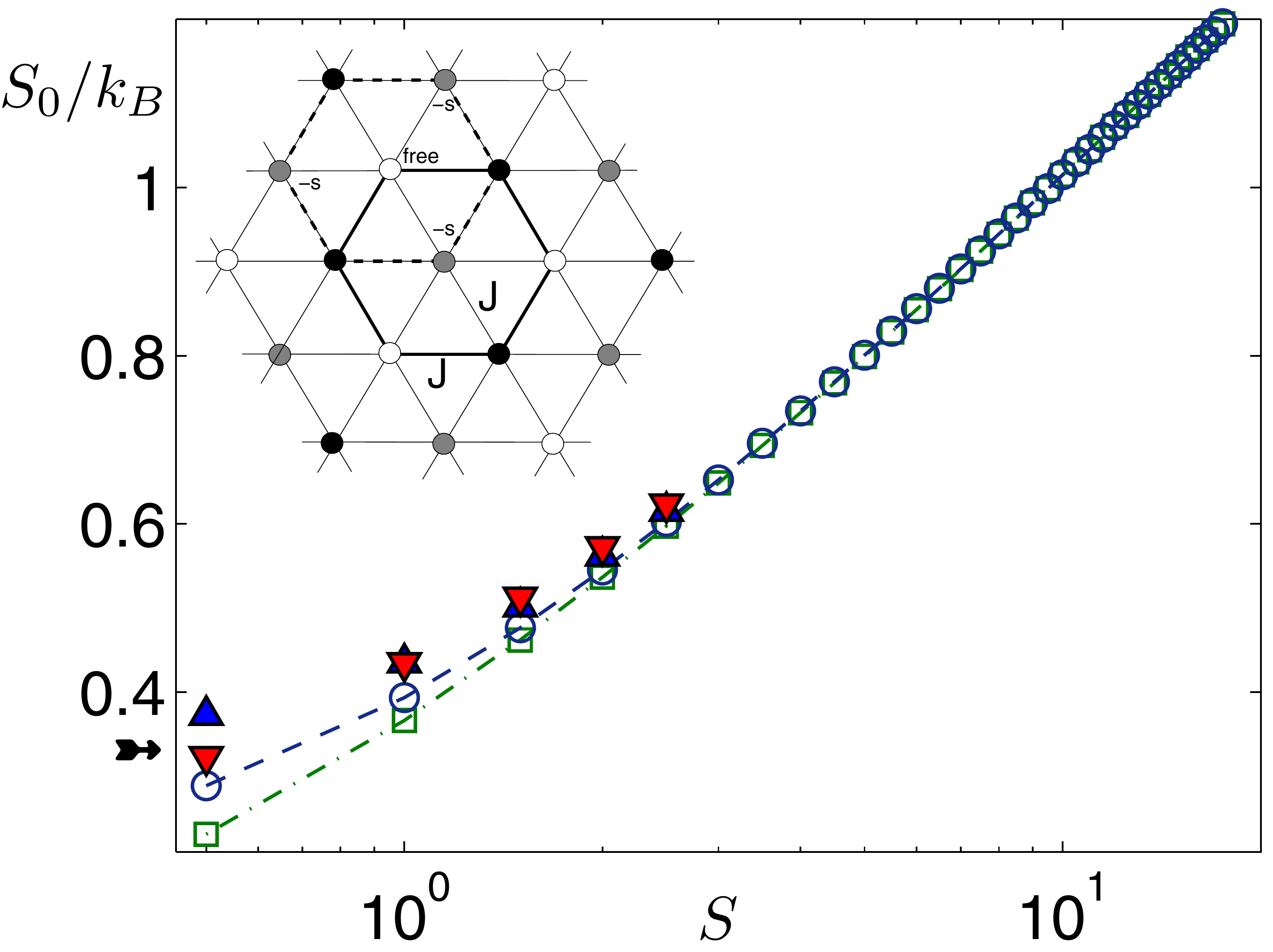}
     \caption{Residual entropy (per site) of the TLIA model as function of spin size S obtained from various numerical methods (symbols) applying finite size scaling. 
     For $S=1/2$ the arrow indicates Wannier's~\cite{wannier:50} exact analytical result $S_0/k_\text B = 0.32306$. Inset shows possible ground state 
     configuration with 1/3 each $\pm S$ (black/white circles) and 1/3 (grey) free (disordered) spins. Dashed hexagon corresponds to 
     alternative configuration (adapted from~\cite{zukovic:13}).}
    \label{fig:triIsing}
\end{figure}

It is instructive to compare the 2D geometrically frustrated n.n. TLIA model with the 2D ANNNI model; the former is defined by (inset of Fig.~\ref{fig:triIsing})
\begin{equation}
{\cal H}_\text{TLIA}=J\sum_{\langle ij\rangle\in\triangle} S_iS_j
\end{equation}
In contrast to the 2D ANNNI model it has no low temperature ordered phase, at least in the case $S=1/2$ and up to $S_\text c\simeq 4$~\cite{nagai:93}. Instead the ground state is known to have a finite residual entropy and algebraic decay of spin correlations. It may be estimated by considering a possible ground state spin configuration  with energy $E_0=(J/3)S^2$ per bond and $1/3$ of the spins in $+S,-S$ state or free ($\pm S$ equally probable), respectively. Then a rough lower bound of the residual entropy per site is $S_0/k_\text B > (1/3)\ln(2S+1)$ or  $S_0/k_\text B> 0.231$ for spin 1/2. It must be larger because `misfit' clusters (dashed lines in the inset) are allowed energetically. For $S=1/2$  its exact value was determined early by Wannier~\cite{wannier:50} to be $S_0/k_\text B=0.32306$. The residual entropy curve as function of general spin $S$ is shown in Fig.~\ref{fig:triIsing} from various numerical methods~\cite{zukovic:13} where the arrow indicates Wannier's value for spin $S=1/2$.

\begin{figure}
    \centering
    \hfill
    \includegraphics[width=.3\columnwidth]{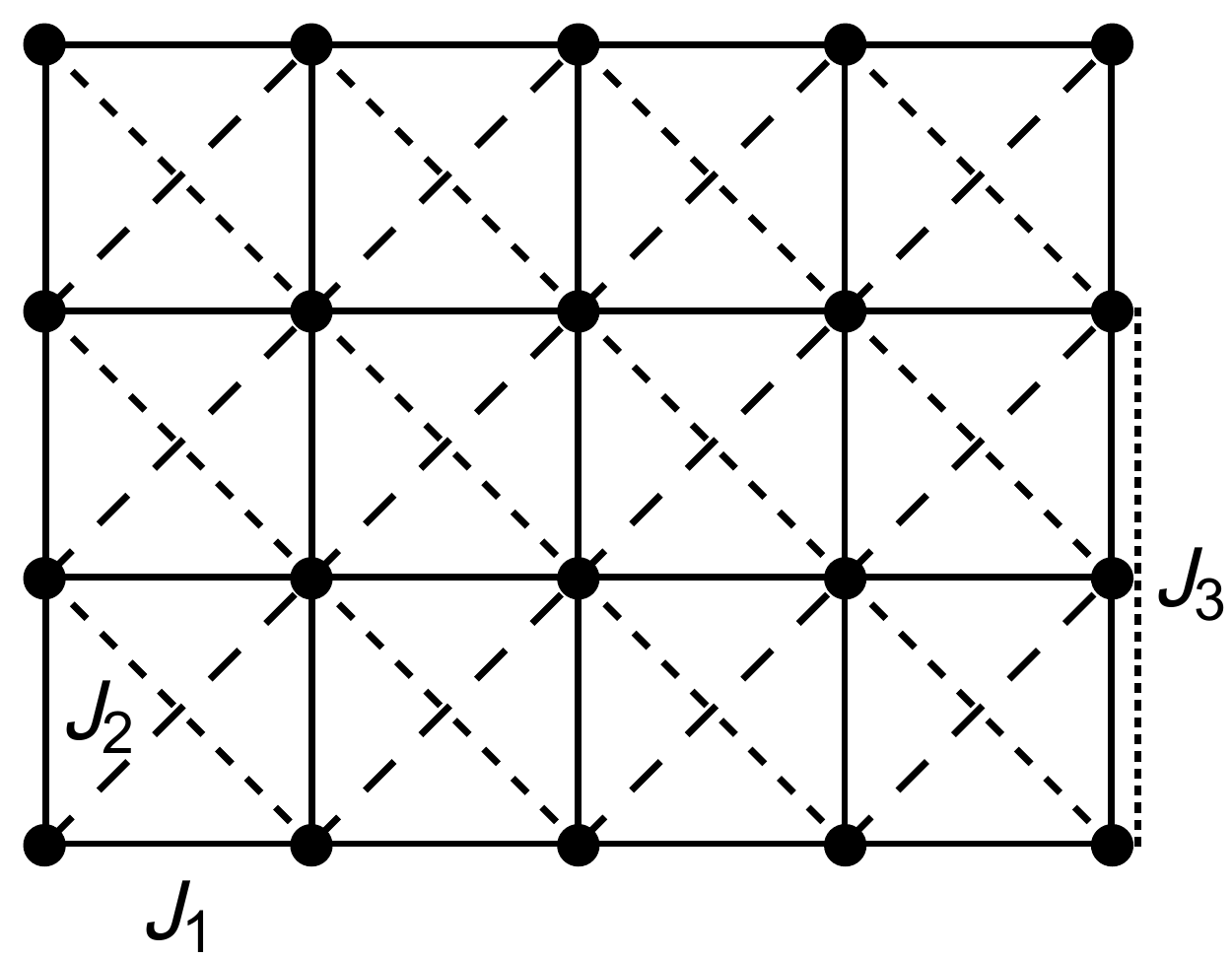}
    \hspace*{.03\columnwidth}
    \raisebox{.015\columnwidth}{\includegraphics[width=.38\columnwidth]{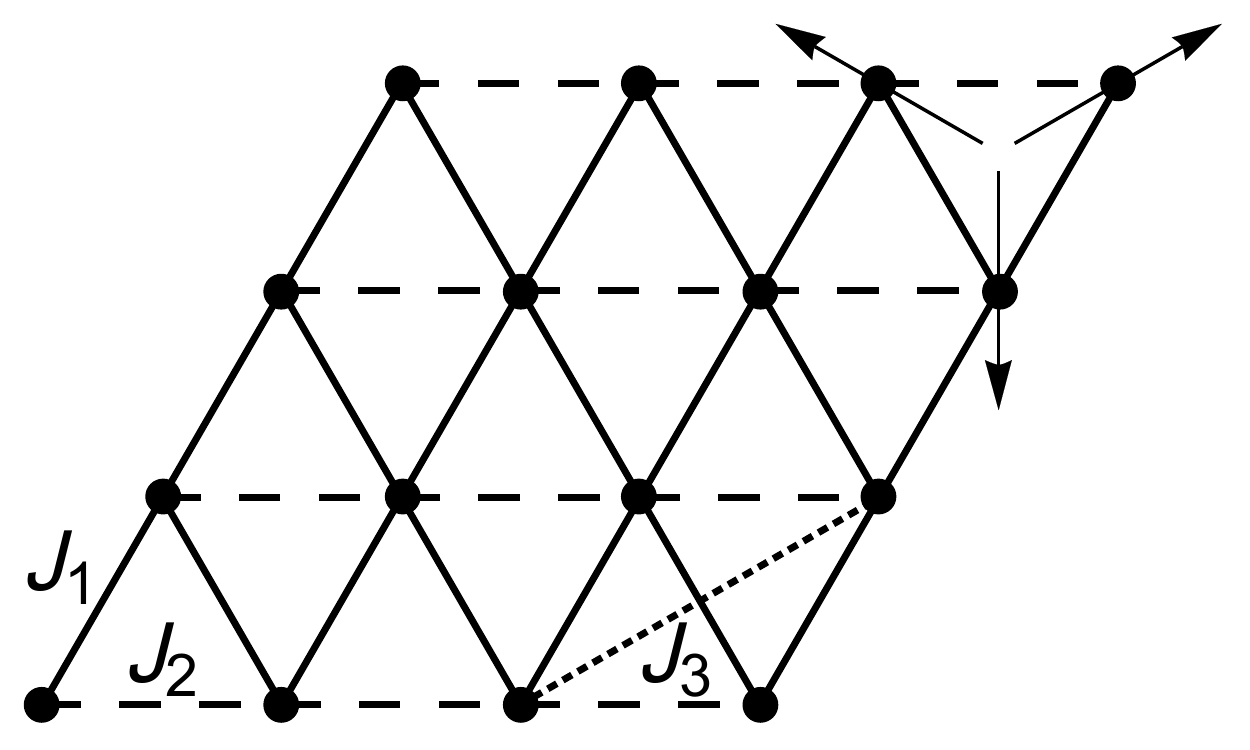}}
    \hfill\null
    \caption{Square or anisotropic rectangular (a) and anisotropic triangular (b) lattice exchange 
    models.\protect\footnotemark[1]
    Exchange bonds $J_1$ (solid lines) and $J_2$ (dashed lines) are indicated. In the rectangular lattice $J_1\rightarrow 
    J_{1a},J_{1b}$ becomes anisotropic.
     The triangular lattice may be obtained from the square lattice model by deleting one set of 
     diagonal $J_2$ bonds (short-dashed lines) and 
     tilting the lattice. Further neighbor interactions $J_3$ are included (dotted lines). 
     They are {\em not\/} equivalent in $\square$ and $\triangle$ lattices. Arrows indicate the frustrated 
     $120^\circ$ spin order of isotropic ($J_1=J_2$) triangular model. }
    \label{fig:lattices}
\end{figure}

We emphasize that the {\em classical\/} $S=1/2$ TLIA model has a disordered ground state with residual entropy whereas the isotropic n.n. {\em quantum\/} Heisenberg model on the same lattice (Fig.~\ref{fig:lattices}) has long range magnetic order in the $120^\circ$  spiral structure (wave vector $\vec Q =(2\pi/3,2\pi/3)$), although it was originally also thought to be in a disordered resonant valence bond (RVB) state. The ordered moment of the spiral structure is, however reduced to $m_{\vec Q}/S = 0.41$ due to quantum fluctuations~\cite{white:07}. The reason for the different ground state is that due to the possible continuous SU(2) spin rotations the frustration in the Heisenberg case may be somewhat relaxed by choosing the compromise  $120^\circ$ spin structure. In fact the microscopic degree of frustration $\kappa$ in the elementary triangular plaquette as defined in Eq.~(\ref{eqn:kappa}) is lower in the Heisenberg model $(\kappa=4/7=0.57)$ than in the Ising model with $\kappa=2/3=0.66$.

More recently geometrically frustrated Ising systems which are generally denoted as `spin ice' compounds have gained a large prominence~\cite{bramwell:01,gingras:14}. This is because they may host a classical realization of an effective magnetic monopole gas in a Coulomb phase with long-range interactions. These cubic compounds of the type R$_2$M$_2$O$_7$ ($\text R=\text{Ho}, \text{Dy}$ and $ \text M=\text{Ti},\text{Sn}$) have the 3D geometrically frustrated pyrochlore structure with a network of corner-sharing tetrahedra with $4f$ moments sitting at their vertices. Again, by the action of a CEF potential their local ground states may be Ising-pseudospin doublets. However, the essential new aspect of spin ice is that the local Ising axis is different at each corner, pointing along one of the four [111] cubic diagonals that meet in the tetrahedron's center. The pyrochlore structure may also be viewed as an alternative stacking of 2D planes with triangular and kagome structure~\cite{hiroi:03}. Due to large moments $(\mu\sim 10\mu_\text B)$ with extreme Ising character not only n.n. exchange interactions but also long range classical dipole-dipole couplings are important. In the elementary tetrahedra all configurations with two spins pointing to the tetrahedron center and two spins pointing away (`2-in, 2-out' structure), leading to zero net spin have the lowest energy. In the corner sharing network then there is a macroscopic number of possible ground state  configurations~\cite{anderson:56}, they can be mapped into each other by applying spin flips along closed hexagon loops connecting a ring of tetrahedrons. The 2-in, 2-out `ice'-state was first proposed for the structurally frustrated cubic water ice where the `spin' corresponds to real  displacement of protons towards or away from the oxygen~\cite{pauling:35}. The estimate for the residual  entropy per site of this macroscopically degenerate manifold of states is $S_0/k_\text B=(1/2)\ln(3/2)$ which is in good agreement with results from specific heat measurements in the spin-ice pyrochlore  Dy$_2$Ti$_2$O$_7$~\cite{ramirez:99,hiroi:03}. They show a Schottky-type peak corresponding to a spin flip energy scale $J/k_\text B\approx 1$K.

\begin{figure}
    \centering
     \includegraphics[width=.45\columnwidth]{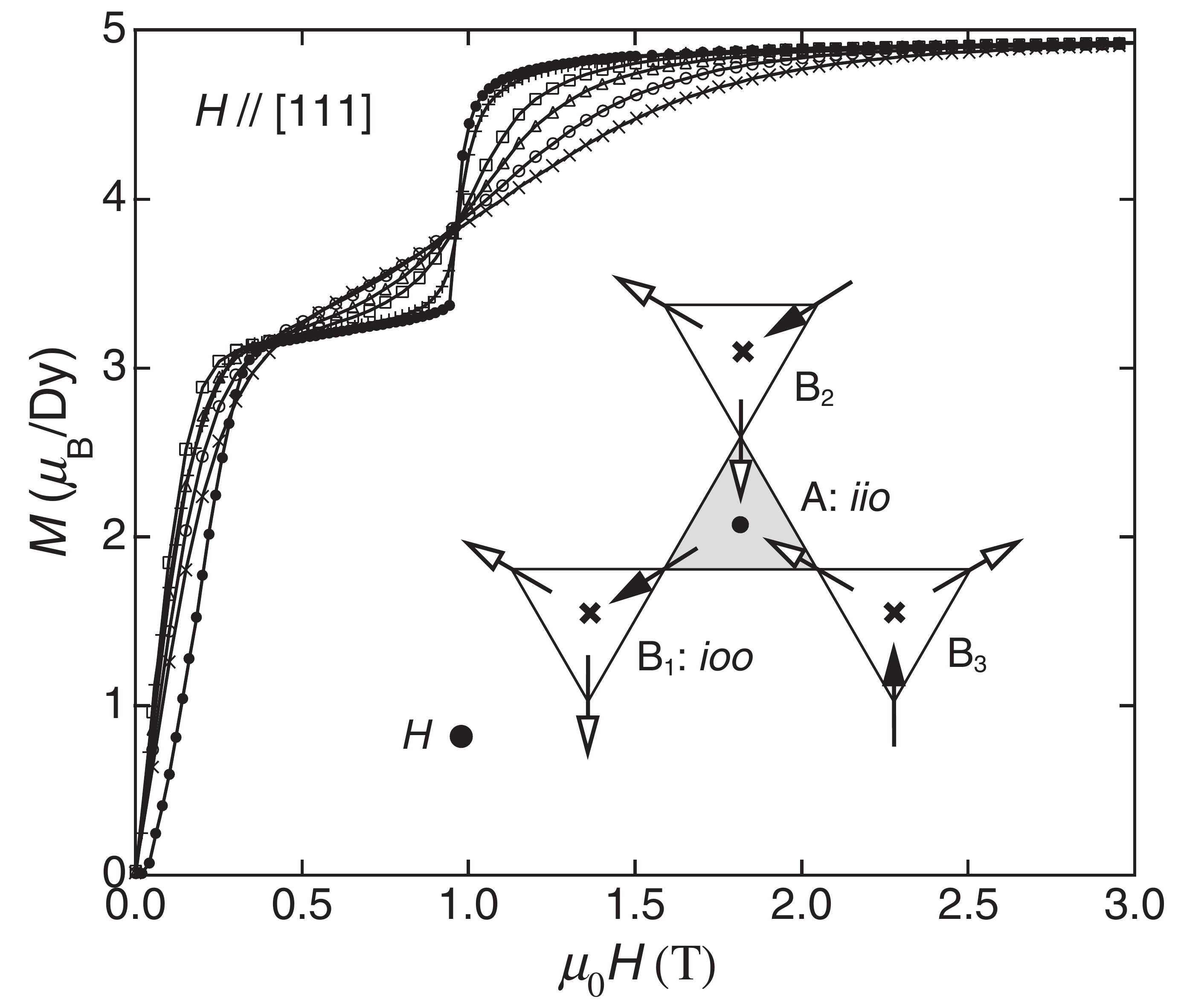}
     \caption{Magnetization curve of 3D pyrochlore spin ice Dy$_2$Ti$_2$O$_7$. The plateau at 
     $M\approx 3\mu_\text B$ is due to the stabilization of 
     2D spin ice state in the frustrated kagome layers with 2-in,1-out (iio) and 1-in,2-out (ioo) spin 
     configurations on corner-sharing triangles. Dark and white arrows indicate
     spins with and without components antiparallel to the out of plane field $\parallel$ [111] and temperatures
     between $0.34$ and $1.65\,\rm K$  (adapted from~\cite{hiroi:03}). }
    \label{fig:kagoice}
\end{figure}

 In a magnetic field along [111]  the 3D tetrahedron spin ice state becomes polarized and finally changes to the 2D kagome spin ice state with 2-in, 1-out (or opposite) spin configuration as depicted in the inset of Fig.~\ref{fig:kagoice} (kagome planes $\perp$ to field direction). It is accompanied by an Ising magnetization plateau in its stability range as shown in the same figure. The residual entropy of this 2D spin ice according to Pauling's estimate~\cite{hiroi:03} is $S_0/k_\text B=\ln(4/3)$ which agrees approximately with the experimental value~\cite{hiroi:03}.

The spin-ice model in pyrochlores is popular for another reason: By representing the classical pointlike dipoles at the tetrahedron links by discrete dipoles with opposite `magnetic charges' sitting at the tetrahedron centers,
the spin ice model may be mapped to a Coulomb gas of magnetic charges~\cite{castelnovo:08}. The ice rule $\sum_a\vec S_a\cdot\hat{\vec z}_a=0$ ($a=1\ldots4$ tetrahedron sites, $\hat{\vec z}_a$ is oriented along local $[111]$ direction to center) has the meaning of a divergence free condition for the spin field $\vec S_a(\vec r_i)$. Coarse grained over tetrahedron centers $i$ the spin field then corresponds to an effective magnetic field whose sources and sinks are the constituent monopoles of the spin-ice dipoles. The monopoles are created by flipping a dipole on the tetrahedron corner, which costs a spin flip energy $J$. At temperatures above $T=J/k_\text B$ the monopoles are deconfined and can diffuse to large separations; the spin ice state with thermally excited spin flips may then be viewed as a gas of magnetic monopoles with long-range Coulomb interactions. Experimental evidence for the existence of  individual monopoles is still controversial~\cite{gingras:14}.

The notion of macroscopic residual entropy in the frustrated spin systems is in conflict with the common interpretation of the third law of thermodynamics which requires that the true ground state of a real material should be unique without a remaining degeneracy. There is evidence that the highly symmetric TLIA model develops quasi-long range order when small perturbations like anisotropy and further neighbor exchange are added~\cite{sato:13}. It is also expected from simulations that the residual entropy in pyrochlore and 2D kagome spin ice is removed by ordering at a temperature scale which is lower by about an order of magnitude compared to the dipolar spin-flip scale where the spin-ice state forms~\cite{bramwell:01}. The order is characterized by a staggered structure of 2-in, 2-out and 2-out, 2-in tetrahedrons or it may be a dipolar spin glass state~\cite{eyvazov:17}.
Spin models may also exhibit continuous accidental degeneracies for specific model parameters which are larger than requested by the global symmetries of the Hamiltonian. In such cases thermal fluctuations (and also quantum fluctuations) which normally work against order may lift these accidental degeneracies and restore order. This mechanism has been termed `order by disorder'~\cite{villain:80,shender:82} and an example is  discussed in Sec.~\ref{sect:quantumphases}.

The classical frustration picture for Ising spins presented so far is dramatically changed when the transverse spin components of quantum models are introduced.  Then effects of frustration and quantum fluctuations have to be treated simultaneously, leading to new physical effects, even in the case of magnetically ordered ground states. In this case instead of thermally excited spin flip excitations with finite energy one has a zero point contribution in the ground state energy resulting from dispersive spin wave excitations. If the spin exchange model  contains a continuous spatial symmetry then Goldstone modes appear which enhance this contribution, in particular in low dimension.

The zero point fluctuations, depending on the frustrated interaction control parameters, may eventually destroy the ordered moment completely in a certain range of these parameters. The ensuing nonmagnetic states without long range moment order are generically called `spin liquids' or `quantum phases'. There are gapped as well as gapless spin liquids characterized by short range (exponential) or long range (algebraic) spin correlations, respectively. The most well known are the spin liquid state in the kagome lattice~\cite{nishimoto:13,norman:16}, the pyrochlore lattice~\cite{hermele:04,gingras:14} or in the square and triangular $J_1$-$J_2$ type Heisenberg models shown in Fig.~\ref{fig:lattices} that are the topic of this review. It is the nonmagnetic quantum phases of these models that have drawn an enormous amount of attention. This is certainly justified from a fundamental theoretical point of view. They offer the possibility to study the emergence of exotic order like valence bond or spin dimer liquids and crystals and  spin nematic states with unconventional excitation spectra and spin correlations. This subject is well covered in existing reviews~\cite{misguich:12,starykh:15,gingras:14,savary:17}.\footnote{In this and all other multi-part figures we use the convention that panels are labeled a, b, c, \ldots from left to right and/or from top to bottom.}

 On the other hand it must be reminded that so far not a single square or triangular spin compound is known to show unanimous clear signature of any such quantum phases. In all triangular (generally anisotropic) or square (rectangular) compound classes investigated so far one eventually finds  magnetic order at low temperatures. The moment formation is stabilized by small symmetry breakings in the exchange terms that reduce the classical degeneracies introduced by frustration and in addition by the effect of the interlayer coupling $J_\perp$. Even though the latter may be smaller by several orders of magnitudes than the in-plane n.n. coupling $J$ (for the pure N\'eel case on the square lattice) the ordering temperature is sizeable because it is still proportional to the in-plane scale $J$ and it varies only logarithmically~\cite{yasuda:05} with $J_\perp/J$ according to
 \begin{equation}
 k_\text BT_\text N = aJ/[b-\ln(J_\perp/J)]
 \label{eqn:3DTN}
 \end{equation}
 where $a$, $b$ are numerical constants. This means that from a physical point of view the exclusive attention on quantum phases in the frustrated magnets seems not justified. In this review we therefore choose a different focus. We will mainly review the actual physical effects on frustrated spin systems that do order in one of the conventional AF phases. We discuss the results of systematic variation of ground state energies, magnetization and its quantum-stabilized plateaux, saturation fields, structure factor, susceptibility, specific heat and magnetocaloric effects as function of the frustration and anisotropy control parameters. In particular we study the behavior of ordered magnetic moment and its field dependence on these control parameters which gives the most direct visualization of the frustration effect in quantum magnets. 
 
 As theoretical methods we use the numerical exact diagonalization (ED) for finite size tiles complemented by finite size scaling procedure for the ground state properties. These numerical results are accompanied by analytical linear spin wave (LSW) calculations to cross check the overall validity range of these methods and support the confidence in the results. For thermodynamic properties the finite temperature Lanczos method (FTLM) is employed throughout which gives excellent results for temperature above the finite size gaps. 
 
In reverse the comparison of theoretical and experimental results of these physical quantities is necessary to actually determine the frustration and anisotropy ratios of exchange parameters and to locate the position of a compound in the model phase diagram. In this context we show the results of systematic studies on the dependence of specific heat and susceptibility maxima positions and values as function of control parameters and their connections to the strength of frustration. Furthermore overall fitting of the temperature dependence of thermodynamic quantities using the FTLM will be discussed. We demonstrate that this is an excellent tool to extract the physical exchange parameters of 2D magnets.

This will be summarized for various important classes: oxo-vanadate, Cu-pyrazine and Cu-halide compounds.  Similar methods are also important for organic spin compounds~\cite{kanoda:11}. We show that the results of this analysis is in excellent agreement with direct spectroscopic determination of control parameters. In our view the aforementioned physical topics are the most important in the investigation of frustrated magnets. To make the review self contained however, we also briefly discuss some characteristics of quantum phases and in particular give a discussion why their existence may be prevented in real materials. This necessitates to discuss a few of the many possible generalizations of the generic $J_1$-$J_2$ models that are the central focus of this article.

The review is organized as follows: In Sec.~\ref{sect:frustmodel} we introduce the parametrization of square and triangular Heisenberg exchange models of Fig.~\ref{fig:lattices}  and their classical phase diagrams are presented in Sec.~\ref{sect:classical}. The numerical and analytical methods to investigate ground state properties are explained in Sec.~\ref{sect:quantum} and the numerical technique to compute finite temperature properties in Sec.~\ref{sect:temp}. Sec.~\ref{sect:structurefactor} contains a discussion of the static structure factor measured in magnetic neutron diffraction experiments. The nonmagnetic quantum phases are summarily discussed in Sec.~\ref{sect:quantumphases}. Sec.~\ref{sect:magplateaux} addresses characteristic plateaux formation which may appear in magnetization measurements. The application of theoretical analysis to three distinct classes of magnetic materials in presented in  Sec.~\ref{sect:compounds}. We indicate possible extensions of the generic exchange models and the influence on the (non-) existence of quantum phases in Sec.~\ref{sect:extended} while Sec.~\ref{sect:summary} gives a brief summary of our review. Finally some technical details on the numerical ED and FTL methods that have been used throughout this article can be found in the appendices.

\section{Frustration of Heisenberg exchange model on square and triangular lattice}
\label{sect:frustmodel}

In two dimensions five Bravais lattices exist; square, rectangular, hexagonal (triangular), oblique rectangular and rhombic \cite{dionne:92}. We will discuss the magnetism of the Heisenberg model on the first three lattices which are realized in numerous physical examples. If we restrict to AF spin exchange to next neighbors  only the triangular lattice shows competition between  formation of singlet exchange bonds which is called `geometric frustration'. However, if further neighbor exchange interactions are added competition of the nearest neighbor (n.n.)  and next nearest neighbor (n.n.n.) exchange bonds also appears in the square and rectangular lattice which is termed `interaction frustration'. Physically this distinction is not  important and indeed we will treat both types of frustration within the same theoretical framework. The spin Hamiltonian for the 2D model with spin S comprising exchange  and Zeeman term is then defined by  ${\cal H}={\cal H}_\text{ex}+{\cal H}_Z$, explicitly
\begin{equation}
    {\cal H}
    =
    \sum_{\left\langle ij\right\rangle}
    \vec S_{i}\hat{J}_{ij}\vec S_{j}
    -g\mu_\text B\mu_0\vec H\cdot\sum_i\vec S_i
    \label{eqn:h}
\end{equation}
where the first sum runs over bonds connecting sites $i$ and $j$ $^2$. 
\footnote{Here $\mu_0\vec H$ is the applied field, $g$ the gyromagnetic ratio and $\mu_B$ the Bohr magneton.} The exchange tensor $\hat{J}_{ij}$ consists of a symmetric part which will be restricted to a form with uniaxial (spin-space) anisotropy  and an asymmetric Dzyaloshinskii-Moriya (DM) term, respectively:
\begin{equation}
    {\cal H}_\text{ex}
    =
    \sum_{\left\langle ij\right\rangle}\left[
    J_{ij}\left( S^x_{i}S^x_{j}+S^y_{i}S^y_{j}\right)+
    J^z_{ij} S^z_{i}S^z_{j}+
     \vec D_{ij}\cdot\vec S_i\times\vec S_j\right].
    \label{eqn:hex}
\end{equation}
For most part of this review we consider only the spin-space isotropic ($J=J_z$) Heisenberg part which reads, explicitly
\begin{equation}
    {\cal H}_\text{ex}
    =
   J_1 \sum_{\left\langle ij\right\rangle_1}
    \vec S_i\cdot\vec S_j+
    J_2 \sum_{\left\langle ij\right\rangle_2}
     \vec S_i\cdot\vec S_j
    \label{eqn:hex12}
\end{equation}
with exchange bonds $J_{ij}=J_1$ for $i$, $j$ being nearest neighbors, and $J_{ij}=J_2$ for $i,j$ being next-nearest neighbors on square lattice as shown in Fig.~\ref{fig:lattices}. For the triangular lattice we use the convention that both $J_1,J_2$ are n.n. exchange bonds that may be real-space anisotropic, i.e. $J_1\neq J_2$, later we consider such possibility also in the rectangular lattice where $J_1^a\neq J_1^b$. In addition models with further neighbor interactions $J_3$ will be discussed later. It is obvious from Fig.~\ref{fig:lattices} that the two models are related: The anisotropic triangular case is topologically equivalent to the square lattice model by cutting one set of the diagonal $J_2$ bonds in the latter.

\begin{table}
    \centering
	\begin{tabular}[c]{lll}
	symbol & exchange  & model \\
	\hline
	 $\boxtimes$ & $J_1,J_2$ & general frustrated model \\
	 $\square$ & $J_2=0$ &pure N\'eel, latt. const. $a$ \\
	$\Diamond$ & $J_1=0$ & pure N\'eel, latt. const. $\sqrt{2}a$\\
	\hline
	$\Delta$ & $J_1,J_2$ & anisotropic triangular \\
        $\triangle$ & $J_1=J_2$ & isotropic triangular \\
        $\square$ & $J_2=0$ & pure N\'eel, latt. const. a \\
         $\parallel$ & $J_1=0$  & decoupled 1D chains  \\
	\end{tabular}
	\hfill
	\begin{tabular}[c]{llcc}
	square lattice & triangular lattice & \;\; $\phi/\pi$ & \;\; $J_2/J_1$  \\
	\hline
	 $\square$ & $\square$ & -1 & \;\; 0 \\
	FM/NAF & FM/NAF &\;\; -0.5 & \;\; $-\infty$  \\
	$\square$ & $\square$ & 0 &\;\;0\\
        NAF/CAF & NAF/SP & \;\;0.15 & \;\;0.5 \\
        $\boxtimes$ & $\triangle$ & \;\;0.25 & \;\;1.0 \\
         $\Diamond$ & $\parallel$ & \;\;0.5 & \;\; $\infty$  \\
	CAF/FM &SP/FM & \;\; 0.85 & \;\;-0.5  \\
	$\square$ & $\square$ & \;\; 1 & \;\;0 \\	
	\end{tabular}
    \caption
    {(a) definition of symbols for general and special exchange models for square (top three) 
    and triangular lattices (bottom four). (b) Equivalences of typical values for $J_2/J_1$ ratios and frustration 
    angle $\phi/\pi$ listed in anti-clockwise direction for $\phi$ $(-\pi\leq\phi\leq\pi)$.
    Values correspond to phase boundaries or special points (cf. Figs.~\ref{fig:frustration},\ref{fig:classical_PD}a). 
     }
    \label{tbl:specialvalues}
\end{table}
It is useful to introduce a convenient polar parametrization of the model given by 
\begin{eqnarray}
    J_{1}=J_{\text c}\cos\phi,
    &\quad&
    J_{2}=J_{\text c}\sin\phi,
    \\
    J_{\text c}=\sqrt{J_{1}^{2}+J_{2}^{2}},
    &\quad&
    \phi=\tan^{-1}\left(\frac{J_{2}}{J_{1}}\right)
    \nonumber
    \label{eqn:exchange}
\end{eqnarray}
where $J_\text c$ defines the energy scale and the polar angle $\phi$ is the frustration control parameter of the model that tunes between the different physical regimes. The advantage of using $\phi$ consists in mapping the model to a compact control parameter interval $-\pi\leq \phi \leq\pi$ in contrast to using directly the exchange ratio where $-\infty < J_2/J_1< \infty$. For ease of comparison we give special equivalent values in Table~\ref{tbl:specialvalues}.

The concept of `frustration' is central to the magnetism of 2D magnets, both in classical and quantum case. Therefore we give an intuitive measure which characterizes the strength of frustration effects as function of $\phi$. Obviously for some regions the model is unfrustrated, in particular when both $J_1,J_2 <0$ are FM whereas it is strongly frustrated when e.g. $J_2 \approx J_1 >0$. The measure is provided by the ratio of the exchange energy of the fundamental plaquette (i.e. square or triangle) to the sum of exchange energies of the decoupled dimers and trimers. For example in the triangular case 
the frustration degree $\kappa_\triangle(\phi)$ is given by
\begin{equation}
    \kappa_\triangle(\phi):=1-\frac{E_\triangle}{E_{\text t}+E_{\text d}}.
    \label{eqn:kappa}
\end{equation}
Here $E_\triangle$ is the ground state energy of the frustrated
triangle. $E_{\text t}:=E_{\triangle}(J_{2}=0)$ and $E_{\text
d}:=E_{\triangle}(J_{1}=0)$ are those of its
unfrustrated trimer and dimer parts, respectively where
\begin{equation}
    E_\triangle(\phi)
    :=
    \min\left(
    -\frac{3}{4}J_2,-J_1+\frac{1}{4}J_2,\frac{1}{2}J_1+\frac{1}{4}J_2
    \right),
    \label{eqn:etriangle}
\end{equation}
and the minimum is taken over the three different energy eigenvalues
of the triangle. A similar analysis for the square lattice models with
\begin{equation}
	E_\square(\phi)=\min\left(
	-\frac{3 J_2}{2},
	\frac{1}{2} \left(J_2-4J_1\right),
	-\frac{J_2}{2},
	\frac{1}{2} \left(J_2-2J_1\right),
	J_1+\frac{J_2}{2}
	\right)
\end{equation}
leads to
the identical result $\kappa_\square(\phi)\equiv\kappa_\triangle(\phi)$.

\begin{figure}
    \centering
    \includegraphics[width=.5\columnwidth]{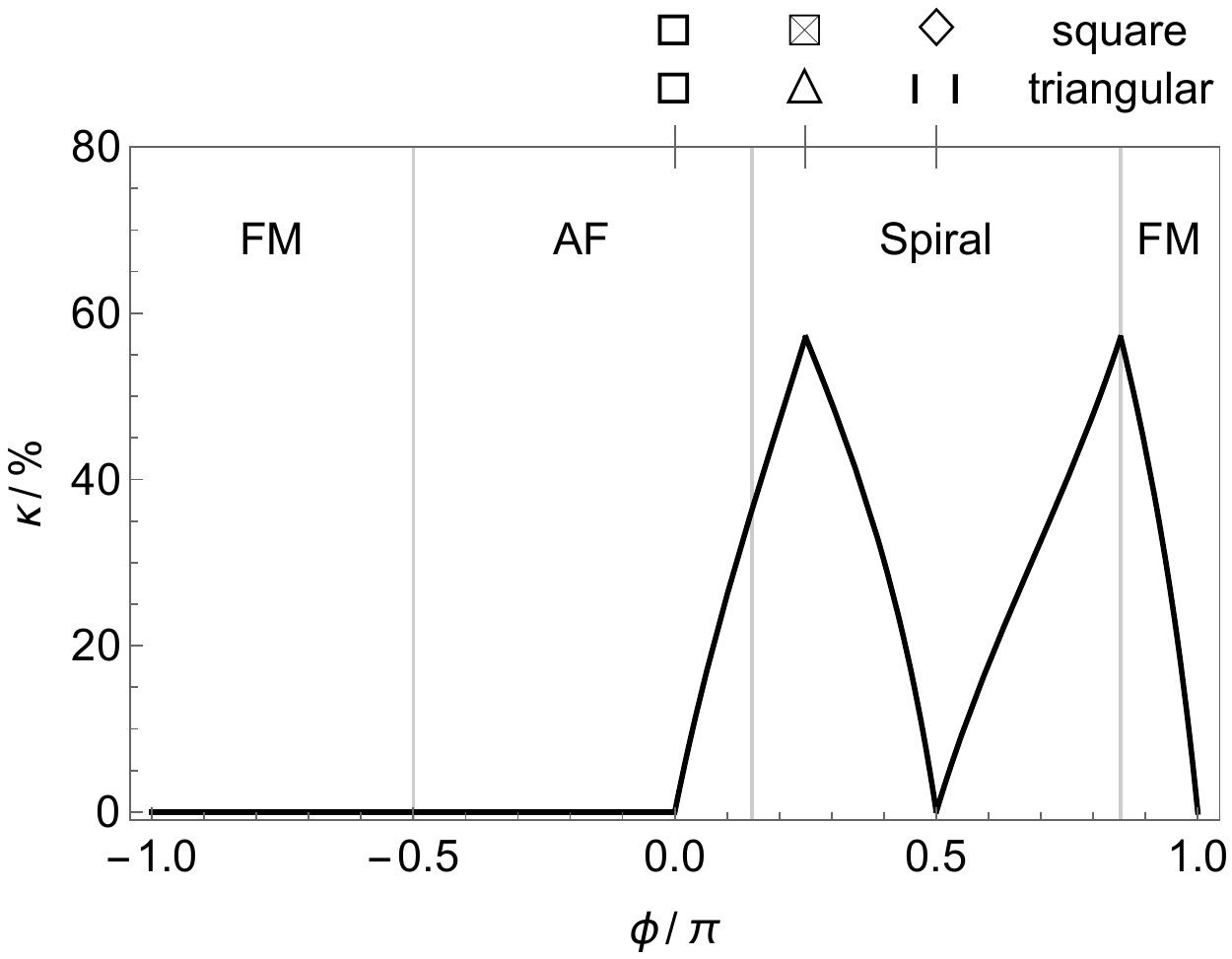}
    \caption{Degree of exchange  frustration in square  and anisotropic triangular lattices in per cent.
     The maximum appears at $J_2=J_1$ ($\triangle$, $\boxtimes$) and at the Spiral (SP)/FM (or 
     CAF/FM for square lattice) phase boundaries 
     for $J_2=-J_1/2$. Here and in following figures the symbols on top are defined by
     Table~\ref{tbl:specialvalues}a.
     }
    \label{fig:frustration}
\end{figure}
The degree of frustration is shown in Fig.~\ref{fig:frustration}. Obviously for $J_2 \leq 0$ the model is unfrustrated and also when $J_1=0$. The maximum frustration $\kappa=4/7$ is reached in the $J_2>0$ sector for the isotropic triangular case $J_1=J_2$ ($\phi=\pi/4$) and at $J_2/J_1=-0.5$ $(\phi=0.85\pi)$ corresponding to a classical phase boundary discussed below.  Close to these regions of maximum frustration its interplay with quantum fluctuation will lead to the strongest suppression of ordered ground state moment.

Various extensions of the above generic 2D exchange model have been proposed by adding further interaction terms like rectangular anisotropy \cite{bishop:08,schmidt:11}, $3^\text{rd}$ neighbor interactions \cite{chandra:90}, Dzyaloshinskii-Moriya (DM) asymmetric exchange \cite{veillette:05} and ring exchange \cite{chubukov:92,lauchli:05} (Sec.~\ref{sect:extended}).


\section{The classical phase diagram}
\label{sect:classical}

\begin{table}
    \centering
    \begin{tabular}{l|ll}
        phase
        &
        $\vec Q (\triangle) $
        &
        equivalent \vec Q
        \\
        \hline
        FM
        &
        $0$
        &
        $\left(\pm2\pi,\pm\frac{2\pi}{\sqrt3}\right)$
        \\
        CAF
        &
        $\left(\pm\pi,\pm\frac{\pi}{\sqrt{3}}\right)$
        \\
        NAF
        &
        $\left(\pm2\pi,0\right)$
        &
        $\left(0,\pm\frac{2\pi}{\sqrt{3}}\right)$
         \\
        SP
        &
        $\left(\pm2\tan^{-1}\left(\sqrt{4\left(\frac{J_{2}}{J_{1}}\right)^{2}-1}\right),0\right)$
        &
        $\left(\pm2\tan^{-1}\left(\sqrt{4\left(\frac{J_{2}}{J_{1}}\right)^{2}-1}\right),\pm\frac{2\pi}{\sqrt{3}}\right)$
	\\
	SP($120^\circ$)
	&
	$\left(\pm\frac{4\pi}{3},0\right)$
        &
        $\left(\pm\frac{2\pi}{3},\pm\frac{2\pi}{\sqrt{3}}\right)$
        \end{tabular}
    \caption{Classical ordering vectors for  triangular $(\triangle)$
    lattice.  For each phase the possible equivalent wave vectors in the first BZ that produce the
    same spin structure are given.}
   \label{tbl:Qvec}
\end{table}
As an instructive reference for further discussion we now introduce the classical phase diagram obtained from minimizing the classical ground state energy $E_{\text{cl}}=NS^2J(\vec Q)$  where $\vec Q$ is the modulation vector of the magnetic phase and $J(\vec k)$ the Fourier transform of the (spin-space isotropic) exchange tensor given by
\begin{equation}
    J(\vec k)=
    \left\{
      \begin{array}{l@{\qquad}l}
	  J_{1}(\cos k_x+\cos k_y)+2J_2\cos k_x\cos k_y&\boxtimes\\
	  2J_{1}\cos\frac{1}{2}k_x\cos\frac{\sqrt{3}}{2}k_y +J_{2}\cos k_x&\Delta
      \end{array}
    \right.
 \label{eqn:exfourier}   
\end{equation}
 in cartesian momentum coordinates.  Minimization of $E_{\text{cl}}$ with respect to $\vec Q$ leads to four possible magnetic phases:  Ferromagnet (FM): $\vec Q=(0,0)$, N\'eel antiferromagnet (NAF): $\vec Q=(\pi,\pi)$  and columnar antiferromagnet (CAF): $\vec Q=(\pi,0)$ or $(0,\pi)$ in the square lattice. A spiral phase (SP) with varying incommensurate $\vec Q(\phi)$ appears as ground state in the triangular lattice where the spiral wave vector interpolates continuously between that of Néel and FM phase, see Table~\ref{tbl:Qvec} for a complete list of ordering vectors. The classical value of the ordered moment for each phase is $m_{\vec Q}=S$.  (Unless otherwise noted, we express magnetic moments in units of $g\mu_{\text B}$.)
  
\begin{figure}
    \centering
    \includegraphics[width=.5\columnwidth]{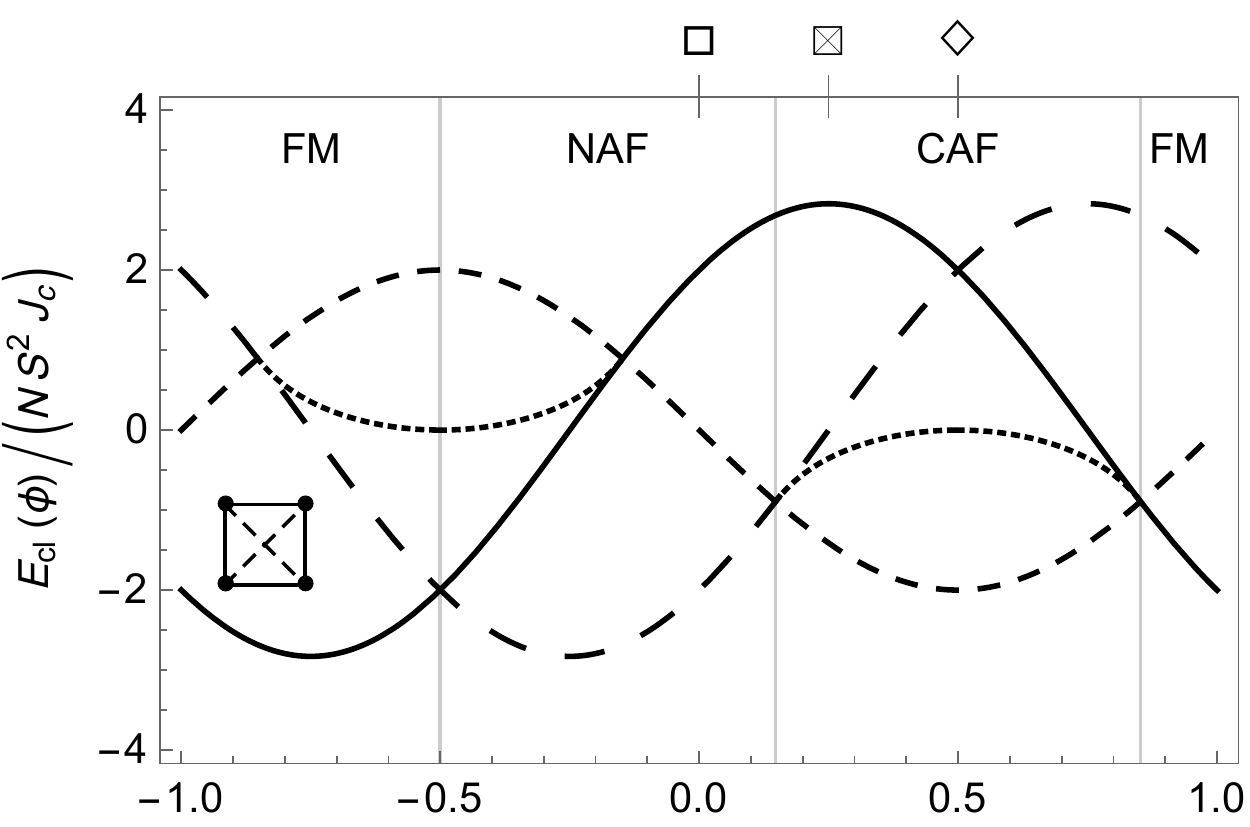}
    \includegraphics[width=.5\columnwidth]{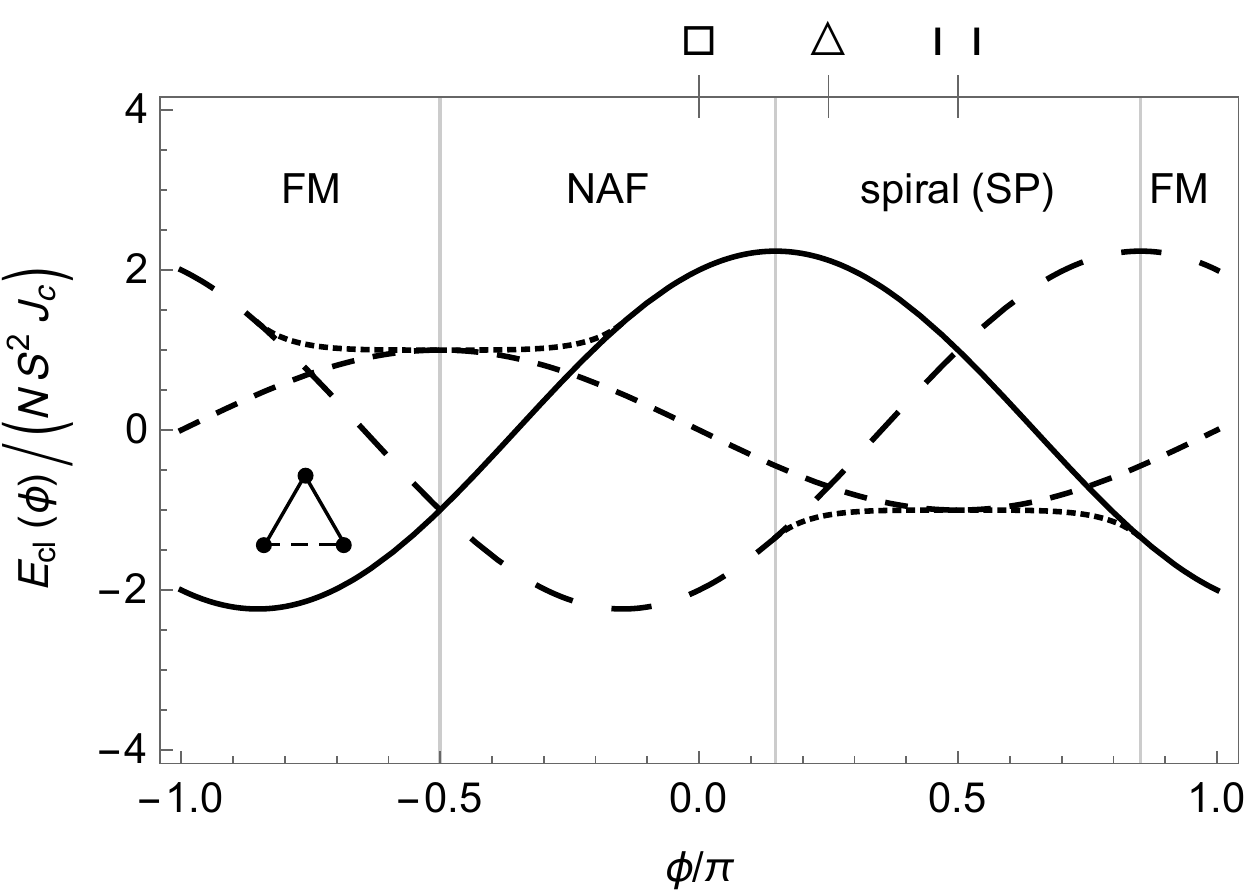}
    \caption{Classical energies of four possible phases as function of control parameter for square (a)  and 
    anisotropic triangular lattices (b).
    Positions of phase boundaries are identical. $E_\text{cl}(\phi)$ for the four possible phases are denoted by full line (FM), 
    long-dashed (NAF), dashed (CAF) and dotted line (SP). }
    \label{fig:classical_gsen}
\end{figure}
From the behavior of the ground state energies as function of control parameter presented in Fig.~\ref{fig:classical_gsen} for the two cases it is obvious that the classical phase {\it boundaries} are identical. However, in one sector for $J_2>0$ the incommensurate spiral (SP) phase appears as ground state in the triangular case instead of the commensurate CAF for the square lattice. From the energy curves we infer that this is caused by the missing $J_2$ bond in the triangular lattice which pushes the CAF energy above that of the SP phase.
\begin{table*}
    \small
    \begin{tabular}{lclll}
        phase
        &
        $E_{\text{cl}}/(NS^{2})$
        &
        ground state
        &
        conditions
        &
        range
        \\
        \hline
        ferromagnet (FM)
        &
        $2J_{1}+(1+\zeta)J_{2}$
        &
        $\boxtimes$ and $\Delta$
        &
        $J_{1}\le0\ \wedge\ \frac{J_{2}}{|J_{1}|}\le\frac 12$
        &
        $\phi\in [0.85\pi,\pi] \cup [-\pi,-\frac{\pi}{2}]$
        \\
        antiferromagnet (NAF)
        &
        $-2J_{1}+(1+\zeta)J_{2}$
        &
        $\boxtimes$ and $\Delta$
        &
        $J_{1}\ge0\ \wedge\ \frac{J_{2}}{J_{1}}\le\frac 12$
        &
        $\phi\in [-\frac{\pi}{2},0.15\pi]$
        \\
        columnar AF (CAF)
        &
        $-(1+\zeta)J_{2}$
        &
        $\boxtimes$ only
        &
        $J_{2}\ge0\ \wedge\ \frac{J_{2}}{|J_{1}|}\ge\frac 12$
        &
        $\phi\in [0.15\pi,0.85\pi]$
        \\
        spiral (SP)
        &
        $-J_{2}\left[1-\zeta+\frac12\left(\frac{J_{1}}{J_{2}}\right)^{2}\right]$
        &
        $\Delta$ only
        &
        $J_{2}\ge0\ \wedge\ \frac{J_{2}}{|J_{1}|}\ge\frac 12$
        &
          $\phi\in [0.15\pi,0.85\pi]$
        \\
        commensurate SP ($120^\circ$)
        &        
        $-\frac{3}{2}J_1$
        &
        $\triangle$ only
        &
        $J_1=J_2 >0$
        &
        $\phi=\frac{\pi}{4}$
    \end{tabular}
    \caption{Classical ground states: energies and conditions. 
    Here $\zeta=0$ for the anisotropic triangular lattice ($\Delta$) and
    $\zeta=1$ for the square lattice ($\boxtimes$).}
    \label{tbl:ecl}
\end{table*}
These results may best be summarized in a compilation shown in Table~\ref{tbl:ecl} valid for both square and triangular lattice. 

\begin{figure}
    \centering
    \hfill
    \includegraphics[width=.43\columnwidth]{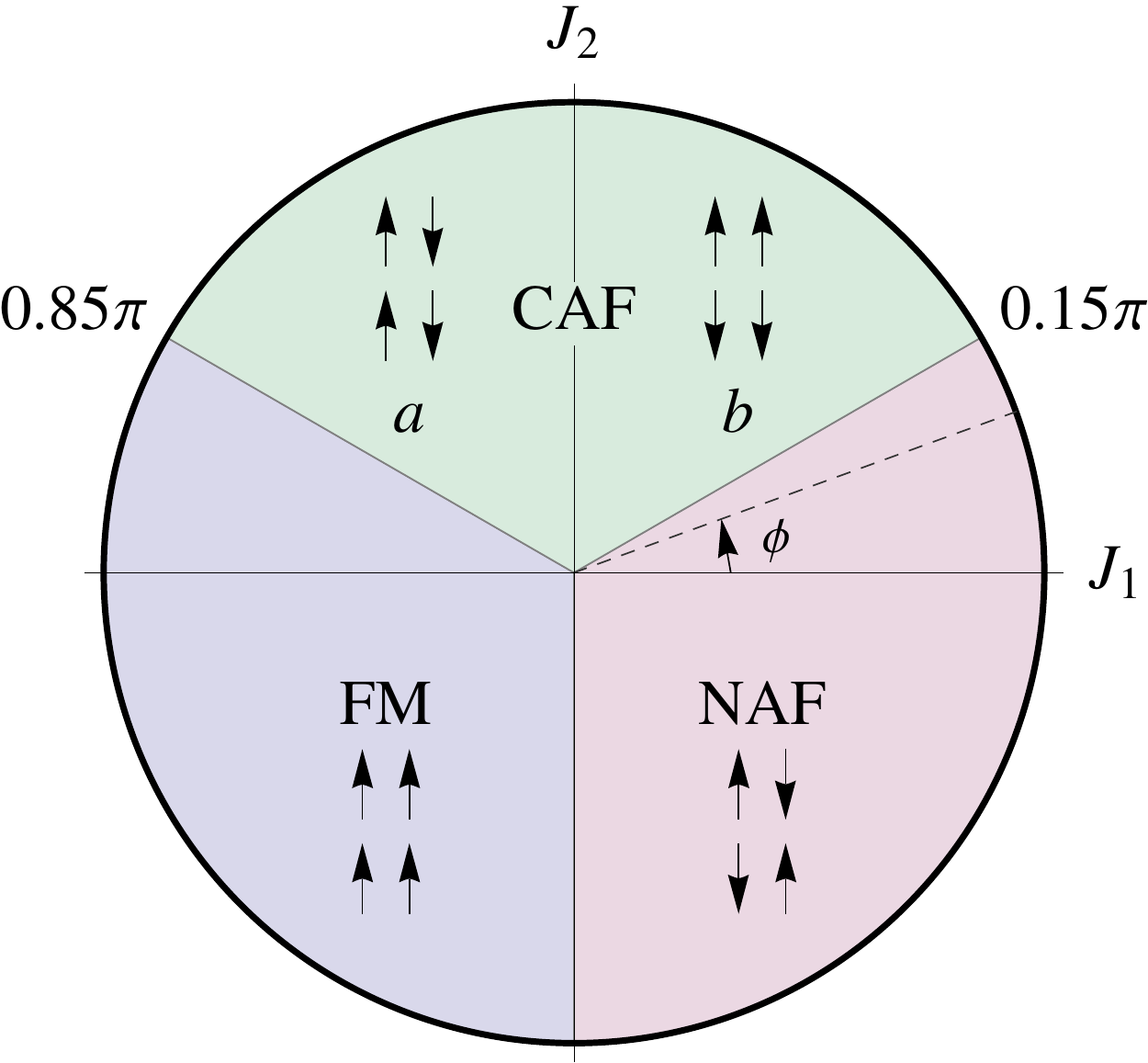}
    \hfill
    \includegraphics[width=.4\columnwidth]{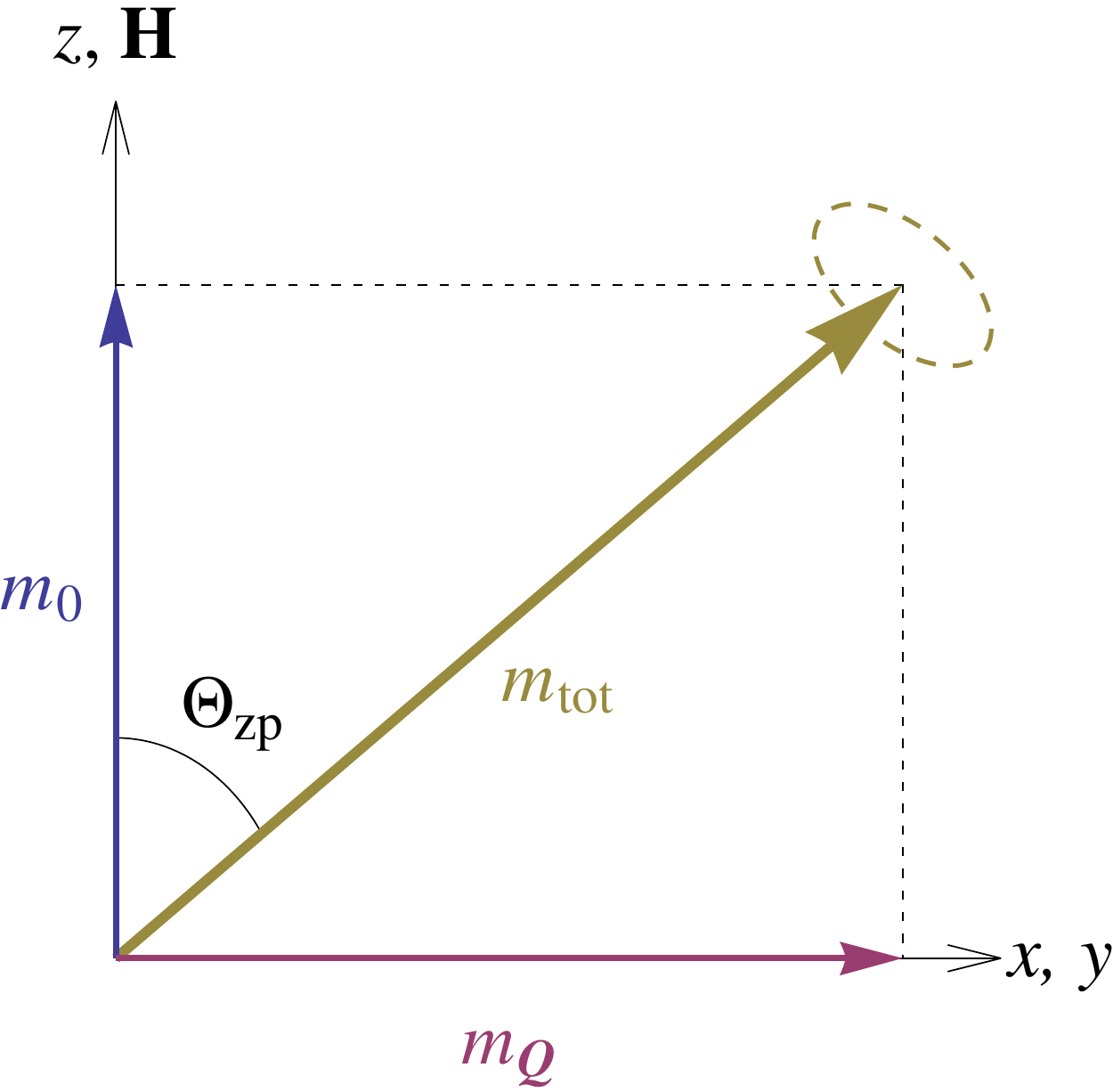}
    \hfill\null
    \caption{(a) Classical phase diagram $(-\pi\leq\phi\leq\pi)$ derived from classical 
    ground state energies in Fig.~\ref{fig:classical_gsen} shown for the 
    square lattice. The CAF phase is twofold degenerate (a,b). In the triangular lattice the CAF 
    phase is replaced by the spiral (SP) phase. (b) Geometry of moments in the field 
    $\vec H\parallel\hat{\vec z}$. Here $m_0=m_\text{tot}\cos\Theta_\text{zp}$, 
    $m_{\vec Q}=m_\text{tot}\sin\Theta_\text{zp}$, $m_\text{tot}$ are homogeneous, 
    ordered (wave vector $\vec Q$) and total moment size respectively. $\Theta_\text{zp}$ is the total 
    moment canting vector counted from z, including quantum corrections. In the classical case
     $m_\text{tot}=S$ and $m_0$, $\Theta_\text{zp}\rightarrow \Theta_{\text{cl}}$ are given by Eq.~(\ref{eqn:enclass}).}
    \label{fig:classical_PD}
\end{figure}
\begin{figure}
    \centering
    \includegraphics[width=.5\columnwidth]{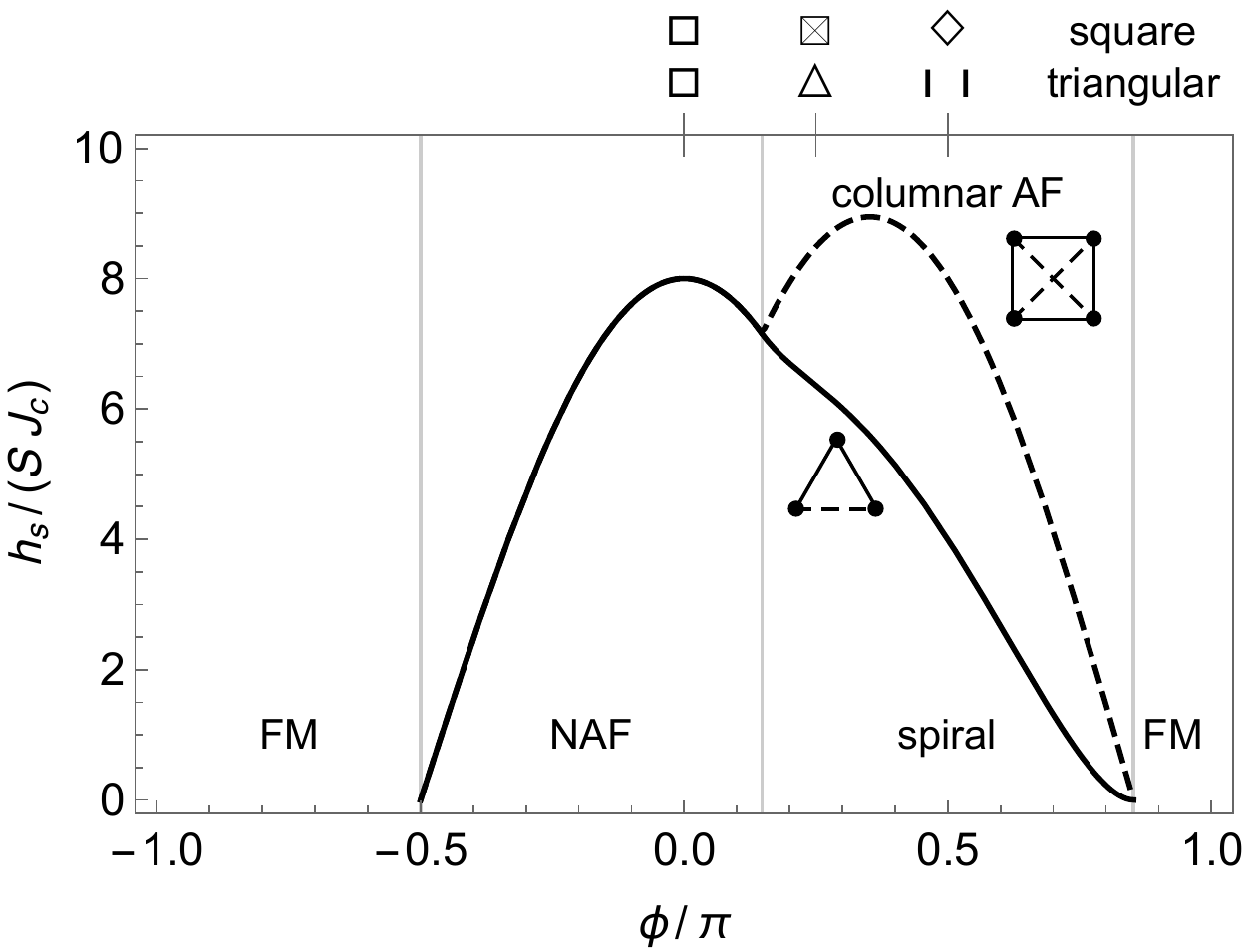}
    \caption{Classical saturation fields for triangular (full line) and square lattice (dashed line) models. They are asymmetric with 
    respect to $\phi=\frac{\pi}{4}$  $(J_1=J_2)$ position. }
    \label{fig:satfield_class}
\end{figure}

The classical analysis can be extended to include an external field $\vec H$ conveniently oriented along $z$ (this is irrelevant for the spin-space isotropic exchange of Eq.~(\ref{eqn:hex12})). In the antiferromagnetic phase the moments will be gradually tilted out of the plane by the field (Fig.~\ref{fig:classical_PD}b). We introduce $\Theta_{\text{cl}}$ as classical canting angle of moments from the z- direction such that $\Theta_{\text{cl}}=0$ for the fully polarized phase above the saturation field $H_\text s$ and $\Theta_{\text{cl}}=\pi/2$ for zero field, corresponding to moments lying in the plane which may be assumed as a consequence of an infinitesimal easy-plane anisotropy. Then the {\it classical} ground state energy for $N$ spins, classical canting angle $\Theta_\text{cl}$ and homogeneous magnetic moment $m_0$ are given by
\begin{align}
	E_{\text{cl}}
	&=
	NS^2\left[
	J(\vec Q)-\cos^2\Theta_{\text{cl}}\left(J(0)-J(\vec Q)\right)
	\right],
	\nonumber\\
	\cos\Theta_{\text{cl}}
	&=
	\frac{h}{2S(J(0)-J(\vec Q))}=\frac{h}{h_{\text s}},
	\nonumber\\
	m_0
	&=
	S\frac h{h_{\text s}}
\label{eqn:enclass}
\end{align}
with $h:=g\mu_{\text B}\mu_0H$. The exchange Fourier transform $J(\vec k)=(1/2)\sum_n J_n\exp(-i\vec k\vec R_n)$ comprises a sum over all bonds connecting an arbitrary but fixed site at position $\vec R_i$ with its neighbors at positions $\vec R_j=\vec R_i+\vec R_n$. Classically the canting increases linearly with the field corresponding  in general to a noncoplanar cone- or umbrella configuration of moments up to the saturation field
\begin{equation}
\frac{h_{\text s}(\phi)}{2S}=J(0)-J(\vec Q).
\end{equation}
Using $J(\vec Q)=E_{\text{cl}}\left/\left(NS^2\right)\right.$ for $\vec H=0$ from Table~\ref{tbl:ecl} with the physical wave vectors $\vec Q$ that minimize $E_{\text{gs}}$ we obtain
\begin{equation}
    \frac{h_{\text s}}{2S}=
    \left\{
      \begin{array}{l@{\qquad}ll}
	  4J_{1}&\mbox{NAF}&\boxtimes\mbox{ and }\Delta\\
	  2(J_{1}+2J_{2})&\mbox{CAF}&\boxtimes\mbox{ only}\\
	  2(J_1+J_2)+\frac{1}{2}\frac{J_1^2}{J_2}&\mbox{SP}&\Delta\mbox{ only}\\
	  \frac{9}{2}J_1 &\mbox{SP}&\triangle\ (120^\circ)
      \end{array}
    \right..
 \label{eqn:satclass2}   
\end{equation}
The curve of classical saturation fields for both models is shown in Fig.~\ref{fig:satfield_class}. The saturation field, together with susceptibility and specific heat is an important tool to determine $J_1,J_2$ for a real compound.

Due to the quantum nature of spin, in particular for $S=1/2$ and the low dimensionality the interplay of quantum fluctuations and frustration will strongly modify the classical picture as far as ground state energy $E_\text{gs}$, magnetization $m_0$ and ordered moment $m_{\vec Q}$ are concerned. The former ($m_0$) may exhibit strongly nonlinear behavior, including plateau formation while the latter ($m_{\vec Q}$) may become unstable in certain regions of the control parameter $\phi$ leading to nonmagnetic states, generically called `spin liquids' or to more exotic order than magnetic. Furthermore in the stable magnetic regime the size of the ordered moment may exhibit nonmonotonic field dependence due to quantum corrections. Finally, in the triangular lattice the classical ordering vector of the spiral phase (Table~\ref{tbl:Qvec}) is generally incommensurate. Its dependence on triangular anisotropy is shown in Fig.~\ref{fig:Q_triangular}.


\section{Methods to treat quantum fluctuations in the ground state }
\label{sect:quantum}

In the classical or mean field picture of magnetic order the spin structure is uniquely determined by minimizing the mean field ground state energy. Then it is expected that contributions of quantum fluctuations of spins to the total ground state energy do not change its structure but renormalize the physical properties like ground state energy, size of ordered moment and magnetization. This topic will be discussed now using the analytical linear spin wave approach and unbiased numerical exact diagonalization treatment. The more subtle question what happens when the classical ground state is not stable due to effect of quantum fluctuations or has a continuous degeneracy will be approached later in Secs.~\ref{sect:quantumphases} and~\ref{sect:magplateaux}.

\subsection{Linear spin wave theory and physical consequences}
\label{subsect:LSW}

We are mainly interested on the quantum effects of frustration in compounds with magnetic order, which comprises, as we shall see the largest part of the phase diagram. Therefore we may apply linear spin wave (LSW) theory based on the Holstein-Primakoff (HP) approximation \cite{keffer:66,manousakis:91,majlis:07}. Formally it corresponds to a $1/S$ expansion around the magnetically ordered phase in terms of bosonic eigenmodes, the magnon excitations. This approximation is valid in the low density, i.\,e. the low temperature limit. The zero point energy of these modes then leads to a correction to the classical ground state energy. Likewise this results in a modification of magnetization and staggered moment which are accessible physical quantities that provide a test for the validity and limitations of LSW theory. In the presence of an external field the HP expansion has to be performed in the local coordinate system whose $z$-axis is aligned with the tilted moment at every site. The details of this approach are described in Ref.~\cite{schmidt:10,siahatgar:12}.

The HP transformation from local spin variables $S^\alpha_i$ ($\alpha=x,y,z$) to bosonic variables $a_i,a_i^\dagger$  at site $i$ is given by $S_i^x\rightarrow\sqrt{S/2}\left(a_i+a_i^\dagger\right)$, $S_i^y\rightarrow-{\rm i}\sqrt{S/2}\left(a_i-a_i^\dagger\right)$,  and  $S_i^z\rightarrow S-a_i^\dagger a_i$. Then, performing the Fourier transform $a_i=(1/\sqrt N)\sum_{\vec k}a_{\vec k}^\dagger\exp(-{\rm i}\vec k\cdot\vec R_i)$ the exchange Hamiltonian Eq.~(\ref{eqn:h}) may be written as a bilinear (harmonic) form in $\hat{a}^\dagger_{\vec k}=\left(a^\dagger_{\vec k}, a_{-\vec k}\right)$ which may be diagonalized (Eq.~(\ref{eqn:hdiagonal})) by the Bogoliubov transformation 
\begin{equation}
    \alpha_{\vec k}
    =
    u_{\vec k}a_{\vec k}+v_{\vec k}a_{-\vec k}^{\dagger},
    \quad
    \alpha_{-\vec k}^{\dagger}
    =
    v_{\vec k}a_{\vec k}+u_{\vec k}a_{-\vec k}^{\dagger}
\end{equation}
to the magnon creation and annihilation operators $ \alpha_{\vec k}^\dagger,  \alpha_{\vec k}$ of spinwave modes given in Eq.~(\ref{eqn:ek}). The transformation coefficients are obtained as
\begin{eqnarray}
    &&u^2_{\vec k}=
    \frac{1}{2}\left[
    \frac{A(\vec k)
    -\cos^{2}\Theta\left(B(\vec k)+2A(0)\right)
    +(h/S)\cos\Theta}{E(h,\vec k)}
    +1\right]
    \nonumber\\
  &&v^2_{\vec k}=
    \frac{1}{2}\left[
    \frac{A(\vec k)
    -\cos^{2}\Theta\left(B(\vec k)+2A(0)\right)
    +(h/S)\cos\Theta}{E(h,\vec k)}
    -1\right]
    \label{eqn:bogol}
\end{eqnarray}
with the sign convention $u_{\vec k}=\mathop{\rm sign}\nolimits B(\vec k)|u_{\vec k}|$, $v_{\vec k}=|v_{\vec k}|$ and $E(h,\vec k)$ denoting magnon energies given below. We note that $\Theta$ in Eq.~(\ref{eqn:bogol}) is still the general canting angle that is determined by minimizing the total ground state energy. Just minimizing the classical part $E_\text{cl}$ leads to $\Theta_\text{cl}$. In the $1/S$ expansion scheme the latter appears, however, within the first $1/S$ quantum correction $E_\text{zp}$.
The final result of the HP transformation  is then the free magnon Hamiltonian
\begin{eqnarray}
{\cal H}&=&E_\text{cl}+E_\text{zp}+S\sum_{\vec k} E_\text{sw}(h,\vec k)\alpha^\dagger_{\vec k}\alpha_{\vec k}\nonumber\\
E_\text{cl}&=&NS^2\bigl(J(\vec Q)-A(0)\cos^2\Theta_\text{cl}\bigr)\nonumber\\
E_\text{zp}&=&NSJ(\vec Q)+\frac{S}{2}\sum_{\vec k}E(h,\vec k)
\label{eqn:hdiagonal}
\end{eqnarray}
Here $E_\text{cl}$ is the (negative) classical ground state energy as before with $A(0)=J(0)-J(\vec Q)$, the second term $E_\text{zp}$ is the (negative) energy of zero point fluctuations of magnon modes and the last term describes the free Hamiltonian of excited magnons. The total ground state energy is  $E_\text{gs}=E_\text{cl}+E_\text{zp}$.  The zero point contribution is of relative order $1/S$ as compared to the classical part. The spin-wave or magnon energy  $\omega(h,\vec k)=SE_\text{sw}(h,\vec k)$ is obtained from the Bogoliubov transformation as 
\begin{align}
E_\text{sw}(h,\vec k)&=
E(h,\vec k)+C(\vec k)\cos\Theta_\text{cl},
\nonumber\\
E(h,\vec k)&=\sqrt{\left[A(\vec k)-B(\vec k)\cos^{2}\Theta_\text{cl}\right]^{2}
   -\left[B(\vec k)\left(1-\cos^{2}\Theta_\text{cl}\right)\right]^{2}},
    \label{eqn:ek}
\end{align}
where intra- and inter-sublattice interactions $A(\vec k)$ and  $B(\vec k)$ as well as the interaction $C(\vec k)$ which is antisymmetric in $\vec k$ (only relevant for the SP phase) are given by
\begin{eqnarray}
A(\vec k)&=&J(\vec k)
    +\frac{1}{2}
    \left[
    J(\vec k+\vec Q)+J(\vec k-\vec Q)
    \right]
    -2J(\vec Q),
    \nonumber\\
     B(\vec k)&=&J(\vec k)
    -\frac{1}{2}
    \left[
    J(\vec k+\vec Q)+J(\vec k-\vec Q)\right],
    \label{eqn:bk}
    \nonumber\\
    C(\vec k)&=&J(\vec k+\vec Q)-J(\vec k-\vec Q).
\label{eqn:swcoeff}    
\end{eqnarray}
We already mentioned that in the spirit of the $1/S$ expansion the appropriate canting angle $\Theta_\text{cl}$ to start from is the {\em classical\/} one of Eq.~(\ref{eqn:enclass}) even though this angle will also exhibit $1/S$ corrections due to zero point fluctuations leading to a modified $\Theta_\text{zp}$ (Eq.~(\ref{eqn:cant_zp}))~\cite{schmidt:10,thalmeier:08}. In zero field $\Theta_\text{cl}=\pi/2$ and Eq.~(\ref{eqn:ek}) reduces to 
\begin{equation}
E(h,\vec k)=
\sqrt{A^2(\vec k)-B^2(\vec k)}
\end{equation}
The total ground state energy $E_\text{gs}$ depends on frustration control parameter $\phi$ and field. The zero-point contribution is determined by the $\phi,h$-dependent spin wave energies which may become singular in the strongly frustrated regions of Fig.~\ref{fig:frustration}. Therefore we should expect that physical properties like ordered moment, magnetization and susceptibility will strongly depend on $\phi$ and show anomalous behavior in the same region. Within the LSW approach we obtain the following quantum corrected physical quantities per site \cite{schmidt:10,schmidt:14,thalmeier:08}:
\begin{figure}
    \centering
    \hfill
     \includegraphics[height=.4\columnwidth]{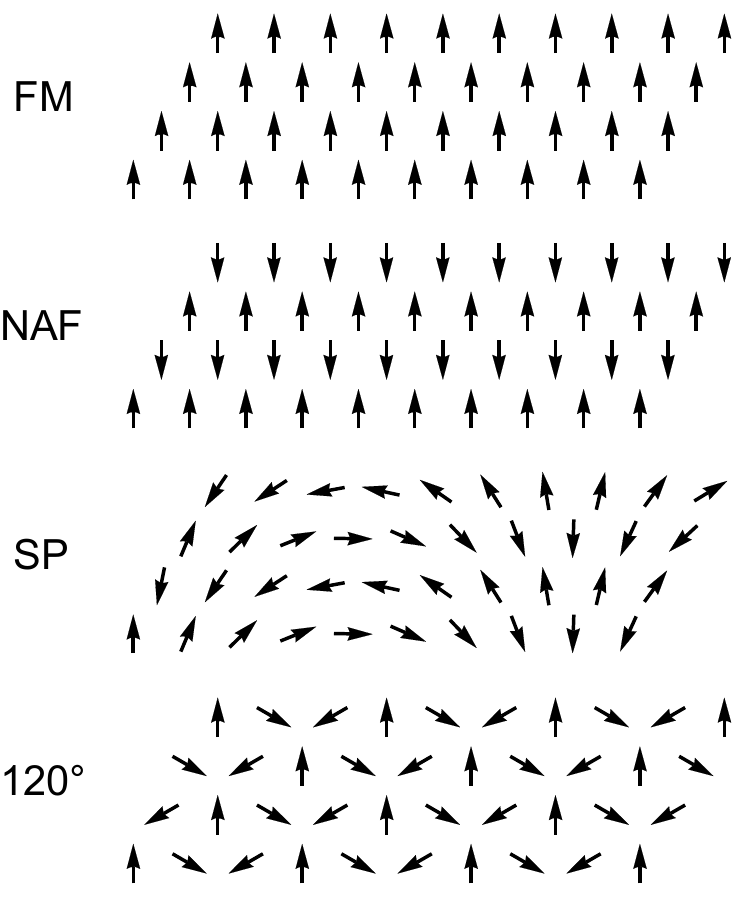}
     \hfill
    \includegraphics[height=.4\columnwidth]{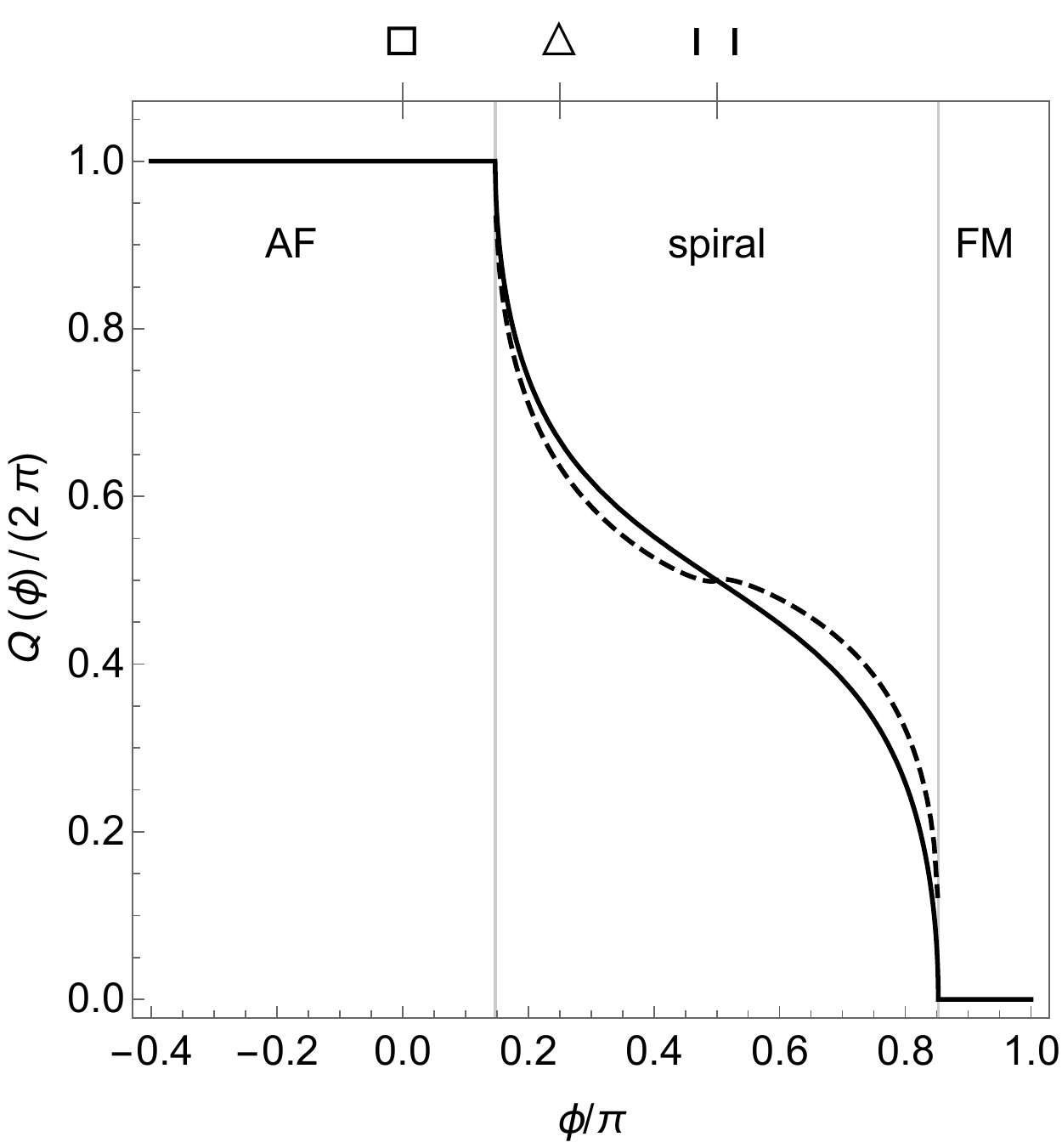}
    \hfill\null
     \caption{(a) Possible magnetic structures in the anisotropic triangular lattice. (b) wave 
     vector of the incommensurate spiral (SP) as function of control parameter $\phi$. Full line:
      classical value (Table~\ref{tbl:Qvec}), broken line: including $1/S$ quantum corrections.}
    \label{fig:Q_triangular}
\end{figure}
\paragraph{Ground state energy}
\begin{eqnarray}
\frac1{NS^2}E_\text{gs}(h)
&=&
\left(1+\frac1S\right)J(\vec Q)-A(0)\left(\frac h{h_\text s}\right)^2
\nonumber\\&\phantom=&
+\frac1{2NS}\sum_\vec k
\sqrt{\left[A(\vec k)-B(\vec k)\right]
\left[A(\vec k)+B(\vec k)\left(1-2\left(\frac h{h_\text s}\right)^2\right)\right]}.
\label{eqn:egsexplicit}
\end{eqnarray}
    
\paragraph{Ordered moment (in $g\mu_\text B$)}
    \begin{equation}
        m_{Q}(h)
        =
        S\sqrt{1-\left(\frac
h{h_{\text{s}}}\right)^{2}}
        \left\{
        1-\frac1{1-(h/h_{\text{s}})^{2}}\frac1{2S}
         \frac{1}{N}\sum_{\vec k}\left[
        \frac{A(\vec k)}{E(h,\vec k)}-1
        +\left(\frac h{h_{\text{s}}}\right)^{2}
        \frac{B(\vec k)}{E(h,\vec k)}
        \left(
        \frac{A(\vec k)-B(\vec k)}{A(0)}-1
        \right)
         \right]\right\}.
    \label{eqn:moment1}
    \end{equation}    
At zero field this reduces to
\begin{equation}
    m_{Q}(0)
    =
    S\left[
    1-\frac{1}{2S}\left(
    \frac{1}{N}\sum_{\vec k}
       \frac{A(\vec k)}{E(h,\vec k)}
    -1
    \right)
    \right].
    \label{eqn:moment2}
\end{equation}    
The numerical reference values for the simple square lattice NAF  ($\square, \phi=0$) are  $m_{\vec Q}/S =0.606$ and for the isotropic triangular lattice ($\triangle, \phi=\pi/4$)  $m_{\vec Q}/S =0.478$ \cite{schmidt:14}. 

\paragraph{Uniform moment (magnetization) (in $g\mu_\text B$)}
Likewise we obtain
\begin{equation}
    m_0(h)
    =
    S\frac{h}{h_\text s}
    \left[
    1+\frac{1}{2S}
    \frac{1}{N}\sum_{\vec k}
    \frac{B(\vec k)\left(A(\vec k)-B(\vec k)\right)}{A(0)
    E(h,\vec k)}
    \right]
    \label{eqn:magnetization}
\end{equation}
\paragraph{Total moment (in $g\mu_\text B$)}
The total moment according to the geometric presentation in Fig.~\ref{fig:classical_PD}b is given by $m^2_\text{tot}(h)=m_0^2(h)+m_{\vec Q}^2(h)$ and is obtained from
\begin{equation}
m_\text{tot} =S-\frac{1}{N}\sum_{\vec k}\langle a^\dagger_{\vec k}a_{\vec k} \rangle
=S-\frac{1}{N}\sum_{\vec k}v^2_{\vec k}.
\label{eqn:mtot1}
\end{equation}
Using the {\em classical\/} $\Theta_\text{cl}$ in Eq.~(\ref{eqn:bogol}) this leads to
\begin{equation}
m_\text{tot}(h) =S\left[
1-\frac{1}{2S}\left(
\frac{1}{N}\sum_{\vec k}
 \frac{B(\vec k)\left(A(\vec k)
    - B(\vec k)\right)}{E(h,\vec k)}
    -1\right)
\right].
\label{mtot2}
\end{equation}
\begin{figure}
    \centering
    \hfill
    \includegraphics[width=.4\columnwidth]{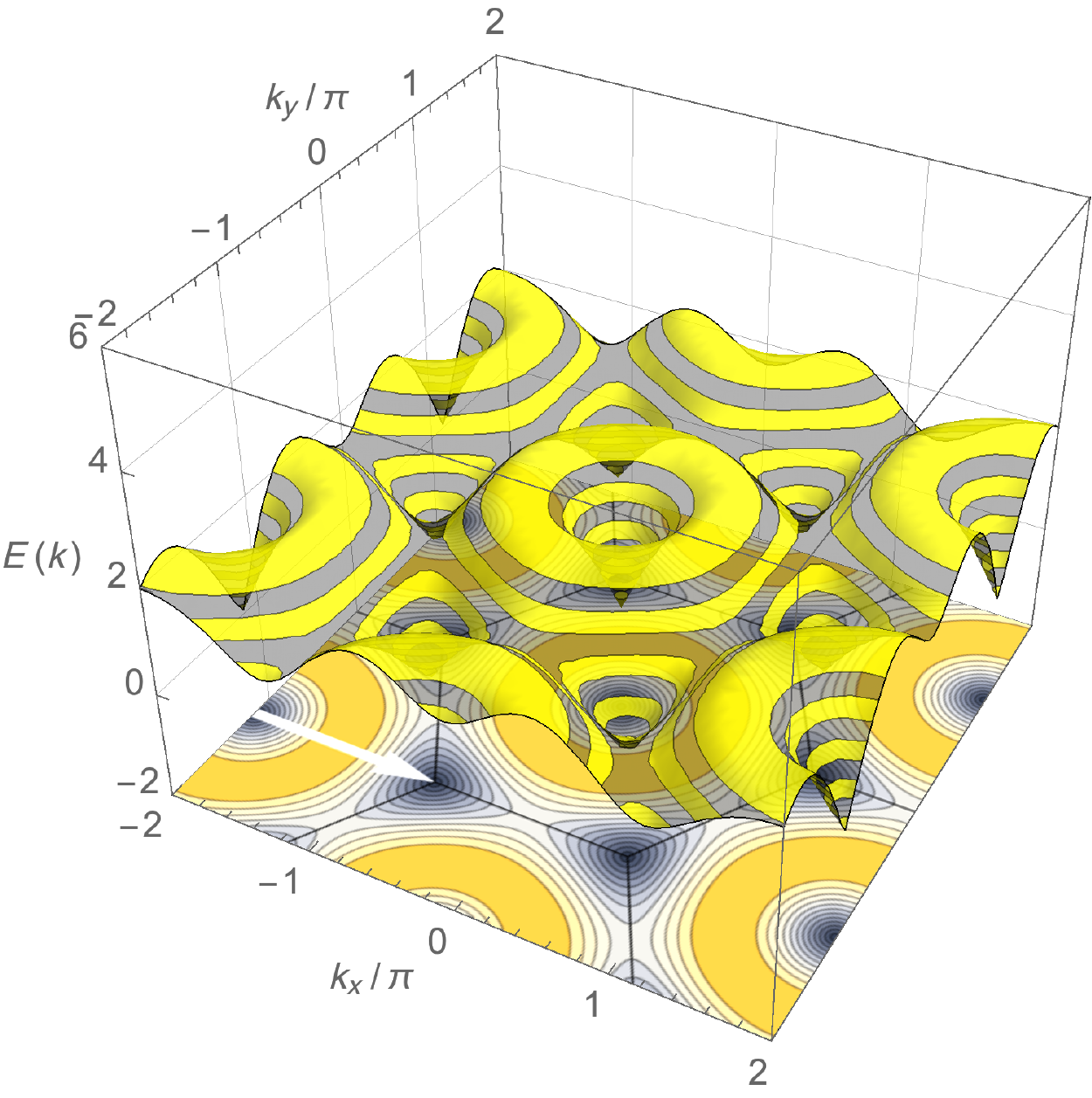}
    \hfill
    \includegraphics[width=.4\columnwidth]{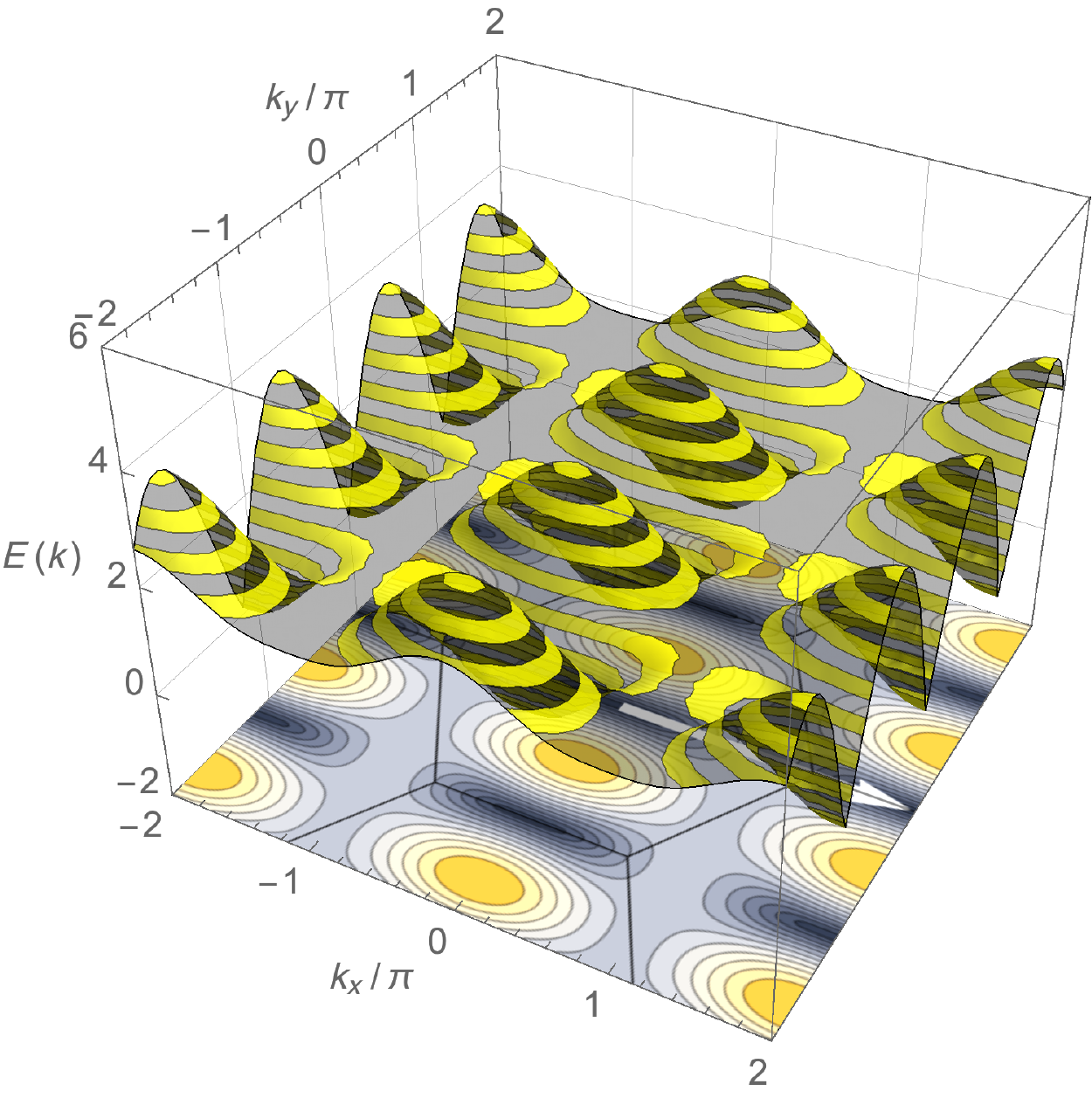}
    \hfill\null
    \caption{3D contour plots of spin wave dispersion $\omega(\vec k)=SE_\text{sw}(\vec k)$ for triangular lattice (a) isotropic 
    ($\triangle$, $J_1=J_2$) case and (b) at AF/Spiral phase boundary ($J_2 =J_1/2$). 
    Yellow and dark 2D contours in the 
    $k_x$-$k_y$ plane indicate the regions of large and small spin wave energy, respectively. }
    \label{fig:sw_triangular}
\end{figure}

\paragraph{Quantum corrected canting angle}
We may also obtain the renormalized canting angle either from $\Theta_\text{zp}=\tan^{-1}(m_0/m_{\vec Q})$ according to Fig.~\ref{fig:classical_PD}b or from minimizing the ground-state energy $E_\text{gs}$ with respect to $\Theta$,
    \begin{equation}
\cos\Theta_\text{zp}=\cos\Theta_\text{cl}\left[
1+\frac{1}{2S}
\left(\frac1N\sum_{\vec k}\frac
{A(0)\left(A(\vec k)- B(\vec k)\cos^2\Theta_\text{cl}\right) + B(\vec k)\left(A(\vec k)- B(\vec k)\right)}
{A(0)\sqrt{\left(A(\vec k)- B(\vec k)\right)\left(A(\vec k)+B(\vec k)(1+2\cos^2\Theta_\text{cl})\right)}}
-1\right)\right]
 \label{eqn:cant_zp}
 \end{equation}  
where $\cos\Theta_\text{cl}=h/h_\text s$.

\paragraph{Quantum corrected uniform differential susceptibility}
Finally we give the expression for $\chi_0(h)=\partial m_0(h)/\partial h$, 
\begin{equation}
    \chi_0(h)=
    \frac{1}{2A(0)}\left[
    1+\frac{1}{2S}\frac{1}{N}\sum_{\vec k}
    \frac{B(\vec k)\left(A(\vec k)-B(\vec k)\right)}%
    {A(0)E(h,\vec k)}   
    +\frac{1}{S}\cos^{2}\Theta_\text{cl}\frac{1}{N}\sum_{\vec k}
    \frac{B^{2}(\vec k)\left(A(\vec k)-B(\vec k)\right)^{2}}%
    {A(0)E^{3}(h,\vec k)}
    \right].
    \label{eqn:chi}
\end{equation}

The physically observable quantities listed above develop non-classical field dependence due to  contributions of  zero point fluctuations. The latter are controlled by the spin wave dispersion $SE(h,\vec k)$. The zero point fluctuations increase when the spin wave energies are small in a large region of the BZ. This depends critically on the frustration control parameter $\phi$.

Therefore we first discuss a few typical examples of spin wave dispersions for various $\phi$. In Fig.~\ref{fig:sw_triangular} it is shown for the triangular lattice  with (a) corresponding  to the isotropic  ($\triangle$) case where maximum local exchange frustration occurs (Fig~\ref{fig:frustration}). Nevertheless the spin wave dispersion is rather normal with localized minima in small BZ areas around zone center and zone boundary symmetry points and ring-like maxima in between. As discussed later this leads to a sizeable but not singular zero point reduction to the ordered moment. On the other hand in (b) we are at the classical NAF/SP phase boundary and the dispersion becomes anisotropic and anomalous in the sense that now large connected areas in the 2D BZ  (dark regions)  with low spin wave energy exist whereas those with high energies have shrunk to small (yellow) pockets. This will lead to singular contribution of quantum fluctuations that pushes  the ordered moment to zero at the NAF/SP phase boundary.
A similar behavior is observed in the square lattice at the NAF/CAF boundary $(\phi/\pi\simeq 0.15)$ where flat spin wave modes appear along directions connecting the two respective ordering vectors $\vec Q$  \cite{shannon:04} , leading again to the destruction of the ordered moment by quantum fluctuations.
\begin{figure}
    \centering
    \includegraphics[width=.5\columnwidth]{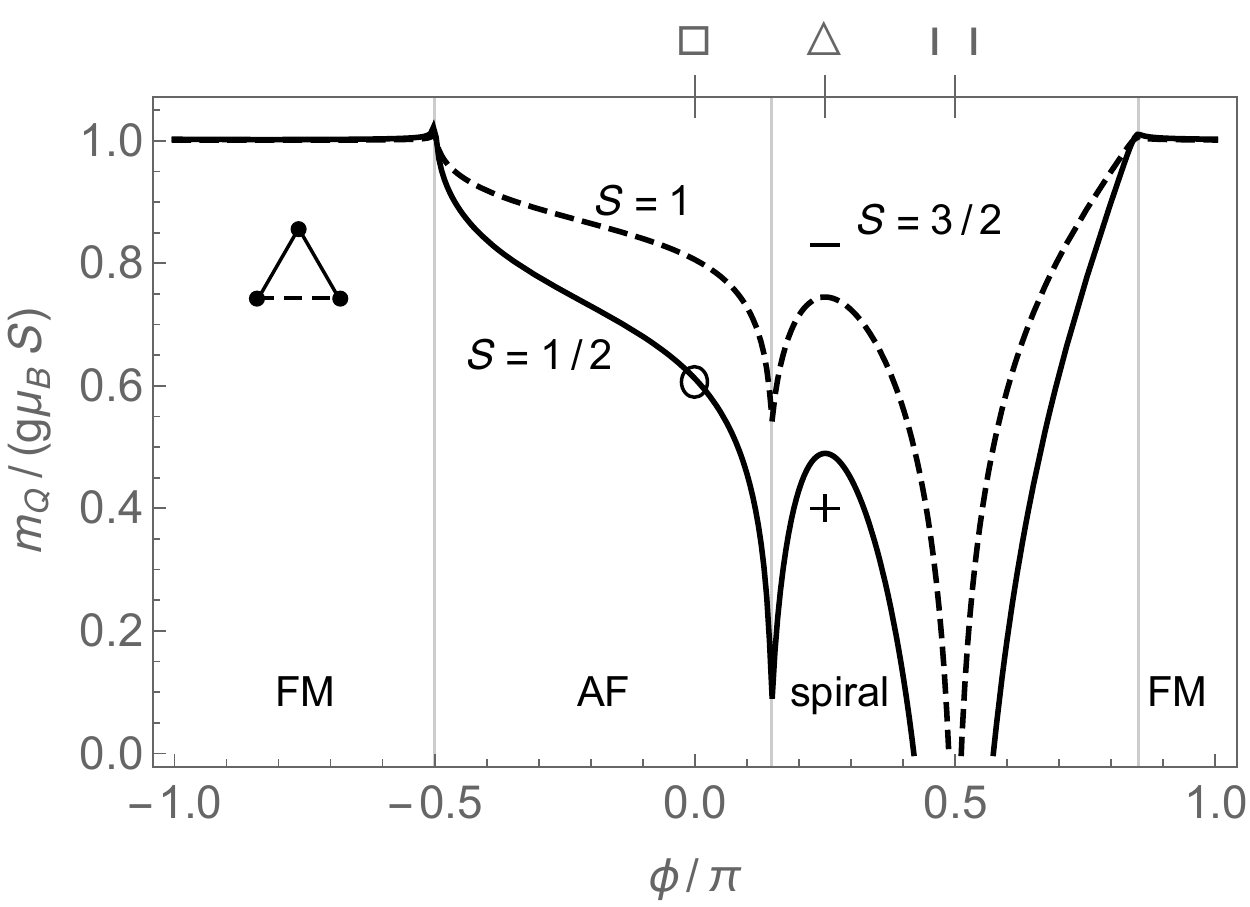}
      \caption{Size of the ordered moment in the triangular model. It vanishes 
    for $J_2=J_1/2$ ($\phi/\pi\approx0.15$) and around the 
    1D case ($J_1=0$, $\pi=\pi/2$). The cross denotes DMRG result \cite{white:07} 
    $m_{\vec Q}(h)/S=0.41$ ( LSW: $m_{\vec Q}(h)/S= 0.478$)
     for isotropic triangular case and
     the circle the unfrustrated HAF ($J_2=0$) ( LSW: $m_{\vec Q}(h)/S= 0.606$). 
     }
    \label{fig:moment_triangular}
\end{figure}

This typical behavior is nicely illustrated by Fig.~\ref{fig:moment_triangular}(a) where the size of the ordered moment for zero field is shown throughout the phase diagram of the triangular lattice.  Consider the quantum spin case $S=1/2$ . For the unfrustrated HAF $(\square, \phi=0)$ the moment is reduced to the well known value $m_{\vec Q}/S =0.606$ (small circle).  At the AF/spiral boundary the quantum corrections due to the anomalous spin wave excitations in Fig.\ref{fig:sw_triangular}(b) destroy the moment, it recovers to a LSW value  $m_{\vec Q}/S = 0.478$ at the maximally frustrated isotropic point $(\triangle)$ which is considerably lower than in the unfrustrated case. The ordered moment is also destroyed in a large region around the quasi-1D $(\parallel)$ case due to diverging 1D quantum fluctuations and recovers at the spiral/FM boundary.\footnote{We note that this is {\em not\/} a frustration effect.} This is quite different  from the square lattice case presented below together with ED results for comparison (Fig.~\ref{fig:groundstate}(b)) where $\phi/\pi=0.5$ $(J_1=0)$ corresponds to two decoupled unfrustrated HAF sublattices with the stable moment as given above.

The linear spin wave method is simple and universally applicable to study quantum effects in magnetism. However it is a biased method due to the underlying assumption of magnetic order, i.\,e. a preferred direction in spin space and it is an approximate method due to the (free) magnon expansion around the ordered ground state which includes only quantum fluctuation effects up to order (1/S). Many attempts have been made to improve the method and approximations by including magnon interactions \cite{igarashi:05,chernyshev:09}, using different boson expansions  \cite{auerbach:94}, imposing self consistency conditions in modified spin wave (MSW) theory \cite{takahashi:89, hauke:11} and employing series expansion method \cite{zheng:06}. This gives some insight into the many-body physics of magnons. However there is no guarantee that inclusion of quantum fluctuation effects in higher order of (1/S) removes singular behavior and improves the results for ground state properties \cite{thalmeier:08}. Therefore in this review we restrict to what can be obtained from the simple LSW approach but whenever possible we will confront its predictions with that of an unbiased purely numerical approach. Such countercheck strategy is very useful to obtain a correct judgment of successes and pitfalls of both methods.

\subsection{Numerical exact diagonalization (ED)  method}
\label{subsect:ED}

Now we review the results from well established numerical exact diagonalization (ED) method used to calculate ground state properties and later also the finite temperature behavior of quantum spin systems \cite{shannon:06,schmidt:15}. It is based on the Lanczos tridiagonalization of Hamilton matrices in spin space for spin clusters of finite size N, supplemented by appropriate, commonly periodic boundary conditions (for an alternative see Ref.~\cite{thesberg:14}). To obtain reliable ground state properties in the thermodynamic limit $N\rightarrow\infty$ that can be compared to analytical spin wave results a finite size scaling analysis has to be performed. Furthermore one has to be careful in the choice of clusters involved in the scaling procedure. Only those which contain the  symmetries of the possible ordered phases in the infinite lattice are useful for the numerics.

\begin{figure}
    \vspace{0.3cm}
    \centering
    \hfill
    \includegraphics[width=.3\columnwidth]{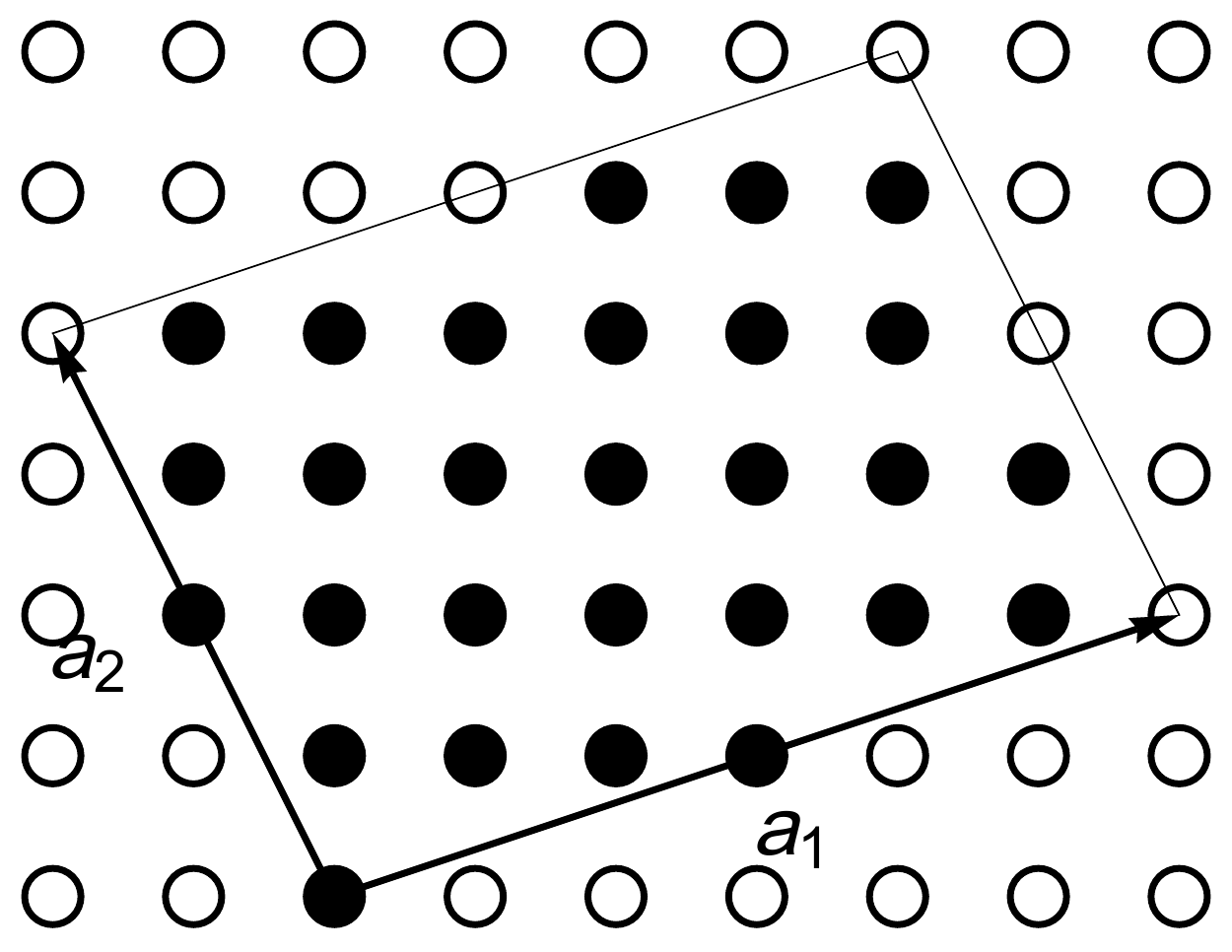}
    \hfill
      \includegraphics[width=.3\columnwidth]{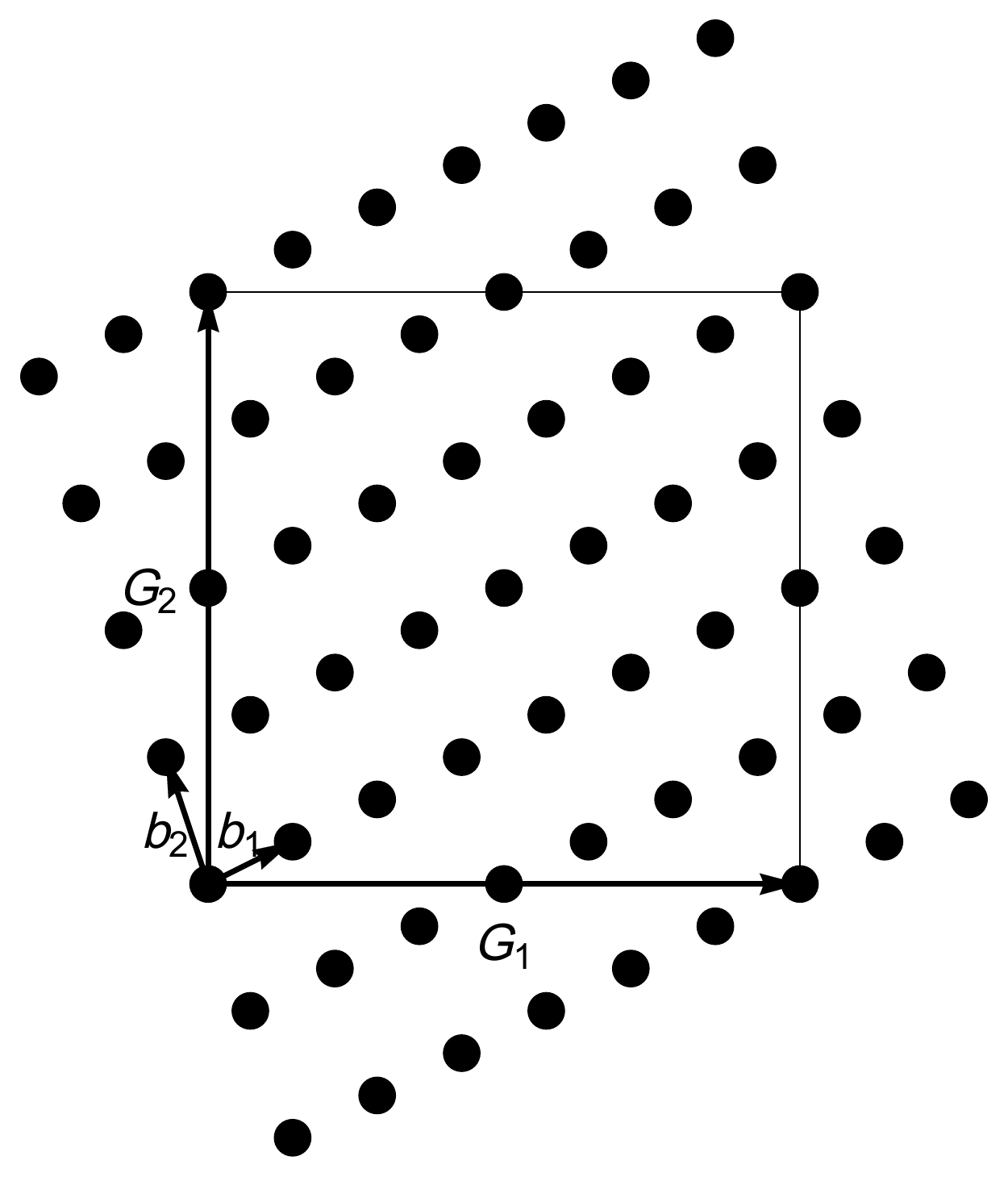}
      \hfill\null
    \caption{Example of the $N=28$ tile 28:2-4 used in ED in direct (a) and reciprocal (b) space with edge vectors $\vec a_{1,2}$ and $\vec b_{1,2}$, respectively. $\vec G_{1,2}$ span the first Brillouin zone with area $(2\pi)^2$.
     }
 \label{fig:tile-14}
\end{figure}
The technical implementation of ED is not an issue here, for more details we refer to \ref{sect:ED} and Refs.~\cite{jaklic:00, schmidt:11,siahatgar:12}. In this method as a first step the Hamiltonian basis for the finite cluster or tile has to be specified. The number of states in Hilbert space is $(2S+1)^N$ which increases exponentially with tile size $N$. This defines the severe restriction of the method to small tiles (here we use $N\leq 32$ tiles) and also implies the necessity of finite size scaling analysis. The Hamiltonian matrix in this space factorizes into sub-blocks when spin rotational symmetry around an axis (z) is preserved with $[S^z_\text{tot},H]=0$ where $S^z_\text{tot}$ is the total spin component along that axis. Furthermore to use periodic boundary conditions such that translational invariance $[t_{\vec R},H]=0$ is satisfied ($\vec R$ is a reciprocal lattice vector and $t_{\vec R}$ the translation operator) Bloch states are chosen as basis vectors.  Due to the finiteness of the tiles the accessible $\vec k$ vectors  are not dense in the BZ but form a reciprocal tile with the same size N. Importantly the reciprocal tile must contain the wave vectors of possible classical ground states or wave vectors close to it. In Fig.~\ref{fig:tile-14} we give an example for a real and reciprocal space tile used in ED. See  \ref{sect:ED}  for details on tile construction and classification~\cite{schmidt:11}.

For tiles with $N > 10$ the iterative  Lanczos tridiagonalization algorithm has to be used.  It has the advantage of requiring only little memory space because at each iteration step only three consecutive recursion vectors have to be stored and the extremal eigenvalues rapidly converge. This is sufficient to calculate ground state and (discussed later) low temperature properties. The drawbacks of the method consisting in spurious eigenvalues due to rounding errors and incorrect ground state degeneracy may be alleviated by using a set of random starting vectors and averaging over the results. An essential question for a successful application of ED to such spin problems is: which tiles should one choose out of the many possibilities (there are 816 different tilings for the square lattice with $8\leq N\leq 32$). A priori, not all of them are useful to be included in the finite size scaling procedure. For that purpose criteria for `optimum tiles' \cite{betts:99,schmidt:11} must be found. For the square lattice this is included in the request of a `maximum squareness'  ratio of the tile \cite{schmidt:11} which is defined by 
\begin{equation}
\rho_\square=\frac{\mbox{area}}{\mbox{(circumference}/4)^2}
\label{eqn:squareness}
\end{equation}
of the tile. If the tile is a true square $\rho_\square=1$ , for all other tiles  $\rho_\square<1$ . Furthermore it is requested that the reciprocal tile contains the four classical ordering vectors $(0,0)$, $(\pi,\pi)$, $( 0,\pi)$, $(\pi,0)$ for FM, NAF and CAF$_a$, CAF$_b$ phases. The latter are two different domains in the square lattice but become different phases in the rectangular lattice (Sec.\ref{subsect:anisotropic}). This leaves us with just a handful of optimal tiles (a list is given in  \ref{sect:ED} and Ref.~\cite{schmidt:11}) which then exhibit smooth scaling behavior as function of $1/N$ for ground state energy, spin correlations etc. from which reliable thermodynamic limit values may be extracted.

The most important quantities  to calculate are the uniform magnetization at finite magnetic field
\begin{equation}
m_0=\frac{1}{\cal N}\sum_i\left\langle S^z_i\right\rangle
\label{eqn:mhom}
\end{equation}
which is the expectation value of z-component of spin $\vec S$ in the {\em global\/} coordinate system,
and the ordered moment
\begin{eqnarray}
    m^{2}_{\vec Q}
    &=&\zeta(\vec Q)\lim_{N\to\infty}S_{N}(\vec Q)\nonumber\\
      S_{N}(\vec Q)&=&\frac{1}{\cal N}\sum_{i,j=1}^{N}
    \left\langle\vec S_{i}\vec S_{j}\right\rangle
    e^{{\rm i}\vec Q\left(\vec R_{i}-\vec R_{j}\right)}
    \label{eqn:mord}
\end{eqnarray}
Here $S_{N}(\vec k)$ is the Fourier transform of the  spin correlation function $\left\langle\vec S_{i}\vec S_{j}\right\rangle$, i.\,e. the spin structure function, of a tile with size $N$ where $\vec k$ belongs to the reciprocal tile (e.g. in Fig~\ref{fig:tile-14}). Furthermore  $\zeta(\vec Q)=1$ for FM and NAF phase and  $\zeta(\vec Q)=2$ for CAF phase due to degeneracy of the latter. At $H=0$, the normalization factor for tiles of size $N$ to achieve the proper thermodynamic limit was shown to be ${\cal N}=N\left(N+1/S\right)$ in Ref.~\cite{schmidt:11}.
\begin{figure}
    \centering
    \includegraphics[width=.48\columnwidth]{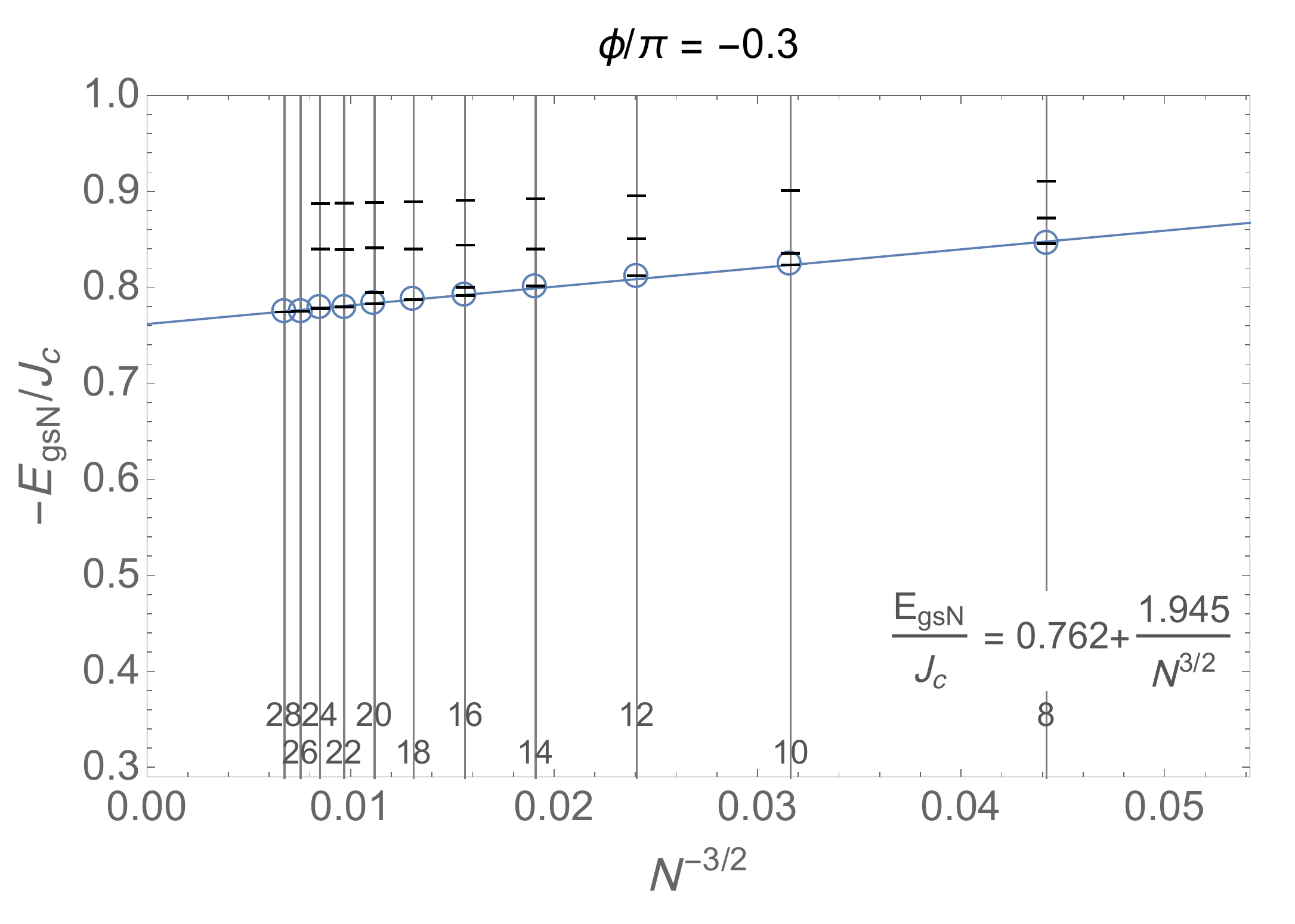}
    \includegraphics[width=.48\columnwidth]{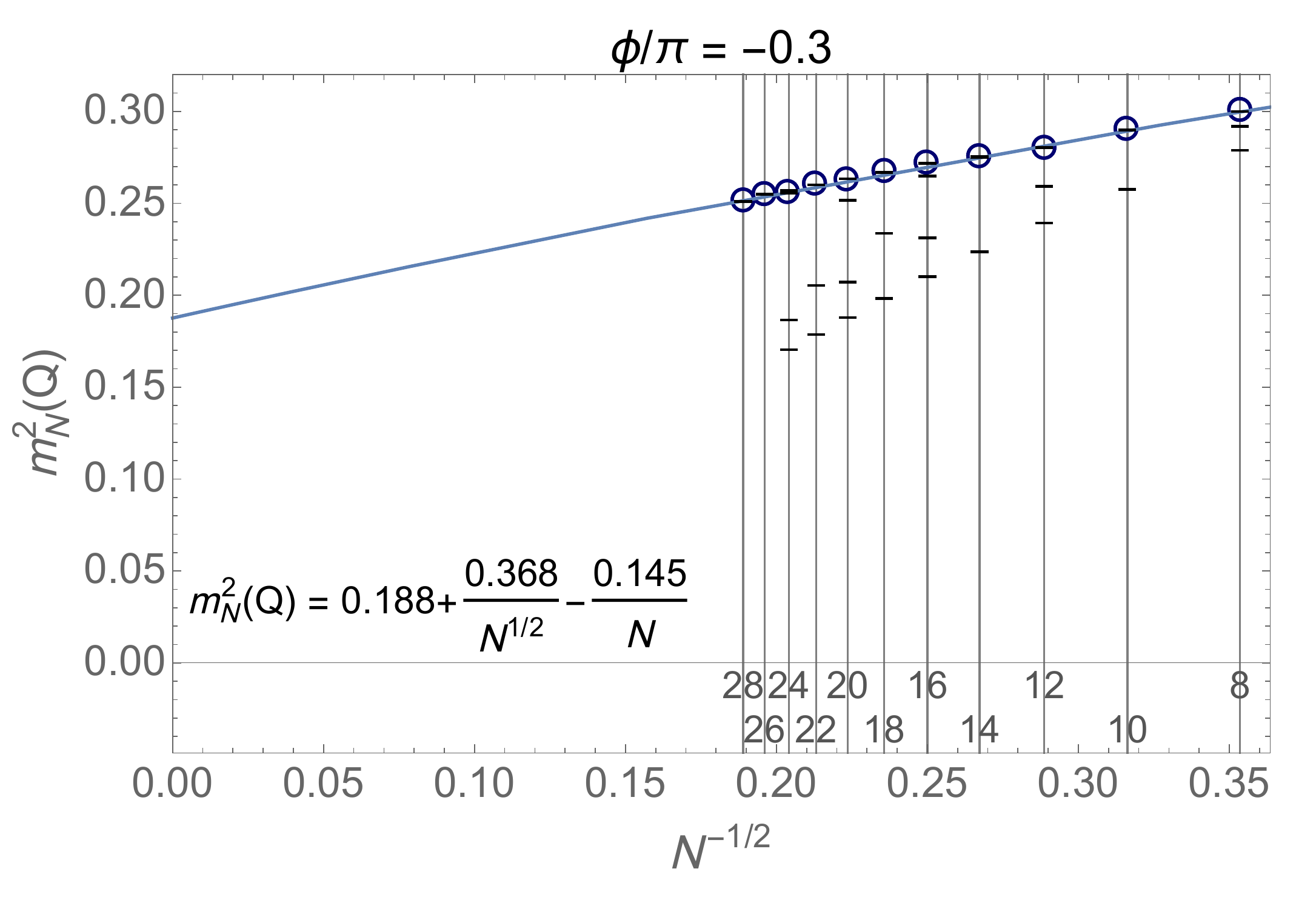}
    \caption{Finite size scaling behavior of ground state energy (a) and ordered moment (b) for $\phi=-\pi/3$ in the 
    N\'eel phase for square lattice. Circles correspond to tiles with maximum squareness, dashes to less optimal tiles. 
    The inset gives the ED scaling laws and coefficients.}
     \label{fig:scaling}
\end{figure}
To derive ground state properties  a scaling procedure to the thermodynamic limit  $N\rightarrow \infty$ is needed. In Refs.~\cite{hasenfratz:93,sandvik:97} their area dependence for the HAF ($J_2=0$) has been derived using chiral perturbation theory. This result is used frequently in  ED ground state investigations \cite{schulz:96,betts:99,schmidt:11}:
\begin{eqnarray}
E_{gsN}&=&E_{gs}+\frac{e_1}{N^{3/2}}+\frac{e_2}{N^{2}}\nonumber\\
m^2_N(\vec Q)&=&m^2_{\vec Q}+\frac{m^2_1}{N^{1/2}}+\frac{m^2_2}{N}
\label{eqn:scaling}
\end{eqnarray}
It is reasonable to assume that the scaling does not depend strongly on the details of short range interactions but only by the universality class defined by spatial $(D=2)$ and order parameter $(n=3)$ dimension. Sometimes scaling laws which involve exponents that are not simple fractions are also invoked \cite{bishop:08}.  In Fig.~\ref{fig:scaling} the scaling with tile size is shown for ground state energy and ordered moment at a particular frustration angle $\phi=-0.3\pi$ (NAF).

It is obvious from this figure that the restriction to tiles of maximum squareness (circles) as described above is essential to achieve a stable scaling to the thermodynamic limit. Less perfect tiles (bars) correspond to states with higher energy; they do not scale to the ground state energy and the scaling of the ordered moment has larger deviations for them. The quality of scaling depends considerably on $\phi$. When frustration is absent as is the case for $\phi < 0$ in the NAF phase of Fig.~\ref{fig:scaling} it is smooth with small error in the result.  In contrast in the frustrated regime (see Fig.~\ref{fig:frustration}) close to the classical NAF/CAF and CAF/FM phase boundaries the accuracy of scaling results decreases \cite{schmidt:11}.

\subsection {Comparative discussion of ground state properties}
\label{subsect:gsprop}

\begin{figure}
    \centering
    \includegraphics[width=.48\columnwidth]{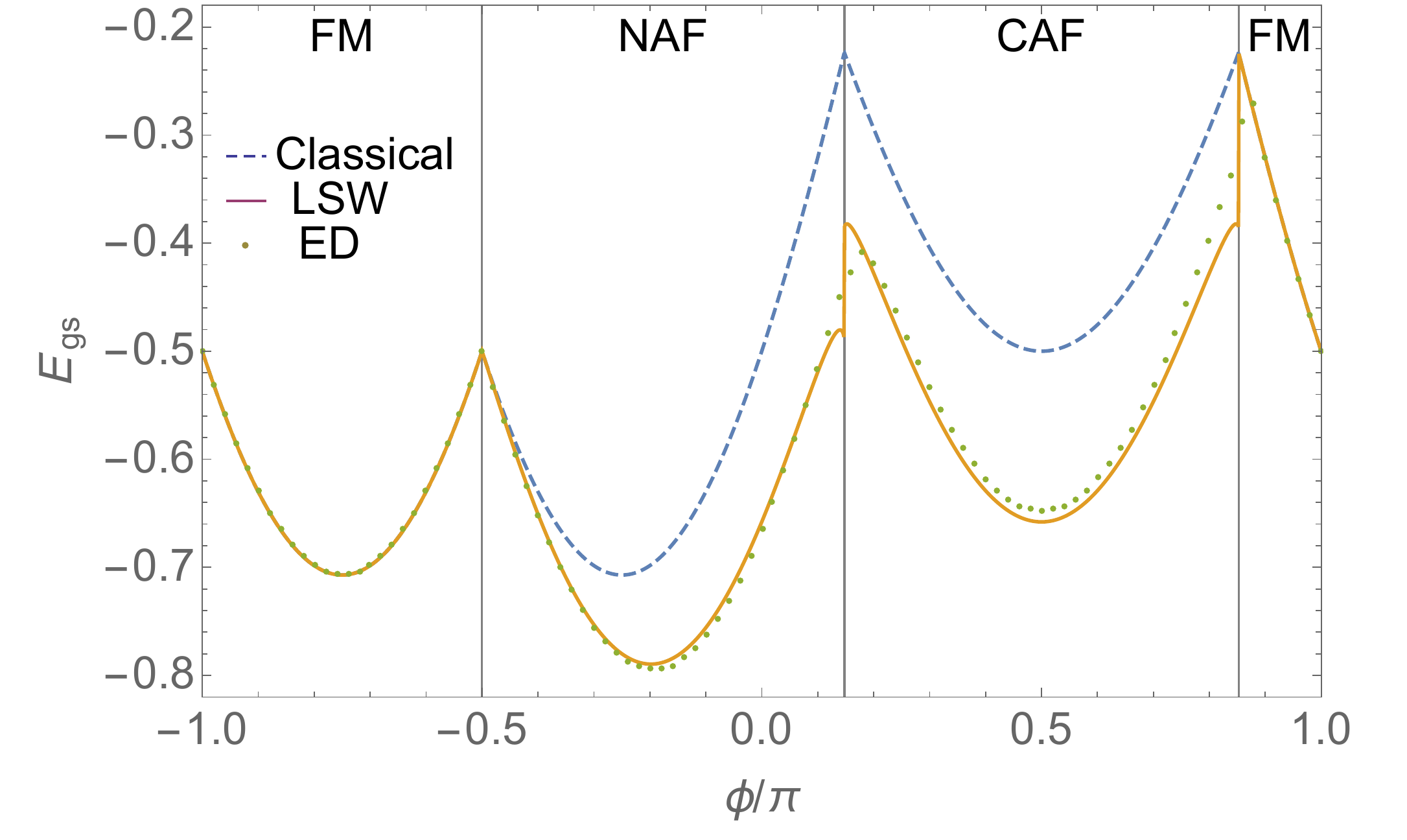}
    \includegraphics[width=.48\columnwidth]{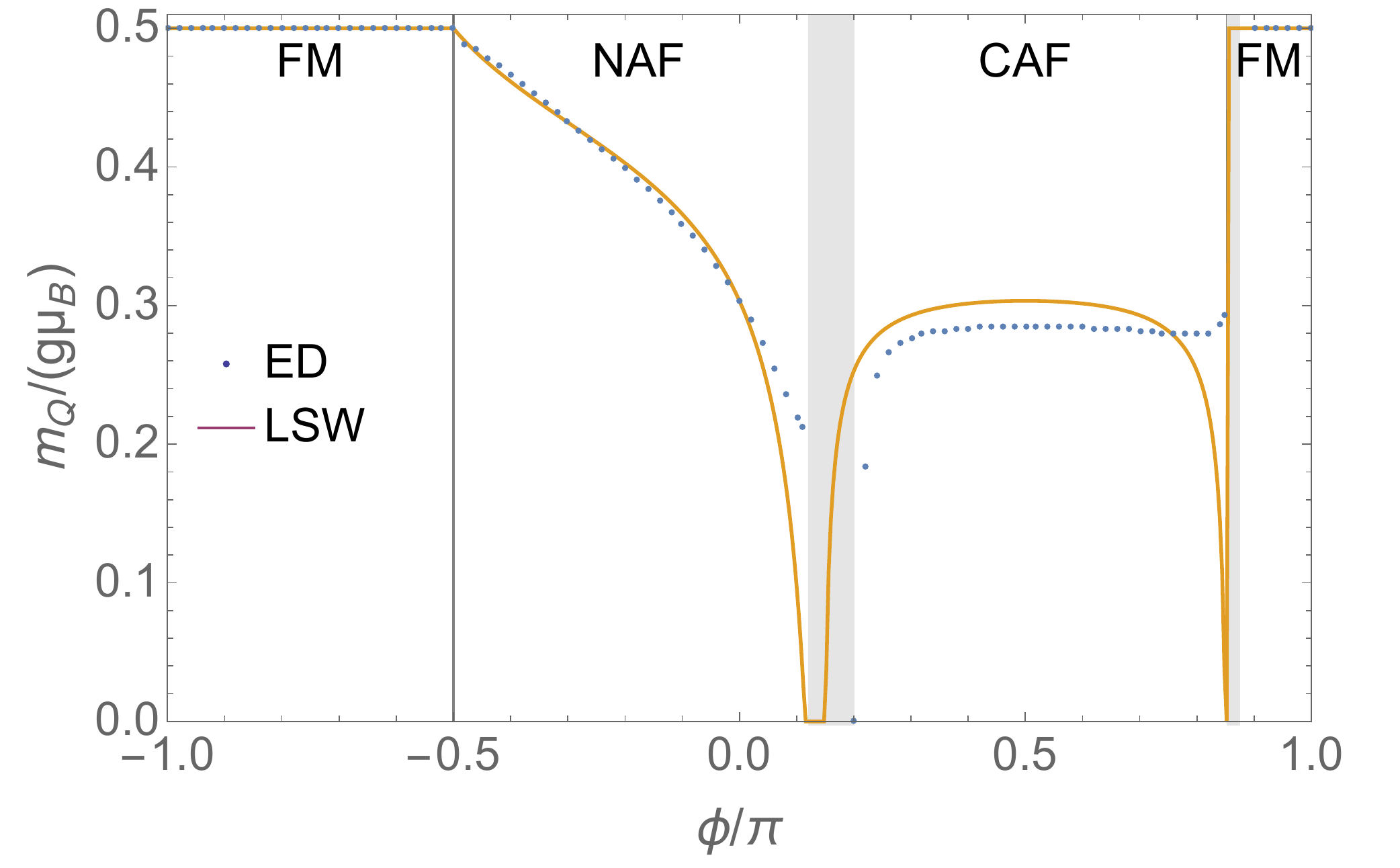}
    \caption{Results of ED finite size scaling analysis in comparison with LSW predictions for (a) the ground state energy
     and (b) the ordered moment of the (isotropic) square lattice model. In the gray areas ED scaling to a magnetic state 
     breaks down indicating the region of a nonmagnetic spin liquid (Table \ref{tbl:spinliquid}).}
     \label{fig:groundstate}
\end{figure}
Here we present the compiled ground state results from the numerical ED scaling procedure in comparison to the analytical LSW results repeated throughout the phase diagram, i.\,e. as function of frustration control parameter $\phi$. This is shown in Fig.~\ref{fig:groundstate} for the square lattice. The overall qualitative variation of $E_\text{gs}(\phi)$ follows that of the classical energy $E_\text{cl}(\phi)$ (dashed line in (a)). In the FM region they are identical because the classical ``all-up'' state is an eigenstate also in the quantum case. In the NAF and CAF region the presence of quantum fluctuations leads to a significant lowering of the ground state energy. It is reassuring that the  LSW results (full red line) from Eq.~(\ref{eqn:hdiagonal}) lead to an excellent agreement with the finite size ED scaling results (dotted line). 

The ordered ground state moment $m_{\vec Q}(\phi)$  shown in (b) is particularly illuminating. In the FM phase it keeps the classical value $m_{\vec Q}=S$ and shows a continuous decrease in the NAF phase towards the well-known value  $m_{\vec Q}/S= 0.606$ for the unfrustrated $J_2=0$ ($\phi=0$) case.  Then for positive $J_2$ ($\phi>0$) the frustration increases and the ordered moment precipitously drops to zero at the classical NAF/CAF boundary as seen from LSW (full line) while ED scaling (dots) no longer converges in the associated gray shaded region. Thus there is a finite range of $\phi$ around $\phi/\pi=0.15$ ($J_2/J_1=0.5$) where the magnetic order is destroyed by  diverging quantum fluctuations. This is correlated with large regions in $\vec k$ space with low energy spin waves that have a flat dispersion on lines in the BZ connecting the wave vectors $(\pi,\pi)$ and $(0,\pi)$ or $(\pi,0)$ of  NAF and CAF phases, respectively  which are degenerate for $\phi/\pi=0.15$. The position and range of the moment instability along $\phi$ axis is slightly different for both methods (Table~\ref{tbl:spinliquid}). The nature of this nonmagnetic `spin liquid'  phase will be discussed further in Sec.~\ref{subsect:VBS}.

\begin{figure}
    \vspace{0.3cm}
    \centering
    \includegraphics[width=.5\columnwidth]{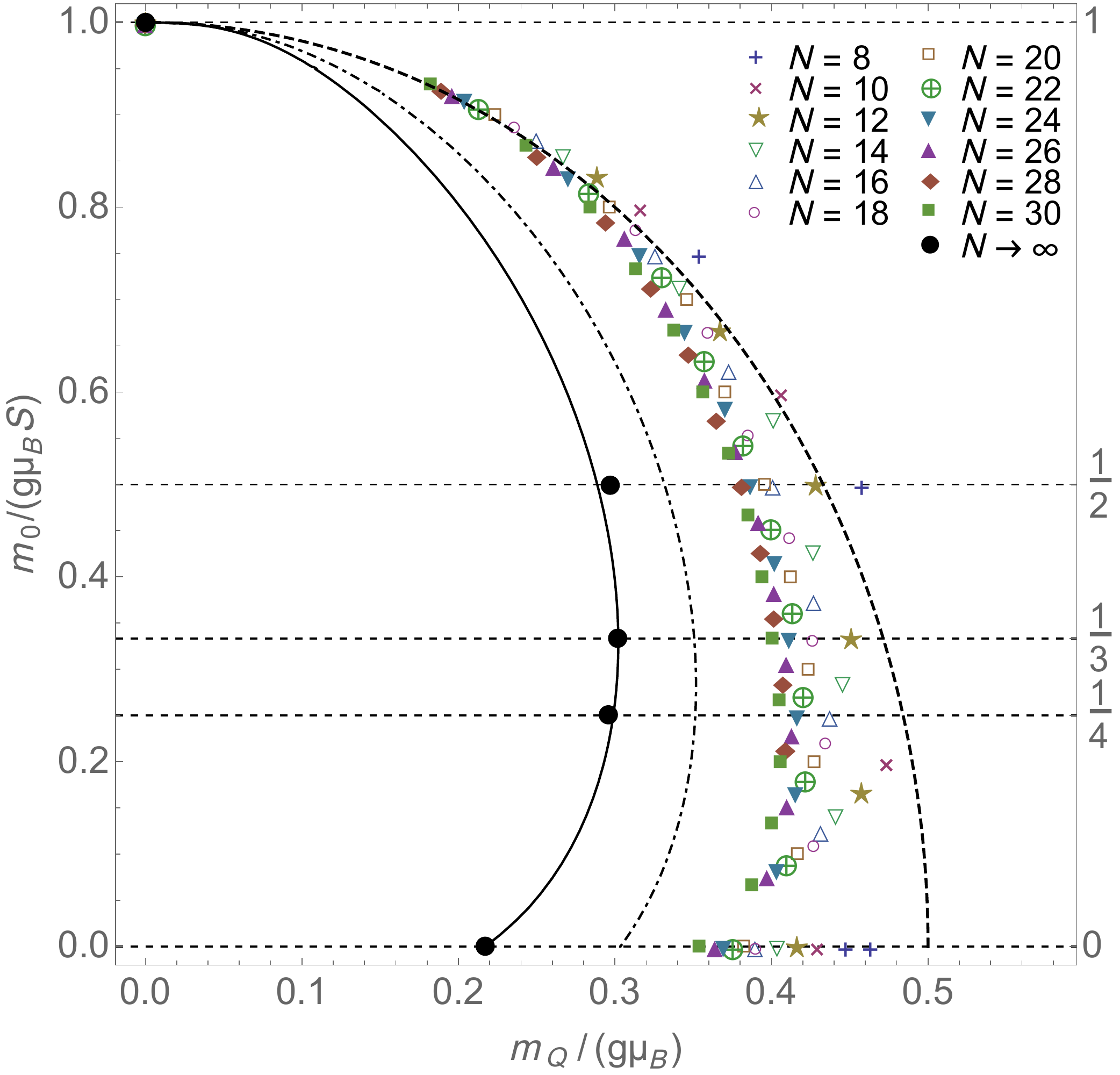}
     \caption{Parametric plot of uniform moment $m_0(h)$ versus ordered (staggered) moment $m_\vec Q(h)$. In this plot $h=0$ corresponds to $m_0=0$ (moments align in a plane perpendicular to the field axis) and $h=h_\text s$ to  $m_0=S$ (moments align parallel to the field axis and gain their full classical value).
     Dashed line: classical ($\phi$-independent) result; full line: LSW result for moderate frustration  ($\phi=0.063\pi$). Symbols: ED results for 
      finite size tiles ($N=8\ldots30$). Filled circles: Finite size scaling results. For comparison the LSW result for the unfrustrated $J_2=0$ N\'eel antiferromagnet is shown as dash-dotted line.}
     \label{fig:parametric}
\end{figure}
The field dependence of the ordered magnetic moments clearly reveals the interplay of frustration and quantum fluctuations. Classically one would expect a very simple behavior (dashed line in  Fig.~\ref{fig:parametric}), namely a canting of the AF ordered moments of fixed size S out of the plane $\perp$ to the field direction whereby the cosine of the canting angle varies linearly with field strength (Eq.~(\ref{eqn:enclass})) as  illustrated in  Fig.~\ref{fig:classical_PD}b. However the inclusion of quantum fluctuations changes the picture dramatically. In the zero-field case they strongly reduce the total moment size. The amount of reduction is determined by the prevalence of frustration according to Fig.~\ref{fig:groundstate}(b). On the other hand at the saturation field the moments are ferromagnetically aligned and quantum fluctuations are therefore turned off, thus the moment reduction is eliminated. This means that for intermediate fields the total moment $m_\text{tot}(h)$ (with uniform $m_0(h)$ and staggered $m_{\vec Q}(h)$ components) will not only rotate but will also increase. The ordered moment $m_{\vec Q}$ which is the projection to the abscissa in Fig.~\ref{fig:parametric} will then first increase with field due to suppressed moment reduction and then decrease again due to the classical geometric effect of canting. Therefore  $m_{\vec Q}(h)$  will depend {\em non-monotonically\/} on field strength. How noticeable this surprising quantum behavior is depends on the size of frustration. For large frustration the starting moment at $h=0$ is strongly reduced and therefore will exhibit a most pronounced nonmonotonic field dependence. 

These astonishing quantum effects are nicely illustrated in the parametric plot of Fig.~\ref{fig:parametric} where classical (dashed line), spin wave (full line) and ED scaling results (black dots) are shown together for comparison for a strongly frustrated case with the initial reduction $m_{\vec Q}(h=0) \approx 0.2$. 
We also show the LSW results near the unfrustrated pure N\'eel case ($\phi=0$ or $J_2=0$) (dash-dotted line) where the ordered moment is already non-monotonic, though less pronounced. The LSW results for the frustrated case (full line) show excellent agreement with the ED scaling results. The pre-scaling results for all cluster sizes used are also presented which illustrates the necessity of a scaling procedure  to obtain useful predictions for the thermodynamic limit. Because we need to scale at constant uniform moment per site (horizontal dashed lines in Fig.~\ref{fig:parametric}), this procedure is only possible for selected groups of finite tiles where the moment is not changed as function of tile size~\cite{siahatgar:11}.

\begin{figure}
    \centering
    \includegraphics[width=.5\columnwidth]{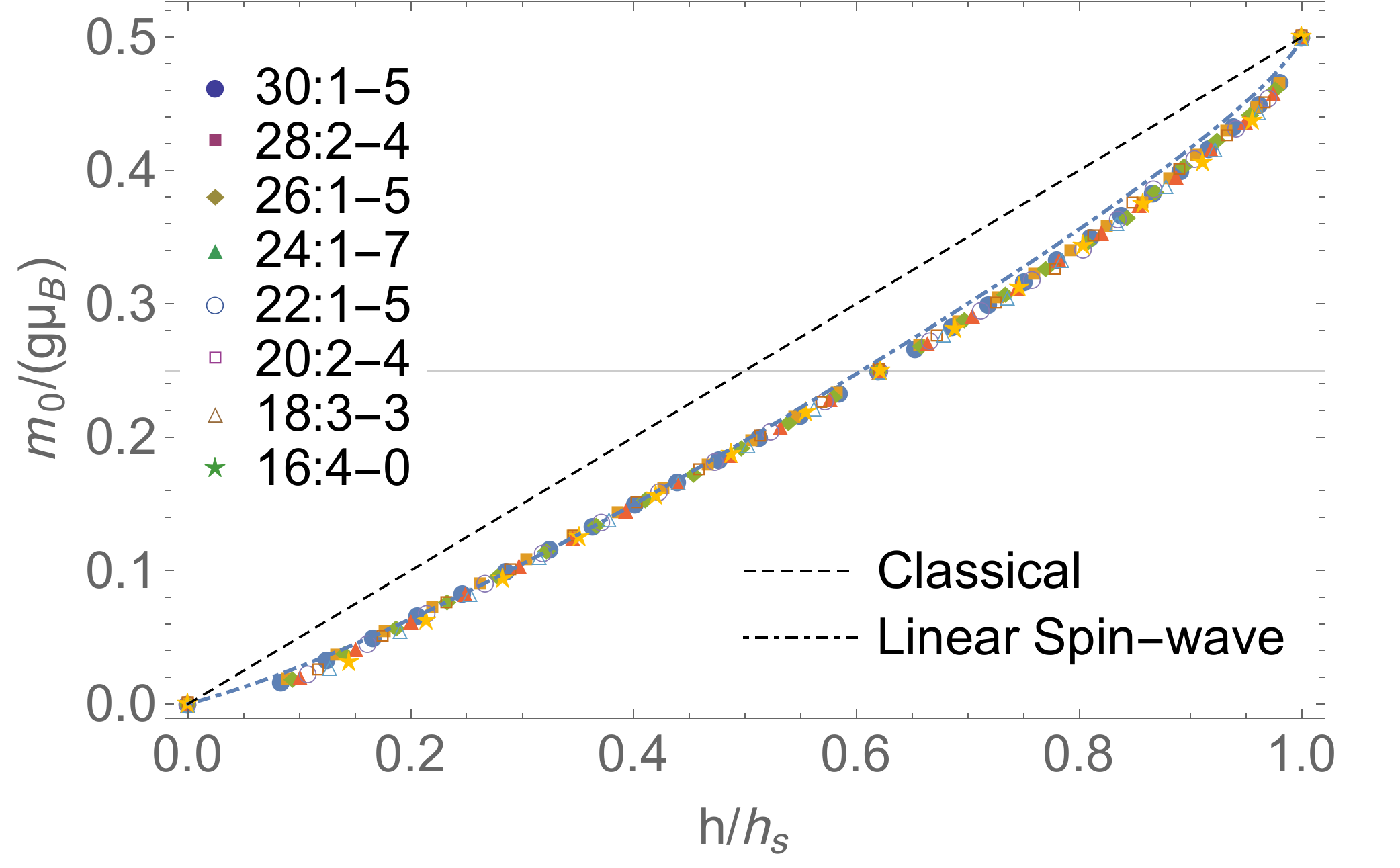}
     \caption{Homogeneous magnetization for the pure N\'eel case ($\phi=0$)  obtained from ED results (symbols)
     of various tile sizes $N=16\ldots30$ of square lattice through the Bonner-Fisher construction.}
     \label{fig:m0_ED}
\end{figure}
Furthermore the homogeneous magnetization $m_0(h)$ itself is instructive for the interplay of frustration and quantum effects. In the ED approach using Eq.~(\ref{eqn:mhom}) it will not be a smooth function of field because the states with a given total $S_z$ quantum number will cross as function of field strength \cite{schmidt:07}. Therefore states of consecutively higher $S_z$ become the ground state. This results in a step and plateau-like magnetization curve for finite tiles. Since the position and widths of these steps varies with cluster size a scaling procedure at fixed field for the magnetization is difficult. A convenient empirical alternative to obtain a smooth magnetization curve is the Bonner-Fisher construction \cite{bonner:64} where the midpoints of steps and plateaux are connected to obtain $m_0(h)$. This is shown in Fig.~\ref{fig:m0_ED} for the pure square lattice N\'eel case ($J_2=0$, $\phi=0$).

\begin{figure}
    \centering
    \includegraphics[width=.5\columnwidth]{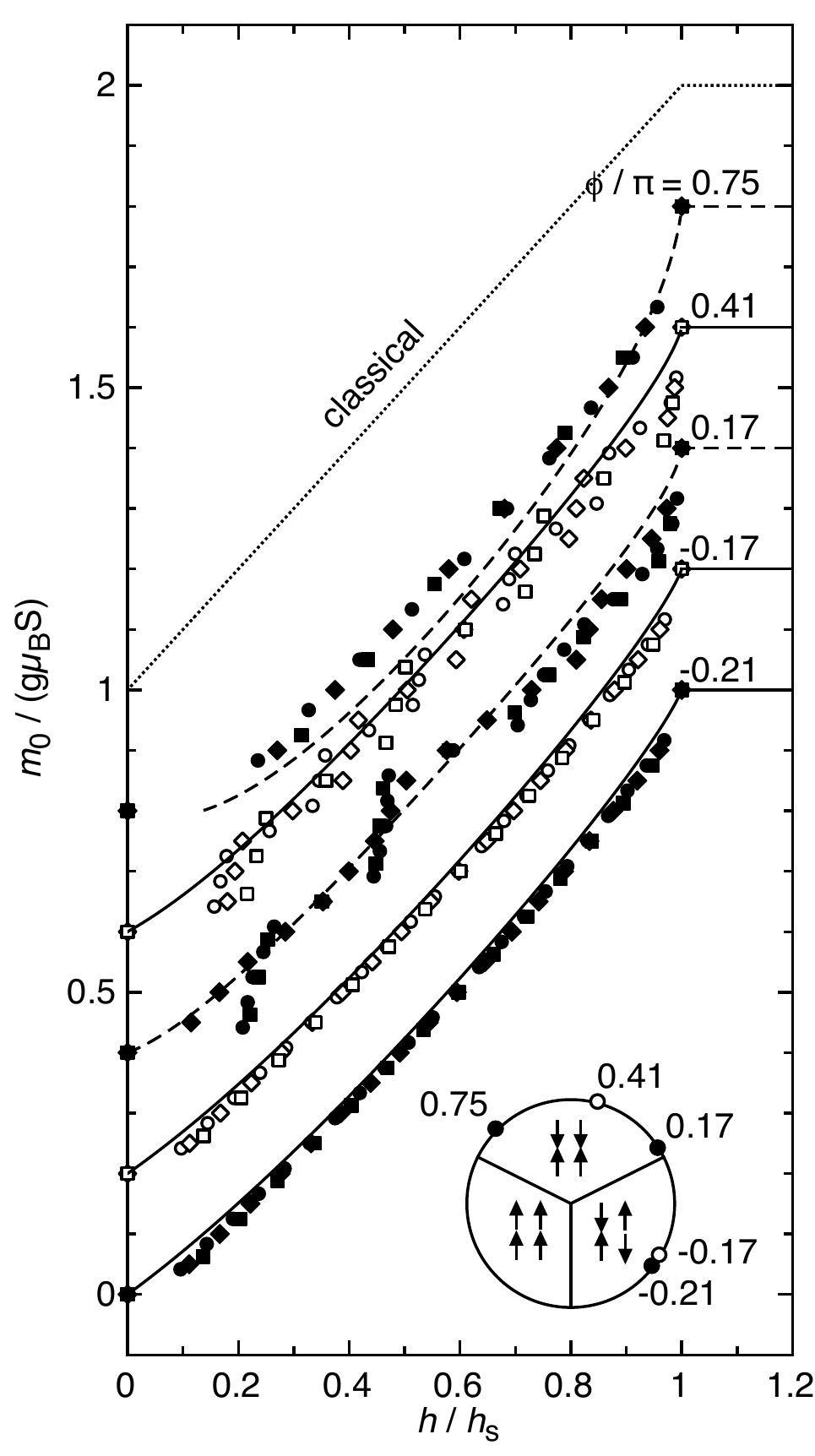}
     \caption{Comparison of $m_0/S$ from LSW and ED (Bonner-Fisher construction)  for various frustration control parameters $\phi$ in the 
     square lattice  phase diagram (inset). Each curve is offset by $0.25$.
     For vanishing $(\phi/\pi=-0.17)$ or moderate  $(\phi/\pi=0.41)$ frustration there is 
     excellent or good agreement. For large frustration strong deviations occur. For AF $J_1$ $(\phi/\pi=0.17)$ plateau formation (ED results) at
     $m_0/S=1/2$ occurs (also Fig.~\ref{fig:mag_plateau} in Sec.~\ref{sect:magplateaux}). For FM $J_1$ ($\phi/\pi=0.75$) LSW 
     results are unstable. }
     \label{fig:m0_EDLSW}
\end{figure}
The combined Bonner-Fisher midpoint data for various cluster sizes reproduce a dense magnetization plot in this figure which gives a  nice illustration for overall quantum effects in the nonlinear magnetization curves. However the smoothing of plateaux and steps also misses some subtle and striking quantum effects because it turns out that in particular cases a plateau in the magnetization may be stable in the thermodynamic limit (Sec.~\ref{sect:magplateaux}). Furthermore a comparison of ED and analytical LSW results of the square lattice $J_1$-$J_2$ model for general frustration parameter $\phi$ confirms the overall observation that the nonlinear deviation from the classical linear behavior is much more pronounced in the frustrated region of the phase diagram (Fig.~\ref{fig:m0_EDLSW}).

This figure shows that for $\phi$ deep within the stable magnetic regions with little frustration there is excellent agreement between both methods. However in the strongly frustrated case close to the classical zone boundaries the ED results show considerable scattering and tendency to large plateau formation whereas the LSW results at the CAF/FM boundary $(\phi/\pi=0.75)$ become unstable (leading even to negative $m_0(h)$ for small fields). An extension including spin wave interactions \cite{thalmeier:08} stabilizes the magnetic moment but does not lead to the physical correct behavior.

The pronounced ED plateau formation at $m_0=1/2$ at the classical NAF/CAF boundary is not just a finite size effect but persists even in the thermodynamic limit \cite{honecker:99} with a plateau width given by $\Delta h/h_\text s=(h_\text c^+-h_\text c^-)/h_\text s \approx 0.16$ for $\phi/\pi=0.17$ (Sec.~\ref{sect:quantumphases}, Fig.~\ref{fig:mag_plateau} (b)) where $h_\text c^\pm$ are the upper and lower critical fields of the plateau state (Sec.~\ref{sect:magplateaux}). Physically it is interpreted as a four spin bound state formation on a square plaquette in a narrow interval $\phi/\pi \in [0.15,0.18]$ \cite{coletta:13} within the spin liquid phase (Table~\ref{tbl:spinliquid}) discussed in Secs.~\ref{sect:quantumphases} and~\ref{sect:magplateaux}.


\section{Finite temperature methods and properties}
\label{sect:temp}

 The previous analysis of ground state properties takes it for granted that the exchange parameters for a specific planar magnet are known. However, this is often not easily achieved. They may be directly obtained by determination of magnetic excitations with inelastic neutrons scattering (INS) and fitting their dispersion to spin wave calculations. A more indirect method is the investigation of finite temperature properties like susceptibility $\chi(T)$, specific heat $c_V(T)$ and magnetocaloric coefficient $\Gamma_\text{mc}(T)$ which also probe these excitations. The characteristics like high temperature asymptotics and peak positions then give information on the exchange constants $J_1$ and $J_2$ involved. This may not be sufficient and ambiguities can remain.  For example the first term in the high-temperature series expansion \cite{oitmaa:06,hehn:17} is given by ($\beta =1/(k_\text BT)$).
\begin{eqnarray}
    \chi
    &=&
    \frac{S(S+1)}3\beta J_{\text c}\left(1-\beta k_\text B\Theta_\text{CW}\right),
    \\
    c_{V}
    &=&
    \frac13\left[S(S+1)\right]^{2}\beta^{2}J_{\text{mc}}^{2}
    \label{eqn:cvht}
\end{eqnarray}
in units of $N_\text L\mu_0(g\mu_\text B)^2/J_\text c$ and $R=N_\text Lk_\text B$, respectively. ($N_\text L$ is the Loschmidt or Avogadro constant.) The Curie-Weiss temperature $\Theta_\text{CW}$ and magnetocaloric energy scale $J_\text{mc}$ (not to be confused with $J_\text c$) for triangular $(\Delta)$ and square  $(\boxtimes)$ lattice models are defined by
\begin{eqnarray}
    \Theta_\text{CW}&:=&\frac{S(S+1)}{3k_\text B}\sum_{n}J_{ii+n}=
    \frac1{k_\text B}
      \left\{
      \begin{array}{l@{\qquad}l}
	 J_{1}+\frac{J_{2}}2 & \Delta\\
	  J_{1}+J_{2} & \boxtimes
      \end{array}
      \label{eqn:thetacw}
    \right.
    \\
    J_{\text{mc}}^{2}&:=&\frac12\sum_{n}J_{ii+n}^{2}=
       \left\{
      \begin{array}{l@{\qquad}l}
	  2J_{1}^{2}+J_{2}^{2}&\Delta\\
	  2(J_{1}^{2}+J_{2}^2)=2J_\text c^2&\boxtimes\\
      \end{array}
    \right.   
     \label{eqn:jth}
\end{eqnarray}
This suggests that one should be able to determine the exchange parameters
$J_{1}$ and $J_{2}$ already from high-temperature fits of the experimental
results to the expressions above.  However, fixing
$J_{\text{mc}}^{2}$ and $\Theta_\text{CW}$ determines $J_1+J_2$ (square lattice) or $J_{1}+(1/2)J_{2}$ (triangular lattice) and
$|J_{1}-J_{2}|$, but the sign of the latter is undetermined.\footnote{More general: For a $S=1/2$ isotropic exchange model with $n_1$ nearest and $n_2$ next nearest neighbors,
$n_1J_1+n_2J_2=4k_\text B\Theta_\text{CW}$ and $\left(J_1-J_2\right)^2=\left[2\left(n_1+n_2\right)J_\text{mc}^2-\left(4k_\text B\Theta_\text{CW}\right)^2\right]/\left(n_1n_2\right)$ are fixed.}
Using $J_{\text c}$ and $\phi$ instead,
this is equivalent to having {\em two\/} possible values $\phi_{\pm}$ for the control
parameter (cf. Fig.\ref{fig:phasedia}).  They can lie in two
different thermodynamic phases with distinct properties. This
ambiguity and its implications were discussed for the square lattice model in
Refs.~\cite{misguich:03,shannon:04}.

The coefficients of the high-temperature expansions for
$\chi(T)$ and $c_{V}(T)$  are polynomial functions of $J_{1}$ and
$J_{2}$ which are known up to at least eighth order~\cite{rosner:03,schmidt:11-1,oitmaa:06,hehn:17}.
Nevertheless the ambiguity persists~\cite{misguich:03},
and it remains difficult to determine $J_{1}$ and $J_{2}$
solely from fits to the high temperature dependence of $\chi$ and
$c_{V}$.  One powerful additional diagnostic on this issue is the investigation
of saturation fields~\cite{schmidt:09,schmidt:10} provided that they
are in an accessible range (see Sec. \ref{subsect:oxovana} and Fig.~\ref{fig:saturation_fields}).

 \subsection{Finite temperature Lanczos method}
\label{subsect:Lanczos}

To overcome these difficulties the more powerful numerical finite temperature Lanczos method (FTLM) \cite{jaklic:00} has been used successfully \cite{shannon:04,schmidt:15}. In this approach the thermodynamic variables are expressed as traces over the statistical operator and averages for finite tiles are evaluated directly, leading to reliable results in the whole temperature range above the finite size gap region. From fits to experimental curves of specific heat and susceptibility the exchange parameters (and g-factors) may then be extracted (Sec.~\ref{sect:compounds}). The evaluation of statistical traces of operators $\langle A\rangle_\beta$  utilizes the eigenvalues and many-body wave functions from numerical ED of Hamiltonians on  finite tiles. It is briefly described in \ref{sect:ftlm}, for more details we refer to Refs.~\cite{jaklic:00,shannon:04,schmidt:15,hanebaum:14}. 

With the internal energy $\bar E=\left\langle{\cal H}\right\rangle_\beta$ and the uniform magnetic moment $\bar\mu=g\mu_{\text B}\left\langle S_z^{\text{tot}}\right\rangle_\beta$ the measurable thermodynamic coefficients associated with these conserved quantities are then written
in terms of second order cumulants:
\begin{align}
    \chi
    &=
    \frac1N
    \left(\frac{\partial\bar\mu}{\partial B}\right)_{T}
    =
    \frac{\beta J_{\text c}}N\left(
    \left\langle\left(S^{\text{tot}}_{z}\right)^2\right\rangle_{\beta}
    -\left\langle S^{\text{tot}}_{z}\right\rangle_{\beta}^{2}
    \right),
    \\
    c_{V}
    &= 
    \frac1N
    \left(\frac{\partial\bar E} {\partial T}\right)_{B}
    =
    \frac{\beta^{2}}N\left(
    \left\langle{\cal H}^{2}\right\rangle_{\beta}
    -\left\langle{\cal H}\right\rangle_{\beta}^{2}
    \right),
    \\
   \Gamma_{\text{mc}}
    &=
    \left(\frac{\partial T}{\partial B}\right)_{S}
   =
   -\frac1N
   \frac T{c_V}\left(\frac{\partial\bar\mu}{\partial T}\right)_{B}
   =
      -\frac1\beta
    \frac{\left\langle{\cal H}S_{z}^{\text{tot}}\right\rangle_{\beta}
    -\left\langle{\cal H}\right\rangle_{\beta}
    \left\langle S_{z}^{\text{tot}}\right\rangle_{\beta}}{
    \left\langle{\cal H}^2\right\rangle_{\beta}
    -\left\langle{\cal H}\right\rangle_{\beta}^2},
    \label{eqn:gammamc}
    \\
    \chi^{(3)}
    &=
    \frac1N
    \left(\frac{\partial^3\bar\mu}{\partial B^3}\right)_T=
    \frac{\beta^3J_\text c^3}{N}\left[
        \left\langle\left(S^{\text{tot}}_{z}\right)^4\right\rangle_{\beta}
    -3\left\langle\left(S^{\text{tot}}_{z}\right)^2\right\rangle_{\beta}^{2}
    \right]
    \label{eqn:thermodynamic}  
  \end{align}
for the molar linear susceptibility, specific heat, magnetocaloric coefficient (the adiabatic temperature change with field) and third order susceptibility (nonlinear field dependence of the magnetic moment) of a lattice tiling with tiles containing $N$ sites, respectively. For zero field quantities one can set $\left\langle S^{\text{tot}}_{z}\right\rangle_{\beta}=0$ in these expressions. $B=\mu_0H$ denotes the applied magnetic field. Unless otherwise noted we express $\chi$ in units of $N_\text L\mu_0\left(g\mu_\text B\right)^2/J_\text c$, $c_V$ in units of $R=N_\text Lk_\text B$, $\Gamma_\text{mc}$ in units of $g\mu_\text B/k_\text B$, and $\chi^{(3)}$ in units of $N_\text L\mu_0\left(g\mu_\text B\right)^4/J_\text c^3$.

\subsection {Discussion of susceptibility and specific heat}
\label{subsect:specsus}

\begin{figure}
    \centering
    \begin{minipage}{.5\columnwidth}
    \includegraphics[height=.6\columnwidth]{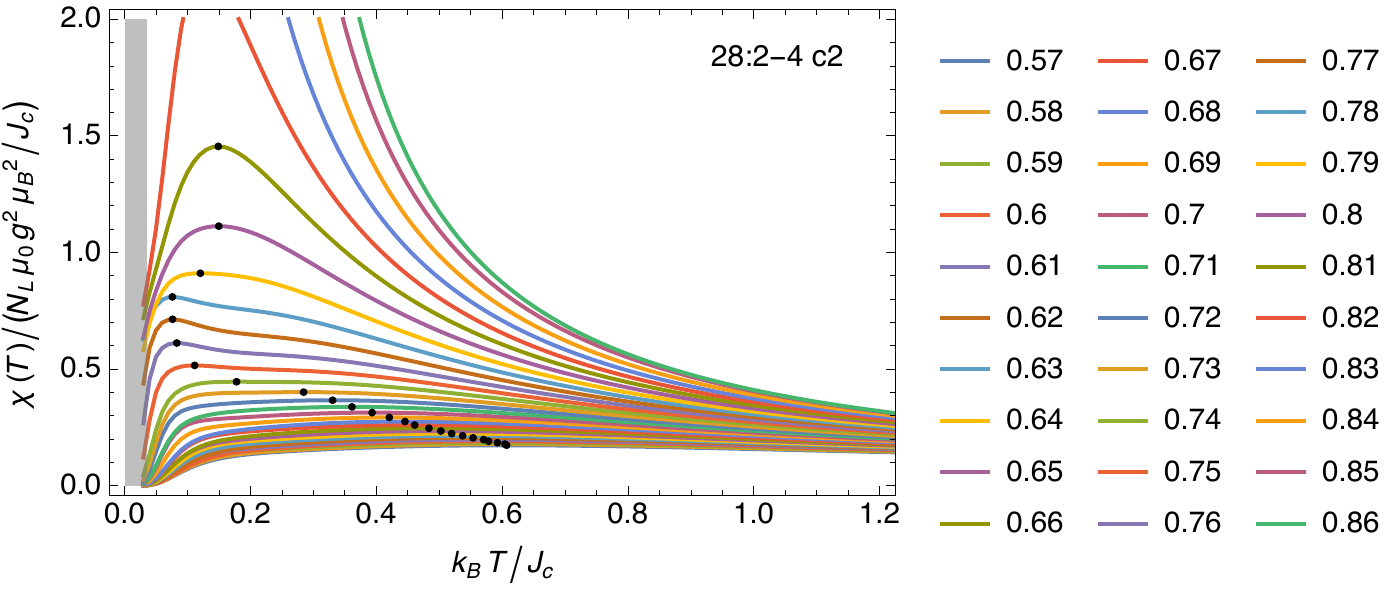}   
     \includegraphics[height=.6\columnwidth]{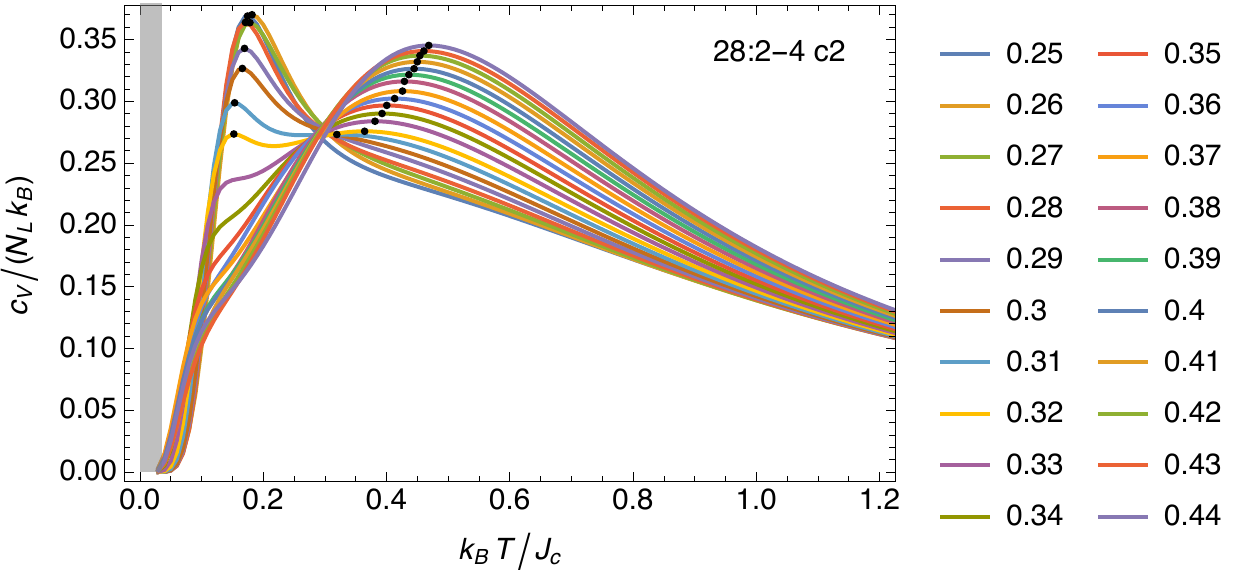}   
     \end{minipage}
    \caption{Susceptibility curves  (a) and specific heat results (b) for various control 
    parameters $\phi/\pi$ in the spiral phase (SP) of triangular 
    model from FTLM results based on tile 28:2-4 (see \ref{sect:ED}).  Dots denote the maximum 
    position and value. Double maximum occurs in $c_V$
     for $\phi$ in limited range of the  SP regime.}
     \label{fig:chicvtri_T}
\end{figure}
\begin{figure}
    \centering
    \includegraphics[width=.5\columnwidth]{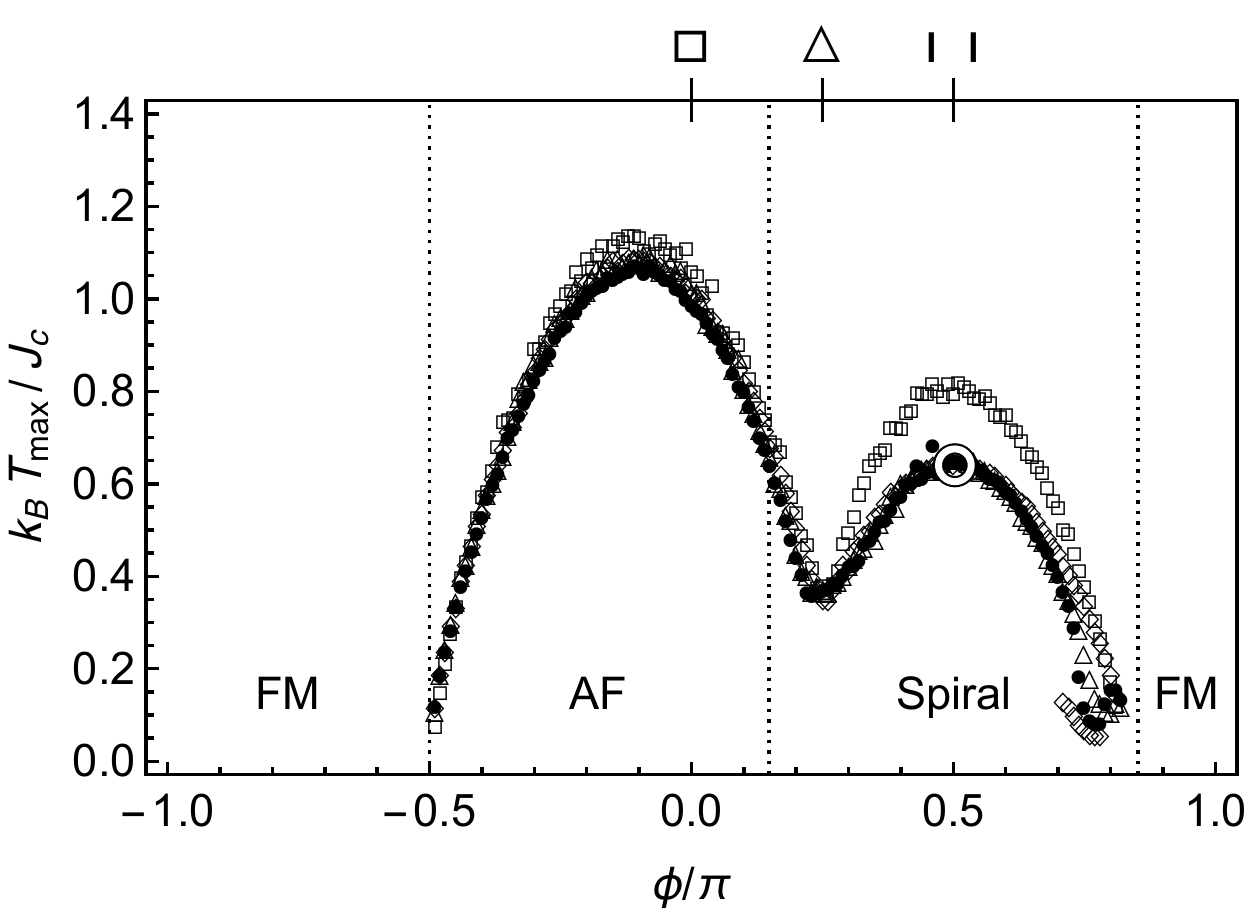}
    \includegraphics[width=.5\columnwidth]{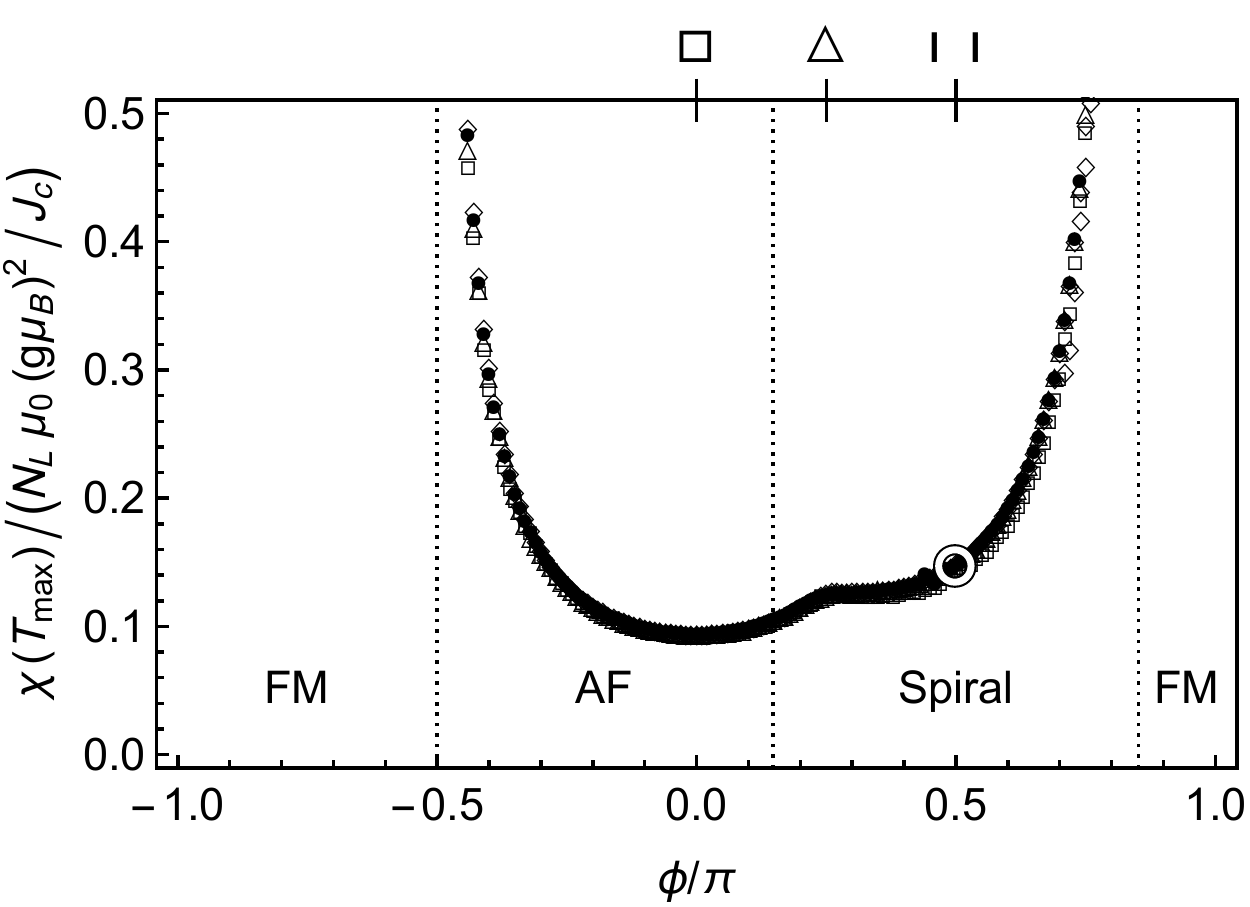}
    \caption{Maximum position (a) and value (b) of FTLM susceptibility as function of $\phi$ for triangular lattice. 
    Centered dot corresponds to special 1D case with $k_\text BT_\text{max}/J_\text c=0.641$ 
    from exact Bethe ansatz result \cite{bonner:64}. 
    Tile sizes are N=16 (squares), 20 (triangles), 24 (diamonds) , 28 (filled circles). 
    }
     \label{fig:chitri_max}
\end{figure}

The susceptibility is the most useful and easily accessible quantity for a determination of parameters in the exchange Hamiltonian.  
The typical temperature dependence obtained from FTLM is shown in Fig.~\ref{fig:chicvtri_T}~(a) for various control parameters in the triangular lattice going from deep inside the spiral phase to the ferromagnet where $\chi(T)$ diverges for $T\to0$ in the thermodynamic limit. A non-monotonic $\phi$ dependence of the susceptibility maximum (black dots) is clearly visible.  The evolution of the maximum temperature $T_\text{max}$ together with the susceptibility value $\chi(T_\text{max})$ is collected in Fig.~\ref{fig:chitri_max}.

Likewise the the specific heat curves are shown in  Fig.~\ref{fig:chicvtri_T}~(b). Here, the range of frustration angles $\phi$ corresponds to the crossover between the 120-degree structure at $\phi/\pi=1/4$ to the nonmagnetic quasi-1D case. In a large part of the AF region (not shown) two distinct maxima exist. 
\begin{figure}
    \centering
    \includegraphics[width=.5\columnwidth]{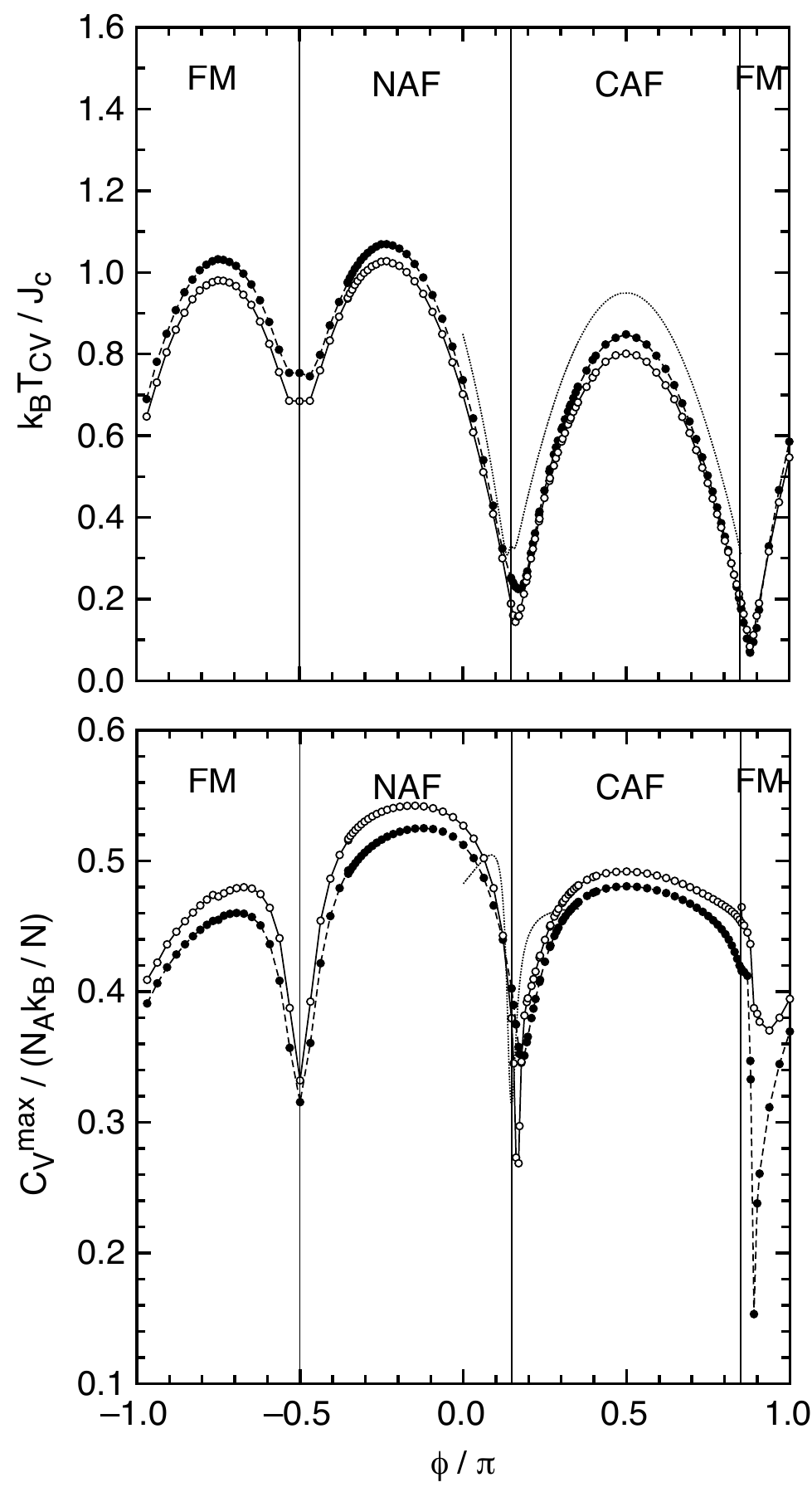}
    \caption{Maximum position (a) and value (b) of the FTLM specific heat $c_V(T)$ as function of $\phi$ for the square lattice.
     Open (solid) circles are for sizes $N = 20$ ($16$). 
     }
     \label{fig:cvsquare_max}
\end{figure}
As function of tile size these quantities converge rapidly to the thermodynamic limit, indeed the 28-site tile results agree with that of exact value from Bethe ansatz solution (centered circle) for the special 1D chain ($\parallel$) case. The absolute maximum of $T_\text{max}(\phi)$ appears in the unfrustrated AF region while the secondary, lower maximum corresponds to the 1D AF chain with quasi-LRO, its value is considerably reduced due to quantum fluctuations. Naturally $T_\text{max}(\phi)$ moves to zero when approaching the FM phase. The local minimum appears at the maximally frustrated isotropic triangular $(\triangle)$ case when the combined effect of frustration and quantum fluctuations leads to a large density of low lying excitations implying a peak in $\chi(T)$ at low temperatures. The $\phi$-dependence of the peak height in the AF and spiral regime is rather flat but increases rapidly when approaching the FM phase (Fig.~\ref{fig:chitri_max}(b)). The qualitative picture for the square lattice model \cite{shannon:04} is similar except that $T_\text{max}(\phi)$ is less asymmetric. The ratio of main to side maximum value is $1.58$ in the triangular lattice in Fig.~\ref{fig:chitri_max}(a) while it is only 1.1  for the square lattice model.

For the specific heat $c_V$ (Eq.~(\ref{eqn:thermodynamic}))  which is the second order cumulant of the energy only eigenvalues of the latter and no matrix elements of operators are needed which renders it  easier to calculate. On the other hand to obtain the experimental magnetic specific heat of a localized spin compound it is necessary to subtract the phonon contribution which is not so easy and may not lead to a unique position and value of the maximum in $c_V(T)$. Their calculated FTLM values scanned through the phase diagram are presented in Fig.~\ref{fig:cvsquare_max} for the square lattice.
For $c_V(\phi)$ the maximum position is now also finite in the FM sector. The temperature is smallest precisely at the classical NAF/CAF and CAF/FM phase boundary where the combined effect  of frustration and quantum fluctuations is large and destabilizes the magnetic order parameters leading two high density of low lying spin excitations in 
this `spin liquid' regime and therefore a specific heat peak at comparatively low temperature. In contrast to the trigonal case (Fig.~\ref{fig:chicvtri_T}) a double maximum structure never appears at any value of $\phi$ for the square lattice. Whether in the former case it survives in the thermodynamic limit is still questionable.

\subsection{Magnetocaloric effects and adiabatic cooling}
\label{magnetocaloric}

The LSW analysis has shown that the effect of frustration  and quantum fluctuations in ground state properties is strongly suppressed by application of a magnetic field. This is due to the change in the spin wave spectrum when the moments become polarized. Therefore one should also expect corresponding changes in thermodynamic quantities with magnetic field, in particular the specific heat and the magnetocaloric coefficient  where the latter measures the adiabatic change of temperature ${\rm d}T_\text{ad}=\Gamma_\text{mc}{\rm d}B$ with field change, i.e. the adiabatic field-cooling rate. The integrated temperature change under adiabatic field variation between $B_0$ and $B_1$  is then given by
\begin{equation}
\Delta T_\text{ad}(T,\Delta B)=-\int_{B_0}^{B_1}{\rm d}B
\frac{T}{c_V(T,B)}\left(\frac{\partial m_0(T,B)}{\partial T}\right)_B.
\label{eqn:fieldcooling}
\end{equation}
\begin{figure}
    \centering
    \includegraphics[width=.4\columnwidth]{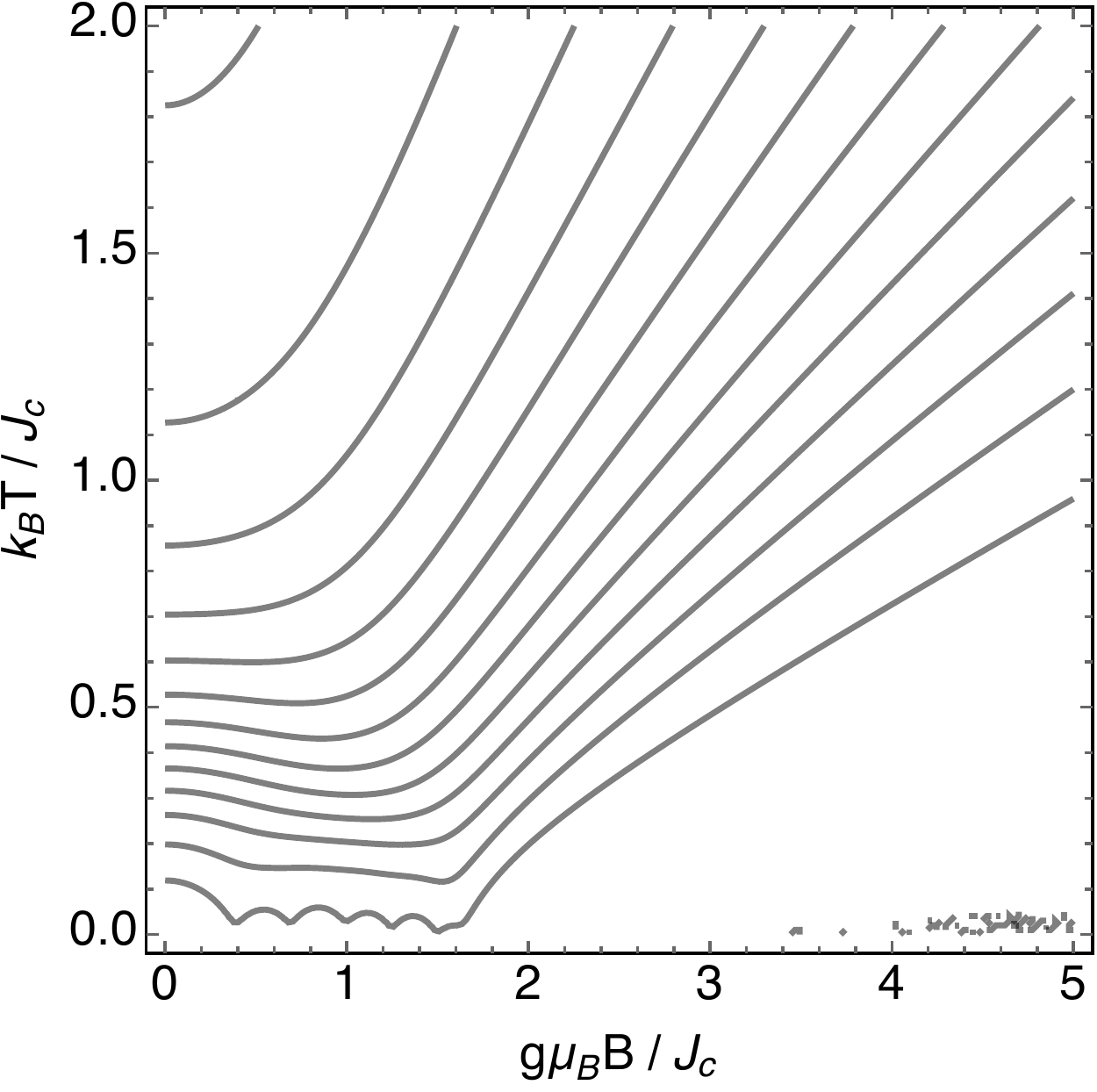}   
    \caption{Isentropics in the CAF phase ($\phi/\pi=0.74$) for tile 24:4-0 (\ref{sect:ED}) in the $B$-$T$ plane. 
    Lowest curve: $S=0.05k_\text B$ (per site), highest curve: $S=0.6 k_\text B$.
    When approaching the saturation field from above (the fully polarized state) at low $T$ the slope of isentropics 
    increases leading to the enhancement of the magnetocaloric coefficient that traces the saturation field  curve as 
    function of frustration parameter $\phi$ (cf. Fig.~\ref{fig:mcal}(b)).
     }
     \label{fig:isentropic_square}
\end{figure}
For practical cases $(\partial m_0/\partial T) < 0$ and  therefore a cooling is achieved when the
field is lowered adiabatically i.e. $B_1 <B_0$.  This effect is of great technical importance for cooling applications and we refer to Refs.~\cite{pecharsky:99,tishin:99} on this issue. The magnetocaloric coefficient of free paramagnetic spins is given by $\Gamma^0_\text{mc}=T/B$ and then $\Gamma_\text{mc}/\Gamma^0_\text{mc}$ defines the enhancement due to spin interaction effects. This qualitative statement is supported by the behavior of isentropics calculated within FTLM (Fig.~\ref{fig:isentropic_square}). It shows that for large fields and temperature the ratio of $T/B = \Gamma^0_\text{mc}$ is constant for constant entropy (cf. Eq.~(\ref{eqn:gammamc})) and therefore is a suitable normalization quantity. The isentropics of the (isotropic) triangular lattice show qualitatively similar behavior \cite{honecker:06}.

\begin{figure}
    \centering
    \includegraphics[width=.48\columnwidth]{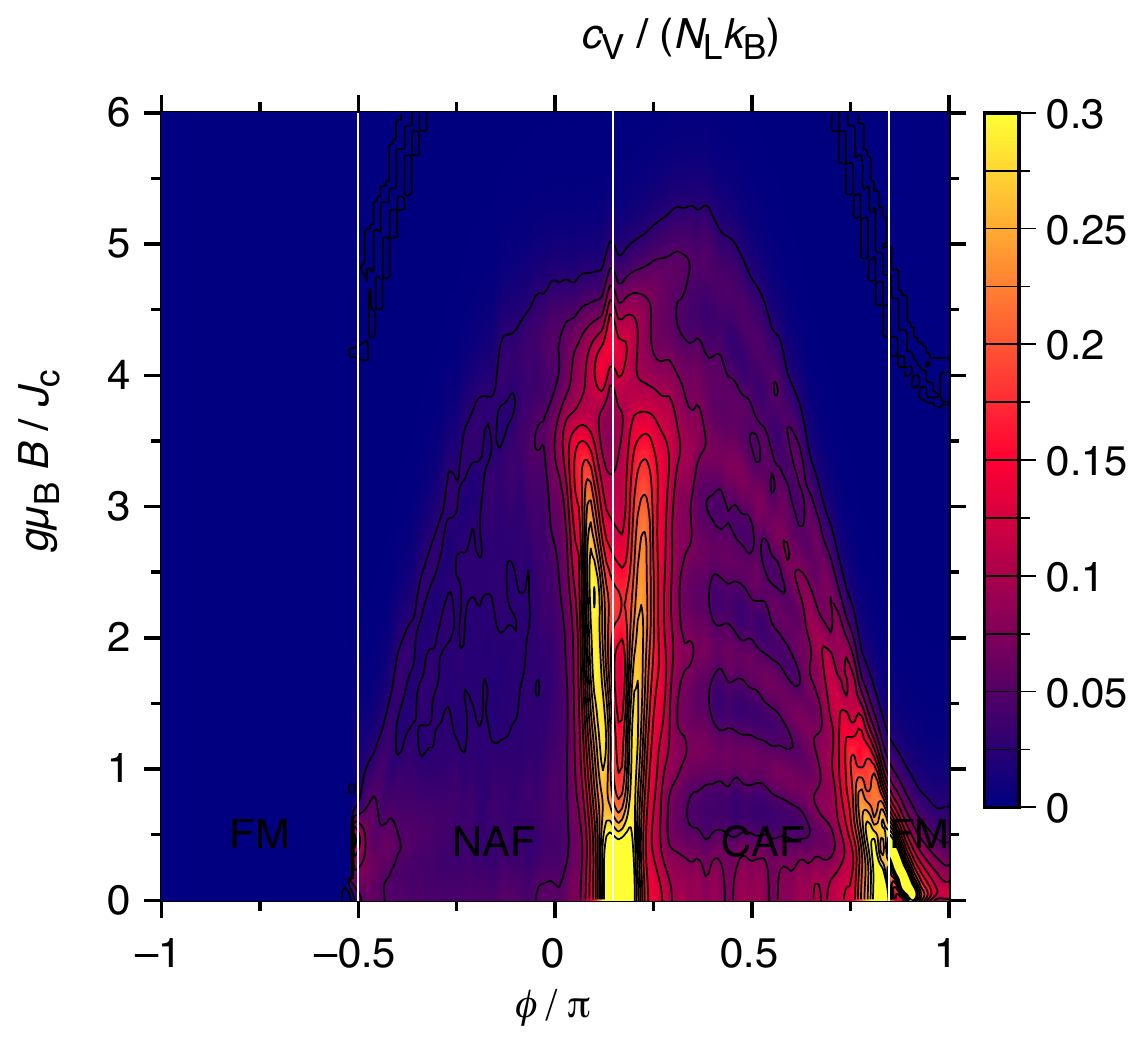}
    \includegraphics[width=.48\columnwidth]{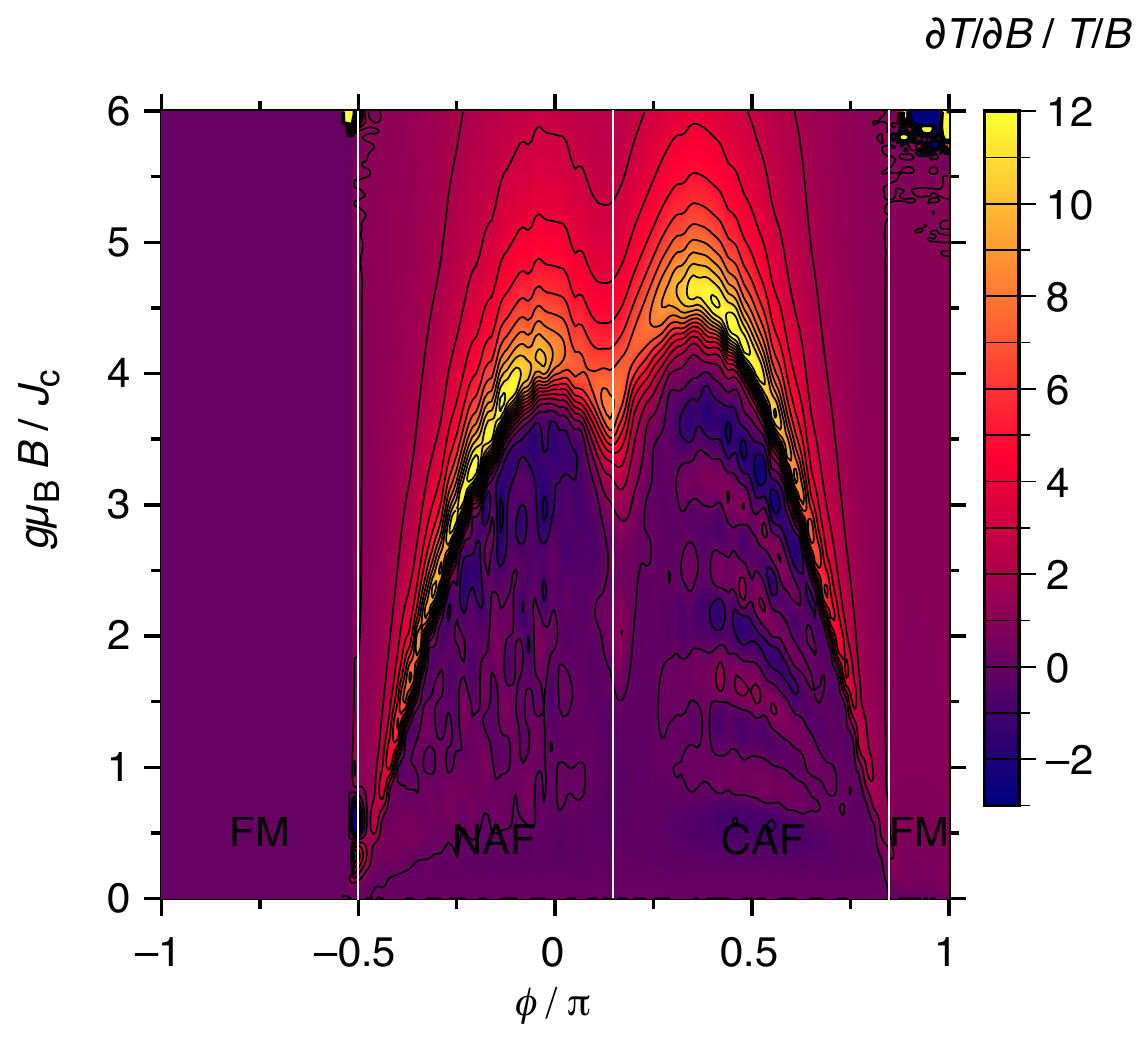}    
    \caption{Contour plots of the specific heat (a) and magnetocaloric enhancement ratio 
    (b) for  the N=24 square lattice tile at fixed 
    $T=0.2J_\text c/k_\text B$ as function of control parameter $\phi$ and field $B$. The specific heat is maximal 
    in the strongly frustrated and low field regimes.
    The magnetocaloric enhancement ratio is maximal in the weakly frustrated regime at saturation field (see Fig.~\ref{fig:satfield_class}). }
     \label{fig:mcal}
\end{figure}
The magnetocaloric FTLM results are presented in Fig.~\ref{fig:mcal} as a contour  plot of $c_V$ (a) and $\Gamma_\text{mc}/\Gamma^0_\text{mc}$ (b) in the $\phi,B$ plane for the square lattice model. This representation has the advantage to reveal immediately how the enhancement of this quantities by interactions is related to the degree of frustration.
Firstly both envelope curves in this figure track the saturation field curve of the square lattice in Fig.~\ref{fig:satfield_class}, in particular for $\Gamma_\text{mc}$ (b).
The specific heat at small fields shows two well localized maxima in the highly frustrated spin liquid regimes around the classical CAF/FM and NAF/CAF  boundaries. In the latter $(\phi/\pi\simeq 0.15)$ the peak splits into a double peak structure at larger fields indicating that in a small $\phi$-region in between another type of order may be stabilized. On the other hand the magnetocaloric cooling rate inside the ordered regime is small because changing the field there leads to little entropy change. The maximum enhancement of the cooling rate is observed when approaching the transition to the ordered regime from above the saturation field. Interestingly the absolute maximum of magnetocaloric enhancement occurs at the saturation field well {\it inside} (as function of $\phi$) the NAF and CAF phases and not in the spin liquid regime at the NAF/CAF or CAF/FM boundary. This is due to the large specific heat peaks there which rather suppresses the cooling rate according to the second expression in Eq.~(\ref{eqn:gammamc}) and also Eq.~(\ref{eqn:fieldcooling}). The maximum enhancement factor obtainable in the square lattice model is about one order of magnitude. The magnetocaloric effect has also been investigated for other types of frustrated lattices using the classical Monte-Carlo method \cite{zhitomirsky:03}.


\section{Neutron scattering and the static magnetic structure factor $S(\vec q)$}
\label{sect:structurefactor}

A detailed understanding of the magnetism of the materials discussed later (Sec.~\ref{sect:compounds}) can be obtained by spectroscopic methods. For properties like magnetic order (or the absence of it) and the magnetic excitation spectrum, neutron scattering experiments belong to the key experimental tools. The magnetic moment of the neutron probes directly the static and dynamic magnetic properties of a sample via the weak dipolar interaction. The weak interaction with matter due to its neutrality also results in a large penetration depth, therefore the true bulk properties of a crystal can be studied. Furthermore it facilitates investigations at low temperatures and high magnetic fields. In a magnetic neutron diffraction experiment the differential scattering cross section measured is proportional to the static structure factor given by the Fourier transform of the spin correlation function:
\begin{equation}
	S_{\alpha\beta}(\vec q)	=
	\frac1{\cal N}\sum_{ij}{\rm e}^{{\rm i}\vec q(\vec R_i-\vec R_j)}
	\left\langle S_i^\alpha S_j^\beta\right\rangle_\beta
	\label{eqn:sq}
\end{equation}
(Here and in the following we omit the time index of the moment operator in equal-time correlation functions and we implicitly assume time-independent, rigid ion positions). The clear distinction between the Néel phase with ordering vector $\vec Q=(\pi,\pi)$ and the columnar antiferromagnet with ordering vector $\vec Q=(\pi,0)$ or $\vec Q=(0,\pi)$ is possible by determination of $S_{\alpha\beta}(\vec q)$ in a diffraction experiment (Sec.~\ref{subsect:compounds_neutrons}). In this way, the ambiguity in determining the exchange constants from measurements of the heat capacity and the susceptibility of a given compound can be resolved. Therefore the evaluation of $S_{\alpha\beta}(\vec q)$ for the $J_1$-$J_2$ model is important. We briefly discuss the latter in the ordered as well as paramagnetic state. 

\subsection{Magnetically  ordered phase}
\label{subsect:neutron_ordered}

Typical interaction times in a diffraction experiment on a magnetically ordered system correspond to evaluating the spin correlation function in the limit of long time $t\to\infty$, where the individual spins loose their correlation. This means that in this case we can make the replacement
\begin{equation}
  \left\langle S_i^\alpha S_j^\beta\right\rangle
  =
  \left\langle S_i^\alpha\vphantom{S_j^\beta}\right\rangle
  \left\langle S_j^\beta\right\rangle,
  \quad i\ne j.
\end{equation}
Adding and subtracting $\left\langle S_i^\alpha\vphantom{S_j^\beta}\right\rangle\left\langle S_j^\beta\right\rangle$ to/from Eq.~(\ref{eqn:sq}) for $t\to\infty$, we can write
\begin{equation}
  S_{\alpha\beta}(\vec q)=
  \frac{1}{\cal N}
  \sum_{i=1}^N
  \left[
  \left\langle S_i^\alpha S_i^\beta\right\rangle
  -
  \left\langle S_i^\alpha\vphantom{S_i^\beta}\right\rangle
  \left\langle S_i^\beta\right\rangle
  \right]
  +
  \frac{1}{\cal N}
  \sum_{i=1}^N\sum_{j=1}^N
  \left\langle S_i^\alpha\vphantom{S_j^\beta}\right\rangle
  \left\langle S_j^\beta\right\rangle
  {\rm e}^{{\rm i}\vec q\left(\vec R_i-\vec R_j\right)},
  \label{eqn:lsw:sfsplit}
\end{equation}
splitting the structure factor into a momentum-transfer independent background (first term above)
plus a coherent scattering part.

For finite systems like those we are discussing with exact diagonalization, the assumption of vanishing intersite correlations and the appearance of spontaneous magnetic order is, of course, not correct. However it is useful to discuss selected limiting cases here to better understand the nature of the scattering process.
We assume a spin-flop phase, i.\,e.~magnetic ordering in the $xy$ plane with ordering vector $\vec Q$, and a uniform magnetic field defining the $z$ direction, and omit the incoherent part in Eq.~(\ref{eqn:lsw:sfsplit}). The cross-section of coherent magnetic diffraction of a magnetically ordered system in a uniform applied magnetic field, ignoring the Debye-Waller-Factor is evaluated as \footnote{Here $g$ is the gyromagnetic ratio, $r_0$ the classical electron radius and $F(\vec q)$ the dimensionless atomic magnetic form factor.}
\begin{align}
\frac{{\rm d}\sigma}{{\rm d}\Omega}
  =
  \frac{N}{\cal N}
  r_0^2\left[\frac{1}{2}gF(\vec q)\right]^2
  \frac{(2\pi)^3}{v_0}\left\langle S\right\rangle^2
  \sum_{\vec G}
  &\left\{
  \frac{1}{4}\sin^2\Theta\left(1+\hat q_z^2\right)
  \left[
    \delta(\vec q+\vec Q-\vec G)+
    \delta(\vec q-\vec Q-\vec G)
  \right]
  \right.
  \nonumber\\
  &\left.\vphantom{\frac14}
  +
    \cos^2\Theta\left(1-\hat q_z^2\right)
    \delta(\vec q-\vec G)
  \right\}
  \label{eqn:sqclass}
\end{align}
The parameter $\Theta$ specifies the canting angle of the magnetic moments relative to the $z$ direction. Apart from the peaks at the nuclear Bragg positions $\vec q=\vec G$ proportional to the square of the induced moment or magnetization $m_0\sim\left\langle S\right\rangle\cos\Theta$ which appear in a finite magnetic field only, additional intensity occurs at the positions $\vec q=\vec G\pm\vec Q$, which is proportional to square of the staggered moment $m_{\vec Q}\sim\left\langle S\right\rangle\sin\Theta$, where $\left\langle S\right\rangle$ is the ``length'' of the ordered moment. This additional intensity is due to the $xx$ and $yy$ components of the structure factor.

\subsection{High temperature approximation}
\label{subsect:neutron_hightemp}

In a similar way as for the uniform susceptibility (Sec.~\ref{sect:temp}) we can expand Eq.~(\ref{eqn:sq}) in powers of the inverse temperature $\beta$. For a crystal with an isotropic exchange, the first two terms
\begin{equation}
	S_{\alpha\beta}(\vec q)\approx\frac{1}{3}\delta_{\alpha\beta}S_{\text{HTSE}}(\vec q),\quad
	S_{\text{HTSE}}(\vec q)=S(S+1)\left(1-\frac{2S(S+1)}3\beta J(\vec q)\right)
	\label{eqn:shtse}
\end{equation}
with $J(\vec q)$ defined in Eq.~(\ref{eqn:exfourier}) in many cases already give a good approximation at temperatures $T\approx J_{\text c}/k_{\text B}$ and above~\cite{shannon:04,skoulatos:08}. For systems which order magnetically at low temperatures, $J(\vec q)$ has minima for $\vec q=\vec Q+\vec G$ where $\vec Q$ is the ordering vector and $\vec G$ a reciprocal lattice vector, therefore we can expect characteristic maxima in $S(\vec q)$ at these magnetic Bragg positions in reciprocal space already above the magnetic ordering temperature.

\subsection{Finite temperature Lanczos method}
\label{subsect:neutron_FTLM}

With the finite temperature Lanczos method, we can directly calculate the spin correlation functions $\left\langle S_m^\alpha S_n^\beta\right\rangle_{\beta}$ contained in Eq.~(\ref{eqn:sq}) in an unbiased way (for details see \ref{sect:ftlm}). Because $\left[{\cal H},S_m^\alpha S_n^\beta\right]\ne0$ in general, the  effort to compute one  of these correlation functions  occurring in the sum~(\ref{eqn:sq}) is about twice as large as for thermodynamic quantities like the heat capacity and the susceptibility where the respective operators commute with the Hamiltonian.
\begin{figure}
\begin{center}
\hfill
\includegraphics[width=.45\columnwidth]{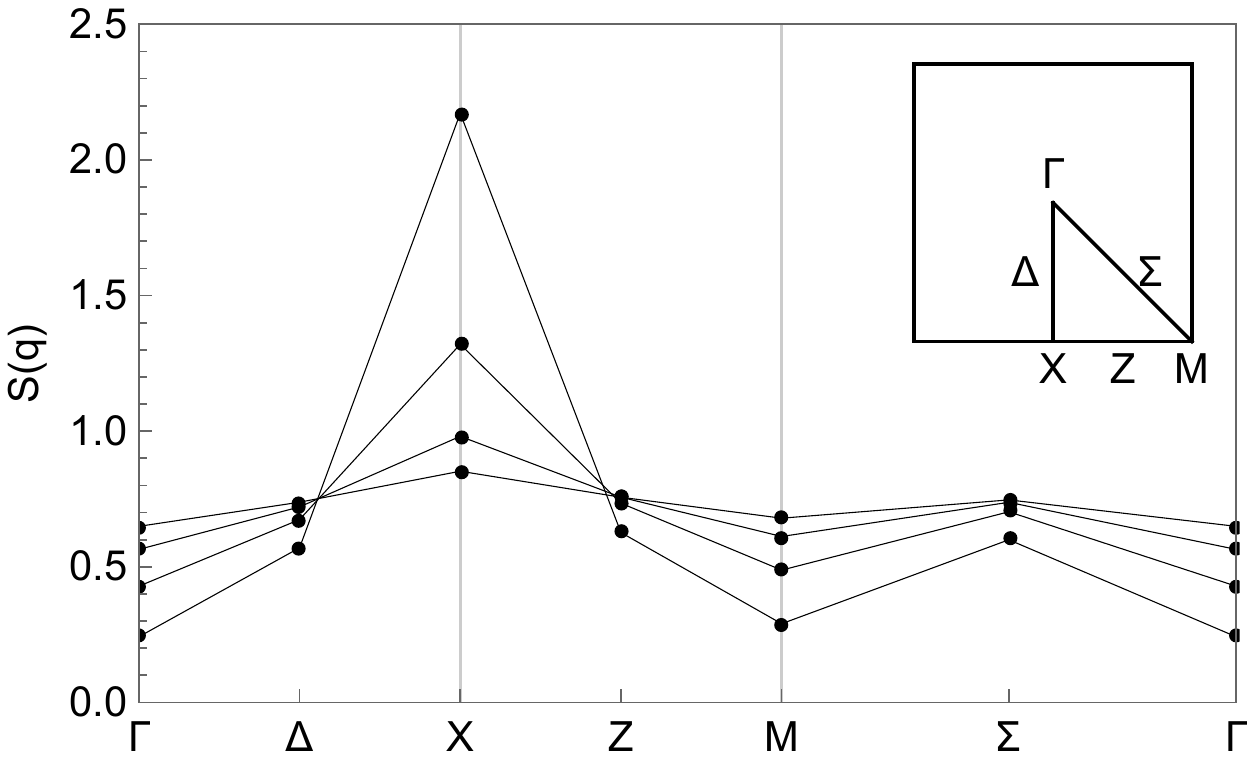}
\hfill
\includegraphics[width=.45\columnwidth]{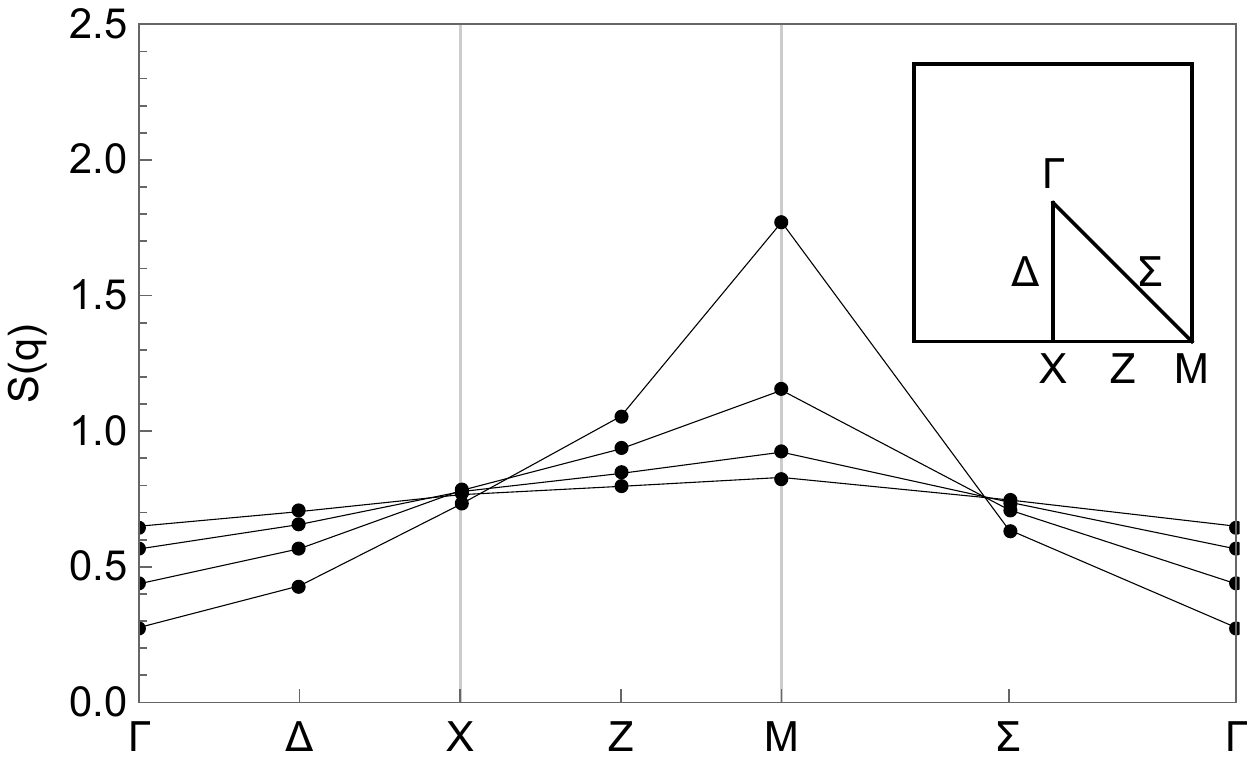}
\hfill\null
\caption{Static spin structure factor $S(\vec q)$ of tile 16:4-0 for the $S=1/2$ square-lattice model in the two antiferromagnetic phases. (a) Columnar phase with $J_1/k_{\text B}=1.25\,\text K$ and $J_2/k_{\text B}=5.95\,\text K$. (b) Néel phase with $J_{1}/k_{\text B}=5.95\,\text K$, $J_{2}/k_{\text B}=1.25\,\text K$. The values chosen for $J_{1}$ and $J_{2}$ correspond to those determined for Li$_{2}$VOSiO$_{4}$~\protect\cite{misguich:03,shannon:04}.  The lines are guides to the eye, the dots denote the numerical results.  The individual curves in each figure, from bottom to top at ${\bf q}=\Gamma$, correspond to fixed temperatures $k_{\rm B}T/J_{\rm c}=1$, $2$, $4$, and $8$.}
\label{fig:sq}
\end{center}
\end{figure}

Fig.~\ref{fig:sq} shows the resulting structure factor $S(\vec q)=S_{\alpha\alpha}(\vec q)$ of tile 16:4-0 for the square-lattice model with spin-space isotropic exchange at the special high-symmetry points in the irreducible triangle of the Brillouin zone. The values chosen for $J_1$ and $J_2$ are those obtained for Li$_{2}$VOSiO$_{4}$ from measurements of the magnetic susceptibility and the heat capacity~\cite{misguich:03,shannon:04}. They correspond to frustration angles $\phi/\pi\approx0.43$ (columnar phase) and $\phi/\pi\approx0.07$ (Néel phase). Already for temperatures $T$ near $J_{\text c}/k_{\text B}$ (which is often of the order of a few Kelvin), a clear distinction between the different antiferromagnetically ordered phases is possible: We find pronounced peaks at the ordering vectors of the respective ground states and we have $S(\vec Q)>S(\vec q)$ for all $\vec q\ne\vec Q$, reflecting the expected behavior from the associated broken symmetries in the thermodynamic limit. 

This relation no longer holds necessarily in the spin liquid regime (not shown): There $S(\vec q)$ is comparatively structureless and has approximately the same values for both $\vec Q=(\pi,\pi)$ and $\vec Q=(\pi,0)$ at temperatures $T\ge J_{\text c}/k_{\text B}$ which is another indication that the strong frustration prevents magnetic order.
\begin{figure}
\begin{center}
	\includegraphics[width=.4\columnwidth]{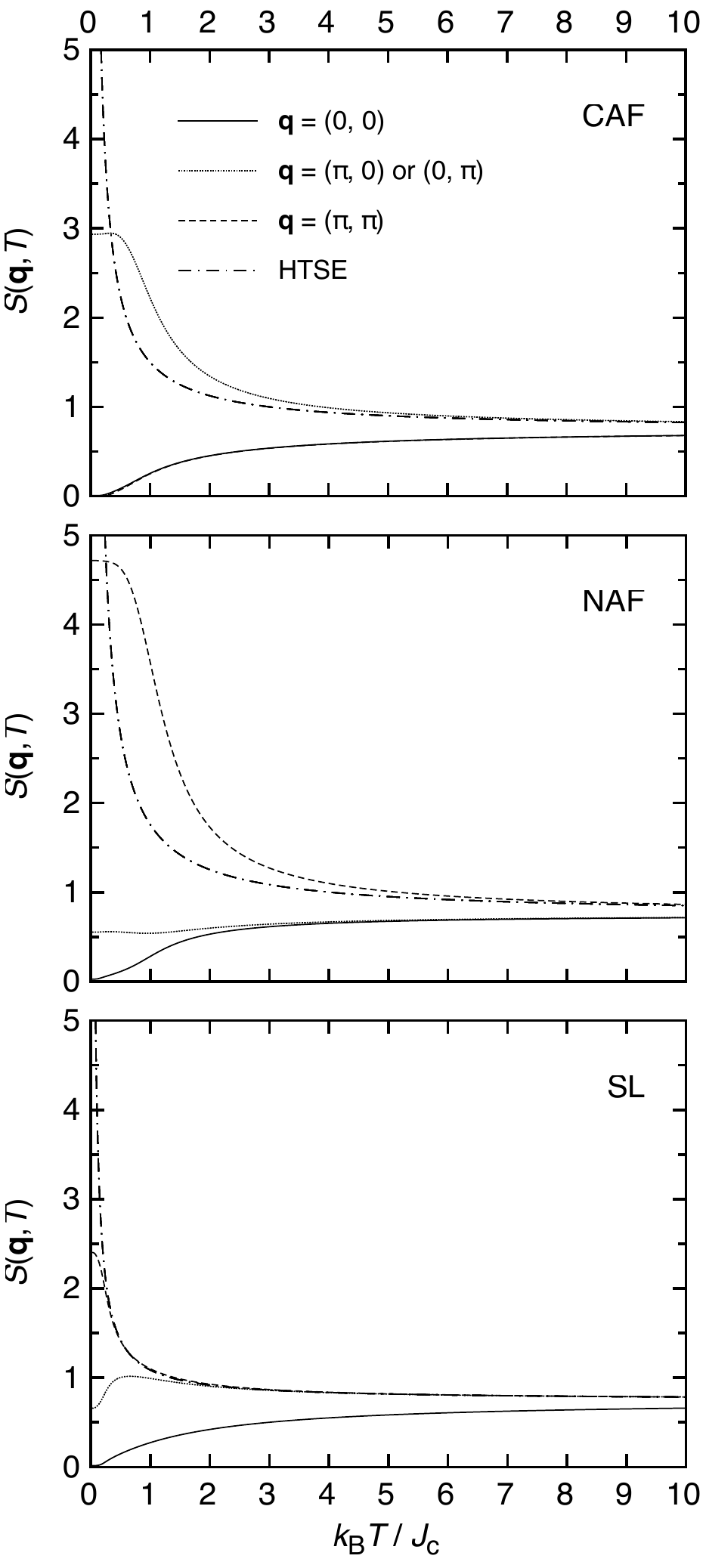}
	\caption{Temperature dependence of the static structure factor $S(\vec q)$ of the square-lattice model for $\vec q=(0,0)$ (solid lines), $(\pi,0)$ or $(0,\pi)$ (dotted lines), and $\vec q=(\pi,\pi)$ (dashed lines). From top to bottom: Columnar phase $(\phi/\pi=0.5)$, Néel phase $(\phi/\pi=-0.15)$, and disordered SL regime $(\phi/\pi=0.15)$. The dash-dotted line in each panel shows the high-temperature expansions (Eq.~(\ref{eqn:shtse})) at the respective value for $\vec q$ where $S(\vec q)$ diverges for $T\to0$.}
	\label{fig:sqt}
\end{center}
\end{figure}

The temperature dependence of the static structure factor is illustrated in the three panels of Fig.~\ref{fig:sqt}. They show $S(\vec q)$ of the square-lattice model for $\vec q=(0,0)$ (solid lines), $(\pi,0)$ or $(0,\pi)$ (dotted lines), and $\vec q=(\pi,\pi)$ (dashed lines). In each panel, the additional dash-dotted line displays $S_{\text{HTSE}}(\vec q)$ given by the high-temperature expansion ({HTSE}) in Eq.~(\ref{eqn:shtse}) at those values for $\vec q$ where it actually diverges for $T\to0$. In both antiferromagnetically ordered phases, $S(\vec q)$ is at maximum for the ordering vector $\vec Q$ which decreases as a function of $T$, eventually approaching the $T\to\infty$ value $S(S+1)=3/4$. For a given temperature, the relation $S(\vec Q)>S(\vec q)$, $\vec q\ne\vec Q$ always holds, and $S(\vec q)\approx0$ for $\vec q\ne\vec Q$ at low temperature. Qualitatively, $S_{\text{HTSE}}(\vec q)$ is a good approximation, however at low temperatures for the ordered cases it under-estimates the numerical results for $T\lesssim 5J_{\text c}/k_{\text B}$.

The behavior of $S(\vec q)$ is quite different in the disordered regime (lower panel in Fig.~\ref{fig:sqt}): Here, $S(\vec q)>S(0)$ for both NAF and CAF ordering vectors even at high temperatures $T\gg J_{\text c}/k_{\text B}$~\cite{shannon:04}. Correspondingly, $S_{\text{HTSE}}(\vec q)$ diverges for low temperatures at both ordering vectors as well.


\section{Non-magnetic quantum phases}
\label{sect:quantumphases}

The calculation of ground state moment with numerical exact diagonalization scaling approach and the analytical linear spin wave approach have both demonstrated that in the square lattice model for $J_2/|J_1|\simeq1/2$; $(\phi\simeq0.15\pi$, $0.85\pi$), $J_2>0$ the magnetic moment breaks down due to large quantum fluctuations (Fig.~\ref{fig:groundstate}~(b)). This signifies the appearance of a `spin-liquid' phase on the CAF/NAF and CAF/FM boundaries. This interpretation is supported by the finite temperature anomalies in the specific heat as discussed above.  In the triangular lattice the instability occurs only close to the SP/NAF boundary and obviously around the quasi-1D region (Fig.~\ref{fig:moment_triangular}) where it is not related to frustration.

Although they are not the main topic of this review we discuss briefly some proposals for their nature which has been scrutinized in an enormous amount of theoretical work (for a review, see~\cite{misguich:12,starykh:15,savary:17}). The generic designation `spin liquid' encompasses many different types of ground states, genuinely disordered as well as exotic, non-magnetically ordered. They are characterized by exponentially or algebraically decaying spin correlations when the spin excitations are gapped or gapless respectively. A particular case of the latter is the quasi-1D U(1) spin liquid with gapless fermionic spinon excitations \cite{karbach:97}.

It must be said clearly, however, that presently there is no real 2D square or triangular lattice material candidate, at least among transition metal insulators, that exhibits such a non-magnetic ground state. Therefore we find it justified not to give too much attention to these special phase regions. Interestingly, the opposite `order by disorder' scenario where quantum fluctuations stabilize magnetic order by selecting among a continuously degenerate classical manifold is also realized in the square lattice model. 

\begin{figure}
    \centering
    \includegraphics[width=.5\columnwidth]{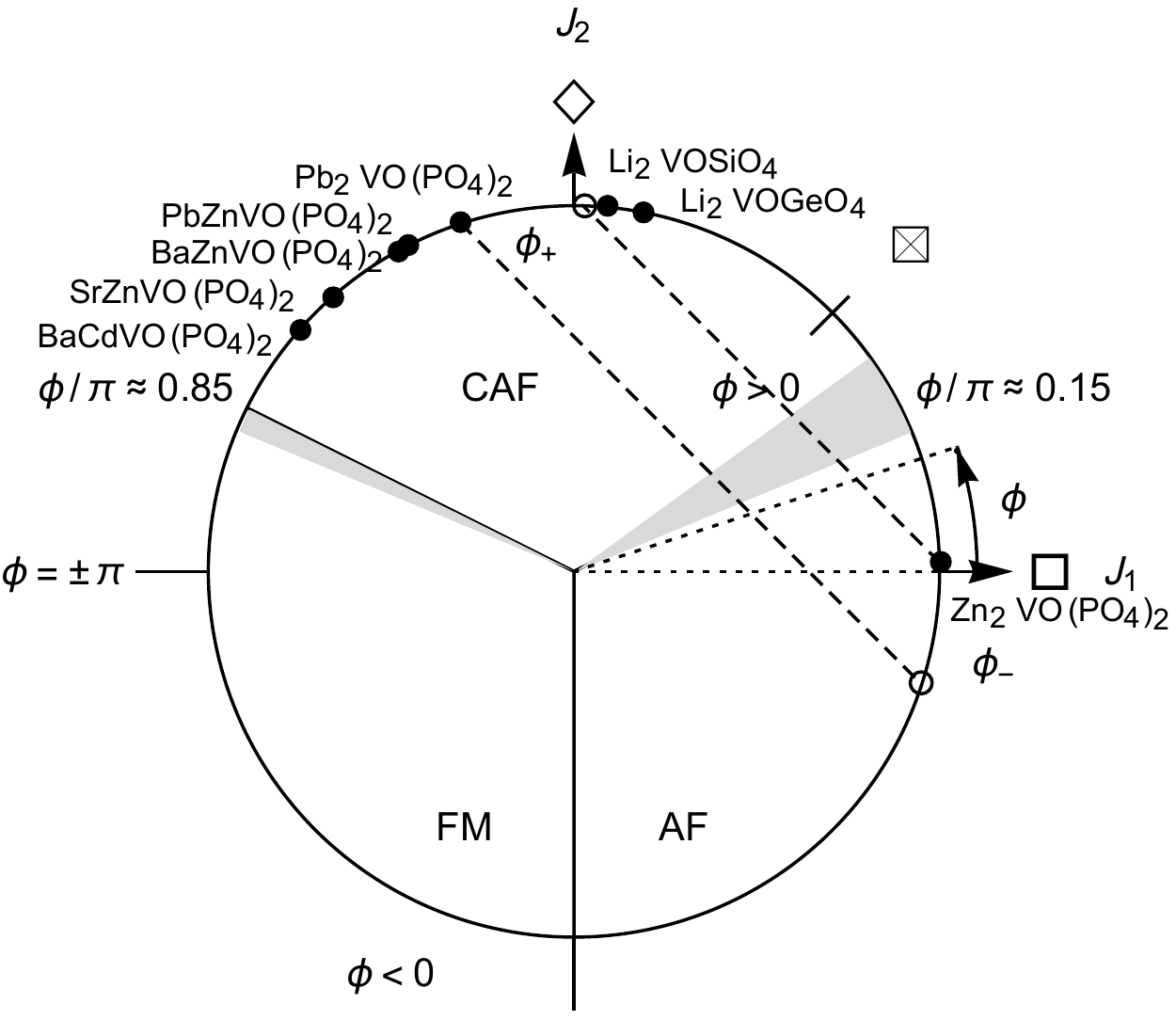} 
    \caption{Phase diagram of square lattice model mapped to the unit circle $(J_\text c\to1)$ 
    with the control parameter 
    $\phi=\tan^{-1}(J_2/J_1)$  ($-\pi\leq\phi\leq\pi$). The upper half ($J_2 >0$, $0\leq\phi\leq\pi$) is generally a 
    frustrated region  while the 
    lower one  ($J_2 <0$, $-\pi\leq\phi\leq 0$) is unfrustrated (cf. Fig.~\ref{fig:frustration}). 
    The positions of $\phi$ for the vanadium compounds listed Table~\ref{tbl:exchange} are 
    indicated with the black dots.
    The dashed lines correspond to the experimental Curie-Weiss temperature 
    $\Theta_\text{CW}=(J_1+J_2)/k_\text B$ for the Pb$_2$ and the Zn$_2$ compounds \cite{kaul:04,kini:06} 
    with two possible values $\phi_-$ (NAF) and $\phi_+$ (CAF). The ambiguity is resolved by neutron diffraction, 
    FTLM fit or saturation field analysis. All compounds are in the CAF sector ($\phi_+$) 
    except for Zn$_2$VO(PO$_4$)$_2$ 
    which is a N\'eel antiferromagnet ($\phi_-$) \cite{kini:06,yusuf:10}. Regions with possible quantum phases as determined by finite-size scaling of ED data (Table~\ref{tbl:spinliquid}) are shaded in grey.}
     \label{fig:phasedia}
\end{figure}
The appearance of special regions in the phase diagram may be understood by rewriting the Hamiltonian in Eq.~(\ref{eqn:hex12}) using block-spins on a square plaquette with sites denoted clockwise  $i=1\ldots4$~\cite{shannon:04}. Since the square lattice is bipartite we may assign diagonal opposite sites $i=1,3$ to the A sublattice and $i=2,4$ to the B sublattice. Then, introducing $\vec S^\text A_\text t=\vec S^\text A_1 +\vec S^\text A_3$ and $\vec S^\text B_\text t=\vec S^\text B_2 +\vec S^\text B_4$ and denoting their combinations by $\vec S_{t\pm}=\vec S^\text A_\text t\pm \vec S^\text B_\text t$ we can rewrite the $J_1-J_2$ Hamiltonian in two equivalent forms $H^\pm$ as \cite{shannon:04}
\begin{eqnarray}
H^\pm
&=&
-2J_2S(S+1)N+\sum_\square H^\pm_\square,
\nonumber\\
H^\pm_\square
&=&
\pm\frac14J_1\vec S^2_{t\pm}+\frac12\left(J_2\mp\frac{1}{2}J_1\right)
\left[\left(\vec S^\text A_\text t\right)^2+\left(\vec S^\text B_\text t\right)^2\right]
\label{eqn:ham_squ}
\end{eqnarray}
Because the original Hamiltonian is symmetric under the transformation $\vec S^\text B\rightarrow -\vec S^\text B$, $J_1\rightarrow -J_1$ we also have $H_\pm\leftrightarrow H_\mp$ under this transformation. This means firstly that the phase diagram on the classical level is symmetric under reflection at $J_1=0$ (Fig.~\ref{fig:phasedia}). It is evident from this form that (i) $J_2/J_1=\pm1/2$ and (ii) $J_1=0$ are special cases where one of the two terms in Eq.~(\ref{eqn:ham_squ}) vanishes. For $J_2/|J_1|=1/2$ the lowest energy is achieved by {\em any\/} state for which the sum or difference of sublattice A, B spins vanishes on {\em each square\/} which leads to a very high classical ground state degeneracy. This is also reflected in the dispersionless spin waves leading to continuous degeneracy along BZ symmetry directions for these exchange ratios. This leads to the breakdown of the ordered moment and advent of exotic spin liquid phases. For $J_2>|J_1|/2$, the classical ground-state energy does not depend on $J_1$ (see Table~\ref{tbl:ecl}), thus the relative orientation of the A and B sublattice spins is arbitrary and again a continuous degeneracy results. For $J_2 <0$ any small $|J_1|$ will lead to either FM ($J_1 < 0$) or NAF  ($J_1 > 0$) phase already on the mean field level with a continuous transition between them. For  $J_2 > 0$ the (negative) contributions of quantum fluctuations to ground state energy will select among the degenerate states and stabilize a collinear CAF order of the two AF sublattices which is termed as `order from disorder' \cite{villain:80}.

\begin{figure}
    \centering
    \includegraphics[width=.6\columnwidth]{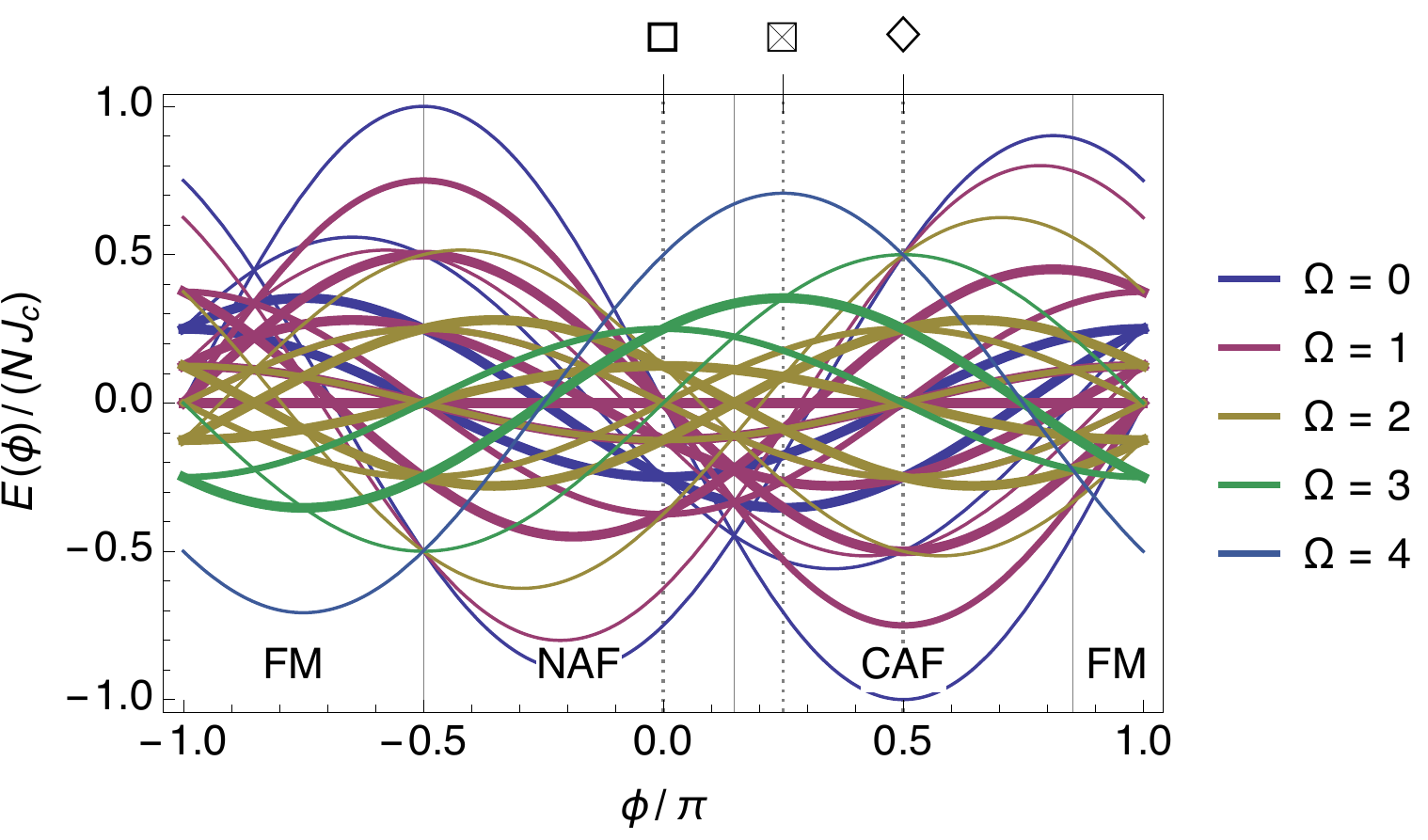} 
    \caption{Energy (per spin, in units of $J_\text c$) of square eight-site cluster levels, classified according to total 
    spin $\Omega=0\ldots4$ as function of $\phi$~\cite{shannon:04}. Line thicknesses indicate the degeneracy of the respective energy level, being either $1$ (thin lines), $2$, or $4$ (thick lines). High degeneracies occur around the strongly frustrated 
    positions ($\phi/\pi=0.15$, $0.85$) and the special unfrustrated points $J_1=0$ ($\phi/\pi=\pm 0.5$) and $J_2=0$ ($\phi/\pi=0,\pm1$).}
     \label{fig:eight_square}
\end{figure}
The large degeneracy of classical states is reflected also in the exact solution of the eight-site quantum model \cite{shannon:04}. Its level scheme is presented in Fig.~\ref{fig:eight_square}. A large number of level crossings can be seen,  leading to degeneracies precisely at the special cases $J_2/|J_1|=1/2$ ($\phi/\pi=0.15$, $0.85$) and $J_1=0$ ($\phi/\pi=\pm 0.5$). 

\subsection{Spin liquid vs. valence bond crystal phase at $J_2/J_1 \simeq1/2$, $J_1>0$}
\label{subsect:VBS}

\begin{figure}
    \centering
    \includegraphics[width=.3\columnwidth]{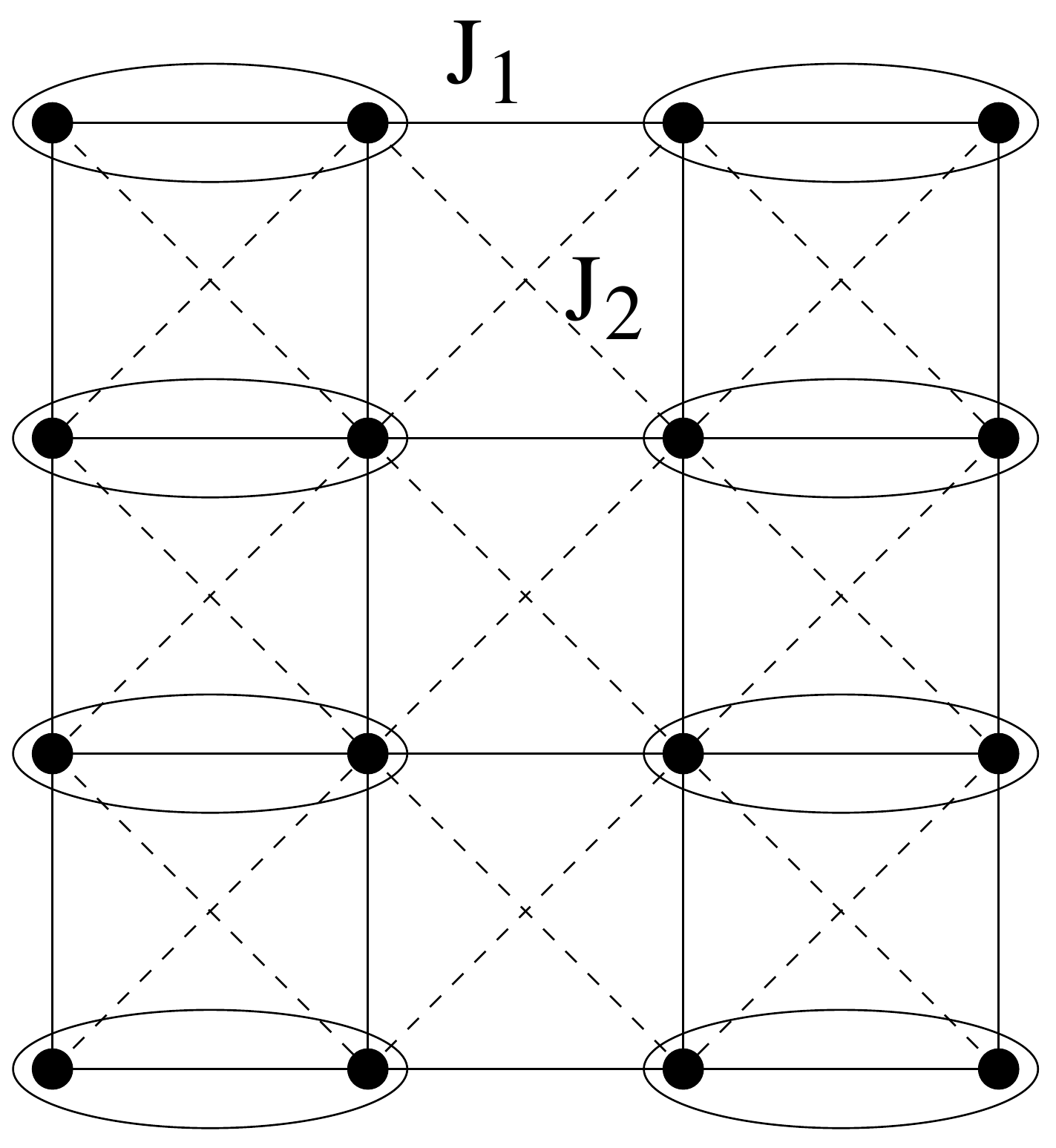} 
    \caption{In the columnar (striped) VBS phase the translational symmetry is broken by the spontaneous formation 
    of stacked spin-singlet dimers (rung ellipsoids) with wave vector $\vec Q=(\pi,0)$ (from Ref.~\cite{sushkov:01}).}
     \label{fig:VBS_col}
\end{figure}
The designation `spin liquid' is used generically for all many body ground states on spin lattices that do not exhibit long range magnetic order of some kind when temperature approaches zero. 
In the square lattice model
the spin wave and ED results in Fig.~\ref{fig:groundstate}~(b) predict a breakdown of the ordered moment in the regime around $J_2/J_1\simeq1/2$ ($\phi/\pi\simeq 0.15$). The estimated boundaries of this region (Table~\ref{tbl:spinliquid}) differ considerably depending on the method; ED leads to an instability region $0.125\leq\phi/\pi\leq0.2$ corresponding to $0.4\leq{J_2}/{J_1}\leq0.7$ where the frustration degree (Fig.~\ref{fig:frustration}) achieves its maximum. Then the question arises whether this intermediate phase is a genuine disordered  `spin-liquid' phase, e.g. a resonating valence bond state (VBS) \cite{anderson:73} with a finite gap for spin excitations and short range spin correlations,  a gapless spin liquid with algebraic spin correlations and fractionalized excitations (spinons) or whether the ground state exhibits some other more exotic order that does not break SU(2) spin-rotation symmetry but possibly translational invariance. The latter has been frequently suggested  in the context of VBS crystals. There a local spin-singlet formation occurs on dimers which are stacked in a staggered fashion such that the unit cell is enlarged. Most  commonly this produces a striped  VBS phase  \cite{kotov:99,singh:99}  with doubling of the unit cell (Fig.~\ref{fig:VBS_col}) along the crystallographic $a$ or $b$ direction.

\begin{table*}    
	\centering
        \begin{tabular}{lccc}
            quantum phase
            &
            $J_2/J_1$ interval
            &
            $\phi/\pi$ interval
            &
            method-Ref.
            \\
            \hline
	   spin dimer $(\boxtimes)$
	   &
            $[0.378,0.509]$
            &
            $[0.115,0.15]$
            &
            LSW \cite{shannon:04}
            \\
            $(J_2/J_1\simeq1/2)$
            &
            $[0.414,0.726]$
            &
            $[0.125,0.2]$
            &
            ED \cite{schmidt:11,siahatgar:12}
            \\
	   &
            $[0.35,0.60]$
            &
            $[0.11,0.17]$
            &
            SE \cite{gelfand:90}
            \\           
            &
            $[0.38,0.62]$
            &
            $[0.11,0.18]$
            &
            BOT \cite{kotov:99}
            \\
             &
            $[0.41,0.62]$
            &
            $[0.12,0.18]$
            &
            DMRG \cite{jiang:12}
            \\[0.3cm]            
             spin nematic $(\boxtimes)$
            &
            $[-0.513,-0.498]$
            &
            $[0.849,0.853]$
            &
            LSW \cite{shannon:04}
	    \\
	    $(J_2/J_1\simeq -1/2)$
	    &
	    $[-0.498,-0.418]$
	   &
	   $[0.853,0.874]$
	   &
	   ED \cite{schmidt:11,siahatgar:12}
	   \\
	   &
	   $[-0.70,-0.41]$
	   &
	   $[0.80,0.88]$
	   &
	   ED \cite{shannon:06,smerald:15}
	   \\ 
	   &
	   $[-0.45,-0.31]$
	   &
	   $[0.865,0.904]$
	   &
	   PFFRG \cite{iqbal:16}\\	     
	    \hline
	    2D spin liquid $(\Delta)$
	    &
	    $0.5$
	    &
	    $0.148$
	    &
	    LSW \cite{trumper:99,schmidt:14}
	    \\
	    $(J_2/J_1\simeq1/2)$
	    &
	    $[0.71,0.91]$
	    &
	    $[0.20,0.23]$
	    &
	    MSW \cite{hauke:13}
	    \\
	    &
	    $[0.68,0.91]$
	    &
	    $[0.19,0.23]$
	    &
	    DSE \cite{zheng:99}
	    \\	    	    
	    &
	    $[0.7,0.8]$
	    &
	    $[0.19,0.21]$
	    &
	    VMC \cite{ghorbani:16}
	    \\[0.3cm]   
	    quasi-1D spin liquid $(\Delta)$
	    &
	    $[3.894,\infty]$
	    &
	    $[0.42,0.50]$
	    &
	    LSW \cite{schmidt:14}
	    \\    
	    ($J_1\approx 0$)
	     &
	    $[-\infty,-5.35]$
	    &
	    $[0.50,0.56]$
	    &
	    LSW \cite{schmidt:14}
	    \\    	    
	    &
	    $[1.54,\infty]$
	    &
	    $[0.32,0.50]$
	    &
	    MSW \cite{hauke:13}
	    \\	    
	    &
	    $[1.67,\infty]$
	    &
	    $[0.33,0.50]$
	    &
	    VMC \cite{ghorbani:16}
	\end{tabular}
    \caption{Intervals of frustration/anisotropy ratio $J_2/J_1$ or  frustration/anisotropy angle $\phi$ for square/triangular 
    lattice  where the classical magnetic order is destroyed or quantum phases are stabilized  around
     $J_2/|J_1|\simeq1/2$ ($\phi/\pi\simeq0.15$, $0.85$). The intervals of the magnetic instability or quantum phase 
     stability regions differ widely depending on method and reference, sometimes having no overlap. Here 
     SE = series expansion, BOT= bond operator theory, DMRG = density matrix renormalization group, 
     PFFRG = pseudo-fermion functional renormalization group,
     VMC = variational Monte Carlo, MSW = modified spin wave theory, DSE = dimer series expansion.}
      \label{tbl:spinliquid}
\end{table*}
Using the bond-operator representation \cite{sachdev:90} the spin dimers in Fig.~\ref{fig:VBS_col} may be represented by bosonic triplon $t_\alpha$ $(\alpha =x,y,z)$ operators creating the dimer excited triplet states out of the singlet ground state. Expressing the spin operators in the $J_1-J_2$ model by the $t_\alpha$ one arrives at an interacting bosonic triplon gas. This problem may be solved within hard-core boson approximation \cite{kotov:99} which results in a spin gap (minimum energy of the single triplon dispersion or two-triplon continuum) phase with broken translational symmetry in the  regime $0.38\leq{J_2}/{J_1}\leq0.62$. The existence of the staggered VBS state has been challenged by unbiased DMRG calculations \cite{jiang:12} on cylinders up to $2\times 100$ sites. There it was concluded that long range VBS order is absent and the ground state is rather a genuine spin liquid in the range  $0.41\leq{J_2}/{J_1}\leq0.62$ without breaking of translational symmetry.
The triplet spin gap in the VBS crystal case achieves a maximum value of $\Delta_\text t/J_1\simeq 0.21$ at $J_2/J_1\simeq 0.6$ $(\phi/\pi=0.17)$ where the maximum frustration  (Fig.~\ref{fig:frustration}) occurs. 

In the triangular model the zero-field 2D spin liquid range around  $J_2/J_1\simeq 0.5$ is less well investigated. In LSW the moment instability occurs only at this point. Dimer series expansion \cite{zheng:99} and MSW theory \cite{hauke:11} stabilize the AF phase and shift it to larger values of $J_2/J_1$ and also lead to a finite interval for the 2D spin liquid regime \cite{schmidt:14}. (Table~\ref{tbl:spinliquid}).

\subsection{The spin nematic phase sector at $J_2/J_1\simeq-1/2$, $J_1<0$}
\label{subsect:nematic}

This special case with FM $J_1<0$ has been studied more recently \cite{shannon:04,shannon:06,smerald:15}. As for AF $J_1$ the ordered moment calculated in LSW approximation breaks down at this special value separating the classical CAF/FM phases. The region of instability on the FM side is very small  in LSW and in zero-field ED scaling analysis (Table \ref{tbl:spinliquid} and Fig.~\ref{fig:groundstate}b). Analytical investigations of the phase diagram \cite{smerald:15,shindou:11} suggest a nonmagnetic phase in the range  $0.41\leq J_2/|J_1|\leq0.62$ rather similar to the AF spin liquid regime (Fig.~\ref{fig:phasedia}). On the other hand ED close to the classical saturation field gives indication about the nature of the nonmagnetic ground state \cite{schmidt:07}. For AF $J_1$ the instability of the fully polarized state below saturation field leads to single spin-flip (one-magnon) $S=1$ states. In contrast for FM $J_1$ the first instability occurs in the two-magnon sector ($S=2$) meaning that two magnon bound states appear first instead of single spin flip states. They have the form $|\phi\rangle=\sum_{ij}\phi_{ij}S_i^-S_j^-|P\rangle$ where P denotes the fully polarized state and $\phi_{ij}$ have d-wave type spatial symmetry. The existence of these two magnon bound states has also been corroborated by analytical methods \cite{zhitomirsky:10}. When the field is lowered they may proliferate and eventually condense into a  non-magnetic ground state which is of the {\em spin nematic\/} type. This order parameter does not break time reversal but the $C_4$ rotational and possibly translational symmetry. 

The designation `spin-nematic' is used here in the genuine sense of a nonmagnetic order parameter based on
$S=1/2$ spin degrees of freedom of bonds introduced first by Andreev and Grishuk in Ref.~\cite{andreev:84}. We exclude its application for the common case of possible spin quadrupoles based on on-site spin variables for $S\geq 1$. In this case the order corresponds to a local molecular field splitting of spin states. The $S=1/2$ spin nematic order is described by the traceless symmetrized tensor \cite{andreev:84,shannon:04}
\begin{equation}
O_{\alpha\beta}^{ij}
=
S_i^\alpha S_j^\beta+S_j^\alpha S_i^\beta
-\frac23\delta_{\alpha\beta}\left\langle\vec S_i\cdot\vec S_j\right\rangle
\label{eqn:two_magnon}
\end{equation}
Numerical evidence for this type of order comes form ED of up to $N=36$ tiles \cite{shannon:06}. 
Since it does not break spin rotational symmetry a Goldstone mode will appear whose signature may be found  in inelastic neutron scattering \cite{smerald:15}. It is primarily associated with the rotation of the spin-nematic quadrupole moment in the plane perpendicular to the field. The two magnon operators can be expressed by the spin nematic tensor operators in Eq.~(\ref{eqn:two_magnon}) as \cite{smerald:15}
\begin{eqnarray}
S_i^-S_j^-|P\rangle=\left[O^{ij}_{x^2-y^2}-iO^{ij}_{xy}\right]|P\rangle
\end{eqnarray}
where $O^{ij}_{x^2-y^2}=\left(O^{ij}_{xx}-O^{ij}_{yy}\right)/2$. This indicates that the nematic order parameter may be interpreted as resulting from a condensation of two magnon bound states when the field is lowered from saturation to zero.

\subsection{The order-by-disorder regime}
\label{subsect:order-disorder}

In the discussion of classical phases and their ground state energies (Sec.~\ref{sect:classical} and Fig.~\ref{fig:classical_gsen}) for the square lattice we tacitly assumed the CAF phase structure as a collinear phase. Actually on the classical level this is not obvious, because the ground-state energy for $J_2/|J_1|\ge1/2$ does not depend on $J_1$ such that the two interpenetrating N\'eel sublattices (lattice constant $\sqrt{2}a$) are completely decoupled and a continuous degeneracy exists with respect to the relative rotation angle of the sublattice moments. In particular this leads to an accidental degeneracy of the $\vec Q =(0,\pi)$  and $(\pi,0)$ columnar states with FM order along $x$ and AF order along $y$ or vice versa. This is also reflected in the accidental degeneracy of LSW excitation spectrum showing Goldstone modes at wave vectors $(0,0), (0,\pi), (\pi,0), (\pi,\pi)$.

Except at $J_1=0$, this degeneracy on the classical level is lifted by the effect of quantum fluctuations which select the proper ground state by lowering its energy~\cite{singh:03}. For finite $J_1$ it is most pronounced when the two sublattices are aligned, i.e. collinear. Indeed Fig.~\ref{fig:groundstate}~(a) shows a significant lowering of the ground state energy of the collinear CAF structure with respect to the classical energy. Therefore for large $J_2/|J_1|$ $(\phi \approx \pi/2)$ the long range magnetic order is actually stabilized by the effect of quantum fluctuations. This behavior has been coined `order-by-disorder' \cite{villain:80,shender:82} and it is opposite to the destruction of the magnetic moment in the spin liquid and spin nematic regimes discussed before.

The degeneracy of the two CAF structures amounts to a hidden discrete Ising variable $\sigma$ in the problem corresponding to $\sigma =+1$ for $\text{CAF}_x$ state and  $\sigma =-1$ for $\text{CAF}_y$ state. Therefore it has been proposed that even in 2D long range order of $\sigma$ may occur at a finite temperature \cite{chandra:90} and this has been corroborated by MC calculations of the classical (large $S$) $J_1$-$J_2$ model \cite{weber:03}. On the other hand for the $S=1/2$ quantum spin case such transition at finite T seems to be suppressed by quantum tunneling between $\sigma=\pm 1$ domains \cite{singh:03}. This may well be the case since neither high-temperature series expansion \cite{singh:03}  nor the present FTLM method for $\chi(T)$ gives any indication of a finite temperature anomaly indicating such a transition.


\section{Magnetization plateaux}
\label{sect:magplateaux}

The nonlinear magnetization curves obtained from spin wave theory based on the classical canted ground state configuration deviated from the strictly linear classical behavior, however, they are still smooth curves (Fig.~\ref{fig:m0_ED}). Depending on the control parameter more pronounced anomalies like magnetization plateaux occur. These are characterized by a constant magnetization with a rational value which persists over a finite field interval~\cite{chubokov:91,coletta:13,coletta:16}. Plateaux states  generally break the translational symmetry of the lattice but their microscopic wave function depends on the model. Aside from the simple Ising-type spin-flip plateaux (Sec.~\ref{sect:introduction}) one distinguishes two types~\cite{hida:05} in the Heisenberg case: i) Quantum plateaux which may be understood as Wigner crystallization of multiplets in a homogeneous background of singlets such as observed in spin ladders or the previously discussed 2D valence bond crystal. Such states do not have a classical analogy.
ii) (Semi-)classical plateaux where the moment structure has a classical analogue are generally of the coplanar or collinear type.   The latter are characterized  by commensurate up ($u$) and down ($d$) spin patterns on the lattice which are stabilized in a finite field range by having comparatively large (negative) zero point fluctuation energy.

\begin{figure}
    \centering
    \includegraphics[width=.5\columnwidth]{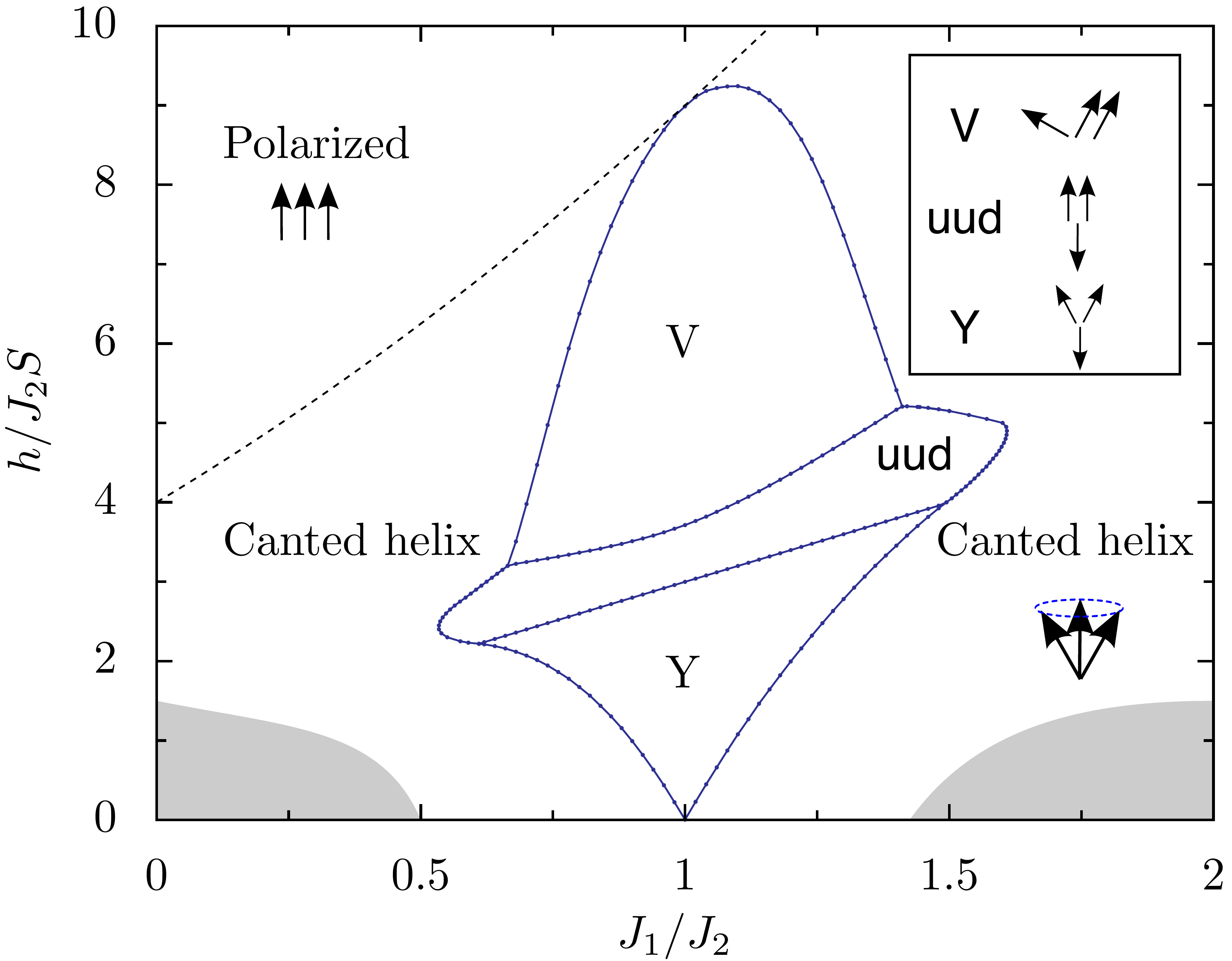} 
    \caption{$h$-$\phi$ phase diagram of the anisotropic triangular model plotted vs. $J_1/J_2=1/\tan^{-1}\phi$. 
    The planar three-sublattice structures are indicated in the inset. 
    The dashed line denotes the saturation field. The $uud$ phase leads to a magnetization plateau at $m_0/S=1/3$. 
    Left shaded area ($\phi/\pi\in [0.5,0.35]$) covers the quasi-1D spin liquid regime. 
    Right shaded area ($\phi/\pi\in [0.18,0.15]$) covers the 2D VBS region. $J_1=J_2$ is isotropic maximally frustrated point. 
    Adapted from Ref.~\cite{coletta:13}.}
     \label{fig:phasedia_triangular_h}
\end{figure}
The most well understood case is the {\it isotropic} triangular model $(J_1=J_2\equiv J)$~\cite{chubokov:91,coletta:16}. As for the square lattice model (Eq.~(\ref{eqn:ham_squ})) its Hamiltonian in Eq.~(\ref{eqn:h}) may be rewritten as a sum over triangular lattice plaquettes p as~\cite{coletta:16}.
\begin{equation}
{\cal H}=\sum_p\frac{J}{4}\bigl(\vec S_{p,1}+\vec S_{p,2}+\vec S_{p,3} -\frac{\vec h }{3J}\bigr)^2
\label{eqn:triplaq}
\end{equation}
where $\vec S_{p,i}$ denotes the spins in a plaquette. Classically when the spin operators are replaced by vectors we notice that the ground state energy is minimized for all spin structures that satisfy $(\vec S_{p,1}+\vec S_{p,2}+\vec S_{p,3}) =\vec h/(3J)$. The corresponding classical ground-state manyfold is therefore continuously degenerate. Previously, when calculating magnetization curves we used as classical starting point the non-coplanar canted helix or 'umbrella' state (Fig.\ref{fig:phasedia_triangular_h}). However there are degenerate coplanar states (all plaquette moments are lying in the plane that contains $\vec H$) like Y, $uud$ (collinear) and V  states depicted within Fig.~\ref{fig:phasedia_triangular_h} that fulfill the above constraint and are classically degenerate with the umbrella state. Because the coplanar states have a larger (negative) zero point contribution to the ground state energy they are stabilized by quantum fluctuations in the above sequence when field increases to the saturation value $h_\text s$~\cite{chubokov:91}. Since the collinear $uud$ state has a constant magnetization $m_0/S=1/3$ in its stability region a magnetization plateau evolves around $h\approx h_\text s/3$. The full magnetization curve was calculated in Ref.~\cite{coletta:16} using variational energies for the coplanar states.

\begin{figure}
    \vspace{0.3cm}
    \centering
    \includegraphics[width=.5\columnwidth]{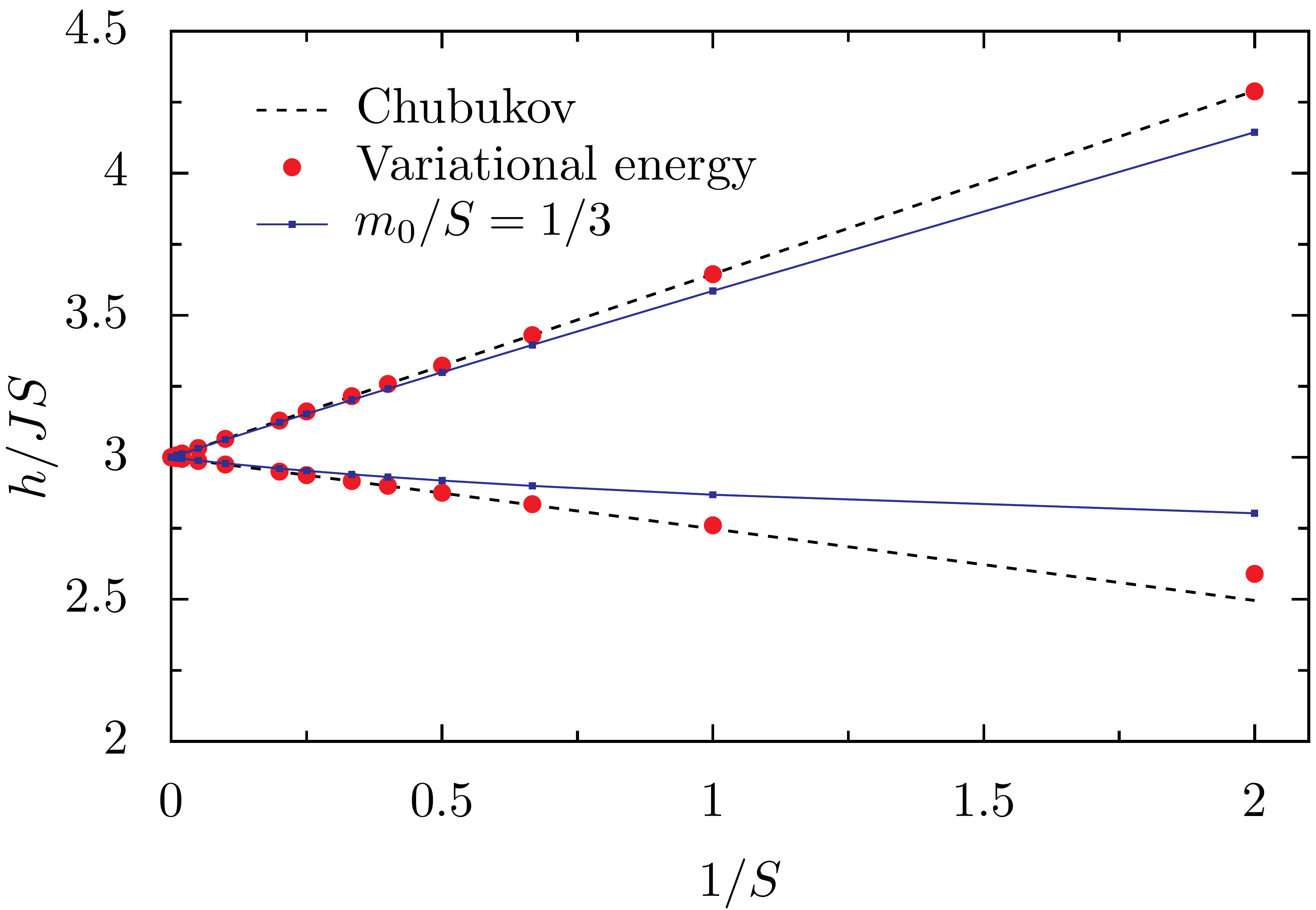} 
    \caption{Evolution of magnetization plateau width of isotropic triangular model with $1/S$. In the field 
    range enclosed by the two curves  $m_0/S=1/3$  is constant. Full line is estimate from LSW and red dots 
    are obtained from variational ground state energy~\cite{coletta:16}. In the classical limit $1/S\rightarrow 0$ 
    the plateau vanishes. Dashed line from Ref.~\cite{chubokov:91}. Adapted from~\cite{coletta:16}.}
     \label{fig:plateaux_width}
\end{figure}
\begin{figure}
    \centering
    \includegraphics[width=.46\columnwidth]{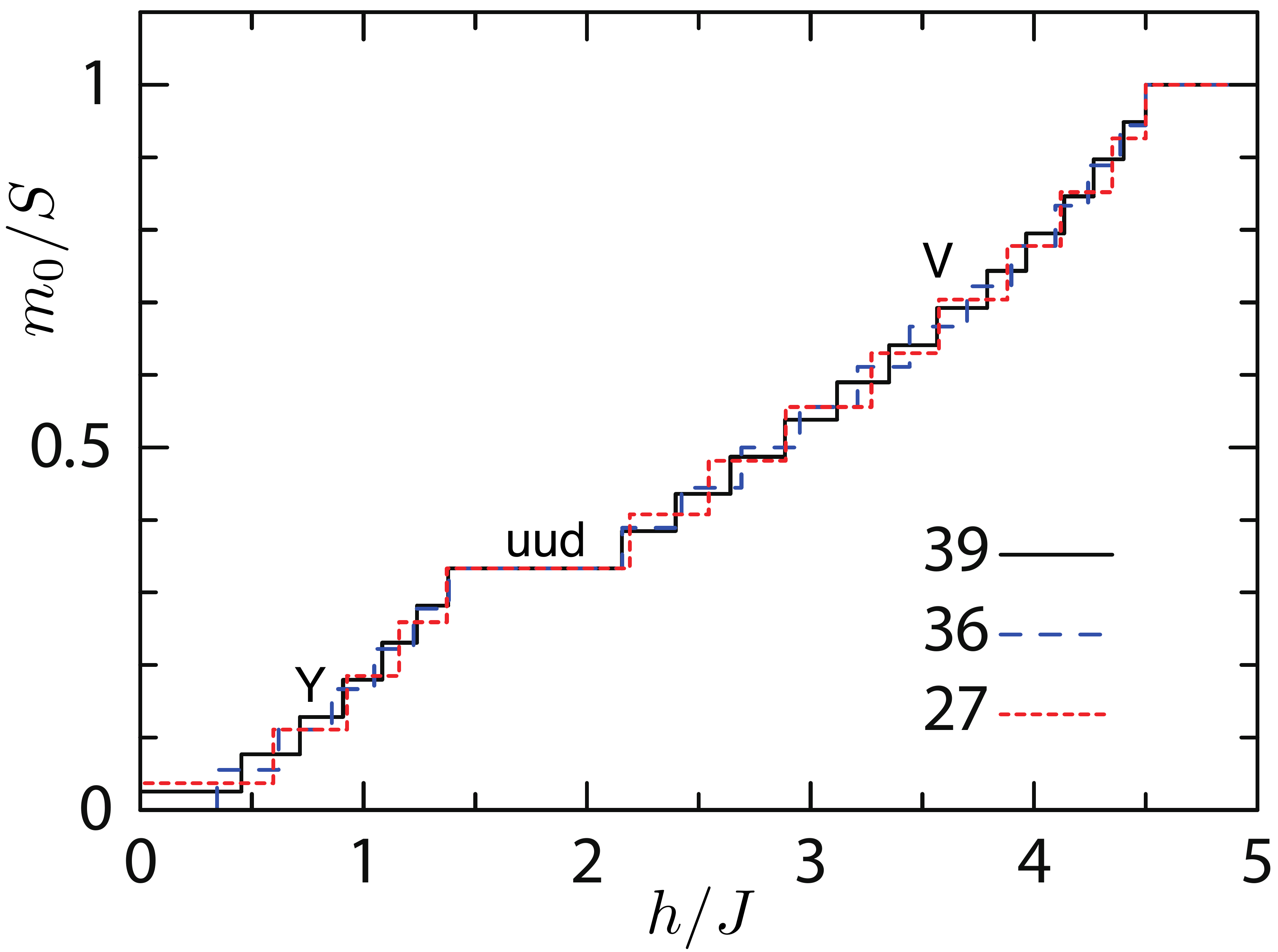}
    \includegraphics[width=.48\columnwidth]{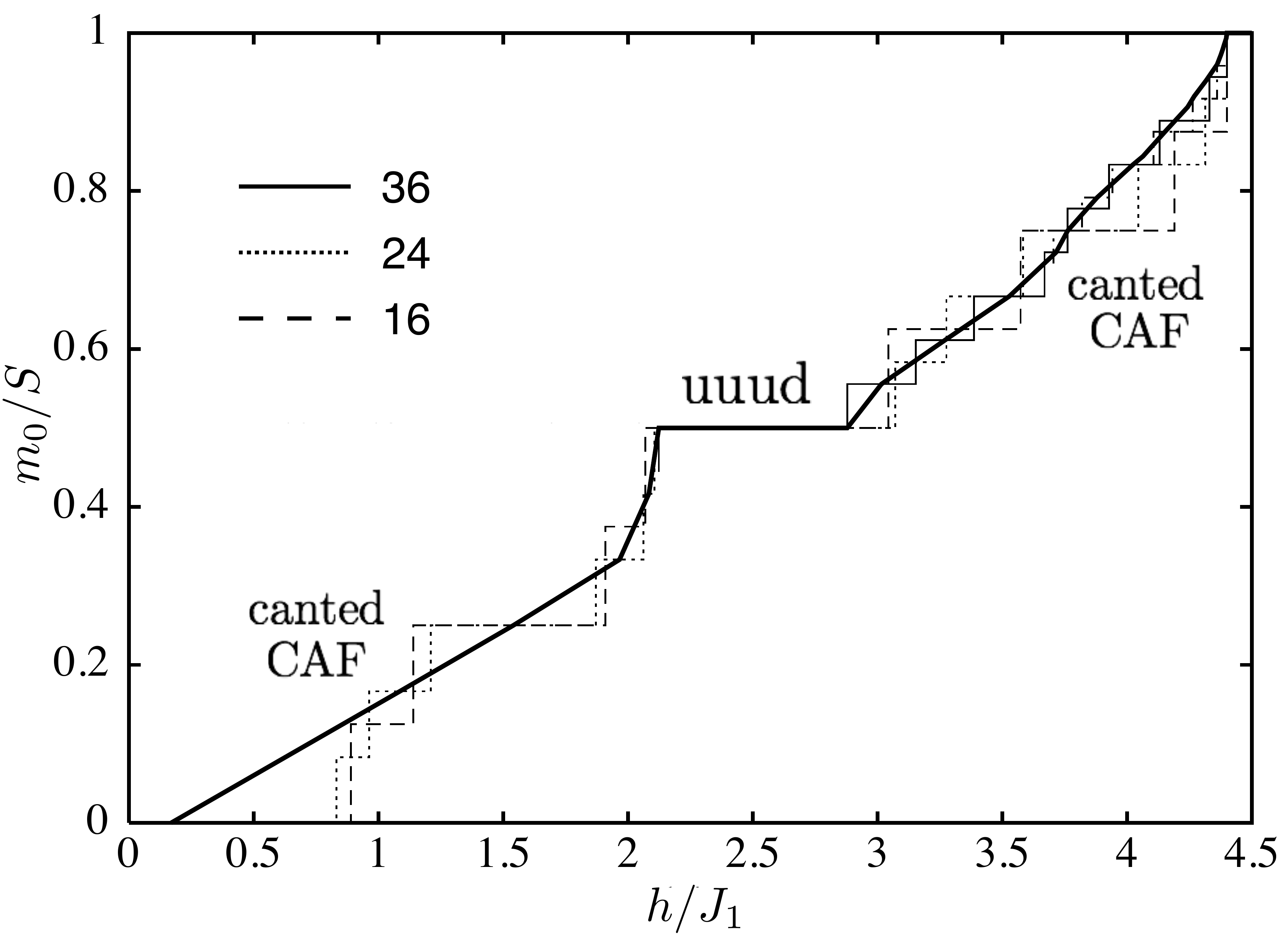}
    \caption{(a) Triangular lattice magnetization and plateau from ED 
    (inset: tile size $N$) at  $m_0/S=1/3$ due to stable $uud$ ($\uparrow\uparrow\downarrow$) 
    structure between canted Y and V structures 
    (see Fig.~\ref{fig:phasedia_triangular_h}). Saturation field $h_\text s=(9/2)J$ ($h_\text s/S=9J$),
     adapted from Ref.~\cite{sakai:11}, see also Ref..~\cite{honecker:99}).
     (b) Magnetization and plateau at $m_0/S=1/2$ in the square lattice model from 
    ED with $J_2/J_1=0.6$ ($\phi/\pi=0.17$) 
    inside the spin liquid region (adapted from Ref.~\cite{honecker:01}). The stable collinear $uuud$ phase 
    ($\uparrow\uparrow\uparrow\downarrow$ spin structure  of a square plaquette) leads to the plateau with
    maximum width for this  $J_2/J_1$ value. The saturation field $h_\text s=2J_1(1+2J_2/J_1)=4.4J_1$.}
     \label{fig:mag_plateau}
\end{figure}

The resulting plateau width is  shown in Fig.~\ref{fig:plateaux_width} as function of $1/S$. In the classical limit $1/S\rightarrow 0$ the stability region of collinear $uud$  state and hence the $m_0/S=1/3$ plateau width shrinks to zero. The plateau width may also be estimated by using the magnetization curves $m_0(h)=-N^{-1}\partial E_\text{gs}(h)/\partial h$ obtained from the ground state energy for Y and V phases in linear spin wave theory by tracking their crossing with the $m_0/S=1/3$ value of the $uud$ collinear phase~\cite{coletta:16}.
Conventional $1/S$ expansion~\cite{alicea:09} leads to lower $(h_{\text c1})$ and upper $(h_{\text c2})$ critical fields for the $uud$ phase given by:
\begin{equation}
\frac{h_{\text c1}}{S}=3J-\frac{0.5J}{2S};\;\; \frac{h_{\text c2}}{S}=3J+\frac{1.3J}{2S}
\end{equation}
and a corresponding plateau width $\Delta h/S= (h_{\text c2}-h_{\text c1})/S = 1.8J/(2S)$. This agrees reasonably well with the numerical ED results of Fig.~\ref{fig:mag_plateau}(a).

The question of magnetization plateaux is more difficult to understand for the anisotropic triangular $(J_1\neq J_2)$ model. In this case the coplanar phases are not degenerate classically compared with the umbrella phase and therefore are per se not at good starting point for $1/S$ expansion. This situation may be remedied by adding an artificial staggered potential with a variational parameter to the Hamiltonian which stabilizes the coplanar phases. Then again quantum corrections to the ground state energy of all coplanar and canted helix (umbrella) phase can be calculated in spin wave approximation . The resulting phase diagram is shown in Fig.~\ref{fig:phasedia_triangular_h}. The $uud$ phase is stable in a finite field range not just at the isotropic point $J_2=J_1$ but also in a wide range of the anisotropy parameter $J_2/J_1$. Therefore the $m_0/S=1/3$ magnetization plateau also exists for this range of $J_2/J_1$. The critical values $J_1/J_2\simeq 0.55, 1.6$ where $h_{\text c1}$ and $h_{\text c2}$ in Fig.~\ref{fig:phasedia_triangular_h} merge into a point are very special. Analysis beyond the 1/S expansion 
suggests that around this point a finite region in $J_1/J_2$-$h$ plane with an exotic phase exists, either with spin nematic order~\cite{chubukov:13} or a chiral spin liquid phase~\cite{parker:17}.

Furthermore the $J_1/J_2$-$h$ phase diagram for the triangular model in Fig.~\ref{fig:phasedia_triangular_h} is rather  similar to that of a three-leg triangular Heisenberg model on a tube~\cite{chen:13}. The field induced sequence of phases of the quantum model in  Fig.~\ref{fig:phasedia_triangular_h} appears also in the {\it classical} spin model at finite temperature~\cite{seabra:16}. In particular the $uud$ phase which is not stable for classical spins at $T=0$ (Fig.~\ref{fig:phasedia_triangular_h}) acquires a finite stability region due to 'order by disorder' from thermal fluctuations.

Similar considerations hold for the square lattice $J_1$-$J_2$ model~\cite{coletta:13} where the colinear $uuud$ phase competes with the canted NAF and CAF phases considered in Sec~\ref{subsect:LSW}. The $uuud$ associated magnetization plateau at $m_0/S$=1/2 is, however constricted to a narrower region above the special point $J_2/J_1=0.5$ $(\phi/\pi =0.15)$. This agrees with the unbiased ED results for medium sized clusters in Fig.~\ref{fig:m0_EDLSW} for $\phi/\pi=0.17$  ($J_2/J_1= 0.6$) and is shown for the largest possible cluster ($N=40$) again in Fig.~\ref{fig:mag_plateau} for the same parameter. The $0.5\leq J_2/J_1\leq 0.65$ existence range  of the plateau obtained from ED~\cite{honecker:99} is  also in qualitative agreement with the semiclassical analytical estimates~\cite{coletta:13}.
Recent DMRG calculations~\cite{morita:16} on finite cylinders report considerably more magnetization plateaux  aside from the main one at $m_0/S=1/2$. It remains to be seen whether they survive in the thermodynamic limit.

In the square lattice model the  $J_2/J_1$ range of the $m_0/S=1/2$ plateau coincides with the spin liquid regime at zero field. In fact the appearance of the $uuud$ plateau phase may be interpreted as the closing of the spin gap of the spin liquid phase in finite field. This situation is a rather different from the anisotropic triangular case.
There the collinear three-sublattice $uud$ structure which leads to a magnetization plateau at $m_0/S=1/3$ appears mainly centered above the isotropic ($\triangle$) $J_2=J_1$ point with a stable $120^\circ$ AF spiral magnetic structure and not above the 2D spin liquid regime at $J_2/J_1\simeq1/2$ (at  $J_1/J_2\simeq 2$ in  Fig.~\ref{fig:phasedia_triangular_h}).

Experimentally magnetization plateaux  were observed in the 2D triangular \CCB~compound~\cite{ono:03} and were proposed already before in the quasi-1D triangular chain compound \CCCl~\cite{jacobs:93}. More recently they have also been found in the 2D triangular compounds Ba$_3$CoSb$_2$O$_9$ $(S={1/2})$,  Ba$_3$NiSb$_2$O$_9$ $(S=1)$~\cite{koutroulakis:15}  and  RbFe(MoO$_4$)$_2$ $(S={5/2})$~\cite{svistov:06}. The widths of observed $m_0/S=1/3$ plateaux follow well the predictions in Fig.~\ref{fig:plateaux_width}.

Another example of quantum plateau formation is found in an extended $J_1-J_1-J_3$ isotropic triangular model with n.n.n. exchange $J_3$ \cite{ye:17} (see also Sec.~\ref{sect:extended}). For zero field this model has a four-sublattice stripe phase for $J_3/J_1>0.135$ (Fig.~\ref{fig:PD_triangular123}). On the classical level it is infinitely degenerate for all spin structures that minimize Eq.~(\ref{eqn:triplaq}). The proper order is then chosen by the 'order-by-disorder' mechanism. This means that at a given field the phase with the largest decrease of ground state energy due to quantum fluctuations is stabilized. Surprisingly for fields $h/h_s\simeq 1/2$ the collinear uuud phase is stable for a finite field interval. This means the (extended) triangular quantum magnet in the stripe regime can exhibit a $m_0/S=1/2$ plateau like the square lattice model instead of the $1/3$ plateau in the case of the original n.n. ($J_3=0$) triangular model.


\section{Application to 2D $S=1/2$ magnetic compounds}
\label{sect:compounds}

We have emphasized before that we find the effects of frustration and quantum fluctuations in  square and triangular 2D magnets most interesting because they have experimental realizations, unlike in the case of more exotic quantum phases. There are two aspects to consider: Firstly the identification of those typical effects like nonlinear magnetization, plateaux formation, saturation field and ordered moment reduction and its field dependence as well as associated thermodynamic anomalies and magnetic structure factor. Secondly a comparison of  experimental results with theoretical prediction that may actually allow to determine the model parameters and thus locate the compound in the phase diagram. This will be demonstrated in the following for a number of different compound classes.

\subsection{The layered vanadium oxides: magnetization and susceptibility}
\label{subsect:oxovana}

The discovery of two classes of layered vanadium oxides \LiX~
($X=\text{Si},\text{Ge}$)~\cite{millet:98, melzi:00, melzi:01,carretta:04} and
\AAVO~($A,A'=\text{Pb},\text{Zn},\text{Sr},\text{Ba}$)~\cite{kaul:04,kaul:05,kini:06,nath:08,nath:09} provided a variable platform of 2D frustrated quantum magnets with different chemical composition. Nevertheless their magnetism is described universally by the $J_1$-$J_2$ model with $J_2/J_1$ or $\phi$ depending on the specific compound. Each of them features V$^{4+}$ ions with $S=1/2$ surrounded by oxygen polyhedra, forming layers of $J_1$-$J_2$ square lattices with weak interlayer coupling \cite{melzi:01,kini:06}.

These compounds  were experimentally investigated e.g. in Refs.~\cite{kaul:04,kaul:05,skoulatos:07,skoulatos:09}.  We discuss the uniform magnetization, saturation field  and susceptibility and show how they may be used to extract the exchange parameters of the model.
 The temperature dependence of the susceptibility exhibits a pronounced maximum due to 2D spin fluctuations.  The peak temperature is  a measure for the exchange energy scale $J_\text c$.  The advantage of these magnetic compounds is a relatively low scale $J_\text c\sim10\,\text K$ (Table ~\ref{tbl:exchange}) which leads to accessible saturation fields (compare Eq.~(\ref{eqn:satclass2})). At considerably lower temperature the additional sharp peak of the 3D magnetic phase transition occurs due to the interlayer coupling~\cite{kaul:05,kini:06}. 

\begin{figure}
    \centering
    \includegraphics[width=.5\columnwidth]{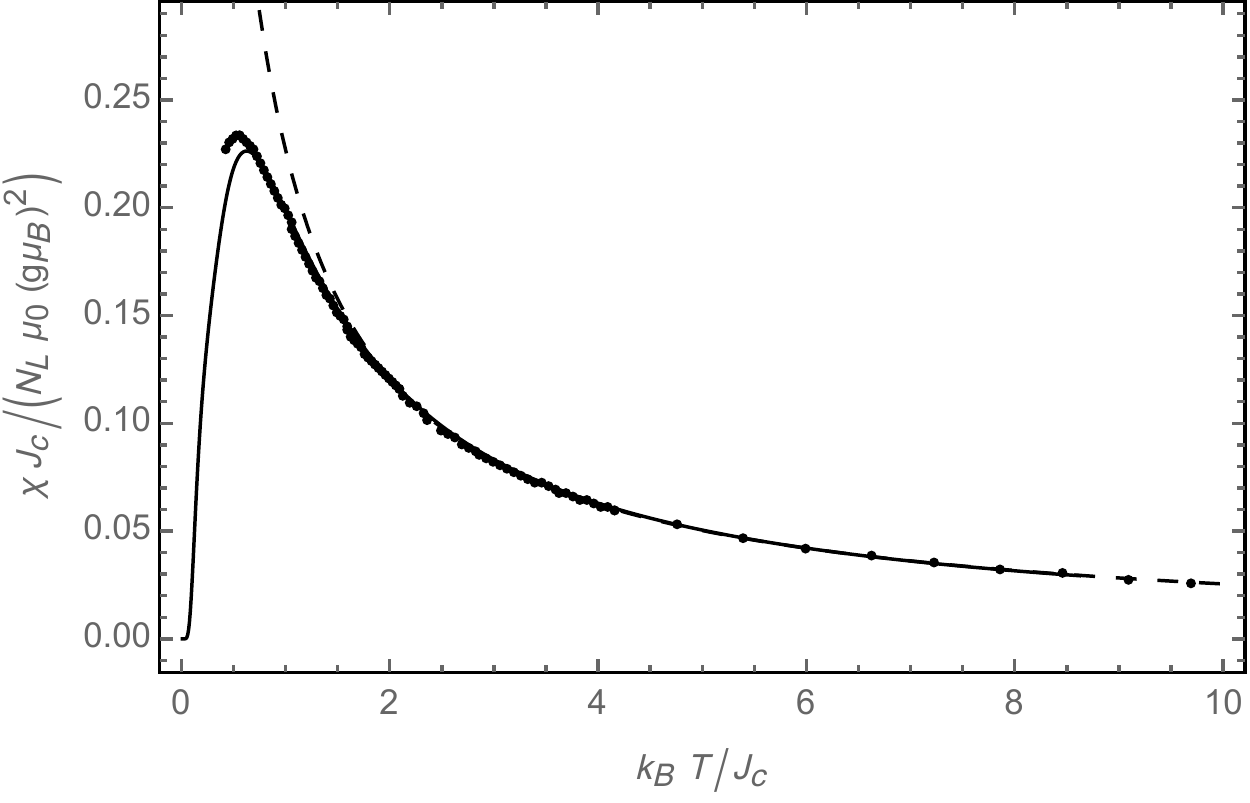} 
    \caption{Magnetic susceptibility curve of \BCVO. The dots represent experimental values~\cite{nath:08}, 
    the dashed line 
    is the high temperature Curie-Weiss fit and the full line gives FTLM results using $J_\text c$ and $\phi$ as 
    fit parameters (Table\ref{tbl:exchange}). }
     \label{fig:chi_bcvo}
\end{figure}
\begin{figure}
    \centering
    \includegraphics[width=.5\columnwidth]{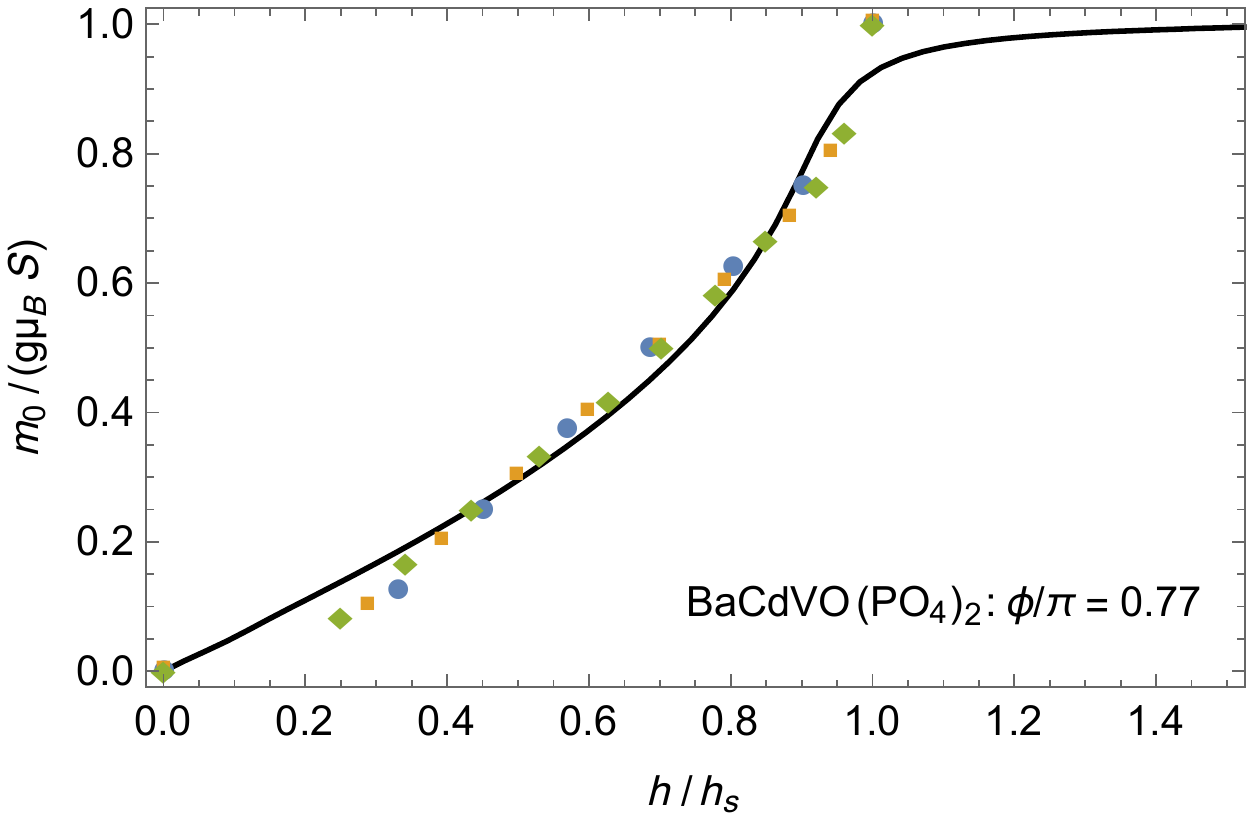} 
    \caption{Uniform magnetization for \BCVO. Solid line denotes experimental data and symbols represent
     ED results ($T=0$) for different cluster sizes $N=16\ldots24$.}
     \label{fig:m0_bacd}
\end{figure}

The previously explained methods have been applied to determine the exchange constants or $J_\text c$ and $\phi$ to this class of compounds. As an example we show the susceptibility curve and theoretical fits for \BCVO~in Fig.~\ref{fig:chi_bcvo}. In particular the FLTM fit gives excellent results and reproduces the spin fluctuation maximum. This comparison leads to $\phi=0.77\pi$ and $J_\text c=4.8\text{K}$. It means that the compound is closest to the instability  of the CAF magnetic phase to the spin nematic sector (Fig.~\ref{fig:phasedia}). Therefore the peak position occurs at rather low value $T_\text{max}\simeq 0.5 \text{K}$ or $k_\text BT_\text{max}/J_\text c\simeq 0.1$. The proximity to the spin-nematic sector also leads to the pronounced nonlinearity of the magnetization curve in Fig.~\ref{fig:m0_bacd}.

\begin{table} 
    \centering
	\begin{tabular}[c]{lcccccccl}
	Compound & $\phi/\pi$ & $J_{\text c}/(k_{\text B}\text K)$ & $J_2/J_1$  & $J_1/(k_{\text B}\text K)$  & $J_2/(k_{\text B}\text K)$ & $\Theta_{\text CW}/{\text K}$ & $\mu_0H_\text s/\text T$  & \;\;Ref.  \\
	\hline
	Zn$_{2}$VO(PO$_{\text{4}}$)$_{\text{2}}$ & 0.008 & 7.9 & 0.025 & 7.91 & 0.2 & 8.11 & 23.5$^*$&   \;\; \cite{kini:06}  \\
	Li$_{\text{2}}$VOGeO$_{\text{4}}$ & 0.44 & 4.2 & 5.0 & 0.82 & 4.1 & 4.92 & 13.4$^*$& \;\; \cite{kaul:04,kaul:05}  \\
        Li$_{\text{2}}$VOSiO$_{\text{4}}$ & 0.47 & 6.3 & 11.25 & 0.56 & 6.3 & 6.86 & 19.6$^*$& \;\; \cite{kaul:04,kaul:05}  \\
        Pb$_{\text{2}}$VO(PO$_{\text{4}}$)$_{\text{2}}$ & 0.60 & 6.8 & -3.25 & -2 & 6.5 & 4.5 & 20.9 & \;\; \cite{skoulatos:07}\\
        Pb$_{\text{2}}$VO(PO$_{\text{4}}$)$_{\text{2}}$ & 0.63 & 8.4 & -2.41 & -3.2 & 7.7 & 4.5 & 20.9 & \;\; \cite{skoulatos:09}\\
        PbZnVO(PO$_{\text{4}}$)$_{\text{2}}$ & 0.65 & 11.27 & -1.92 & -5.2 & 10.0 & 4.8 & 23.4 & \;\; \cite{tsirlin:10,tsirlin:09}\\
        Na$_{\text{1.5}}$VO(PO$_{\text{4}}$)$_{\text{2}}$F$_{\text{0.5}}$ & 0.65 &7.1& -2.0 & -3.2 & 6.3 & 3.1 &15.4 & \;\; \cite{tsirlin:09} \\
        BaZnVO(PO$_{\text{4}}$)$_{\text{2}}$ & 0.66 & 10.5 & -1.86 &-4.99 & 9.26 & 4.27 & 20.1$^*$ &\;\; \cite{kaul:05}  \\
	Pb$_{\text{2}}$VO(PO$_{\text{4}}$)$_{\text{2}}$ & 0.66 & 10.7 & -1.84  & -5.1 & 9.4 & 4.3 & 20.9 &\;\; \cite{tsirlin:09}  \\
	Pb$_{\text{2}}$VO(PO$_{\text{4}}$)$_{\text{2}}$ & 0.67 & 11.5 & -1.63 & -6 & 9.8 & 3. 8 & 20.9 & \;\; \protect\cite{kaul:04}  \\
	SrZnVO(PO$_{\text{4}}$)$_{\text{2}}$ & 0.73 & 12.2 & -1.07 & -8.3 & 8.9 & 0.6 &14.2 & \;\; \cite{tsirlin:09}  \\
        BaCdVO(PO$_{\text{4}}$)$_{\text{2}}$ & 0.77 & 4.8 & -0.89 & -3.6 & 3.2 & -0.4 & 4.3 & \;\; \cite{nath:08,tsirlin:09} \\
        \hline 
	\end{tabular}
    \caption[Exchange interactions constants for vanadium oxide compounds]
    {Exchange interactions constants for various vanadium oxide compounds, ordered with increasing $\phi$,
     i.e. approaching the CAF/FM boundary in Fig.~\ref{fig:phasedia} counterclockwise.
    Results are obtained mostly from susceptibility $\chi(T)$ and magnetization $m_0(h)$ analysis, except for fourth and fifth 
    row which are deduced from neutron diffraction.
    Here $\Theta_{\text CW} =(J_1+J_2)/k_{\text B}$ is the Curie-Weiss temperature. 
    The saturation fields $\mu_0H_\text s$ are experimental values whenever known; values with asterisk are 
    estimated from Eq.~(\ref{eqn:satclass2}) using approximate $g=2$. All compounds
    except the first are in the CAF sector (Fig.~\ref{fig:phasedia}). The example of  
    Pb$_{\text{2}}$VO(PO$_{\text{4}}$)$_{\text{2}}$ shows that a certain variation in exchange parameters 
    as determined by different methods and in different references occurs.}
    \label{tbl:exchange}
\end{table}
A compilation of exchange energy scales $J_\text c$ and frustration angles $\phi$ for the compound family is presented in Table \ref{tbl:exchange}. In most cases the entries have been determined just from high temperature fits like the dashed line in  Fig.~\ref{fig:chi_bcvo}. As explained before this leads to an ambiguity $\phi_\pm$ in the frustration angle, see Fig.~\ref{fig:phasedia}. It can be resolved in the most direct way by verifying whether the low temperature magnetic order is CAF $(\vec Q=((\pi,0), (0,\pi))$ or NAF $(\vec Q=(\pi,\pi))$.

\begin{figure}
    \centering
    \includegraphics[width=.5\columnwidth]{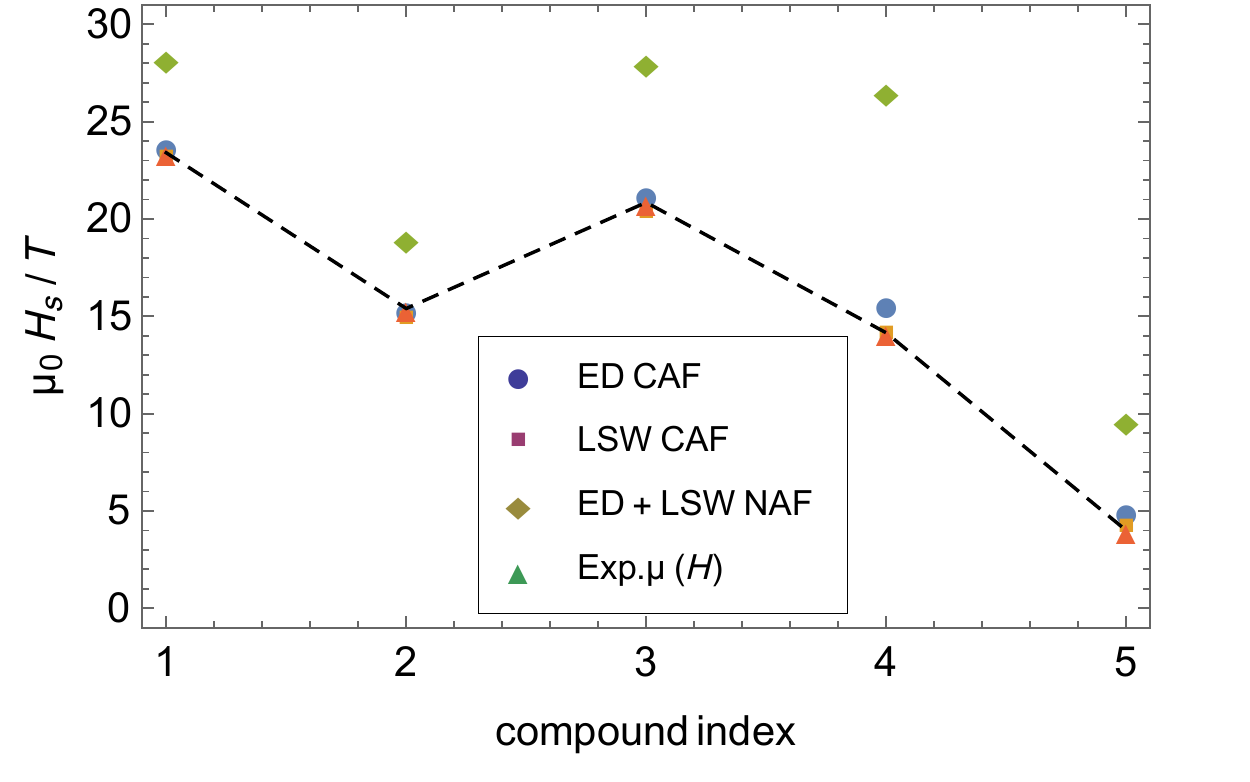} 
    \caption{Comparison of saturation fields form ED, spin-wave theory for CAF $(\phi_+)$ and NAF $(\phi_-)$ structures
    with experimental results. The compounds are denoted by (1) PbZnVO(PO$_{\text{4}}$)$_{\text{2}}$, (2) Na$_{\text{1.5}}$VOPO$_{\text{4}}$F$_{\text{0.5}}$,  (3) Pb$_{\text{2}}$VO(PO$_{\text{4}}$)$_{\text{2}}$, (4) SrZnVO(PO$_{\text{4}}$)$_{\text{2}}$ and (5)  BaCdVO(PO$_{\text{4}}$)$_{\text{2}}$ . The agreement with theoretical results proves that all 
    compounds are in the CAF sector of Fig.~\ref{fig:phasedia} and $\phi_+$ is realized. Dashed line is guide to the eye.}
     \label{fig:saturation_fields}
\end{figure}

A more ingenious method is the determination of the saturation field $H_\text s$. Because of the CAF/NAF asymmetry of the latter for $\phi_\pm$, according to Fig.~\ref{fig:satfield_class}, a quantitative comparison with the experimental value of $H_\text s$ may distinguish which value is the correct one. A comparison of experimental values and LSW/ED results for the saturation field is given in Fig.~\ref{fig:saturation_fields}. In all cases considered in this figure clearly the $\phi_+$ value characteristic for the CAF phase is realized. They are indicated in Fig.~\ref{fig:phasedia}  and included in Table~\ref{tbl:exchange} as corresponding to CAF phase, among others.

Microscopically the values of the exchange parameters in Mott insulators are of the superexchange-type $J_{ij}=4t_{ij}^2/U$  where U is the Coulomb repulsion of 3d electrons and $t_{ij}$ their effective (e.g. tight binding) hopping elements. Therefore they are fixed by the lattice structure and chemical bonding. Nevertheless a limited tuning of the exchange parameters $J_1$, J$_2$ might seem possible by applying hydrostatic pressure which changes bonding lengths and angles and therefore the effective hopping parameters. Such tuning has been attempted in \LiS{} and changes in the ratio $k_\text BT_\text{max}/(J_1+J_2)$ with pressure were observed from the susceptibility~\cite{carretta:04} but the ratio $J_2/J_1$ as function of pressure was not extracted. Another possibility to tune $J_2/J_1$ systematically is chemical substitution which will be discussed for another compound class in Sec.~\ref{subsect:halides}.

\subsection{Structure factor and neutron diffraction: Experimental results for oxovanadates}
\label{subsect:compounds_neutrons}

With thermodynamic experiments alone it is difficult to determine the exchange constants $J_1$ and $J_2$ unambiguously: From measurements of the magnetic susceptibility we can extract the Curie-Weiss temperature $\Theta_\text{CW}$ (Eq.~(\ref{eqn:thetacw})), heat capacity measurements give the magnetocaloric energy scale $J_{\text{mc}}$ (Eq.~(\ref{eqn:jth})). With only $J_\text{mc}$ and $\Theta_\text{CW}$ available an ambiguity (Fig.~\ref{fig:phasedia})  remains about the relative sign of the two exchange constants. As demonstrated in the previous Sec.~\ref{subsect:oxovana} FTLM fits to susceptibility and use of saturation field analysis can help to clarify the ambiguity. Another more direct method is neutron scattering measurements of the static structure factor $S(\vec q)$ to resolve this issue. 

For a large enough single crystal, neutron diffraction gives a characteristic pattern of Bragg peaks. If the magnetic moments are not fully ordered, some of the scattering intensity appears as quasielastic diffuse scattering, coexisting with the Bragg peaks but typically with much weaker intensity. The diffuse magnetic scattering intensity is characterized by the nonzero correlations between the local moments, and by integrating over energy transfer one obtains $S(\vec q)$. To extract the small magnetic component, polarized neutron scattering is required, making use of the basic fact that if the polarization direction of the neutrons is aligned with the momentum transfer $\vec q$, then all magnetic scattering is spin-flip scattering~\cite{skoulatos:08}. Therefore, taking the difference of two measurements with the neutron polarization parallel and perpendicular to $\vec q$ yields the desired magnetic component. In addition, because the wave functions of the electrons carrying the magnetic moment are extended in space, an incoherent background proportional to the magnetic form factor $F^2(\vec q)$, is observed.

In the following we discuss three compounds representing three different locations in the phase diagram of the $J_1$-$J_2$ model (Fig.~\ref{fig:phasedia}, Table~\ref{tbl:exchange}): Pb$_2$VO(PO$_4$)$_2$ and SrZn$_2$VO(PO$_4$)$_2$ represent columnar antiferromagnets with ferromagnetic $J_1<0$. The former is located in the center of $J_1 <0$ stable CAF phase, the latter  nearer to the CAF/FM phase boundary or spin nematic sector thus having a substantially reduced ordered moment. The third compound, Li$_2$VOSiO$_4$, turns out to be in the `order by disorder' regime of the CAF phase, with a very small antiferromagnetic $J_1>0$.
\begin{figure}
\begin{center}
	\includegraphics[width=.5\columnwidth]{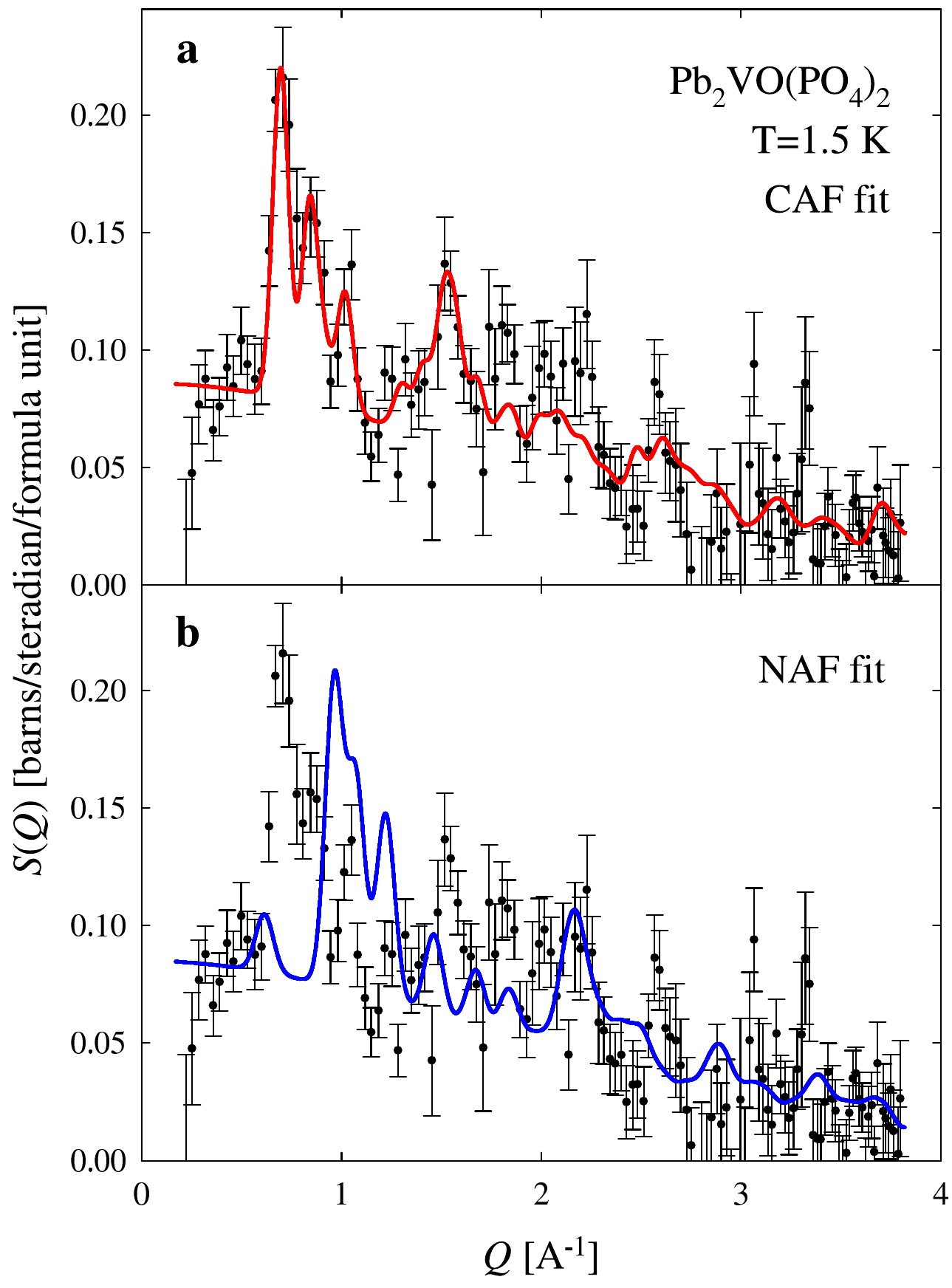}
\caption{Diffraction (elastic) cross section of Pb$_2$VO(PO$_4$)$_2$ at $T=1.5\,\text K$ below $T_\text N$~\cite{skoulatos:08,skoulatos:07,skoulatos:09}. 
Sharp peaks develop on top of the incoherent background. Here (a) shows the measured data (dots with error bars) together with the calculated cross section 
compatible with the columnar AF phase. In (b) the same data are displayed together with a fit (Eq.~(\ref{eqn:sqclass})) assuming a Néel-ordered phase which 
gives no agreement. This measurement resolves the $\phi_\pm$ ambiguity  (Fig.~\ref{fig:phasedia}) in favor of $\phi_+$ (CAF).}
\label{fig:sqpb2lowt}
\end{center}
\end{figure}
At room temperature in the paramagnetic phase the static structure factor of all (powder) samples show a Gaussian-type background determined by the directionally averaged form factor  $\bar{F^2}(|\vec q|)$ \cite{skoulatos:08,skoulatos:07,skoulatos:09}.
For lower temperatures, magnetic correlations start to develop on top of this incoherent background. Fig.~\ref{fig:sqpb2lowt} displays the measured data for Pb$_2$VO(PO$_4$)$_2$ at $T=1.5\,\text K$ below  $T_\text N=3.53\, \text K$ together with two fits based on  Eq.~(\ref{eqn:sqclass}). We note that the finite 3D ordering temperature is due to the interlayer coupling $J_\perp$. The panel (a) assumes a columnar arrangement of the magnetic moments with ordering vector $\vec Q_{\text{\sc caf}}=(\pi,0)$ (red curve) while the panel (b) displays the same data but with a fit assuming a Néel-ordered arrangement with $\vec Q_{\text{\sc naf}}=(\pi,\pi)$ in units of the reciprocal lattice constant $a\approx a_{\text{unitcell}}/2$. The columnar magnetic order clearly yields the better fit to the diffraction data: With $a_{\text{unitcell}}\approx9\,\AA$, we have $|\vec Q_{\text{\sc caf}}|\approx\pi/(a_{\text{unitcell}}/2)\approx0.7\,\AA^{-1}$ for the former. For the latter (Néel order), we have $|\vec Q_{\text{\sc naf}}|=\sqrt2|\vec Q_{\text{\sc caf}}|\approx1\,\AA^{-1}$. Therefore the main peak of $S(\vec Q)$ in the NAF model disagrees with the experimental position (Fig.~\ref{fig:sqpb2lowt} (b)).
From the fit, one can furthermore deduce the reduced size of the ordered moment $\mu_Q\approx0.5\mu_{\text B}$ or  $m_{\vec Q}/S\approx0.5$. This is in agreement with the approximate value 0.6 in the center of the CAF region (Fig.~\ref{fig:groundstate}b).

\begin{figure}
\begin{center}
	\includegraphics[width=.5\columnwidth]{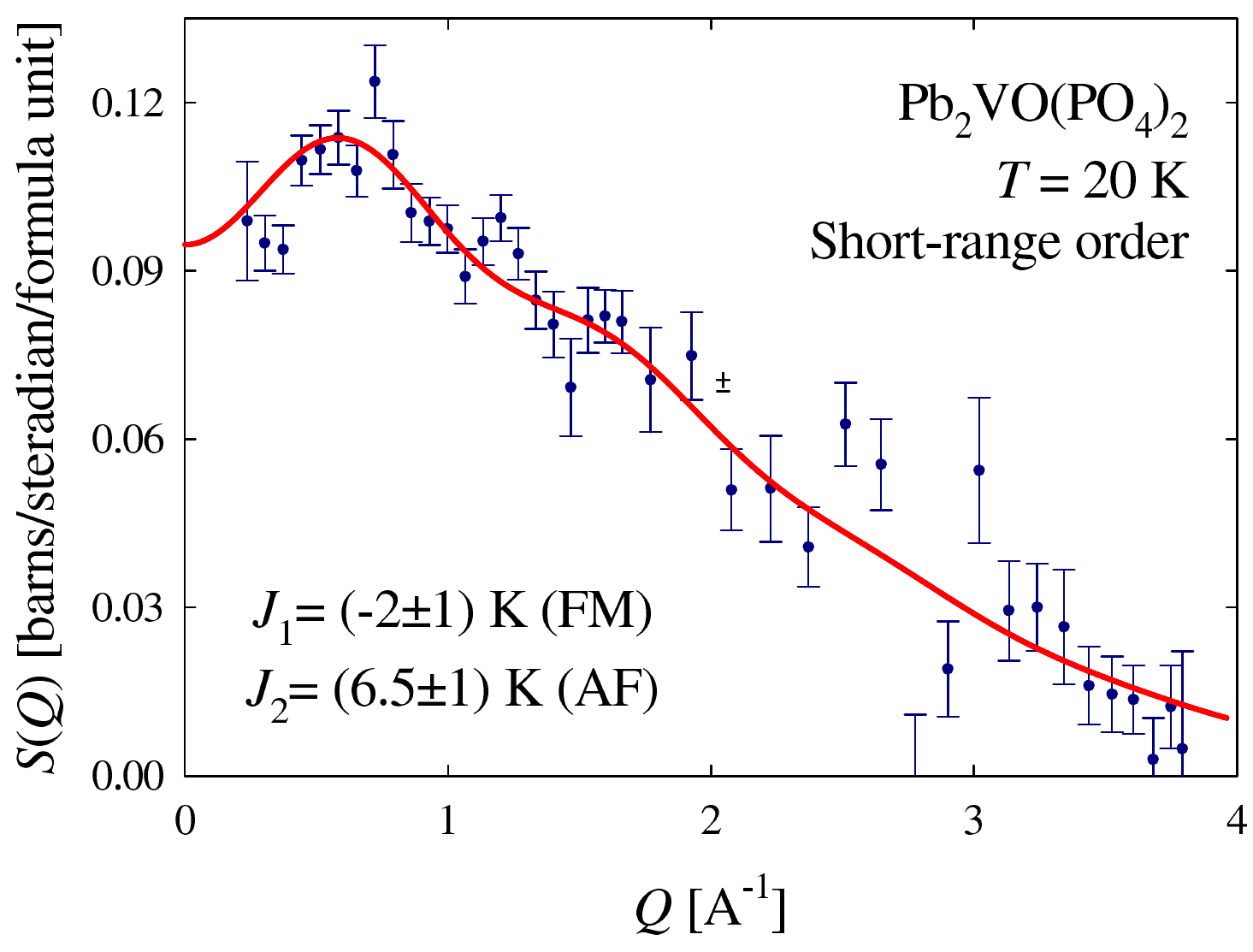}
	\caption{Scattering cross section of Pb$_2$VO(PO$_4$)$_2$ at $T=20\, \text K$ above $T_\text N$
	~\cite{skoulatos:08,skoulatos:07,skoulatos:09}. The dots are experimental data, the curve displays a 
	first-order {\sc htse} fit according to Eq.~(\ref{eqn:shtse}). The obtained values $J_1,J_2$  of exchange 
	constants correspond to $J_\text c=6.8 \text K$ and $\phi/\pi=0.6$  (Table~\ref{tbl:exchange}) in the center of 
	the CAF regime with ferromagnetic $J_1<0$.}
	\label{fig:sqpb220k}
\end{center}
\end{figure}
At intermediate temperatures $T>J_{\text c}/k_{\text B}$, the first-order {\sc htse} expansion of $S(\vec q)$ given in Eq.~(\ref{eqn:shtse}) may be used to achieve a good description of the  scattering cross section, see Fig.~(\ref{fig:sqpb220k}). The values for $J_1$ and $J_2$ determined in this way are in good agreement with those extracted from the thermodynamic measurements.

\begin{figure}
\begin{center}
	\includegraphics[width=.5\columnwidth]{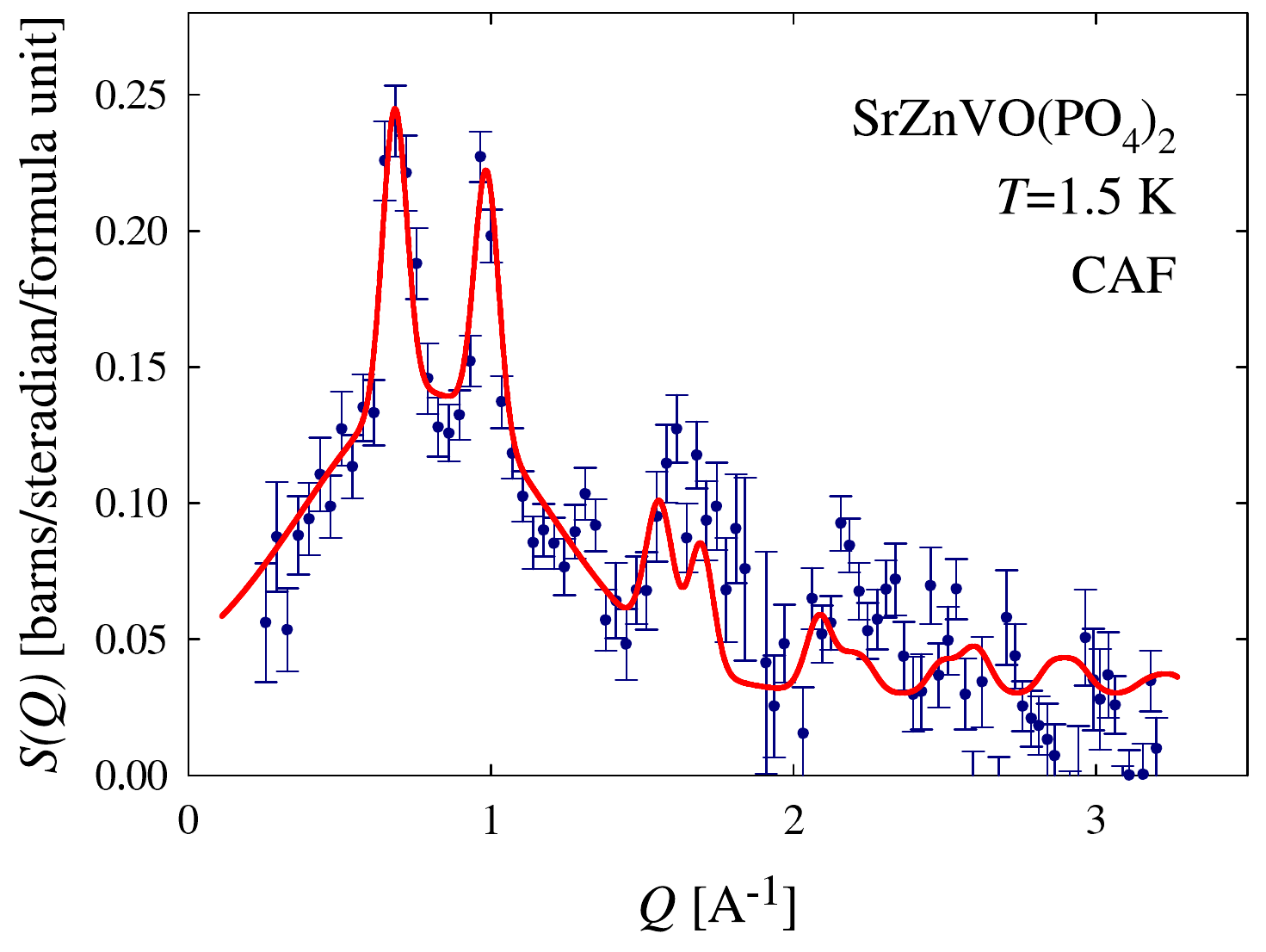}
\caption{Diffraction cross section of SrZnVO(PO$_4$)$_2$ at $T=1.5\,\text K$~\cite{skoulatos:08}. 
Sharp peaks coexist on top of the incoherent background. The fit to the measured data 
(dots with error bars) yields again a columnar arrangement of the magnetic moments but with $J_1,J_2$ values closer to the spin nematic range   (Table~\ref{tbl:exchange}, Fig.\ref{fig:phasedia}).}
\label{fig:sqsrznlowt}
\end{center}
\end{figure}
So far we have only discussed Pb$_2$VO(PO$_4$)$_2$ in the central $(J_1<0)$ CAF region. A similar investigation has been performed on SrZnVO(PO$_4$)$_2$, again confirming a ground state with columnar magnetic order~\cite{skoulatos:08,skoulatos:09}, see Fig.~\ref{fig:sqsrznlowt}. For this compound, the ordered moment $\mu_Q\approx0.42\mu_{\text B}$ is even smaller, bringing it closer to the disordered spin nematic regime in accordance with the results from thermodynamic measurements.

\begin{figure}
\begin{center}
\includegraphics[width=0.5\textwidth]{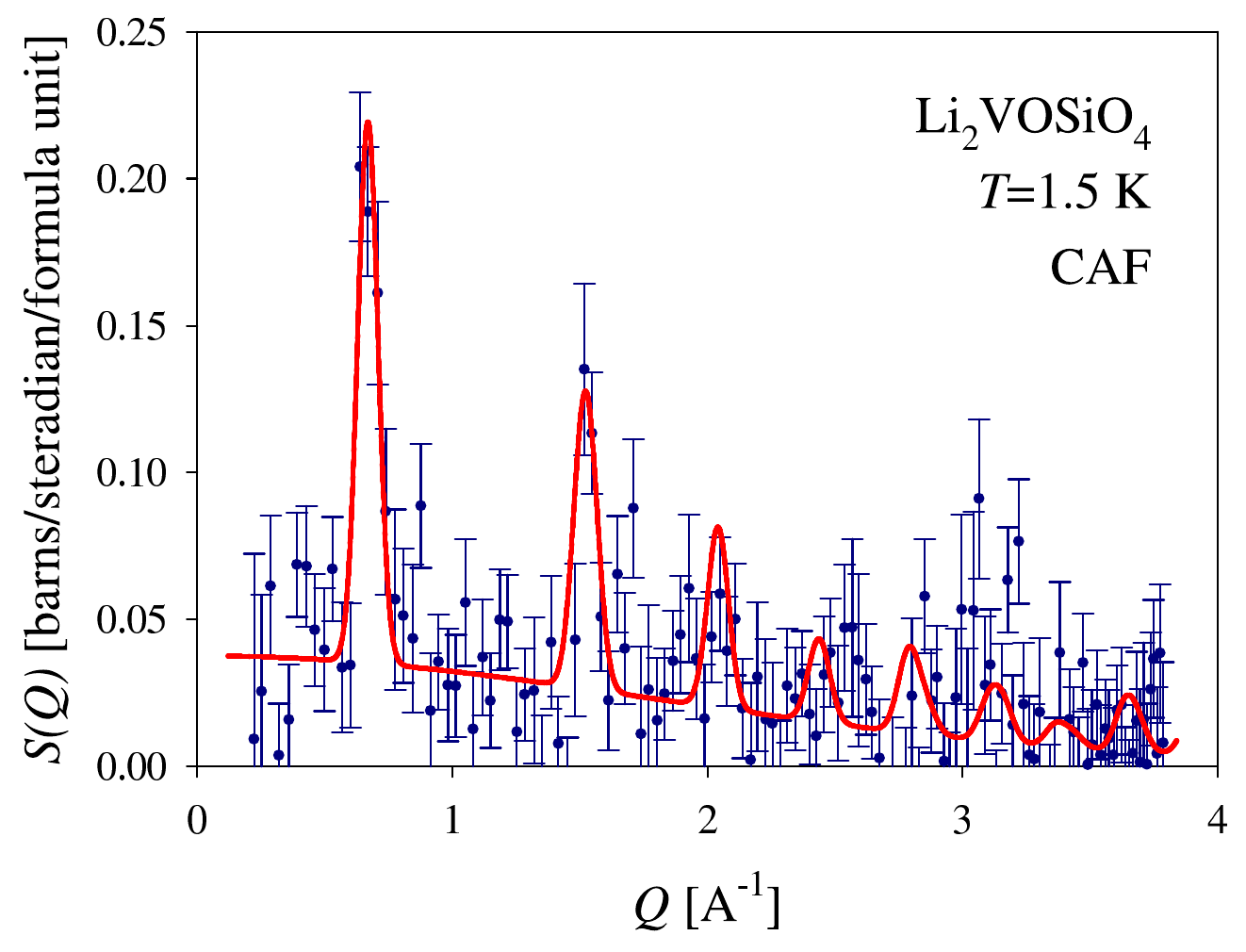}
\caption{Diffraction (elastic) cross section of Li$_2$VOSiO$_4$ at $T=1.5\,\text K$~\cite{skoulatos:08}. Fit with Eq.~(\ref{eqn:sqclass}) (full line) leads to almost vanishing $J_1\approx 0$ (Table~\ref{tbl:exchange}, Fig.\ref{fig:phasedia}) meaning the CAF phase is of the `order from disorder' kind.}
\label{fig:sqli2lowt}
\end{center}
\end{figure}
In case of Li$_2$VOSiO$_4$, a columnar magnetic ground state is obtained as well~\cite{skoulatos:08} with an ordered moment $\mu_Q\approx0.55\mu_{\text B}$, see Fig.~\ref{fig:sqli2lowt}. In contrast to the compounds mentioned before, the diffusive background of the diffraction pattern is substantially smaller, indicating a position in the phase diagram with less frustration. This is in accordance with thermodynamic measurements yielding a frustration angle $\phi\approx0.47\pi$ (see Table~\ref{tbl:exchange}). Both exchange constants are positive with $J_1\ll J_2$ which indeed puts the sample near the unfrustrated point $J_1=0$ in the phase diagram (Fig.~\ref{fig:frustration}). Actually this compound corresponds well to the regime of `order by disorder'  magnetism (Sec.~\ref{subsect:order-disorder}).
\subsection{Field-induced ordered moment stabilization in Cu-pyrazine}
\label{subsect:Cupz}

\begin{figure}
    \centering
    \includegraphics[width=.5\columnwidth]{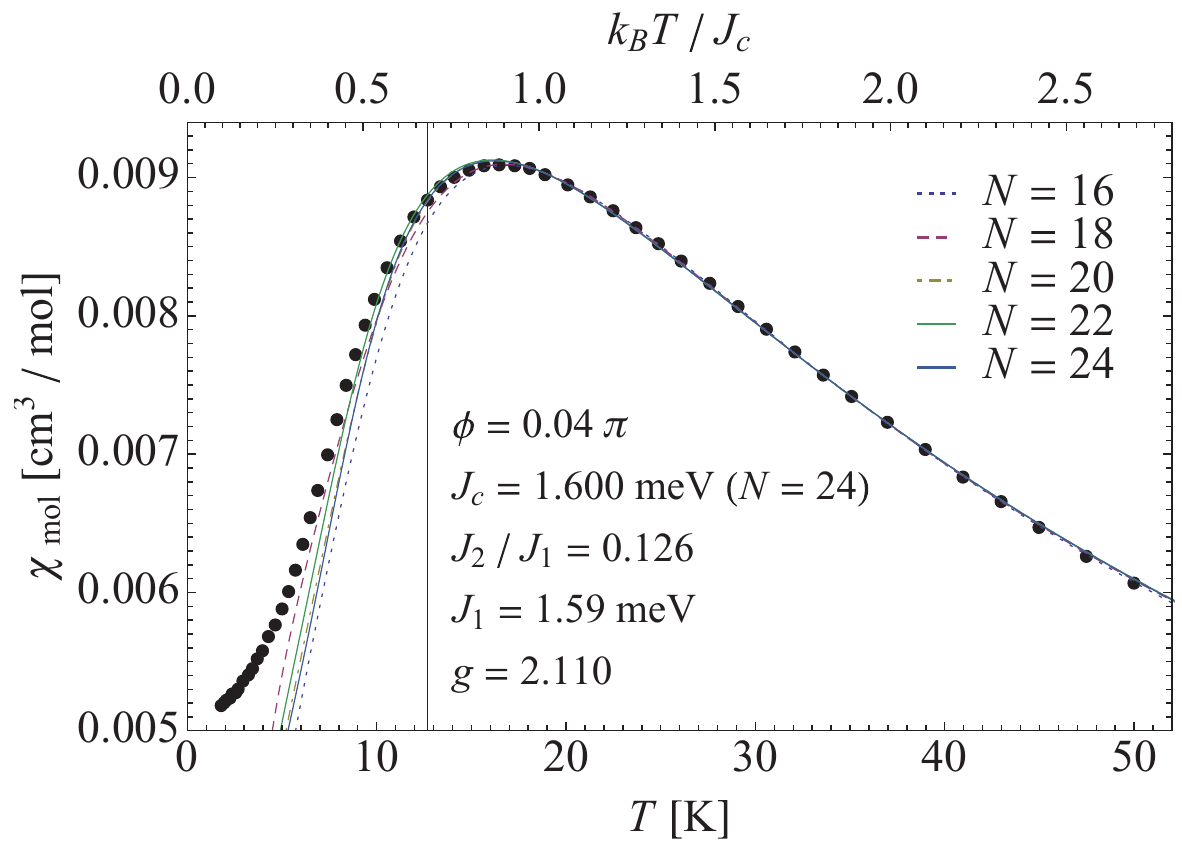} 
    \caption{FTLM susceptibilities for various cluster sizes fitted to the experimental 
    values of  \Cupz~\cite{xiao:09} (dots) to extract the 
    square lattice exchange model parameters as well as the $g$ factor.}
     \label{fig:sus_Cupz}
\end{figure}
Now we discuss an exceptionally interesting field dependence of the interplay of frustration and quantum fluctuations in the  2D magnet \Cupz{} (copper pyrazine)~\cite{siahatgar:11}. It consists of a square lattice of Cu$^{2+}$ ($\text{S}=1/2$) ions with superexchange paths primarily along legs provided by the pyrazine molecules. The ClO$_{\text{4}}$ tetrahedra act as spacer layers~\cite{tsyrulin:10}.

\begin{figure}
    \centering
    \includegraphics[width=.5\columnwidth]{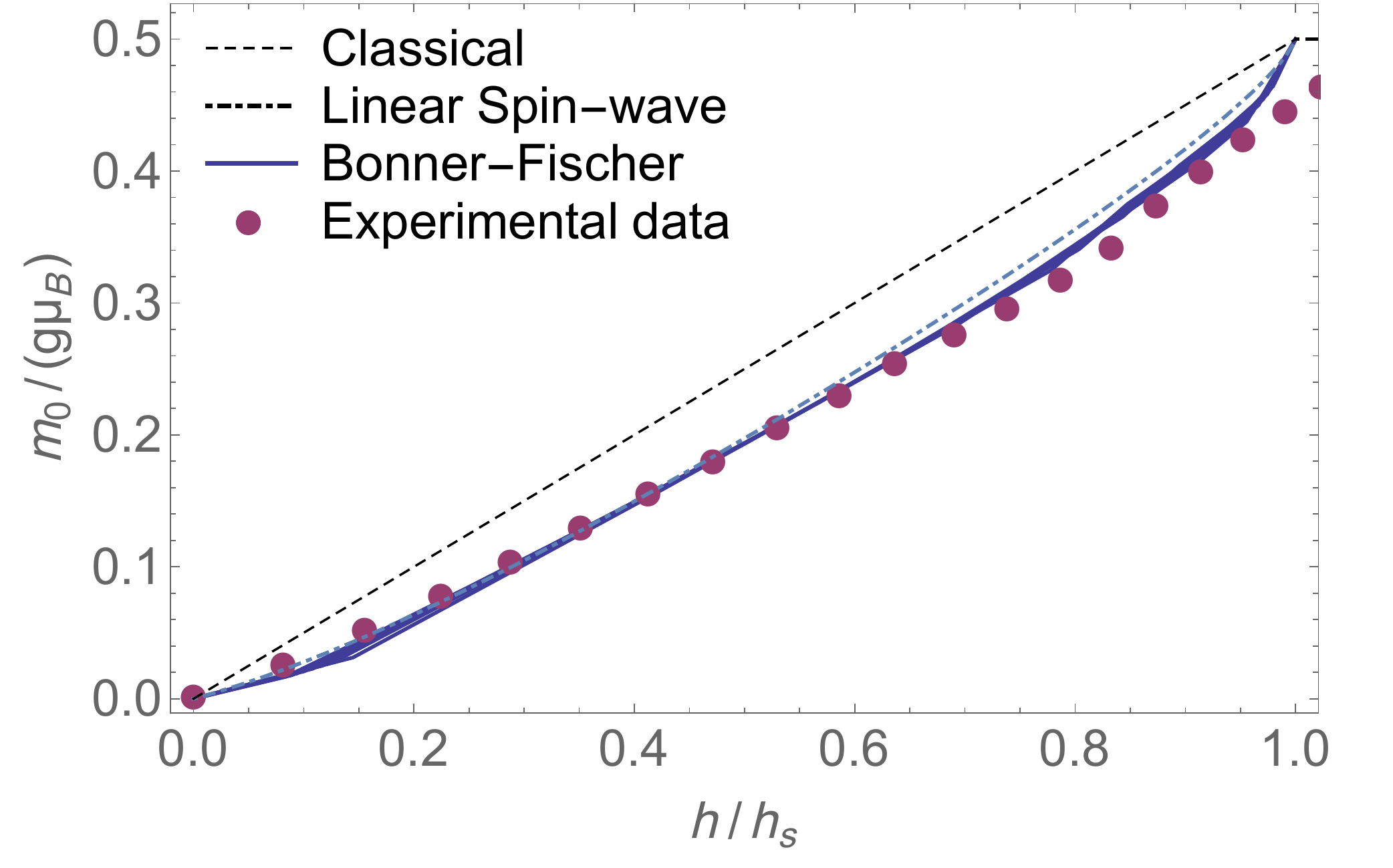} 
    \caption{Experimental magnetization curve ($\mu_0H_\text s=4.9\,\text T$~\cite{xiao:09}) of \Cupz (dots) as compared to spin wave 
    and ED results from Bonner-Fisher construction calculated for $\phi=0.04\pi$.}
     \label{fig:m0_Cupz}
\end{figure}
In addition to the uniform magnetization we also turn our attention to the staggered moment, i.e., the order parameter. As discussed before a highly non-classical and non-monotonic field dependence of the order parameter is predicted both from spin wave theory and from ED scaling results (Fig.~\ref{fig:parametric}). This type of behavior has indeed partly been found in \Cupz~\cite{tsyrulin:09,tsyrulin:10} by determination of the ordered moment in an external field using neutron diffraction. As a starting point the exchange parameters may again be determined by fitting the unbiased FTLM calculations to the experimental data of the susceptibility curve~\cite{siahatgar:11}. The result is shown in Fig.~\ref{fig:sus_Cupz} leading to a moderate frustration ratio of $J_2/J_1=0.126$ or $\phi/\pi=0.04$. Using this value the calculated magnetization curve~\cite{siahatgar:12} is also in good agreement with the experimental results for \Cupz~(Fig.\ref{fig:m0_Cupz}).

\begin{figure}
    \centering
    \includegraphics[width=.5\columnwidth]{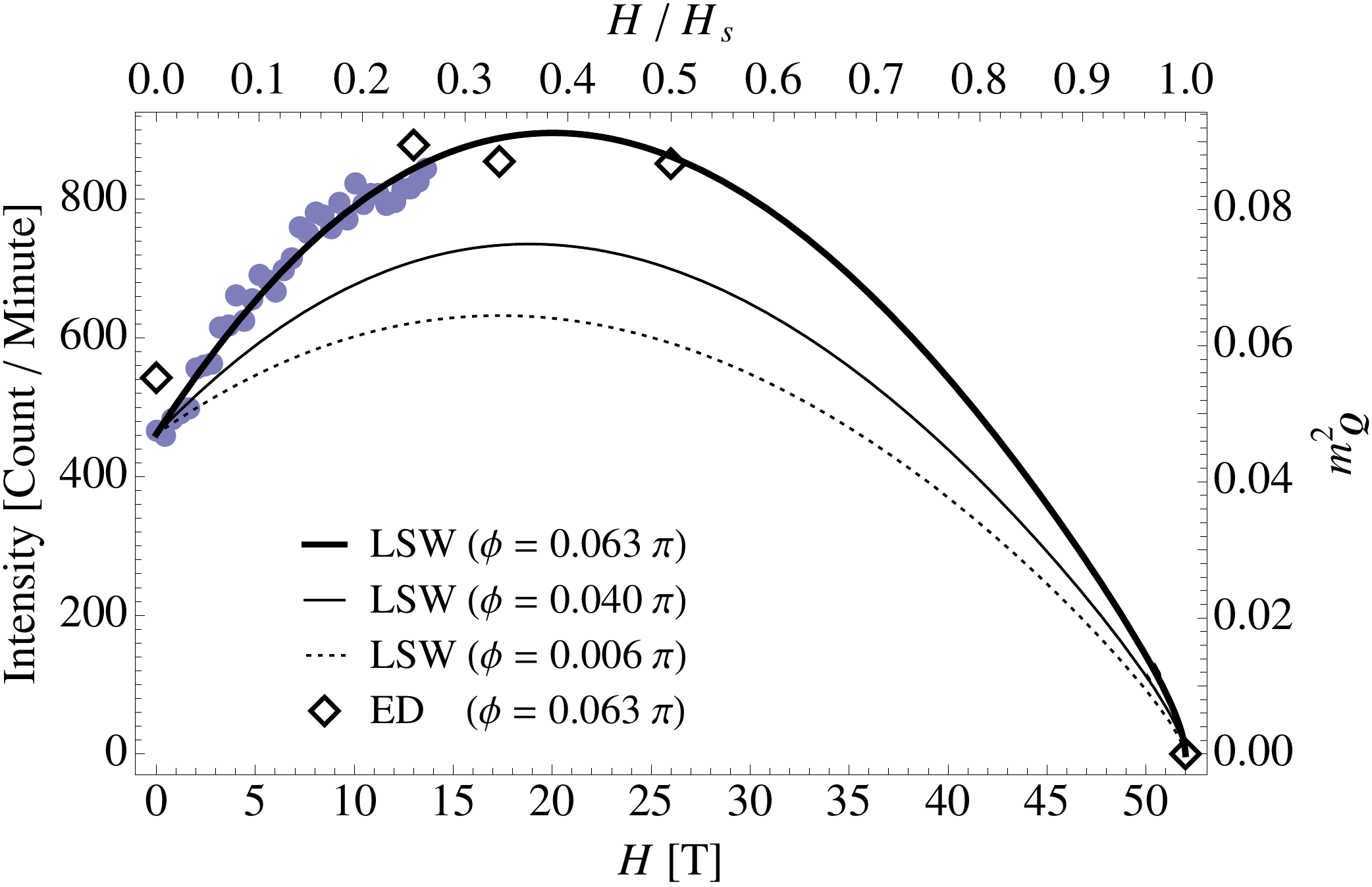} 
    \caption{Field dependence of $m_Q^2 \sim$ scattering intensity in \Cupz~(full circles: neutron diffraction experiment~\cite{tsyrulin:10}). LSW 
    results (lines) fit optimally for $\phi/\pi=0.063$. Diamonds indicate ED scaling results for same $\phi$.}
     \label{fig:mQ_Cupz}
\end{figure}
Alternatively it was shown in Sec.~\ref{sect:quantum} that the field dependence of the ordered moment exhibits a peculiar non-monotonic behavior which is very sensitive to the frustration ratio and allows the determination of $\phi$. This is possible in the rare cases when the field dependence of the staggered moment $m_{\vec{Q}}(h)$ (its square is proportional to the intensity of magnetic Bragg reflections in neutron diffraction) has been determined. The experimental results for \Cupz{} together with LSW and ED calculations of the ordered moment are shown in Fig.~\ref{fig:mQ_Cupz}. The original slope of $m_{\vec{Q}}(h)$  is considerable. This can only be explained if there is some amount of frustration present which suppresses the zero-field moment sufficiently. For the unfrustrated NAF $(\phi\approx 0)$ the suppression is not sufficient and therefore the original slope would be too small. Fitting the LSW results in Eq.~(\ref{eqn:moment1}) for the optimal frustration leads to $\phi/\pi=0.063)$ somewhat larger than obtained from FTLM calculation above. Curves for smaller $\phi$ are also shown in the figure. The results from numerical ED scaling analysis (represented as diamonds) which can only be obtained for a few field values (Sec.\ref{subsect:gsprop}, Fig.~\ref{fig:parametric}) are shown for comparison and agree quite  well with LSW predictions. Unfortunately the estimated saturation field of $\mu_0H_{\text{s}}\simeq 52\,\text{T}$ is not accessible. Therefore the experimental results which stop around $\mu_0H\simeq 14\,\text{T}$ are not sufficient to achieve the maximum and turnaround in the field dependence of the ordered moment. Such experiments might therefore be even more interesting in compounds with smaller saturation fields such as the layered vanadates discussed in the previous section.

In molecular field theory of 3D magnets the transition temperature $T_{\text{N}}$ is proportional to the square of the ordered moment. 
Although for planar magnets it strictly vanishes due to the Mermin-Wagner theorem, any finite interlayer coupling $J_\perp$ immediately leads to a finite $T_{\text{N}}$ (Eq.~(\ref{eqn:3DTN})). One would therefore expect that even in the present very anisotropic quasi-2D case $T_{\text{N}}$ should increase with the size of the ordered moment and consequently with field strength. This behavior is indeed observed in \Cupz.
The transition temperature for the quasi-2D system is obtained from the vanishing total moment calculated in self-consistent RPA approximation~\cite{majlis:92} extended to finite fields~\cite{siahatgar:11,schmidt:13}. One starts from the observation that the moment is renormalized due to thermally excited spin waves according to
\begin{eqnarray}
\langle S\rangle&=&S-\frac{1}{N}\sum_{\vec k}\langle a^\dagger_{\vec k} a_{\vec k}\rangle
\equiv S-\Psi \nonumber\\
\Psi&=&\frac{1}{N}\sum_{\vec k}\bigl[v^2_{\vec k}+(1+2v^2_{\vec k})n_{\vec k}\bigl]
\end{eqnarray}
The coefficient $v_\vec k$ is defined in Eq.~(\ref{eqn:bogol}), and $n_\vec k=\left[\exp(\beta E(h,\vec k))-1\right]^{-1}$. This may be considered as the small $\Psi$ expansion of the selfconsistent RPA equation~\cite{majlis:92,majlis:07} $\langle S\rangle = (1/2)/(1+2\Psi)$ and $T_\text N$ is then determined by requiring $\langle S\rangle =0$. Here in the expression for $\Psi$ the spin wave energies are written in selfconsistent form including the $T$-dependent ordered moment $\langle S\rangle$
in $\langle S\rangle E(h,\vec k)$. In the vicinity of $T_\text N$ where $\langle S\rangle \ll S$, $\Psi$ is given by
\begin{equation}
\Psi=\frac{k_\text BT}{\langle S\rangle}\frac{1}{N}\sum_{\vec k}
\frac{1+2v^2_{\vec k}}{E(h,\vec k)}-\frac{1}{2}
\end{equation}
Using the selfconsistent expression for $\langle S\rangle$ and setting $\langle S\rangle=0$ we obtain the quasi-2D critical temperature for $h\ll h_\text s$
\cite{siahatgar:11,schmidt:13}
\begin{equation}
    T_{\text N}(h)=
    \left(4k_{\text B}
    \frac{1}{N}\sum_{\vec k}
    \frac{A(\vec k)-B(\vec k)\cos^{2}\Theta_\text{cl}}{E^{2}(h,\vec k)}
    \right)^{-1}
    \label{eqn:tnh}
\end{equation}
Here the sublattice exchange interactions $A(\vec k)$, $B(\vec k)$ are again given by Eq.~(\ref{eqn:swcoeff}) and $E(h,\vec k)$ by  Eq.~(\ref{eqn:ek}). Note that the exchange Fourier transform now contains the interlayer coupling $J_\perp$  according to 
$J_{\vec k}=J_\perp\cos k_{z}+J_{1}\left(\cos k_{x}+\cos k_{y}\right)+2J_{2}\cos k_{x}\cos k_{y}$, where we assume that z is the direction perpendicular to layers. Now there is an additional spin wave dispersion perpendicular to layers of order $ J_\perp$. It is very important because it leads to a finite integral in Eq.~(\ref{eqn:tnh}) resulting in a finite  $T_{\text{N}}$. For $J_\perp\rightarrow 0$ the integral diverges and the N\'eel temperature will be zero for the strictly 2D case.

\begin{figure}
    \centering
    \includegraphics[width=.5\columnwidth]{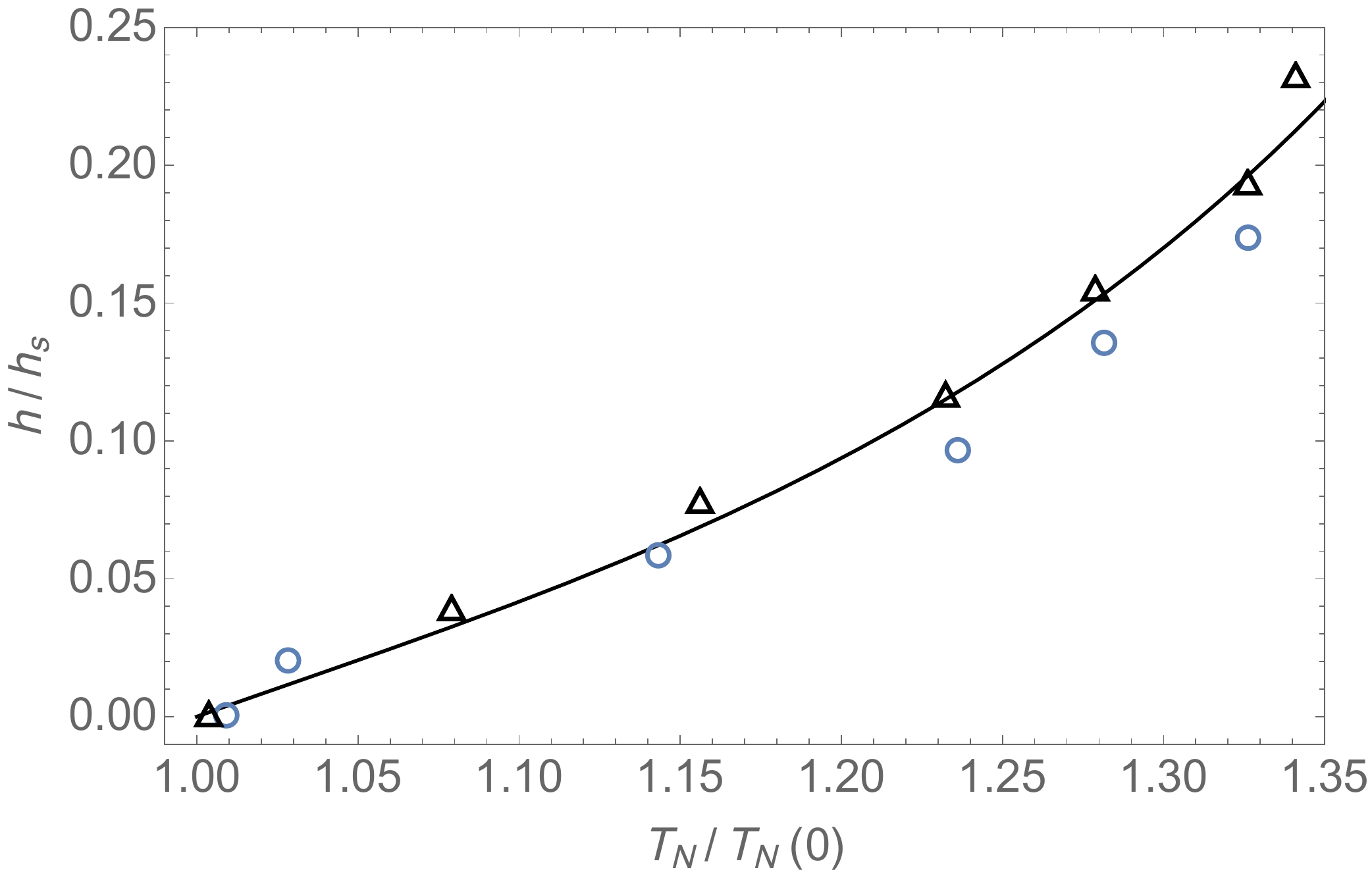} 
    \caption{Increase of the N\'eel temperature of  \Cupz~with field. Experimental results~\cite{tsyrulin:10} 
    are obtained from heat capacity  ($\circ$)
     and neutron diffraction ($\triangle$) . The theoretical full line is obtained for an interlayer 
     coupling $J_\perp/J_1=0.0064$ using an intralayer  
     coupling $J_2/J_1 \simeq 0.2$ ($\phi/\pi=0.063$) corresponding to optimal values in  
     Fig.~\ref{fig:mQ_Cupz}.}
     \label{fig:neeltemp}
\end{figure}
The predicted field dependence of the normalized   $T_{\text{N}}(h)/T_{\text{N}}(0)$ is shown in Fig.~\ref{fig:neeltemp} together with the experimental results. The 2D exchange parameters are the same as the optimal ones of Fig.~\ref{fig:mQ_Cupz} and $J_\perp$ is obtained from a fit to the experimental $T_\text N(h)$. In its essence the increase of $m_{\vec{Q}}(h)$ and $T_{\text{N}}(h)$ have the same origin, namely the field-induced suppression of quantum fluctuations in the frustrated magnet. For the unfrustrated quasi-2D antiferromagnet ($J_2=0, J_\perp >0$) more advanced treatments of the zero-field critical temperature based on results from the nonlinear $\sigma$-model and QMC simulations are available~\cite{yasuda:05}.

\subsection{Strongly anisotropic triangular exchange compounds \CCC{} and \CCB}
\label{subsect:halides}

As a genuine example of frustrated triangular quantum spin systems the compounds \CCC{} ($T_\text N=0.62\,\text K$) and \CCB{} ($T_\text N=1.42\,\text K$) have been investigated intensively by thermodynamic~\cite{carlin:85,cong:11}, high-field~\cite{ono:03,radu:05}, neutron diffraction~\cite{coldea:96} and also spectroscopic methods like INS~\cite{coldea:02,veillette:05-1} and ESR~\cite{zvyagin:14}. They are, however, quite distant from the ideal isotropic case and rather close to the quasi-1D model in Fig.~\ref{fig:moment_triangular} , denoted by $\triangle$ and $\parallel$, respectively. Again the finite ordering temperatures are due to interlayer coupling. 

\begin{figure}
    \centering
    \includegraphics[width=.5\columnwidth]{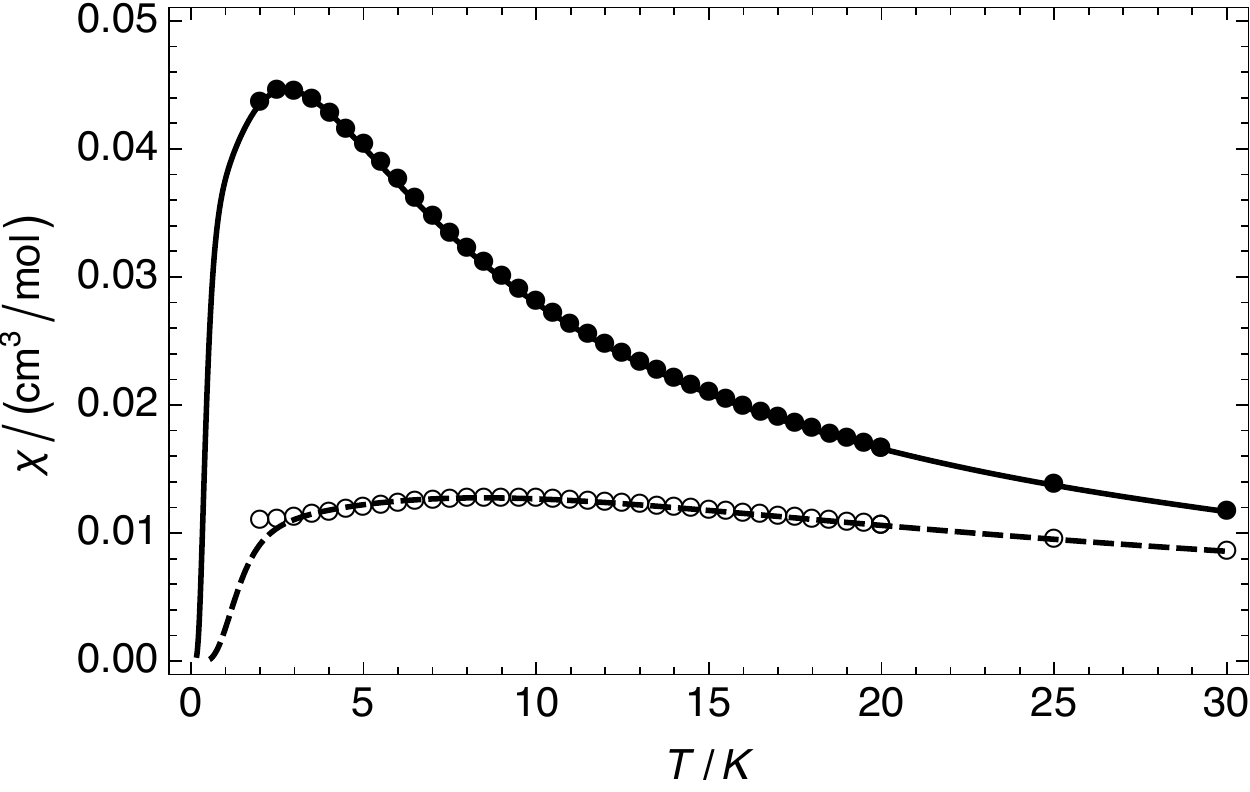}
    \includegraphics[width=.5\columnwidth]{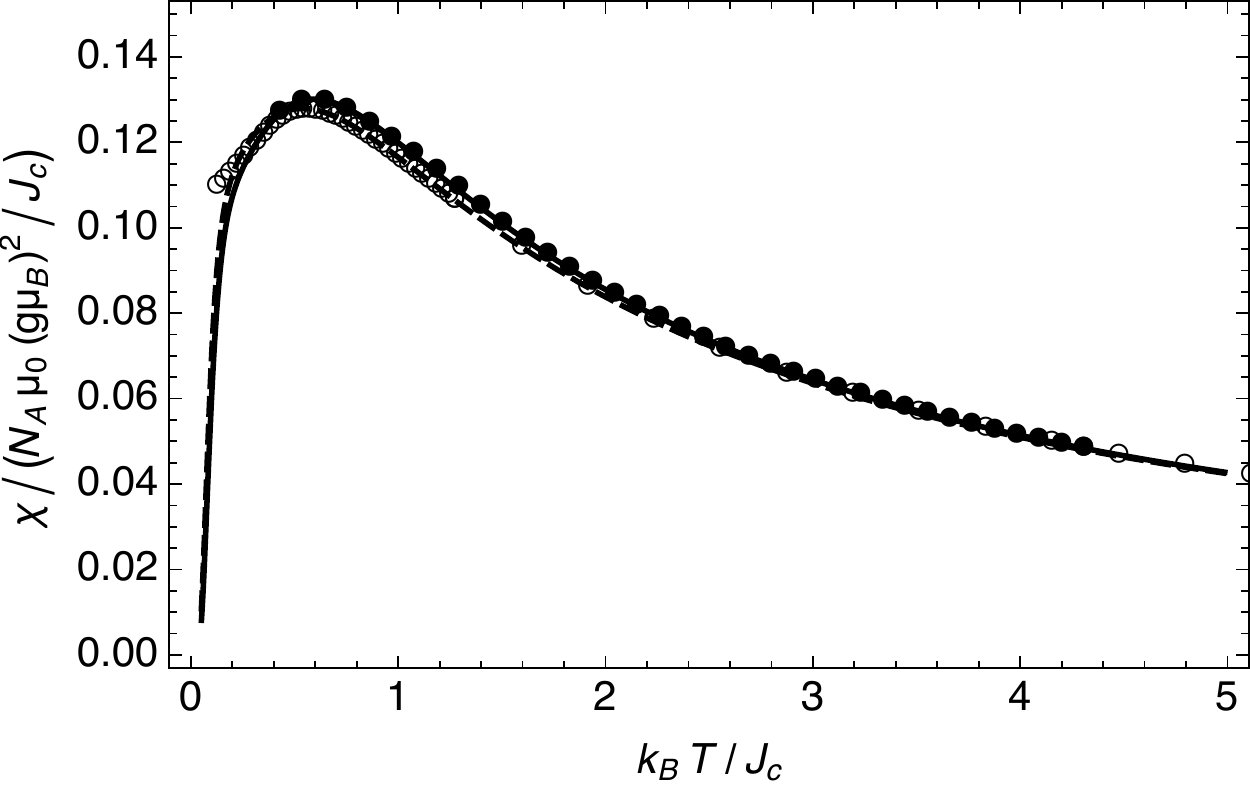}
    \caption{Temperature dependence of the magnetic susceptibility
    $\chi(T)$ of Cs$_{\text 2}$CuCl$_{\text 4}$ (dots and solid line) with $T_\text{max}=2.8\,\text K$
    and Cs$_{\text 2}$CuBr$_{\text 4}$ (open circles and dashed line) with $T_\text{max}=8.8\,\text K$.
    Here (a) displays the experimental data (symbols, taken from
    Ref.~\cite{cong:11}) together with the fits of our FTLM data
    (lines, fitted values see Table~\ref{tbl:compound}).  Plot (b)
    shows exactly the same data, this time in dimensionless
    units using $J_{\text c}$, $\phi$, and $g$ for the two compounds.}
    \label{fig:chicscucl}
\end{figure}
As for the square lattice case a comparison of FTLM results~\cite{schmidt:15} and experimental susceptibility curves presented in  Fig.~\ref{fig:chicscucl} allows to extract the triangular exchange model parameters $(J_\text c,\phi)$.
In this figure dots and
open circles denote the experimental data according to
Ref.~\cite{cong:11}, the solid and dashed lines present
corresponding fits with FTLM data~\cite{schmidt:15}.  
For clarity  the data and the curves are shown in two
different ways: Fig.~\ref{fig:chicscucl} (a) displays
$\chi(T)$ in electromagnetic (CGS) units which is used to reproduce
the values of FTLM model parameters in
Table~\ref{tbl:compound}.  Using these fitted values for $J_{\text
c}$, $\phi$, and $g$,  the same data are shown again in
dimensionless units in of Fig.~\ref{fig:chicscucl}~(b).
It demonstrates that both compounds have almost the same 
temperature dependence when using scaled quantities.

The main effect of replacing Cl with Br is an increase of the exchange
energy scale $J_{\text c}$ by a factor $3.4$, whereas the anisotropy
angle $\phi$ changes only by about $5\,\%$.  Therefore the large
change in $J_{\text c}$ is  exclusively responsible for the decrease and broadening of the maximum
in $\chi(T)$ as well as its shift towards higher temperatures.  The anisotropy ratio $J_2/J_1$ $(\phi)$ and
therefore the position of Cs$_{\text 2}$CuBr$_{\text 4}$ in the phase
diagram is almost identical for Cs$_{\text 2}$CuCl$_{\text
4}$, it is rather close to the quasi-one-dimensional 
regime around $\phi\simeq\pi/2$ or $J_1\simeq 0$.  The original  measurement of $\chi(T)$ of Cs$_{\text
2}$CuCl$_{\text 4}$ led to somewhat different
values~\cite{carlin:85}, however the analysis was performed over
a limited temperature range assuming weakly coupled one-dimensional
chains.  Following Sec.~\ref{sect:temp}, the rapid decrease of the computed $\chi(T)$
for $T\to0$ in both cases is due to the finiteness of the FTLM tiling.
Assuming a finite-size gap
$\Delta\approx J_{\text c}/N$, we expect finite-size effects to dominate below
$T_{\Delta}\approx0.2\,\text K$ for Cs$_{\text 2}$CuCl$_{\text 4}$ and
$T_{\Delta}\approx0.6\,\text K$ for Cs$_{\text 2}$CuBr$_{\text 4}$.
Experimentally, the lowest temperature used was
$T_{\text{min}}\approx2\,\text K$, well above the finite-size gaps.
\begin{table*}
\centering\small
        \begin{tabular}{ccccccccccc}
	    compound 
	    &
            method
            &
            $J_1/\text{meV}$
            &
            $J_{2}/\text{meV}$
            &
            $J_{\text c}/\text{meV}$
            &
            $J_{2}/J_{1}$
            &
            $J_{1}/J_{2}$
            &
            $\phi/\pi$
            &
            $g$
            &
            $\Theta_\text{CW} /\text{K}$
            &
            Ref.
            \\
            \hline
	    Cs$_{\text 2}$CuCl$_{\text 4}$
	    &
            FTLM
            &
            $0.11$
            &
            $0.38$
            &
            $0.40$
            &
            $3.45$
            &
            $0.29$
            &
            $0.41$
            &
            $2.06$
            &
            $3.48$
            &
            \cite{schmidt:15}
            \\
	    &
            INS
            &
            $0.128$
            &
            $0.374$
            &
            $0.40$
            &
            $2.92$
            &
            $0.34$
            &
            $0.40$
            &
            $2.19$
            &
            $3.65$
            &
            \cite{coldea:02}
            \\
	    &
            ESR
            &
            $0.12$
            &
            $0.41$
            &
            $0.43$
            &
            $3.42$
            &
            $0.29$
            &
            $0.41$
            &
            $2.08$
            &
            $3.77$
            &
            \cite{zvyagin:14}
	    \\
	    \hline
	    Cs$_{\text 2}$CuBr$_{\text 4}$
	    &
	    FTLM
	    &
	    $0.5$
	    &
	    $1.26$
	    &
	    $1.35$
	    &
	    $2.52$
	    &
	    $0.40$
	    &
	    $0.38$
	    &
	    $2.04$
	    &
	    $13.11$
	    &
	    \cite{schmidt:15}
	    \\
	    &
	    ESR
	    &
	    $0.53$
	    &
	    $1.28$
	    &
	    $1.38$
	    &
	    $2.44$
	    &
	    $0.41$
	    &
	    $0.38$
	    &
	    $2.09$
	    &
	    $13.57$
	    &
	    \cite{zvyagin:14}
	\end{tabular}
    \caption{Comparison of exchange parameters for Cs$_{\text
    2}$CuCl$_{\text 4}$ and Cs$_{\text 2}$CuBr$_{\text 4}$ as
    determined from thermodynamic FTLM-fit and direct spectroscopic
    INS (Cs$_{\text 2}$CuCl$_{\text 4}$ only) and ESR methods at
    $H>H_{\text{s}}$.  Exchange constant $J_2$ corresponds to the
    crystallographic $b$ direction and $J_1$ to the zigzag bonds in
    the $bc$ plane. The experimental saturation fields are $\mu_0H_\text s\approx8.4\,\text{T}$ 
    for Cs$_{\text 2}$CuCl$_{\text 4}$~\cite{coldea:02}  and  
    $\mu_0H_\text s\approx 30\,\text{T}$ for Cs$_{\text 2}$CuBr$_{\text 4}$~\cite{ono:03}. 
    Curie Weiss temperature is $\Theta_\text{CW}=(J_1+J_2/2)/k_B$. }
      \label{tbl:compound}
\end{table*}
The results from the FTLM fit to susceptibility data are compared in
Table~\ref{tbl:compound} to parameter values from direct spectroscopic methods:
Inelastic neutron scattering (INS)~\cite{coldea:02} (Cs$_{\text
2}$CuCl$_{\text 4}$ only) and electron spin resonance
(ESR)~\cite{zvyagin:14}, both for $H>H_\text s$~\cite{schmidt:14}. Here (Eq.~(\ref{eqn:satclass2}))
\begin{equation}
    \mu_{0}H_{\text{s}}=\frac{2S}{g\mu_{\text B}}
    \frac{(J_{1}+2J_{2})^{2}}{2J_{2}},
\end{equation}
leads to  $\mu_{0}H_{\text{s}}\approx8.4\,\text T$ for Cs$_{\text
2}$CuCl$_{\text 4}$ and $\mu_{0}H_{\text{s}}\approx30\,\text T$ for
Cs$_{\text 2}$CuBr$_{\text 4}$. Using spectroscopic data at high
fields has the advantage of a fully polarized ground state,
corresponding to a single  spinor product state
$|\text{FM}\rangle$ containing only up spins.
The excitations out of the ground state are the $N_\text s$
orthonormal single-particle states
$|\psi_{i}\rangle=(1/\sqrt{2S})S_{i}^{-}|\text{FM}\rangle$, $i=1\ldots
N_\text s$ with $N_\text s={\cal O}(N)$.  Because $\left[{\cal
H},S_z^\text{tot}\right]=0$, the Hamiltonian cannot create more than
these one-spin-flip states, and therefore its spectrum may be exactly
obtained by Fourier transform.

The agreement between the different methods is excellent, for
exchange parameters as well as g-factors.  As explained before in
Ref.~\cite{schmidt:14} if Cs$_{\text 2}$CuCl$_{\text 4}$ is
interpreted as a purely 2D system these exchange parameters put
the compound very close to the quasi-1D spin liquid regime around
$\phi=0.5\pi$ (in LSW theory).  In reality, however, the incommensurate spiral magnetic order with wave vector
$\vec Q=(0,0.535,0)$ in r.l.u.  is stabilized
below $T_{\text N} = 0.62$ K by a finite DM exchange $D\approx 0.020\, \text{meV}$ and an inter-plane coupling 
$J_\perp\approx0.017\,\text{meV}$ along the crystallographic $a$
direction~\cite{coldea:03}.

\begin{figure}
    \centering
    \includegraphics[width=.5\columnwidth]{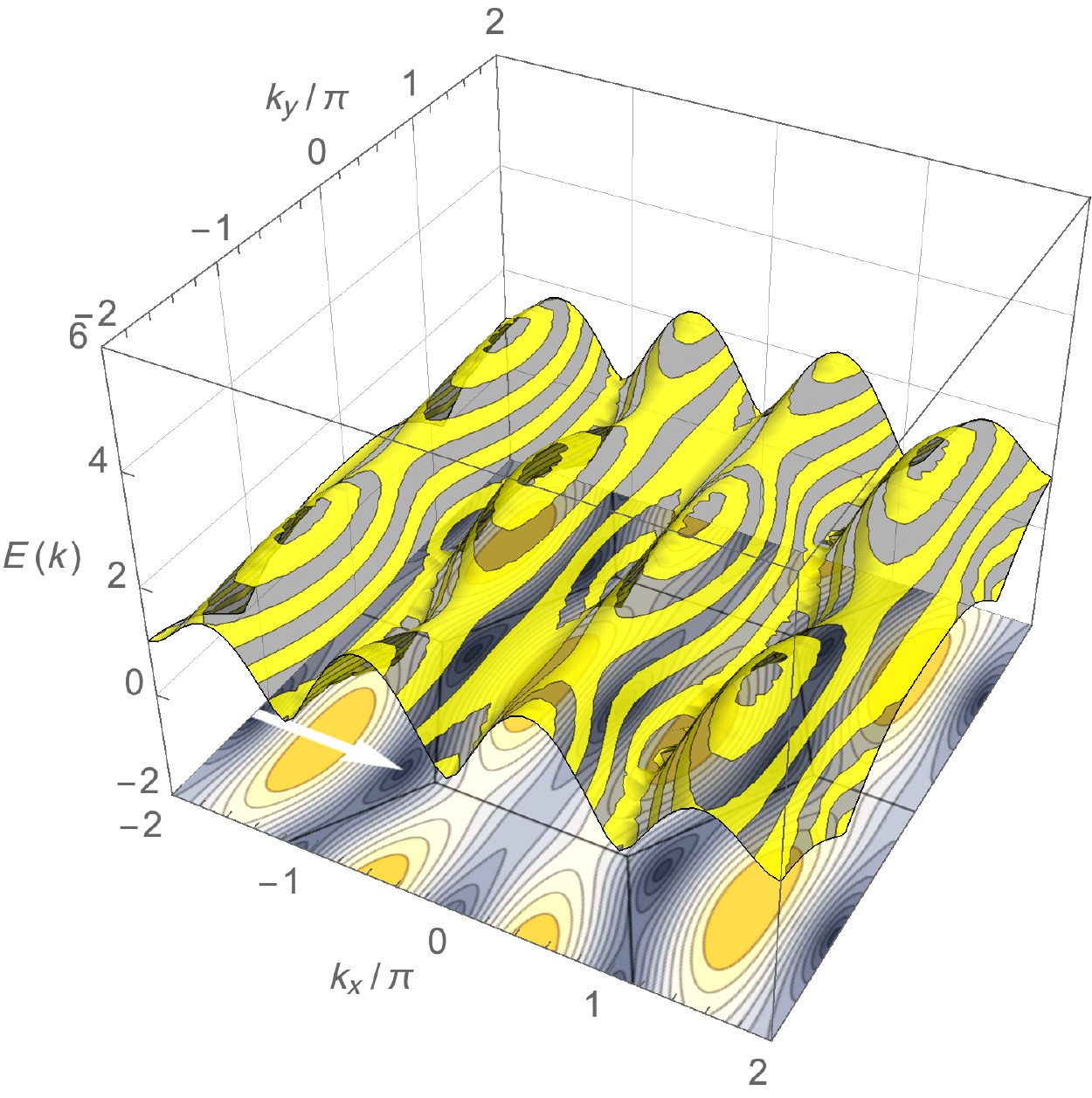} 
    \caption{Spin wave dispersion in triangular model with $\phi/\pi=0.398$ or $J_2/J_1=3$ 
    valid approximately for \CCC~and \CCB. The quasi-1D character is obvious
    from the pronounced dispersion anisotropy.}
     \label{fig:swcscucl}
\end{figure}

The quasi-1D character of the compounds is also visible directly 
from the spin wave dispersions (Sec.~\ref{subsect:LSW}) obtained in Ref.~\cite{coldea:03}.
The calculated dispersion  for the anisotropy ratio $J_2/J_1=3$ 
corresponding approximately to both compounds  is shown in Fig.~\ref{fig:swcscucl}.
Furthermore, although these compounds
are 2D triangular with only quasi-1D character indications of the typical spin fractionalization
of sharp spin wave excitations into a two-spinon continuum known from truly
1D spin chains have been found. For the detailed discussion and analysis of spin excitations in Cs$_{\text 2}$CuCl$_{\text 4}$ from inelastic neutron scattering we refer to 
Refs.~\cite{coldea:01,coldea:02,coldea:03,veillette:05,veillette:05-1,fjaerestad:07}.
For Cs$_{\text 2}$CuBr$_{\text 4}$ there are deviations of the
experimental susceptibility data from FTLM result at the lowest temperatures.  These
are caused by impurities~\cite{cong:11,cong:14}, but may be considered
as an indication for a tendency towards magnetic order in this
compound.  The overall energy scale $J_{\text c}$ is more than three
times larger than for Cs$_{\text 2}$CuCl$_{\text 4}$, however the
estimate for the anisotropy angle $\phi$ is essentially the same.  Therefore 
Cs$_{\text 2}$CuBr$_{\text 4}$ exhibits spiral order with $\vec Q=(0,0.575,0)$
at a higher ordering temperature $T_{\text N} = 1.42$ K than its Cl counterpart.

The agreement of FTLM results with direct spectroscopic
results suggests  that the method may be applied to the
analysis of the whole Cs$_{\text 2}$CuCl$_{\text 4-x}$Br$_{\text x}$
substitutional series~\cite{cong:11,cong:14} to find out the systematic variation
of the triangular exchange model parameters $(J_\text c,\phi)$ in the  series.

The magnetization curve of  Cs$_{\text 2}$CuBr$_{\text 4}$ shows a narrow plateau at $m_0/S=1/3$
for $\vec H$ within the anisotropic triangular plane, however not for field perpendicular to it~\cite{ono:03}. This is confirmed by magnetocaloric experiments~\cite{tsujii:07} which suggest the $uud$ plateau phase between $\mu_0H_\text{c1}=12.9\,\text T$ and  $\mu_0H_\text{c2}=14.3\,\text T$. Interestingly the $J_2/J_1$ ratio (Table~\ref{tbl:compound}) of this compound is slightly outside the region where a $1/3$ magnetization  plateau can still occur in Fig.~\ref{fig:phasedia_triangular_h}. Together with the in-/out-of-plane anisotropy this indicates that further effects (like DM interaction and interlayer coupling ) may play a role in its appearance.
From additional magnetocaloric and torque measurements it was proposed that magnetization plateaux at higher fractions of $m_0/S$ may exist~\cite{fortune:09}. There is, however, so far no theoretical justification for such plateaux in the genuine 2D triangular $J_1$-$J_2$ model.

For Cs$_{\text 2}$CuCl$_{\text 4}$ which has larger $J_2/J_1$ (Table \ref{tbl:compound}) one is further to the quasi-1D (smaller $J_1/J_2$ in Fig.~\ref{fig:phasedia_triangular_h}) regime and no zero field magnetization plateau appears. A direct continuous transition from the low-field umbrella (or cone) phase to the fully polarized FM phase above $H_\text s$ takes place. However, the finite temperature properties around $\mu_0H_\text s=8.4$ T are very interesting. From specific heat and magnetocaloric effect~\cite{radu:05,radu:06} it was concluded that the FM to cone transition may be interpreted as a Bose-Einstein condensation (BEC) of FM magnons whose field induced excitation gap $\Delta$ vanishes at $H_\text s$. The condensate density is then proportional to the transverse spiral order parameter in the crystallographic (trigonal) $bc$ plane  that appears below $H_\text s$. The BEC approach is applicable as long as the U(1) symmetry around the field axis ($\parallel$ to crystallographic $a$ axis) is preserved (for a review of magnon BEC see Ref.~\cite{zapf:14}).

To obtain a finite transition temperature and the proper low energy excitations it is necessary to introduce the extension to a quasi-3D model including inter-triangular plane exchange coupling and in-plane (bc) (staggered along a) Dzyaloshinskii-Moriya (DM) term~\cite{kovrizhin:06} which has to be added to the 2D model (Eq.~(\ref{eqn:hex12})). The additional term is given by
\begin{equation}
{\cal H}_{ex}'=\sum_{n,i} \left[J_\perp\vec S_{n,i}\vec S_{n+1,i} + (-1)^n\vec D\cdot\vec S_{ni}\times(\vec S_{ni_{1}}+ \vec S_{ni_{2}})\right]
\end{equation}
where $n$ is the layer index and $i_1$, $i_2$ are the in-plane n.n.{} positions to $i$ which form a $120^\circ$ angle (Fig.~\ref{fig:lattices}).
Furthermore $J_\perp$ is the interlayer coupling and $\vec D= (0,0,D)$ the DM vector oriented along in-plane c-direction. For  Cs$_{\text 2}$CuCl$_{\text 4}$  intra-chain coupling $J_2$ is the dominating term and  $J_\perp/J_2=0.045$,  $D/J_2=0.053$~\cite{coldea:02,kovrizhin:06}. Then, using the Holstein-Primakoff transformation, ${\cal H}={\cal H}_{ex}+{\cal H}_{ex}'+{\cal H}_Z$ may be transformed to a bosonic representation. The low energy ($E< J_\perp$) bosonic modes are located around the spiral phase wave vector $\vec Q$ and are described by a quadratic 3D dispersion $E_{\vec q}$ with minimum gap $\Delta$ at $\vec q =\vec Q$ which closes at $H=H_\text s$.  The number of bosons (or spin flips in the FM state) slightly below $H_\text s$ is controlled by the bare chemical potential $\mu= g\mu_\text B\mu_0(H_\text s-H)$ of the dilute magnon Bose gas. In a standard procedure~\cite{nikuni:00} a hard core boson constraint to restrict local boson occupation numbers to the physical value is implemented in mean field approximation~\cite{kovrizhin:06} with respect to the effective bosonic interaction strength $\Gamma\approx 0.85 J_2$. This shifts the chemical potential to an effective value 
\begin{equation}
\mu_\text{eff}=\mu-2\Gamma n(T); \;\;\; n(T)=N^{-1}\sum_{\vec q}n_\text B(E_{\vec q})
\end{equation}
where $n_\text B$ is the Bose occupation function and $n$ the total number of bosons. The phase boundary $T_\text c(H)$ between FM and cone (umbrella) phase in the $H$-$T$ plane is then determined by the condition $\mu_\text{eff}=0$ where BEC of magnons occurs signifying the onset of the transverse (in-plane) magnetic order. The ordering temperature of the transverse moment is then obtained from the condition
$g\mu_\text B\mu_0(H_\text s-H)=2\Gamma n(T_\text c)$.

\begin{figure}
    \vspace{0.3cm}
    \centering
    \includegraphics[width=.5\columnwidth]{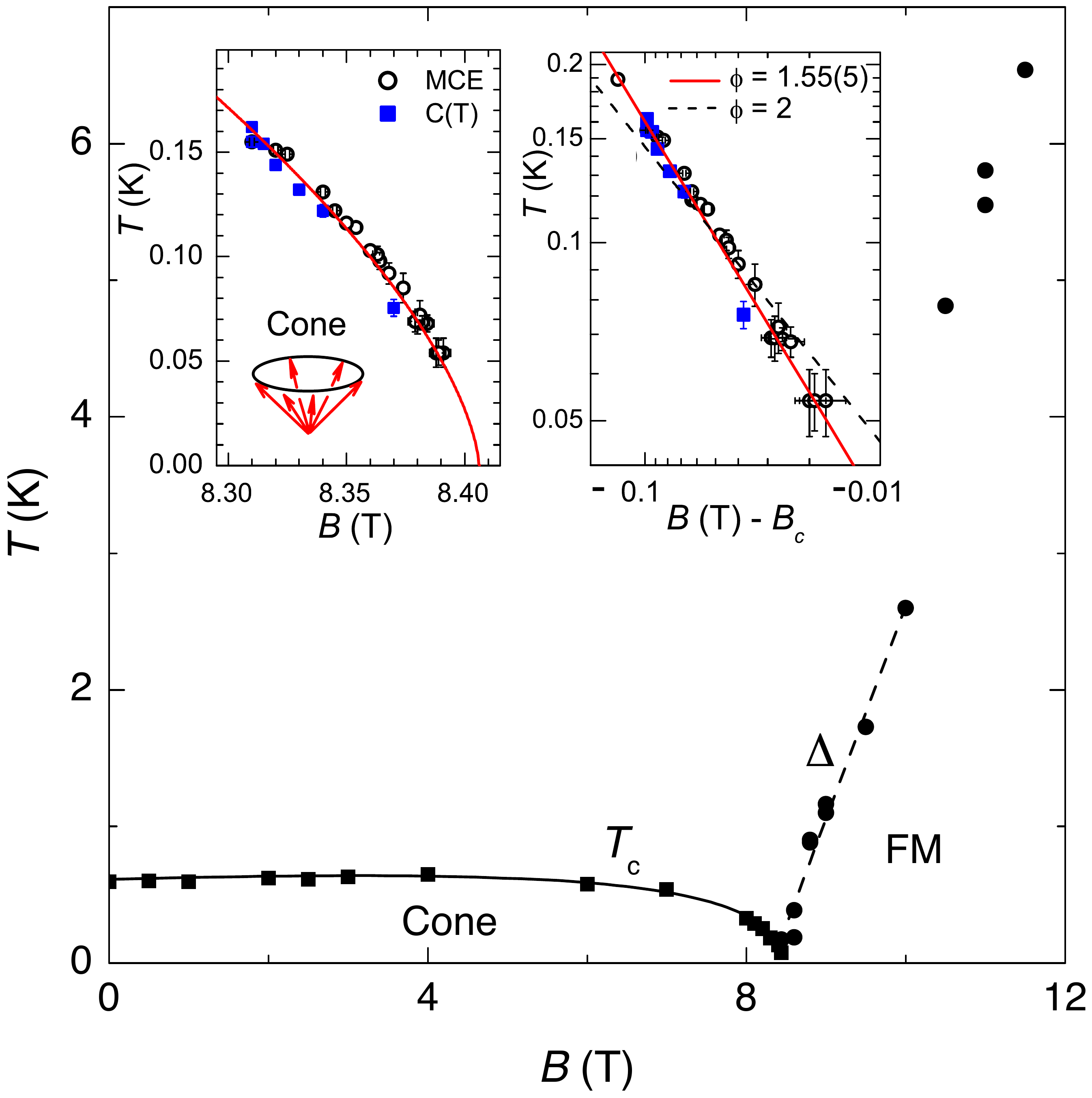} 
    \caption{$B$-$T$ phase diagram of  Cs$_{\text 2}$CuCl$_{\text 4}$ ($B=\mu_0H$). Here critical field is $\mu_0H_\text s=8.4$ T 
    and scaling exponent is $\phi=1.55\pm0.05$ close to the BEC value $3/2$. For comparison scaling with
    $\phi=2$ is also given in the inset. $\Delta$ is the field induced FM magnon gap.
    (Adapted from Refs.~\cite{radu:05,radu:06}).}
     \label{fig:BEC}
\end{figure}
In the asymptotic regime where mainly the 3D quadratic magnon spectrum contributes to $n(T)$ one obtains the universal scaling law characteristic for BEC:
\begin{equation}
T_\text c(H) \sim (H_\text s-H)^{1/\phi}; \;\;\; \phi=\frac{3}{2}
\end{equation}
 where  $\phi$ is the universal BEC scaling exponent. The fitting to specific heat and magnetocaloric cooling rate of  Cs$_{\text 2}$CuCl$_{\text 4}$ in Fig.~\ref{fig:BEC} reveals that this scaling law is very well realized. Therefore the FM to cone transition in this triangular frustrated magnet is a prime example of magnon Bose-Einstein condensation.


\section{Extended variants of the $J_1$-$J_2$ models}
\label{sect:extended}

The genuine square and anisotropic triangular $J_1$-$J_2$ models have obtained such widespread attention because they
 have an attractive property: A single control parameter can tune between ground states
 of quite different nature that are both theoretically interesting and partly realized in
 actual compounds. However, as many of them are 3d-insulators which may have very
 subtle structural distortions from the ideal square or trigonal symmetry it is clear that the 
 simple form of this model is rarely fully adequate. In particular lattice distortions may lead to further
 real-space anisotropies of  the exchange interactions. The spin-orbit interactions can cause spin-space 
 anisotropies of the symmetric exchange and new, smaller terms like Dzyaloshinskii-Moriya (DM) asymmetric
 exchange. Likewise for $S\geq 1$ it may create  a single-ion anisotropy potential for  the spin orientation.
 In addition further neighbor exchange may always be present and even multiple-sites ring exchange can contribute.
From this plethora of possible extensions beyond the genuine models we will discuss only the physically most important
and address in particular the question how they may destabilize the spin liquid formation and reestablish magnetic order.

\subsection{The real-space anisotropic rectangular lattice $J_{1a}$-$J_{1b}$-$J_2$ model}
\label{subsect:anisotropic}

 The introduction of different exchange bonds along the x,y axes of the square lattice (Fig.~\ref{fig:lattices}) lead to the most obvious generalization of the model. The anisotropic n.n. $J_{1a}$-$J_{1b}$ model was discussed first within interacting spin wave theory in Ref.~\cite{igarashi:05}. The frustrated model with nonzero n.n.n.{} exchange $J_2$
has later been much investigated in context of the magnetism of Fe-pnictide parent compounds. This is motivated by two facts: firstly electronic structure calculations \cite{han:09} show that the ordered moment should be considerably larger, up to a factor of two,  than experimentally observed. This led to the early speculation that a magnetic frustration in the context of a local moment $J_1$-$J_2$ square lattice model is the underlying origin of the moment suppression. Secondly from calculations \cite{han:09} as well as experiments \cite{diallo:09} it was concluded that the n.n. exchange along a,b axes should be very anisotropic, i.e. $J_{1a}\neq J_{1b}$ even allowing for different signs, despite the small orthorhombic distortion. Therefore the  $J_{1a}$-$J_{1b}$-$J_2$ model seemed to be an appropriate starting point \cite{schmidt:10}. However, it was demonstrated that within this model the moment reduction by frustration and quantum fluctuations cannot explain the small ordered moment \cite{schmidt:10,schmidt:11}. Subsequently it became clear that Fe-pnictides are not a good realization of the local moment frustrated anisotropic Heisenberg model. A more itinerant and multi-orbital approach is needed \cite{eremin:10}. In fact the existence of low energy longitudinal magnetic excitations \cite{wang:13} is in conflict with the simple local moment picture and advocates an itinerant model. Nevertheless the original $J_{1a}$-$J_{1b}$-$J_2$ model is an obvious and useful extension to describe frustrated local moment magnetism in insulating rectangular lattice compounds. It is described by the Hamiltonian
 \begin{equation}
 \mathcal{H} = 	J_{1a} \sum_{{\langle ij \rangle}_{a}}^N  \mathbf S_i \cdot \mathbf S_j + 
J_{1b} \sum_{{\langle ij \rangle}_{b} }^N \mathbf S_i \cdot \mathbf S_j + 
J_{2}    \sum_{{\langle ij \rangle}_2}^N \mathbf S_i \cdot \mathbf S_j,
\label{eq:j1j2:ham}
\end{equation}
where ${\langle ij \rangle}_{a}$ and ${\langle ij \rangle}_{b}$ are  nearest neighbor bonds along $a$ and $b$ directions, respectively. It is more convenient to define another angle $\theta$, representing the degree of anisotropy between nearest neighbor interactions along $a$ and $b$. This leads to the compact parametrization in terms of frustration and anisotropy angles $-\pi\leq \phi,\theta \leq \pi$ according to
\begin{eqnarray}
J_{1a} &=& \sqrt{2} J_{\text c} \cos\phi \cos\theta, \nonumber \\
J_{1b} &=& \sqrt{2}J_{\text c} \cos\phi \sin\theta, \\
J_{2} &=& J_{\text c} \sin\phi.  \nonumber 
\label{eqn:param_anis}
\end{eqnarray}
Here $\tan \theta=J_{1a}/J_{1b}$ describes the anisotropy ($\theta=\pi/4$ is the previous isotropic case), $\phi$ stands for the frustration of the average interactions of first and second n.n.{} and $J_{\text c} = \sqrt{\left.\left(J_{1a}^{2}+J_{1b}^{2}\right)\right/2 + J_{2}^{2}}$ is the overall exchange energy scale of the anisotropic model. The classical ground state energies defining the phase diagram in the $\phi,\theta$-plane and as function of field are given by 
\begin{equation}
    \frac{E_{\text{cl}}}{NS^2}=\left\{
      \begin{array}{l@{\qquad}l}
	  J_{1a}+J_{1b}+2J_{2}  &  \mbox{FM}\\
	  2 J_{2} - \left(J_{1a} + J_{1b} \right) - 2 \left(J_{1a} + J_{1b}  \right)	\cos^{2}\Theta_\text{cl}  &  \mbox{NAF}   \\
	  J_{1b}  - \left(J_{1a} + 2J_{2} \right) - 2 \left(J_{1a} + 2 J_{2} \right)	\cos^{2}\Theta_\text{cl}  &  \mbox{CAFa} \\
	  J_{1a}  - \left(J_{1b} + 2J_{2} \right) - 2 \left(J_{1b} + J_{2}    \right)	\cos^{2}\Theta_\text{cl}  &  \mbox{CAFb} \\
      \end{array}
    \right.
    \label{eqn:ecl_anis}
\end{equation}
Here  $\mbox{CAFa,b}$ denote columnar AF  phases with ordering vector $\vec Q$ oriented along different axes which are no longer degenerate due to the anisotropic n.n exchange.  Furthermore  $\cos\Theta_\text{cl}=h/h_{\text s}$ and the saturation fields are given by
\begin{eqnarray}
    \frac{h_{\text s}}{2S} =     
    \left\{
      \begin{array}{l@{\qquad}l}
	  2 \left(J_{1a}+J_{1b}\right) & \mbox{NAF}\\
	  2 \left(J_{1a}+2 J_{2}\right) & \mbox{CAFa}\\
	  2 \left(J_{1b}+2 J_{2}\right) & \mbox{CAFb}
      \end{array}
    \right.
\end{eqnarray}
The quantum corrections $E_\text{zp}$ to the total ground state energy $E_\text{gs} = E_{\text{cl}} + E_{\text{zp}}$ may be calculated using again Eq.~(\ref{eqn:hdiagonal}), now with the $J(\vec k)= J_{1a}\cos k_x+ J_{1b}\cos k_y+2J_2\cos k_x\cos k_y$ exchange function for the anisotropic model. As before  $E_{\text gs}$ may also be obtained from unbiased numerical ED of finite tiles and using finite size scaling.

\begin{figure}
\vspace{0.3cm}
\centering
\includegraphics[width=.5\columnwidth]{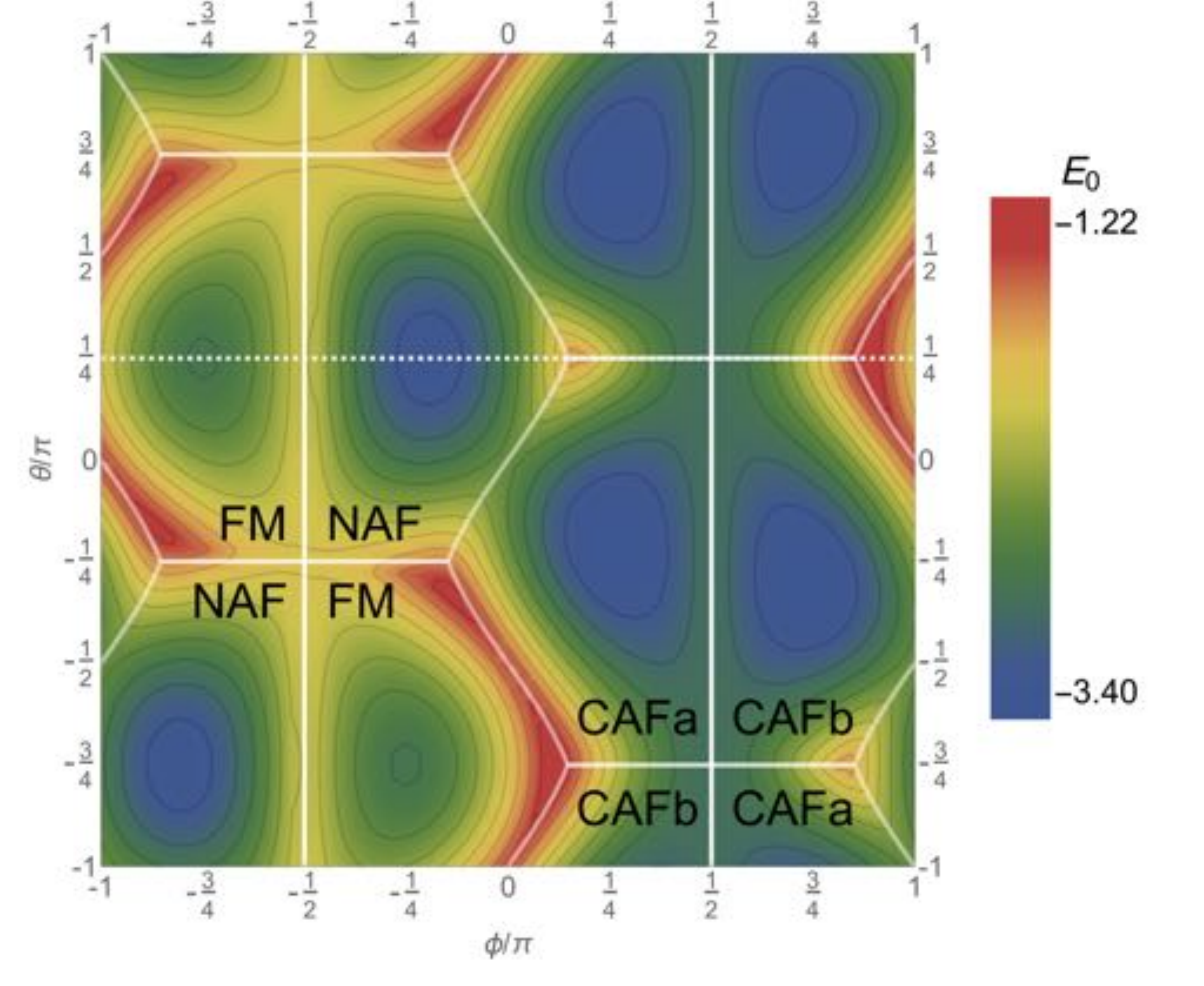} 
 \caption{Contour plot of the ground-state energy from ED as a function of anisotropy ($\theta$) and frustration ($\phi$) parameters for tile 20:2-4. The white lines show the boundaries between the four classical phases, CAFa, CAFb, NAF and FM of the $J_{1a}$-$J_{1b}$-$J_2$ model. The isotropic $J_1$-$J_2$ case corresponds to a cut at $\theta/\pi=1/4$ (dashed line).}
 \label{fig:phasedia_anis}
\end{figure}
The result for tile 20:2-4, overlaid on the classical phase diagram is shown in Fig.~\ref{fig:phasedia_anis}. The regular appearance of the phase diagram is due to the invariance of the zero-field ground state energy $E_\text{gs}(\phi,\theta)$ under various symmetry operations:

i) reflections at the lines $\theta=\frac{\pi}{4}$ and $-\frac{3\pi}{4}$ and inversion at the points $(\phi,\theta)=(\pm\frac{\pi}{2},\frac{3\pi}{4})$ and $(\pm\frac{\pi}{2},-\frac{\pi}{4})$. Both lead to $(J_{1a},J_{1b}) \rightarrow (J_{1b},J_{1a})$ with $J_2$ unchanged. This corresponds to an interchange of the columnar CAFa/b phases, with FM and NAF phases left unchanged.  Thus, it is sufficient to consider only the parameter range $-\pi \leqslant \phi \leqslant \pi$ and $0 \leqslant \theta \leqslant \frac{\pi}{4}$, which can be mapped onto the whole phase diagram applying discrete symmetry operations under which the Hamiltonian in Eq.(~\ref{eq:j1j2:ham}) is invariant.

ii) In the isotropic case ($\theta=\frac{\pi}{4}$ and $-\frac{3\pi}{4}$), CAFa and CAFb are degenerate. By moving away from this symmetry line, one of the two phases is selected. Cutting along a line for fixed anisotropy gives a phase diagram in polar presentation, similar to the isotropic one  in Fig.~\ref{fig:phasedia}.
\begin{figure}
\centering
\includegraphics[width=.6\columnwidth]{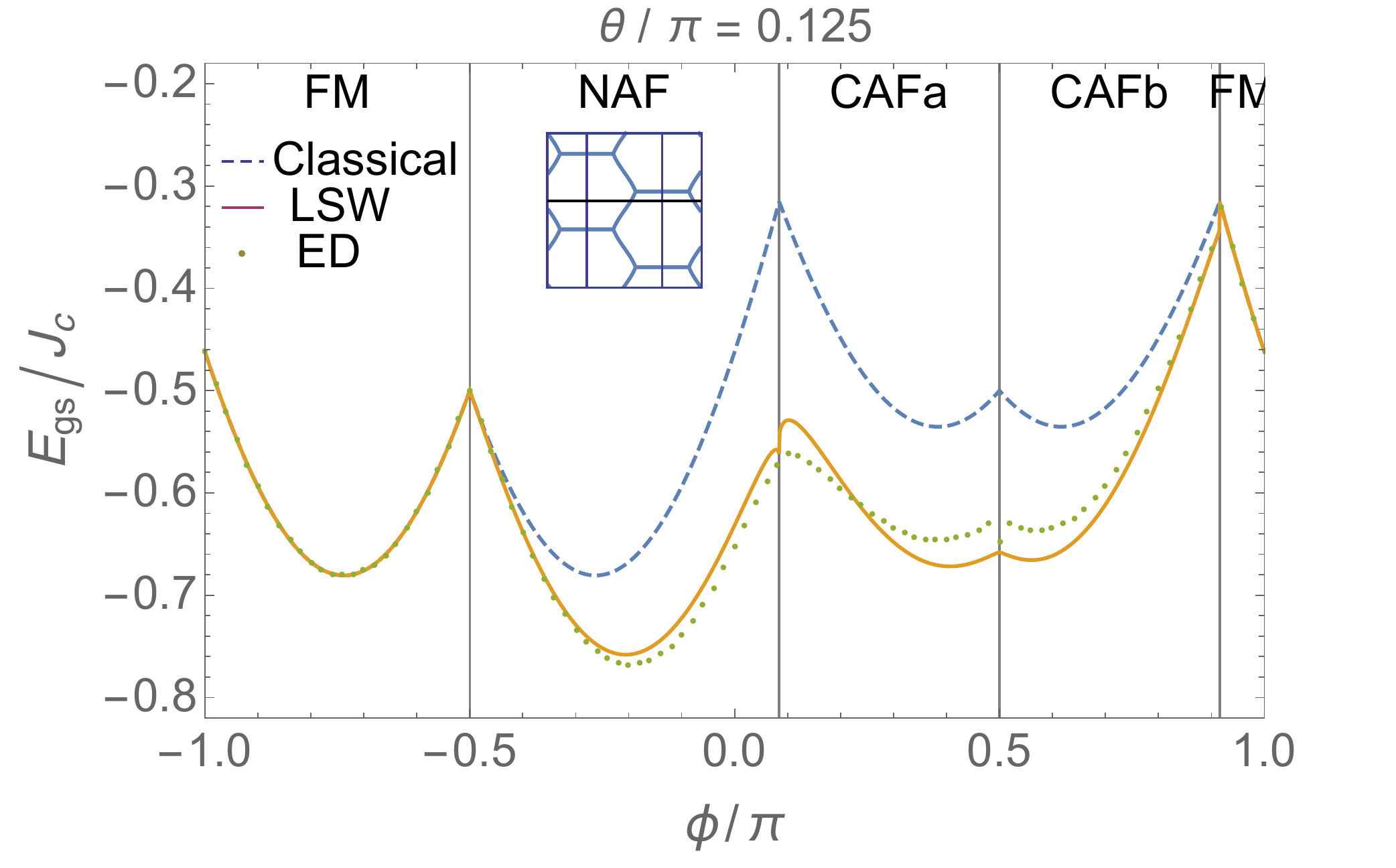} 
 \caption{The ground-state energy as function of the frustration angle $\phi$, for the model with intermediate anisotropy with $\theta = \pi/8$ $(J_{1a}/J_{1b}=0.41)$. It corresponds to a cut shown in the inset where abscissa and ordinate range are same as in Fig.~\ref{fig:phasedia_anis}. This should be compared to the isotropic case $(\theta=\pi/4)$ in Fig.~\ref{fig:groundstate}a.}
 \label{fig:egs_anis}
\end{figure}

As for the isotropic case (Fig.~\ref{fig:groundstate}) it is instructive to compare the ground state energy in the thermodynamic limit obtained from finite size scaling of ED results with that of the analytical LSW treatment. For the intermediate anisotropic case with $\theta/\pi=1/8$ or $J_{1a}/J_{1b}=0.41$ we show $E_\text{gs}(\pi/8,\phi)$ in Fig.~\ref{fig:egs_anis}. A comparison of both methods shows that they give consistent results for the whole range. Apart from the ferromagnetic state, which is an eigenstate to the Hamiltonian, the quantum corrections stabilize the classical ground-state, i.\,e., the zero-point energy (Eq.~\ref{eqn:hdiagonal}) is negative for all values of $\phi$ (Fig.~\ref{fig:egs_anis}). These observations holds for all other anisotropy ratios as well, i.e. for the complete phase diagram  in the $(\theta,\phi)$-plane.

\begin{figure}
\centering
\includegraphics[width=.6\columnwidth]{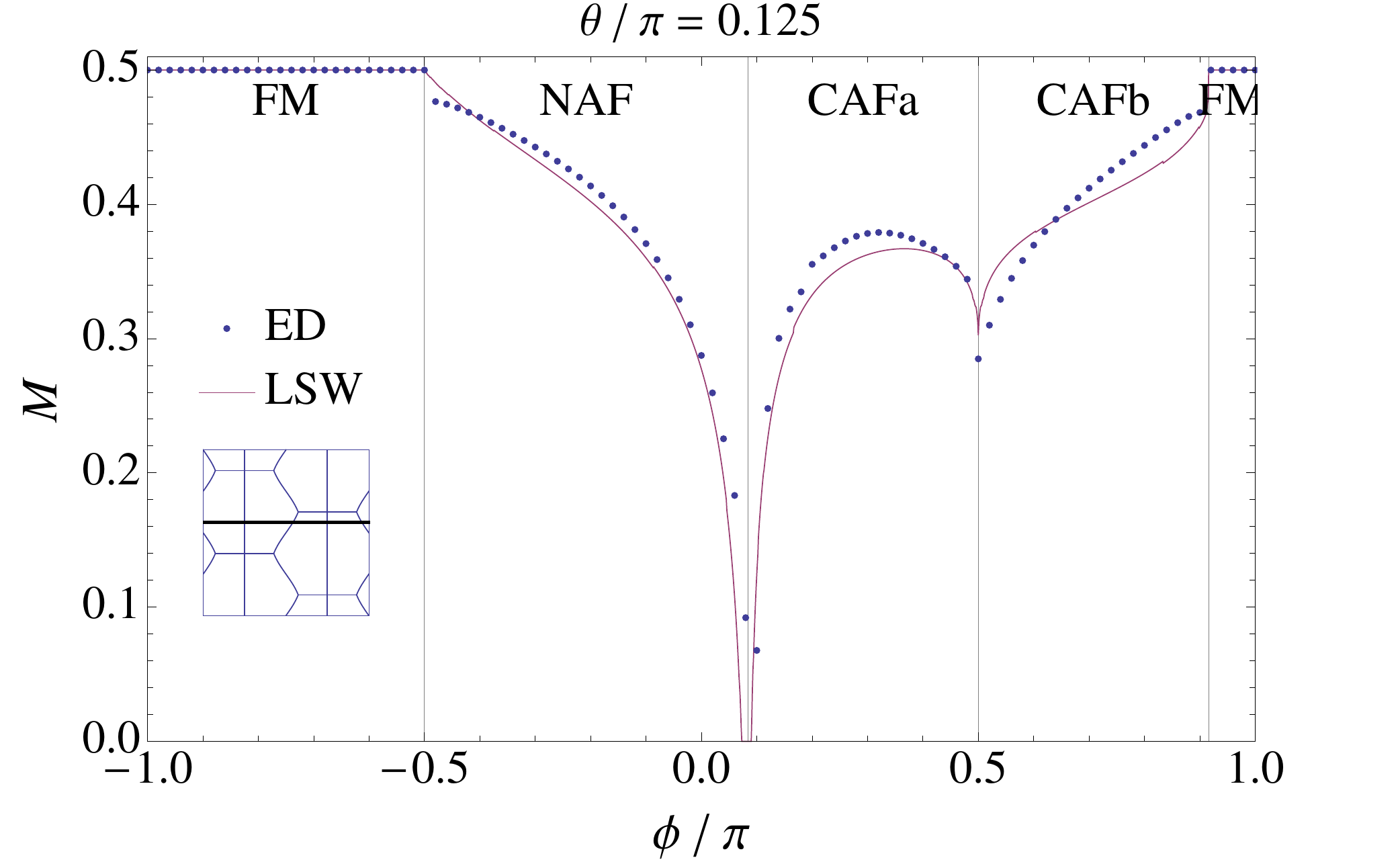} 
 \caption{The ordered moment  $m_\vec Q=M$ as function of the frustration angle $\phi$, for the model with intermediate anisotropy  $\theta = \pi/8$.  The corresponding isotropic case $(\theta=\pi/4)$ is shown in Fig.~\ref{fig:groundstate}(b).The anisotropy generally stabilizes the moments by reducing quantum fluctuations, thereby removing the moment instability in the (spin nematic) regime around $\phi/\pi\simeq 0.85$.}
 \label{fig:moment_anis}
\end{figure}
Both methods may also be applied to investigate the size of the ordered moment. For the intermediate anisotropic case results are shown in  Fig.~\ref{fig:moment_anis}. It is observed that generally in the stable moment regime the moment reduction by quantum fluctuations is less than in the isotropic case (Fig.~\ref{fig:groundstate}b). Apparently the zero point fluctuations are reduced by the anisotropy. This is is also obvious from the reduced width of the instability regimes at NAF/CAFa (spin liquid regime) and CAFb/FM (spin nematic regime). The latter is actually absent for the $\theta/\pi=1/8$ anisotropy, i.e. the spin nematic phase discussed previously for the isotropic model (c.f. Fig.~\ref{fig:groundstate}b) is easily destroyed (the CAFb moment is stabilized) for finite anisotropy. 

\begin{figure}
\centering
\includegraphics[width=.5\columnwidth]{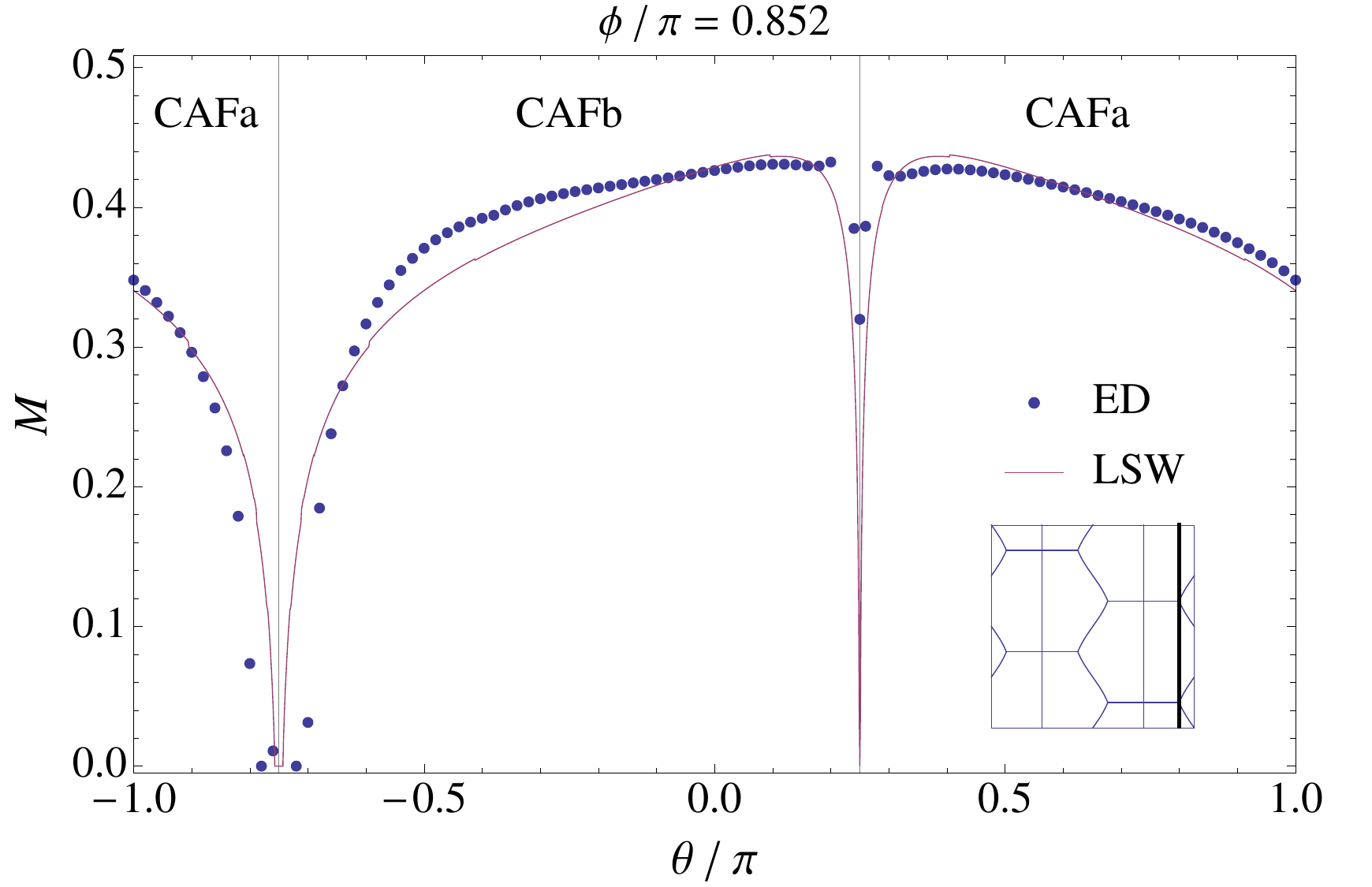} 
 \caption{Ordered moment $m_\vec Q=M$ for $\phi/\pi=0.852$ as a function of the anisotropy parameter $\theta$ (see thick line cut in the inset). The frustration angle $\phi$ is chosen such that at the isotropic point ($\theta=\pi/4$ or $J_{1a}=J_{1b}$), the system is in the disordered (spin nematic) regime at the CAFa/b-FM corner.}
 \label{fig:moment_nematic}
\end{figure}
This is even more strikingly seen in Fig.~\ref{fig:moment_nematic} where we show the anisotropy ($\theta$) dependence of the ordered moment for a fixed $\phi/\pi=0.852$ that corresponds to the spin nematic region at the CAF/FM boundary of the {\em isotropic\/} ($\theta/\pi=1/4$) case  in Fig.~\ref{fig:groundstate}b. Obviously Fig.~\ref{fig:moment_nematic} demonstrates that even a tiny deviation from the isotropic case immediately stabilizes the moment i.e. destroys the spin nematic state. In order that the latter should be realized, two model parameters $\theta,\phi$ controlling anisotropy and frustration should be fine-tuned to a very narrow regime. Therefore it may well be that this phase will rarely be found in a real compound.

\subsection{Spin-space anisotropic xxz triangular exchange model}
\label{subsect:triangular_xxz}

So far we discussed the triangular model with real space anisotropy which is due to anisotropic bond distances and angles. A complementary model is {\em isotropic\/} ($J_1=J_2=J$) in real space, however, has {\em anisotropic\/} exchange constants $J_z\neq J$ for spins oriented along the $z$ and $x$, $y$ axis, respectively in the Hamiltonian of Eq.~(\ref{eqn:h}). This comprises the classical easy plane ($J>J_z$) and easy axis ($J<J_z$) regime. The interesting phase diagram of this model was derived in Ref.~\cite{yamamoto:14} using both numerical large-size cluster mean-field method with scaling scheme (CMF+S) as well as analytical hard-core boson expansion close to the saturation field. It was also investigated using the 2D DMRG method~\cite{sellmann:15} and ED method~\cite{yamamoto:17}. We briefly discuss the hard core boson approach because it qualitatively indicates already how the classical phase diagram is influenced by quantum fluctuations. The representation of $S=1/2$ magnetically ordered states by a Bose condensate in the easy plane case ($J>J_z$)  is, e.g., reviewed in Ref. \cite{zapf:14} and proceeds via the bosonic mapping $S^+_i=a_i$, $S^-_i=a^\dagger_i$, and $S^z_i=S-a_i^\dagger a_i$ of spin operators. The relevant magnetic structure may be built from the condensate amplitude of $a_{\vec k}=\sum_ia_i\exp({\rm i}\vec k \vec R_i)$ given by $\psi_{\pm\vec Q}=\left\langle a_{\pm\vec Q}\right\rangle$. Here $\pm\vec Q=\left(\pm4\pi/3,0\right)$ are possible wave vectors of the (real-space isotropic) triangular $120^\circ$ degree structure which are located on opposite corners of the hexagonal Brillouin zone. In terms of these amplitudes the ground state energy per site  in the dilute boson limit ($0<h_\text s-h\ll h_\text s$) is given by
\begin{eqnarray}
E_\text{gs}/N&=&-(h_\text s-h)(|\psi_{\vec Q}|^2 + |\psi_{-\vec Q}|^2)+
\Gamma_1(|\psi_{\vec Q}|^4 + |\psi_{-\vec Q}|^4)+\nonumber\\
&&2\Gamma_2|\psi_{\vec Q}|^2 |\psi_{-\vec Q}|^2+
2\Gamma_3|\psi_{\vec Q}|^3 |\psi_{-\vec Q}|^3\cos 3\phi
\end{eqnarray}
where $h_\text s=(3/2)J+3J_z$ is the saturation field and $\phi=\mathop{\rm arg}(\psi_{\vec Q}/\psi_{-\vec Q})$ is the relative phase between the two condensate amplitudes which is fixed by the last term of $E_\text{gs}/N$. Depending on the effective interaction parameters $\Gamma_1$-$\Gamma_3$ one can have three cases: i) $\Gamma_1>\Gamma_2$ and $\Gamma_3<0$; both amplitudes are nonzero and $\phi=0$ (0-coplanar structure). ii)  $\Gamma_1>\Gamma_2$ and $\Gamma_3>0$; both amplitudes nonzero and $\phi=\pi$ ($\pi$-coplanar structure). iii)  $\Gamma_1<\Gamma_2$; one of the amplitudes is zero (umbrella structure). The calculation of $\Gamma_i$ is the hard part and in the dilute Bose limit may be performed by calculating the four-particle scattering in ladder approximation \cite{yamamoto:14}. This technique was pioneered before by Nikuni and Shiba \cite{nikuni:95} for the quasi-1D quantum magnets.

\begin{figure}
\centering
\includegraphics[width=.5\columnwidth]{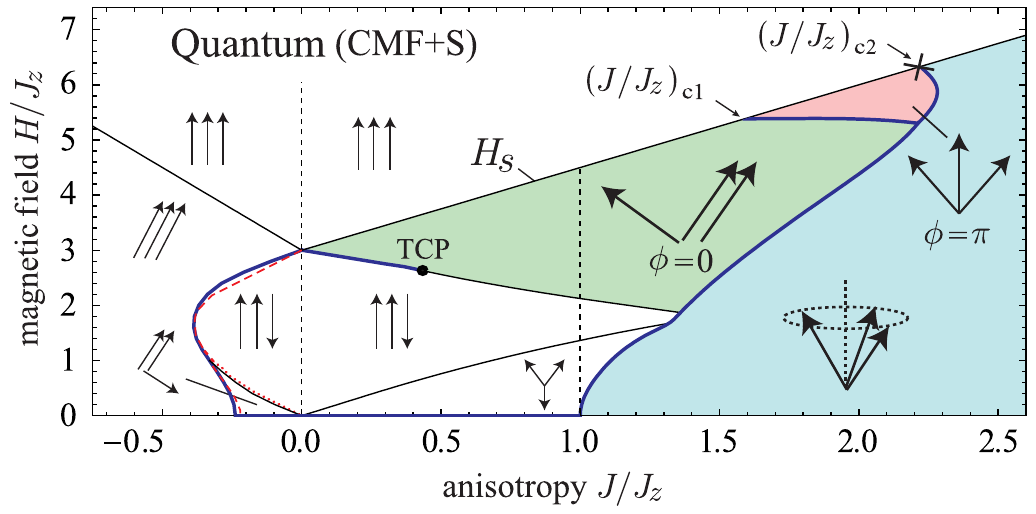} 
 \caption{Phase diagram of the real-space isotropic $(J_1=J_2=J)$ but spin-space-anisotropic $(J_x=J_y=J\neq J_z)$ triangular model. Dashed line at $J_z/J=1$ is classical phase boundary and in this case the whole region for $(J/J_z>1)$ corresponds to the cone (umbrella) state. Coplanar (non-umbrella) phases for $J_z/J>1$ are stabilized by quantum fluctuations. TCP is a tricritical point between first (left) and second (right) order transition lines. (Adapted from Refs.~\cite{yamamoto:14,yamamoto:14-1}).}
 \label{fig:triangular_PDxxz}
\end{figure}
The resulting phase diagram for the present model is then shown in Fig.~\ref{fig:triangular_PDxxz}. Most importantly the introduction of $J/J_z$ anisotropy does not lead to any spin liquid phase, only magnetically ordered ground states are realized. Nevertheless, quantum fluctuations do have an important influence on the phase diagram. In the classical phase diagram the umbrella phase (blue) covers the whole sector to the right $(J/J_z > 1)$ of the dashed line $J/J_z=1$. In the quantum phase diagram this phase boundary  becomes field dependent and reaches essentially to $J/J_z\approx 2.2$ thus widely extending the stability region of coplanar Y, $uud$ and V phases. 

Furthermore a new quantum phase, the coplanar $\phi=\pi$ state (pink region) appears in a small sector for large field between the boundaries $(J/J_z)_{\text c1}=1.588$ and $(J/J_z)_{\text c2}=2.22$. The fully isotropic (real space and spin space) point $J_z/J=1$ corresponds to the point $J_1/J_2=1$ in Fig.~\ref{fig:phasedia_triangular_h} and shows the same sequence of phases Y-$uud$-V for increasing field. The magnetization plateau for $J_z/J=1$ ($h_\text s/JS=9$) in  Fig.~\ref{fig:triangular_PDxxz} has the width $\Delta h/h_\text s =0.171$ which is close to the value  $\Delta h/h_\text s =0.19$ (for $S=1/2$) in Fig.\ref{fig:plateaux_width} and also to the numerical value from ED in Ref.~\cite{sakai:11}. Naturally for Ising type anisotropy $J/J_z<1$ the $uud$ plateau width increases at the expense of Y and V phases.
Until in the pure Ising case $J/J_z=0$ the magnetization develops an uud Ising plateau $m_0/S=1/3$ in the whole field range of the type discussed already in Sec.~\ref{sect:introduction}.

More complicated than uniaxial/easy-plane spin-space exchange anisotropies including asymmetric and non-diagonal exchange terms were considered for triangular YbMgGaO$_4$. This compound has an effective (Yb$^{3+}$-Kramers doublet) pseudo-spin 1/2 and was proposed as a candidate for a triangular spin liquid from ESR~\cite{li:15} and INS~\cite{shen:16}, but there is also compelling evidence for a disorder-induced spin glass state~\cite{ma:17}.

\subsection{Addition of further neighbors: the $J_1$-$J_2$-$J_3$-type  models}
\label{subsect:123}

This is an obvious generalization of the original isotropic square or anisotropic triangular  $J_1$-$J_2$ model which is obtained by adding additional exchange bonds $J_3$ (Fig.~\ref{fig:lattices}).
For the square lattice model depicted in Fig.~\ref{fig:lattices}~a it corresponds to the  $3^\text{rd}$- nearest neighbors along cartesian axes at positions $(\pm 2a,0)$ and $(0,\pm 2a)$. This introduces an additional  2a-periodicity into the exchange function $J(\vec k)$. Therefore in the classical phase diagram, in addition to the FM and commensurate CAF, NAF state generally incommensurate 2D spiral phases characterized by wave vectors $\vec Q=(Q_x,Q_y)$ are possible \cite{chandra:90,seabra:16}. In the special case when $J_1+2J_2-2J_3=0$ ($J_1<0$, $0<2J_3<|J_1|$), only one component is nonzero and has the commensurate value $2\pi/3$ corresponding to a 1D spiral with three-sublattice structure. In this case the exchange Hamiltonian may be mapped to that of an isotropic triangular model \cite{seabra:16} with an effective exchange $J=(1/3)(J_1+2J_2+J_3)$. Indeed it was found by MC simulation that the classical $H$-$T$ phase diagram exhibits phases which are reminiscent to those of the triangular quantum model at zero temperature  \cite{seabra:16}.

\begin{figure}
\centering
\includegraphics[width=.5\columnwidth]{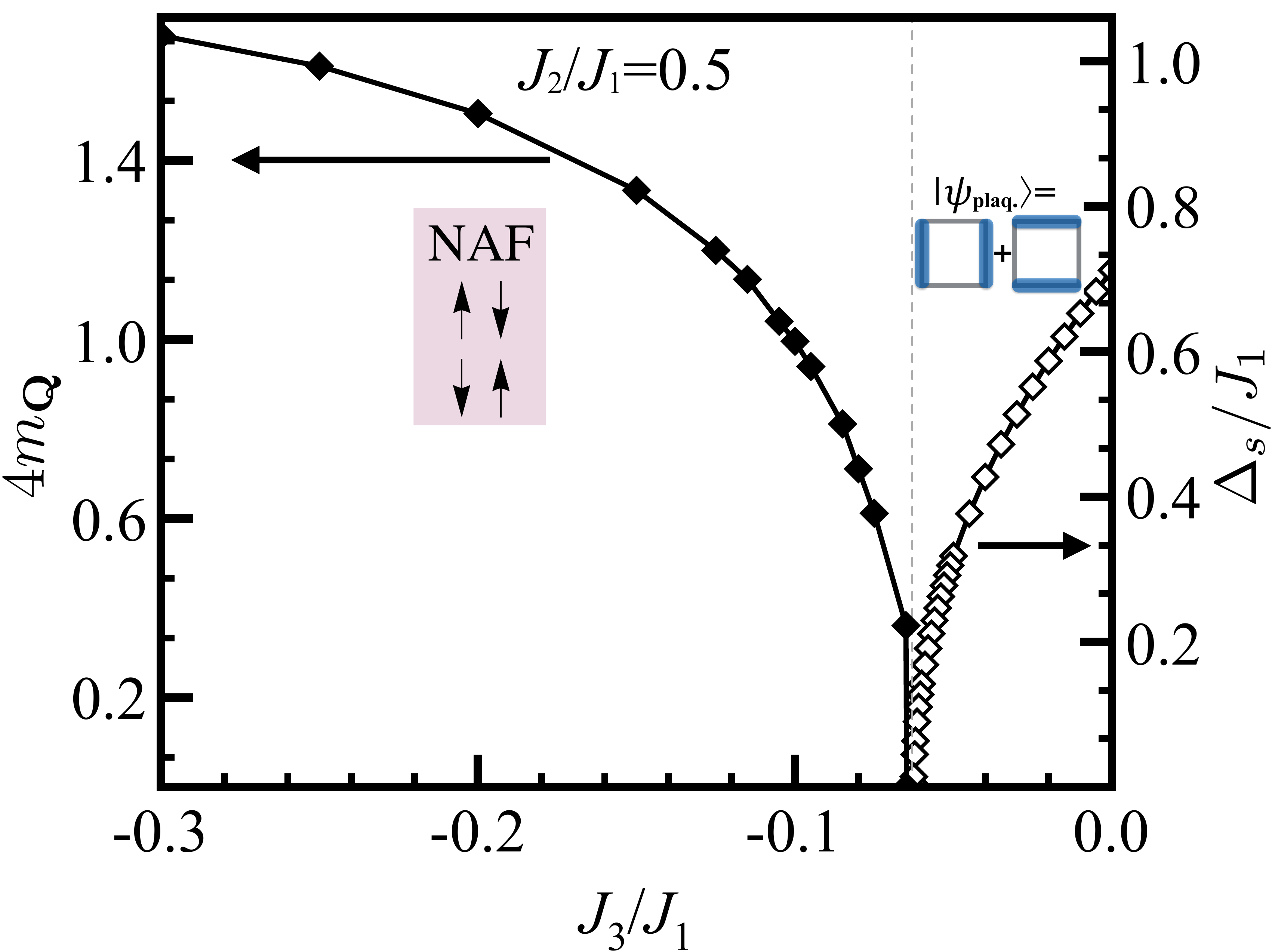} 
 \caption{Spin gap in the dimer phase (right ordinate) and ordered moment in the NAF phase (left ordinate) as function of 
 $3^\text{rd}$-neighbor coupling $J_3$. The nonmagnetic quantum phase is already destroyed for FM $J_3$ at $J_3/J_1\simeq -0.065$
(Adapted from Ref.~\cite{danu:16}).}
 \label{fig:gapmoment_123}
\end{figure}
The most interesting aspect here is the influence of the $J_3$ bond on the quantum phases and we briefly discuss the spin liquid phase around $J_2/J_1=0.5$ $(\phi/\pi=0.15)$ within the spin dimer approach \cite{danu:16}. The spin excitation gap in the dimer phase may be estimated analytically using a singlet-triplon bosonic representation of the model  and a plaquette-factorized variational dimer wave function. The spin gap $\Delta_\text s$ at  $J_2/J_1=0.5$ as function of a FM $J_3 <0$ is shown in Fig.~\ref{fig:gapmoment_123}. The FM $J_3$ coupling favors the moment alignment of third nearest neighbors. Therefore the dimer gap is gradually reduced to zero until triplon condensation is achieved at $J_3/J_1\simeq -0.065$ \cite{danu:16}. For even larger $|J_3|$ the magnetic Neel order is rapidly re-established. This is also the case for all other values of $J_2/J_1$ lying in the spin liquid sector of Fig.~\ref{fig:phasedia}.

\begin{figure}
\centering
\includegraphics[width=.5\columnwidth]{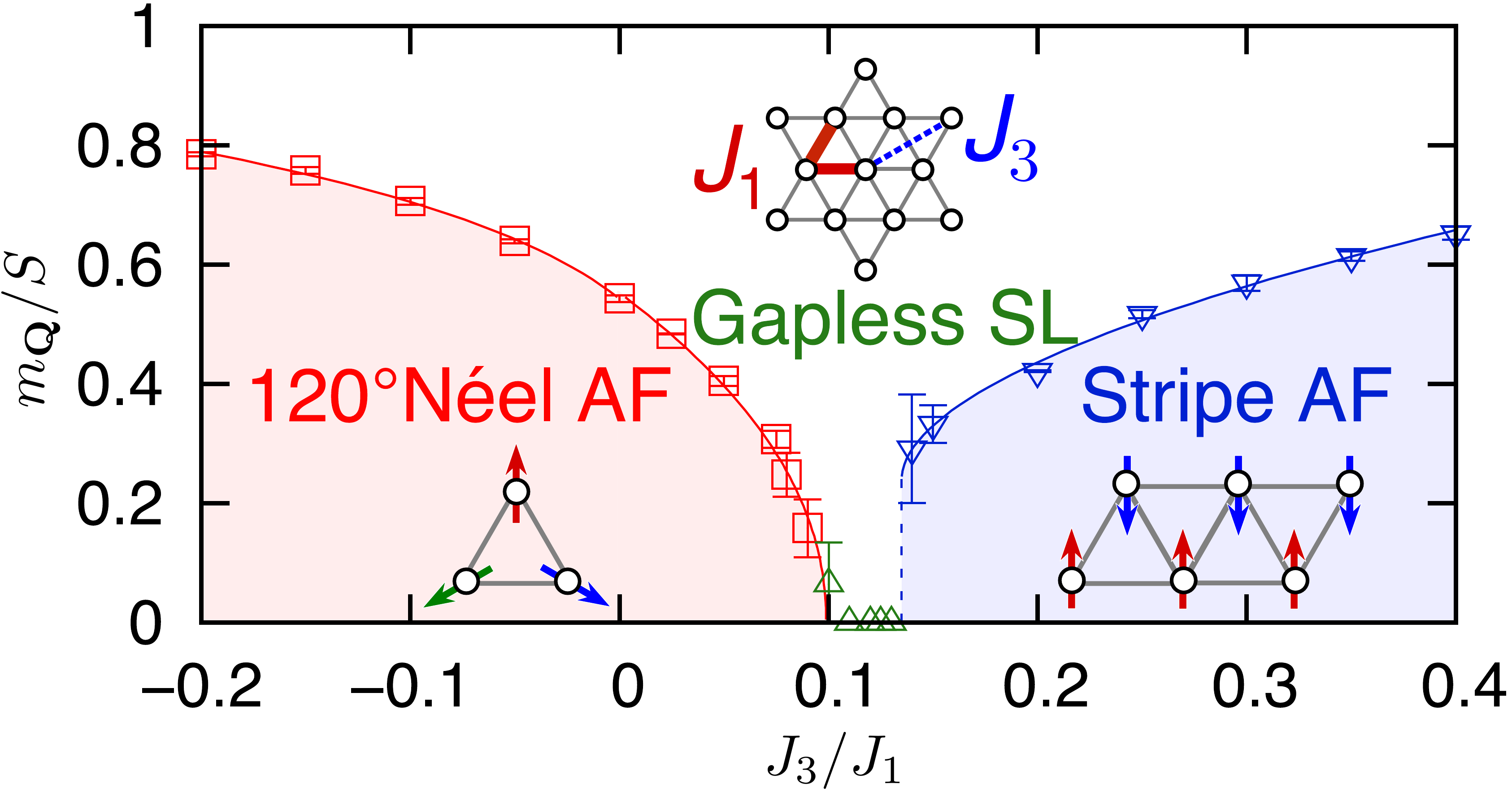} 
 \caption{Ordered moment for isotropic $(J_1=J_2)$ triangular model with next neighbor $J_3$ coupling. A gapless algebraic spin liquid phase appears in a narrow interval above $J_3/J_1\approx 0.1$ between the magnetic ground states.
(Adapted from Ref.~\cite{kaneko:14} ).}
 \label{fig:PD_triangular123}
\end{figure}

As in the square lattice model one may introduce further neighbor $J_3$ interactions in the triangular model along the diagonals between symmetry axes.  (Fig.~\ref{fig:lattices}). The general model was investigated with modified spin wave theory \cite{hauke:11} and 
the isotropic n.n. model ($J_1=J_2$) with variational Monte Carlo methods \cite{kaneko:14, iqbal:16-1}. In the latter case when $J_3=0$  one starts from the stable $120^0$ commensurate spiral structure; the spin liquid phase originally suspected does not exist in this case. To achieve the triangular spin liquid one either has to fine tune the anisotropy of exchange parameters to  $0.7\leq J_2/J_1<0.9$ (Table \ref{tbl:spinliquid}). Or, if one keeps isotropic n.n. exchange $J_2=J_1$ then the next nearest neighbor exchange $J_3$ has to be tuned to the narrow interval  $0.1\leq J_3/J_1\leq 0.135$ to achieve a nonmagnetic state (Fig.~\ref{fig:PD_triangular123}). It was proposed as a gapless spin liquid state with algebraic decay of spin correlations \cite{kaneko:14}.

Recapitulating the effect of interlayer-coupling, $J_{1a}$-$J_{1b}$ anisotropy and additional $J_3$ exchange bonds in the square lattice it is clear that these modifications of the original $J_1$-$J_2$ model are all detrimental to the stability of the nonmagnetic quantum (spin-dimer, spin nematic) phases. Even when small in magnitude compared to the $J_\text c=(J_1^2+J_2^2)^\frac{1}{2}$ energy scale they may quickly reestablish the conventional magnetic phases. Therefore spin liquid phases in the square and also triangular lattices may only be achieved by fine tuning exchange parameters and their anisotropies to narrow intervals. This observation can explain why so far no materials have been found which clearly exhibit the quantum phases in these simple lattices, though some of them may seem to be close to them.

\subsection {Further 2D model extensions and modifications}
\label{subsect:furthermodel}

A variety of additional extended variants of the $J_1$-$J_2$ model that have been investigated will only be mentioned without further discussions. We refer to a selection of the literature. A very interesting extension consists in adding a four spin (cyclic or ring) exchange to the model \cite{chubukov:92,lauchli:05}. This model was discussed for a while in context with the anomalous spin wave excitations in tetragonal La$_2$CuO$_4$ \cite{coldea:01,toader:05} which is an unfrustrated ($J_2=0$) square lattice NAF compound. The ring exchange also opens the way to stabilizing staggered dimer phases \cite{lauchli:05}. A similar extension of triangular models including chiral three-site interactions has been investigated which may support a chiral spin liquid phase \cite{wietek:17}.
A further popular extension for $S\geq 1$ exchange models is the addition of biquadratic terms in the Hamiltonian \cite{lauchli:06,tsunetsugu:06} and possibly a single-ion anisotropy potential. This may lead, in addition to the magnetic and bond-nematic phases to the obviously possible local spin-quadrupole phases. They do indeed have numerous experimental realizations, in particular in Ni ($S=1$) compounds \cite{nakatsuji:10}, in contrast to the true bond spin nematic compounds for $S=1/2$ where none has been found so far. 

Instead of increasing the size of the spin $S$ one may form the product space of spin $\vec S$ and pseudo-spin $\vec T$ which describes e.g. degenerate $e_g$ orbital degrees of freedom when the orbital angular momentum is not completely quenched. The intersite exchange interactions may then be expressed in terms of $S_\alpha T_\beta$ $(\alpha,\beta=x,y,z)$ product operators. If their  intersite-couplings are identical one obtains a supersymmetric spin-orbital model with SU(4) symmetry \cite{ohkawa:85,shiina:97}. Such models were also investigated on the triangular lattice \cite{penc:03}. Due to the enhanced importance of combined spin-orbital fluctuations they exhibit a genuine spin-liquid phase for only (isotropic $J_1=J_2$) nearest neighbor interaction $J_1$, contrary to the pure spin model which has the $120^\circ$ magnetic  order.  However, when n.n. interaction $J_3$ is included even in the  SU(4) model the spin liquid state is rapidly suppressed and for $J_3/J_1 >0.12$ a four-sublattice ordered state with yet unspecified order parameter appears \cite{penc:03} .
%


\section{Summary and Outlook}
\label{sect:summary}

In this review physical effects in quasi-2D quantum magnets have been discussed which are due to simultaneous action of magnetic frustration and low dimensional quantum fluctuations. We focused on the generic $J_1$-$J_2$ or  $J_{1a}$-$J_{1b}$-$J_2$ type Heisenberg models in the square and (anisotropic) triangular or rectangular  lattice geometry, respectively which describe frustrated magnetism for $J_2>0$ ($0<\phi<\pi$).  They have  the advantage that a single control parameter, the frustration (square lattice) or anisotropy (triangular lattice) ratio $J_2/J_1$ allows to tune across a rich phase diagram containing the ferromagnet (FM), collinear commensurate antiferromagnets (NAF, CAF), incommensurate spiral phases (SP) and in narrow regions exotic non-magnetic quantum phases. Furthermore there are  numerous  magnetically ordered compound classes that correspond approximately to these simple models.

A convenient way to discuss the model was found in its compact polar representation using the overall energy scale $J_\text c=(J_1^2+J_2^2)^\frac{1}{2}$ and the frustration control parameter $\phi=\tan^{-1}(J_2/J_1)$ $(-\pi\leq \phi\leq \pi)$.  The phase diagram and boundaries as obtained from the classical ground state energy provide a useful orientation. It contains only magnetic phases (Sec.~\ref{sect:classical}).  The a~priori degree of frustration may be obtained from the ratio of ground state energies of frustrated fundamental plaquettes to those of dimer and trimer constituents (Fig.~\ref{fig:frustration}) (Sec.~\ref{sect:frustmodel}). It shows that the frustration is generally absent for $J_2<0$ and has maxima at the classical phase boundaries NAF/CAF and CAF/FM. 

A prominent effect of frustration and quantum fluctuations is therefore the breakdown of magnetism at these classical phase boundaries due to diverging influence of zero point fluctuations. Subsequently stabilization of possible exotic quantum (`spin liquid') phases like dimer crystal or spin nematic phases without long range magnetic correlations may occur. The latter have received enormous amount of attention in the literature. However to our knowledge there are no unambiguous physical realizations of these hypothetical quantum ground states in the simple square and triangular lattices.
We have argued that they may be easily destabilized by deviations from the generic $J_1$-$J_2$ model due to anisotropies and further interactions (Sec.~\ref{sect:extended}).  In particular the spin-nematic order at the CAF/FM boundary in the square lattice (which spontaneously breaks $C_4$ symmetry)  is extremely sensitive to anisotropies of the n.n. interaction $J_1$ (Fig.\ref{fig:moment_nematic}). Likewise the (dimer-) spin liquid state at the NAF/CAF boundary can easily be destroyed by  $3^{rd}$ neighbor couplings (Fig.\ref{fig:gapmoment_123}). In both cases NAF or CAF magnetic order will be reestablished. Therefore we have mostly restricted ourselves to review the physically measurable effects of frustration and quantum fluctuations in the magnetically ordered parts of the phase diagram.

The influence of quantum fluctuations on the ground state can be approximately treated within the LSW which includes effects to first order in $1/S$ (Sec.~\ref{subsect:LSW}). We have discussed the general results for LSW approach in an external field using a local coordinate system for the $1/S$ expansion. The frustration $(\phi)$ and field dependence of quantum corrections  to physical quantities like ground state energy, ordered moment, uniform moment and canting angle have been calculated. The essential ingredient for the corrections is the field dependent spin wave dispersion and its behavior at symmetry points and along lines as function of frustration control parameter.
As an alternative approach  to calculate quantum effects and as a cross-check to LSW method the unbiased numerical exact diagonalization (ED) technique for finite tiles, supplemented by finite size scaling was employed (Sec.~\ref{subsect:ED}). To obtain reliable results powerful criteria for the proper selection of tiles for the  scaling procedure were derived (\ref{sect:ED}).

A comparative evaluation of results for the ground state was discussed in Sec.~\ref{subsect:gsprop}. Excellent agreement  for the lowering of ground state energies in the whole phase diagram occurs. The ordered moment is strongly reduced by quantum fluctuations, depending on the degree of frustration.
The LSW and ED methods give very good agreement in the regions of stable magnetic phases, although they show differences when approaching the strongly frustrated region around the classical phase boundaries where the quantum phases are expected to emerge. There the moment breaks down in a small finite region of frustration parameter $\phi$. The position and width differs between LSW and ED approach and also between other methods (Table~\ref{tbl:spinliquid}).

The uniform magnetization shows similar behavior (Sec.~\ref{subsect:gsprop}). The nonlinearity of magnetization increases with degree of frustration and together with the saturation field may in principle be used to extract the frustration ratio for the stable magnetic regions. Inside the dimer quantum phase region of the square lattice a pronounced magnetization plateau appears at $m_0/S=1/2$ due to the stabilization of colinear {\em uuud\/} structure. The similar phenomenon happens in the triangular case in a wider range of control parameter around the maximally frustrated isotropic point where the {\em uud\/} collinear phase with $m_0/S=1/3$ occurs. These colinear and also coplanar phases are stabilized by the effect of quantum fluctuations which are maximal when the sublattice moments are parallel (Sec.~\ref{sect:magplateaux}). 

A most direct measure of frustration  is the field dependence of the ordered (staggered) moment (Secs.~\ref{subsect:gsprop}, \ref{subsect:Cupz}). Its starting value at zero field is considerably suppressed by zero point fluctuations with a value that strongly depends on the frustration degree. When the field is turned on quantum fluctuations will be reduced and the ordered moment is enhanced until it reaches a maximum from where it decreases again due to the classical effect of sublattice moment canting. The initial slope of ordered moment increase is an excellent measure for the frustration ratio and it has been vindicated by neutron diffraction study of the Cu-pyrazine compound. We think this is a widely applicable technique with great potential for frustrated magnets. The field dependence of the quasi-2D N\'eel temperature (Sec.~\ref{subsect:Cupz}) is a complementary effect. $T_{\text N}$ increases with field due to the resulting reduction of quantum fluctuations. Because they depend on the degree of frustration $T_{\text N}(H)$, like the ordered moment, contains information on the frustration ratio. Furthermore, for large fields approaching saturation the ordered moment has to vanish quasi-2D quantum magnets may easily exhibit reentrance behavior in the $T$-$H$ plane. For Cu-pyrazine these fields are so far out of reach but reentrance could possibly be observable for some of  the oxovanadates, in particular those close to the spin nematic region (Fig.~\ref{fig:phasedia}).

\begin{figure}
    \centering
     \includegraphics[width=.65\columnwidth]{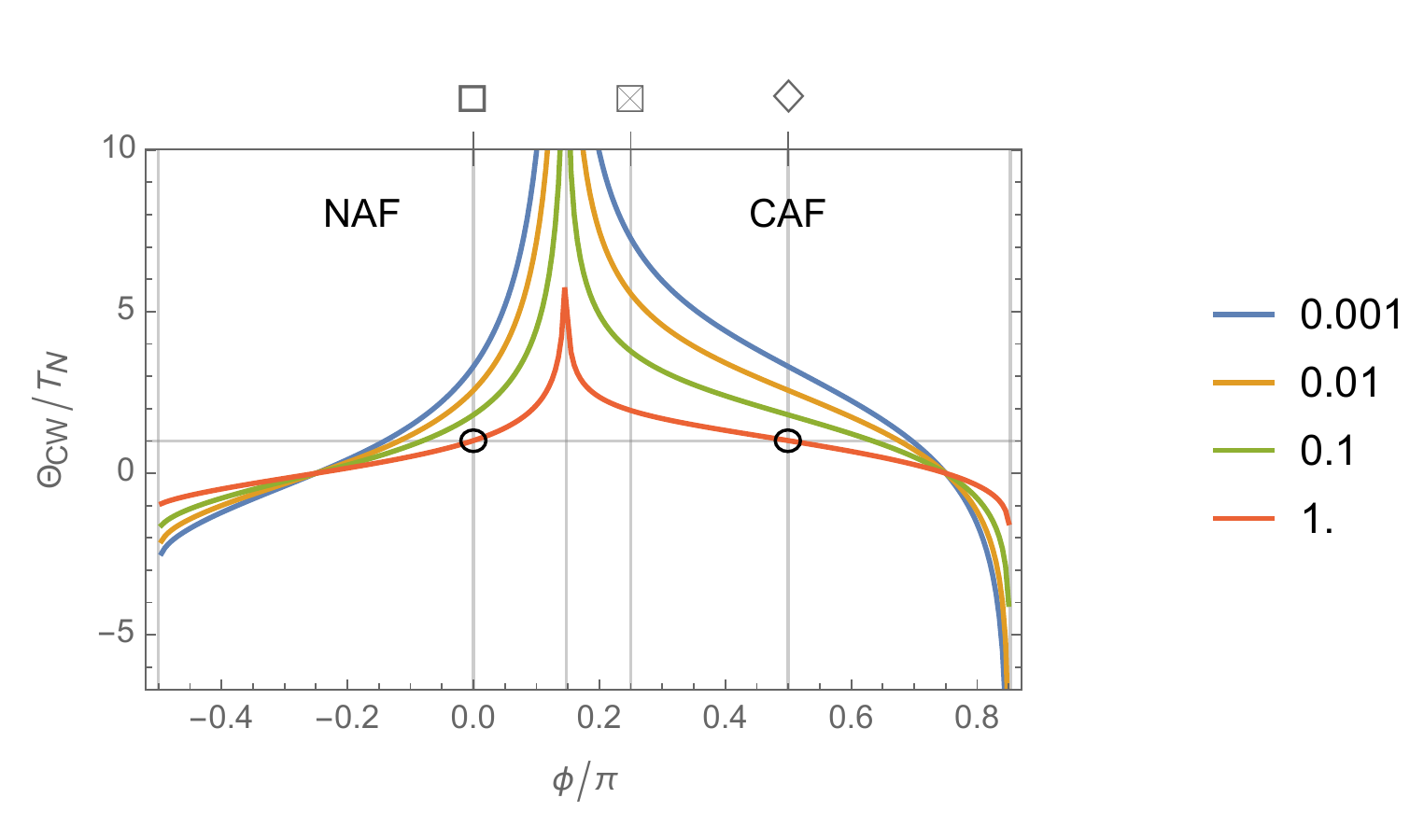}
     \caption{Empirical frustration ratio $f=\Theta_\text{CW}/T_\text N$ as function of $\phi$ for various interlayer coupling 
     strengths $J_\perp/J_\text c$. For 3D isotropic ($J_\perp/J_\text c=1$) pure N\'eel cases (circles) $f=1$. For quasi-2D case
      $(J_\perp/J_\text c\ll 1)$ large enhancement $f\gg 1$ around $\phi/\pi\simeq 0.15$ and $-f\gg 1$ around $\phi/\pi\simeq 0.85$ 
      correlates with the strong frustration in these regimes (c.f. Fig.~\ref{fig:frustration}). Here, $\Theta_\text{CW}=J_1+J_2$.}
    \label{fig:square_f}
\end{figure}

In the introduction it was mentioned that the ratio $f=\Theta_\text{CW}/T_{\text N}$ is frequently used in experimental work as a simple empirical criterion to evaluate the importance of frustration effects. In a common 3D mean-field type magnetism this ratio would be of order one. It is expected that the N\'eel temperature of quasi-2D magnetic order caused by interlayer coupling is strongly suppressed by in-plane frustration leading to a ratio $f$ much larger than one. Using a slight quasi-2D RPA generalization of the LSW in Sec.~\ref{subsect:Cupz} the ordering temperature (Eq.(\ref{eqn:tnh})) and therefore $f$ may be calculated, it is shown in Fig.~\ref{fig:square_f}. The enhancement of the calculated $f$ is clearly observed at the NAF/CAF boundary (dimer spin liquid region, $\phi/\pi\approx 0.15$), similar to the enhanced microscopic frustration degree $\kappa$ in  Fig.~\ref{fig:frustration}. For $\phi/\pi>0.75$ $\Theta_\text{CW}$ becomes negative. Therefore close to the CAF/FM boundary (spin nematic region $\phi/\pi\approx 0.85$) $f$ also acquires negative values and diverges due to the vanishing $T_{\text N}$. This means that $f\gg1$ close to strongly frustrated spin liquid and $-f\gg1$ close to strongly frustrated spin nematic regime.
For the oxovanadates the $\phi$ values (Fig.~\ref{fig:phasedia}) are commonly in between the two regions with $f>1$ only slightly enhanced. However the sign change of $\theta_\text{CW}$ and therefore of $f$ occurs between the last two compounds in Table~\ref{tbl:exchange}.

At finite temperature the unbiased numerical  finite temperature Lanczos method (FTLM) offers many possibilities to investigate the thermodynamic properties of frustrated quantum magnets (Sec.~\ref{sect:temp}). Specific heat, magnetic susceptibility and magnetocaloric cooling rate are the quantities most easily accessible. The former has the experimental drawback that the unknown lattice part must be subtracted, giving an uncertainty to the magnetic part. The latter has so far been only sparsely investigated experimentally for frustrated magnets. 

The most useful quantity is the susceptibility which may be considered in two ways. Firstly the position and height of the susceptibility maximum shows a characteristic variation with the frustration ratio, leading to a low maximum position in the strongly frustrated regimes and likewise a maximum at relatively high temperature for unfrustrated case.
This criterion already gives a preselection for the frustration ratio, however it is not unique by itself because there are always two possible values corresponding to CAF or NAF phase region. It was demonstrated that together with the additional diagnostic tool of the saturation field asymmetry the frustration ratio $J_2/J_1$ or $\phi$ can be uniquely determined.
The magnetocaloric cooling rate calculated for the square lattice model shows considerable enhancement at the saturation field. Surprisingly the maximum enhancement does not occur in the nonmagnetic spin liquid and spin nematic regimes but well inside the stable NAF and CAF sectors. This is due to the enhanced specific heat in the former regions which suppresses the cooling rate coefficient. A larger enhancement may be obtained in more frustrated systems like the kagome lattice \cite{zhitomirsky:03}.

The above mentioned ambiguity in the thermodynamic determination of frustration ratios may also be directly resolved by comparing  the static structure factor from neutron scattering (Sec.~\ref{sect:structurefactor})  with predictions from the ED, FTLM and high temperature expansion. In the quasi-2D ordered magnet the pattern of sharp Bragg peak positions for $T<T_{\text N}$  and their intensity is a finger print for the underlying magnetic structure and shows a characteristic variation with the frustration control parameter. Furthermore, even the broad scattering peaks above $T_N$ due to short range correlations contain enough information to extract the exchange model from a comparison with high temperature series expansion of the structure factor. This program has been successfully carried out for some of the oxovanadate compounds (Sec.~\ref{subsect:Cupz}).

A more complete and most accurate method to determine $J_1$ and $J_2$ uses the fitting of FTLM results to the full experimental susceptibility curve. The strength of FTLM relies on the fact that it still gives a reliable fit for temperatures below the susceptibility maximum but above the finite size gap.  This leads to a high accuracy for the simultaneous determination of the total exchange energy scale $J_\text c$, frustration angle $\phi$  and Landé $g$-factor. Using this procedure for the Cu-chloride and -bromide compounds a highly accurate determination of these parameters in excellent agreement with spectroscopic methods like ESR and INS is possible. This is particularly noteworthy as these anisotropic triangular compounds are close to the disordered quasi-1D spin liquid case. This proves FTLM to be a versatile and efficient method to analyze the thermodynamics of 2D quantum spin systems. When applied to the recent susceptibility data obtained for the whole substitutional series  \CCCB~ \cite{cong:11,cong:14} the systematic trends of exchange constants and frustration ratio with composition may be extracted. Such procedure is a prerequisite tool if one wants to identify a future route to fine tuning the exchange constants in order to approach the narrow intervals of nonmagnetic  quantum phases starting from the magnetically ordered phase regions.

\clearpage
\section*{Acknowledgments}

We want to thank Karlo Penc, Luis Seabra, Nic Shannon, Philippe Sindzingre and Mike Zhitomirsky  for collaboration. Our particular acknowledgments and thanks go to Mohammad Siahatgar whose doctoral thesis has provided many of the results reviewed here. We are also grateful to Christoph Geibel, Enrique Kaul, Michael Lang, Markos Skoulatos, Alexander Tsirlin and Bernd Wolf for sharing their experimental results.

\begin{appendices}

\section{Finite temperature Lanczos method (FTLM)}
\label{sect:ftlm}

In this appendix we give a brief description how partition functions and thermal averages of operators in spin space are
computed in the FTLM method based on ED. The thermodynamic average of an arbitrary time-independent operator $A$ is given by
\begin{eqnarray}
    \left\langle A\right\rangle_{\beta}
    &=&
    \frac{1}{\cal Z}
    \sum_{n=1}^{N_{\text{st}}}\left\langle n\left|
    {\rm e}^{-\beta {\cal H}}A\right|n\right\rangle
    \label{eqn:lancz:trace},
    \\
    {\cal Z} &=& \sum_{n=1}^{N_{\text{st}}}\left\langle n\left| {\rm
    e}^{-\beta {\cal H}}\right|n\right\rangle,
    \label{eqn:thermav}
\end{eqnarray}
where $\cal Z$ is the canonical partition function and $N_{\text{st}}$ is the
dimension of the Hilbert space spanned by the spinor product basis
$\left\{\left|n\right\rangle:n=1\ldots N_{\text{st}}\right\}$
(e.g. $N_{\text{st}}\approx2.7\cdot 10^8 $ for $N=28$).  In each
symmetry-invariant subspace of the full Hilbert space, the eigenvalues
and normalized wave functions are determined by the Lanczos procedure.
To sample an as large as possible part of the Hilbert space, we start
$N_{\text R}={\cal O}(100)$ iterations with different random wave
functions or starting vectors $|r\rangle$, such that eventually we use
no more than ${\cal O}\left(10^{4}\right)$ eigenvalues and wave
functions per Hamiltonian block.  It may be shown~\cite{jaklic:00}
that this procedure yields an asymptotically exact result. 

If the operator $A$ is a conserved quantity with $\left[{\cal
H},A\right]=0$, one can replace $A$ by its quantum numbers
$A_{j}^{r,s}$, and Eqs.~(\ref{eqn:lancz:trace}, \ref{eqn:thermav}) are evaluated as
\begin{eqnarray}
    \langle A\rangle_{\beta}
   & \approx&
    \frac1{\cal Z}\sum_{s}
    \frac{N_{\rm st}^s}{N_{\text R}^s}\sum_{r=1}^{N_{\text R}^s}
    \sum_{j=0}^{M_{\text R}^{s}}
     e^{-\beta\epsilon_{j}^{r}}
     A_{j}^{r,s}
    \left|\langle r_{s}|\psi_{j}^r\rangle\right|^{2}
    \\
        {\cal Z}
    &\approx&
    \sum_{s}
    \frac{N_{\rm st}^s}{N_{\text R}^s}\sum_{r=1}^{N_{\text R}^s}
    \sum_{j=0}^{M_{\text R}^{s}}
    e^{-\beta\epsilon_{j}^{r}}
      \left|\langle r_{s}|\psi_{j}^r\rangle\right|^{2}.   
    \label{eqn:as}
\end{eqnarray}
Here $s$ denotes summation over the independent symmetry sectors. Within each sector containing $N_\text{st}^s$ states, $r$ runs over the $N_\text R^s$ Lanczos procedures with different starting wavefunctions $\left|r_s\right\rangle$. For each tuple $(s,r)$, the iterator $j$ runs over the set of the corresponding Lanczos eigenvectors $\left\{\left|\Psi_j^r\right\rangle\right\}$ with eigenvalues $\left\{\epsilon_j^r\right\}$.
In summary the FTLM replaces the thermal averaging over all microscopic states by an averaging over the low energy Lanczos eigenstates and a simultaneous averaging over the randomly chosen starting vector used to obtain these eigenstates. The most important thermodynamic quantities that may be computed with this procedure are listed in Sec.~(\ref{subsect:Lanczos}).

 An important application where the operator $A$ does not in general commute with the Hamiltonian is the static spin correlation function. Now $A$ refers to a product of spin operators and the expectation value is given by
\begin{align}
	\left\langle S_m^\alpha S_n^\beta\right\rangle_{\beta}
	&\approx
	\frac1{\cal Z}\sum_{s}
	\frac{N_{\rm st}^s}{N_{\text R}^s}\sum_{r=1}^{N_{\text R}^s}
	\sum_{j=0}^{M_{\text R}^{s}}
	e^{-\beta\epsilon_{j}^{r}}
	\left\langle r_{s}|\psi_{j}^{r}\right\rangle
	\left\langle\psi_{j}^{r}| S_m^\alpha S_n^\beta|r_{s}\right\rangle,
	\end{align}
with the same meaning of the symbols as above. The Fourier transform of this product expectation value is the static magnetic structure factor as determined in neutron scattering (Sec.~\ref{sect:structurefactor}).

\section{Technical implementation of exact diagonalization method}
\label{sect:ED}

Exact-diagonalization (ED) methods provide an unbiased way to compute properties of quantum spin models. This is important as it gives a solid check for $T=0$ spin-wave calculations and it extends and complements the results of finite-temperature methods like high-temperature series expansions. The single drawback of ED is the exponential growth of the underlying Hilbert space with increasing system size $N$, restricting us to comparatively small tiles to cover the infinite lattice. In this appendix we describe to some detail the important technical issues connected with the unbiased exact diagonalization method on finite tiles. 

\subsection{Generating and classifying all possible tilings of the square and triangular lattice}
\label{sect:tilings}

At zero temperature ED is useful only when combined with a proper finite-size scaling analysis. This requires clear selection criteria for which tiles should be included in the scaling analysis. At finite temperatures, a finite-size scaling analysis is not possible, however the same principles must be applied to select the tiles for the evaluation of the partition function and the thermodynamic traces involved. Several authors use different approaches to address the tile selection problem: Haan et al.~\cite{haan:92} use helical boundary conditions and define an asymmetry parameter comprising the ratio of difference and sum of the lengths of the edge vectors of the tiles considered. This favors diamond-shaped tiles including perfect squares. Avoiding any non-square-shaped tiles~\cite{schulz:92} appears attractive but leaves only a few coverings. ``Topological imperfection'' is yet another possible selection criterion~\cite{betts:99}, however in the final analysis the authors manually choose tiles to be included.

These examples are in no way a complete list but illustrate the main problems: (i) Only very few coverings might withstand the selection criteria, (ii) the procedure contains some arbitrariness which would lead to selecting different tiles for different Hamiltonian parameters, (iii) given just the edge vectors $\vec a_{i}$ of a particular parallelogram, this is not a unique way to cover a lattice: In general, there are many possible generator matrices $M=(\vec a_1,\vec a_2)$ of the same lattice tiling $\Lambda_{M}$. For completeness and due to the importance of this rather technical issue, we recapitulate the essential ingredients of the procedure we are using to overcome these deficiencies. For details, see Refs.~\cite{schmidt:11,lyness:91,domich:87,schrijver:98}.

First, let us specialize the concept of unitary matrices of dimension 2 to unitary matrices with integer coefficients.  These are called {\em unimodular\/} matrices $U$, which are defined as integer matrices with determinant $\pm1$ and form a (non-abelian) group $U_2$ under matrix multiplication. It has been proven~\cite{schrijver:98} that (a) an arbitrary integer matrix $M$ can be uniquely decomposed into a unimodular matrix $U$ and a Hermite normal form (HNF) matrix $H$ such that
\begin{equation}
    U\cdot M=H,
    \label{app:eqn:umh}
\end{equation}
and (b) that for two HNF matrices $H$ and $H'$ generating the lattice
tilings $\Lambda_{H}$ and $\Lambda_{H'}$, we have
\begin{equation}
    \Lambda_{H}\equiv\Lambda_{H'}\leftrightarrow H=H'.
\end{equation}
For $2\times2$ matrices, this enables us to generate lists of all possible tilings of the square lattice with a given area $N$. The two-dimensional HNF matrices $H$ have the form
\begin{equation}
    H=\left(
    \begin{array}{cc}
        h_{11}&h_{12}\\
        0&h_{22}
    \end{array}
    \right),
    \quad
    h_{11},h_{22}\ge1
    \ \wedge\ 
    0\le h_{12}<h_{22}
    \label{app:eqn:hnf2}
\end{equation}
representing tiles with the special edge vectors $\vec h_{1}=\left(h_{11},h_{12}\right)$ and $\vec h_{2}=\left(0,h_{22}\right)$ and fixed area $N=h_{11}h_{22}$. For numerical purposes, we implement an algorithm using Eq.~(\ref{app:eqn:hnf2}) alone for building the list of possible tilings.

We want to utilize the concept of the ``squareness'' or ``compactness'' of a tile for selecting the proper tiles for finite-size scaling.  For an integer $2\times 2$ matrix $M$ with row vectors $\vec m_1$ and $\vec m_2$, we define a parameter
\begin{equation}
    \rho(M)=
    4
    \frac{\mathop{\rm Det}(\vec m_1,\vec m_2)}{\left(
    \left|\vec m_1\right|+\left|\vec m_2\right|
    \right)^2}
    \label{app:eqn:squareness}
\end{equation}
which measures the ``compactness'' or squareness of the parallelogram spanned by the row vectors of $M$: We have $0\le\rho(M)\le1$, and $\rho(M)=1$ if and only if $M$ describes a square, which we regard as the ``most compact'' tiling of a two-dimensional lattice (see Eq.~(\ref{eqn:squareness}) for an interpretation).

\begin{figure}
    \centering
    \includegraphics[width=.5\textwidth]{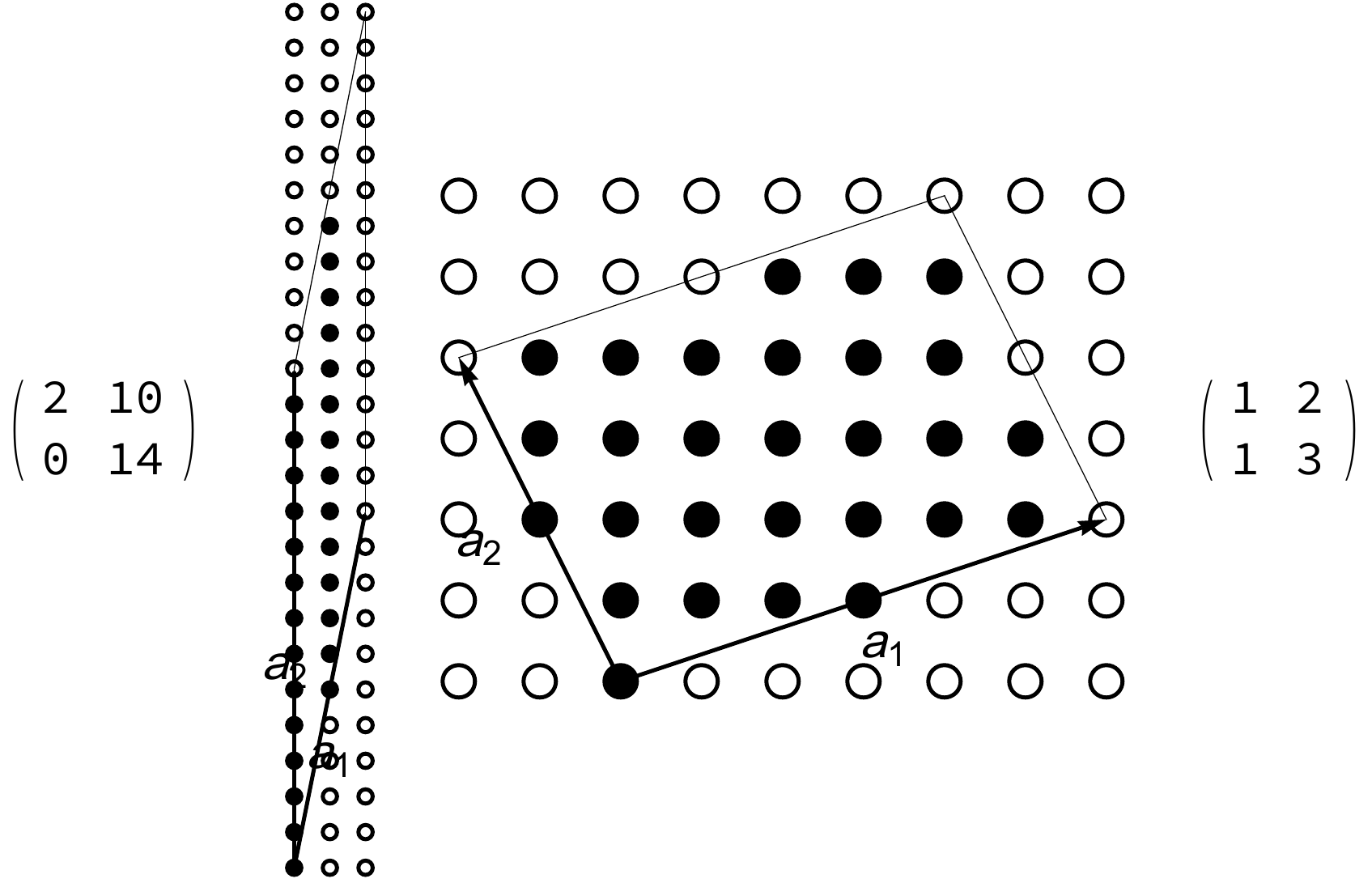}
    \caption{Lattice tiling 28:2-10.  Shown are the HNF matrix $H$, a graphical
    sketch of the HNF tile, a graphical sketch of the compact tile representing $M_\text c$,
    and the matrix $U^{-1}$ mapping $H$ onto $M_{\text c}$.}
    \label{fig:hnf8}
\end{figure}
Calculating $\rho(H)$ for the HNF representation of a lattice tiling
is not very useful: According to Eq.~(\ref{app:eqn:umh}), a single HNF
matrix $H$ represents a whole class ${\cal C}_{H}$ of tiles
with matrix representation $M$ which all lead to an identical
lattice tiling $\Lambda_{H}=\Lambda_{M}$.  But in general,
two matrices $M\ne M'$ with $\Lambda_{M}=\Lambda_{M'}$ have
$\rho(M)\ne\rho(M')$.  From each ${\cal C}_{H}$, we therefore have to
choose a tile which has the maximum compactness of all tiles of its
class,
\begin{equation}
    M_{\text c}=M:\rho(M)=\max_{{\cal T}\in{\cal 
    C}_{H}}\left(\rho(M_{{\cal T}})\right)
\end{equation}
and assign $\rho(M_{\text c})$ to the HNF representation $H=UM_{\text  c}$ of its class ${\cal C}_{H}$. This criterion is convenient to implement numerically, and we use it directly in our tile selection procedure.  To label the tiles in a unique way, we use the scheme $N$:$h_{11}$-$h_{12}$. As an example, Figure~\ref{fig:hnf8} illustrates the lattice tiling 28:2-10.  Shown are the HNF matrix $H$  together with its graphical representation, the corresponding compact tile representing $M_\text c$, and the unimodular matrix $U^{-1}$ needed to map $H$ onto $M_\text c$.

\subsubsection*{Wave vectors for arbitrary parallelogram tiles:}
Assume the edge vectors of a tile have the form
\begin{equation}
    \vec a_{1}=a_{11}\vec e_{x}+a_{12}\vec e_{y},\quad
    \vec a_{2}=a_{21}\vec e_{x}+a_{22}\vec e_{y}
\end{equation}
in cartesian coordinates. For the corresponding basis vectors of the 
reciprocal lattice
\begin{equation}
    \vec b_{1}=b_{11}\vec e_{x}+b_{12}\vec e_{y},\quad
    \vec b_{2}=b_{21}\vec e_{x}+b_{22}\vec e_{y}
\end{equation}
we get from the orthogonality condition
$
    \vec a_{i}\cdot\vec b_{j}=2\pi\delta_{ij}
$
the formulas
\begin{eqnarray}
    b_{11}=\phantom{-}\frac{2\pi}{N}\,a_{22},
    &&
    b_{12}=-\frac{2\pi}{N}\,a_{21},
    \label{eqn:b1}
    \\
    b_{21}=-\frac{2\pi}{N}\,a_{12},
    &&
    b_{22}=\phantom{-}\frac{2\pi}{N}\,a_{11}.
    \label{eqn:b2}
\end{eqnarray}
The denominator in the equations above is just the number of sites of
the tile, or the area of the parallelogram spanned by $\vec a_{1}$ 
and $\vec a_{2}$, 
\begin{equation}
    N=
    \left|\mathop{\rm Det}\left(\vec a_{1},\vec a_{2}\right)\right|=
    \left|a_{11}a_{22}-a_{12}a_{21}\right|.
    \label{eqn:n}
\end{equation}

For a translation vector $\vec r=x_{1}\vec a_{1}+x_{2}\vec a_{2}$ and
a wave vector $\vec k=k_{1}\vec b_{1}+k_{2}\vec b_{2}$, we then have 
$
    \vec k\cdot\vec r=2\pi\left(k_{1}x_{1}+k_{2}x_{2}\right)
$
by construction.
The coefficients $x_{1}=n_{1}/N$ and $x_{2}=n_{2}/N$ with integer
$n_{1}$ and $n_{2}$ are the projections of the lattice points of a
tile onto the edge vectors, and $k_{1}$ and $k_{2}$ are integers.  As
an example, Fig.~\ref{fig:tile-14} illustrates the real and
reciprocal lattices for tile number 28:2-10.

\subsubsection*{Constructing the Brillouin zone for a finite lattice}
Here we will discuss one possible way to determine the points in 
reciprocal space which make up the Brillouin zone, using 
the fact that a translation in $\vec k$ space by a reciprocal lattice 
vector will not change the phase of a wave function, as required by 
Bloch's theorem.

We construct the first
Brillouin zone such that the origin is located in a corner of it.  This is different from the Wigner-Seitz construction for infinite lattices, where the origin is in the center of the
Brillouin zone.  We cannot apply the Wigner-Seitz construction
directly to an arbitrary finite lattice, because we require the wave
vector $\vec k=0$ to be part of the set of $\vec k$ points generated (which corresponds to fixing a global phase of the wave functions),
and in general the center of a given tile does not have a wave vector
associated with it.

For a reciprocal lattice
vector
\begin{equation}
    \vec G=g_{1}\vec b_{1}+g_{2}\vec b_{2},
    \quad g_{1},g_{2}\in{\mathbb Z},
    \quad \left|\vec G\right|=2\pi
    \label{eqn:app:g}
\end{equation}
we need to find the coefficients $g_{i}$.  For a given translation
vector
\begin{equation}
    \vec r=x_{1}\vec a_{1}+x_{2}\vec a_{2},\quad
    x_{i}=\frac{n_{i}}{N},\quad n_{1},n_{2}\in{\mathbb Z}
    \label{eqn:app:r}
\end{equation}
in direct space, the requirement that the phase of a wave function does not change is equivalent to the condition
\begin{equation}
    \vec G\cdot\vec r=\frac{2\pi}{N}
    \left(n_{1}g_{1}+n_{2}g_{2}\right)
    \equiv2\pi m,\quad m\in{\mathbb Z}.
    \label{eqn:g}
\end{equation}
If we express $\vec r$ explicitly in the components of $\vec a_{1}$  and $\vec a_{2}$ (see Eq.~(\ref{eqn:app:r})), we have
\begin{eqnarray}
    r_{x} & = & \frac{n_{1}}{N}a_{11}+\frac{n_{2}}{N}a_{21}
    \equiv m_{x},\quad m_{x}\in{\mathbb Z},
    \\
    r_{y} & = & \frac{n_{1}}{N}a_{12}+\frac{n_{2}}{N}a_{22}
    \equiv m_{y},\quad m_{y}\in{\mathbb Z}.
\end{eqnarray}
First reciprocal lattice vector: Setting $m_{x}=1$, $m_{y}=0$, we obtain
\begin{equation}
    n_{1}=a_{22},\quad n_{2}=-a_{12},
    \label{eqn:a}
\end{equation}
and Eq.~(\ref{eqn:g}) reads
\begin{equation}
    \vec G\cdot\vec r=\frac{2\pi}{N}
    \left(a_{22}g_{1}-a_{12}g_{2}\right)
    \equiv2\pi m,\quad m\in{\mathbb Z}.
    \label{eqn:ga}
\end{equation}
The smallest finite value for $m$ is $m=1$, and using Eqs.~(\ref{eqn:n}) and~(\ref{eqn:app:g}) this leads to
\begin{equation}
    a_{22}g_{1}-a_{12}g_{2}\equiv a_{11}a_{22}-a_{12}a_{21}
    \quad\rightarrow\quad
    g_{1}=a_{11},\quad g_{2}=a_{21}
\end{equation}
for the components of the first reciprocal lattice vector, and eventually we have
\begin{equation}
    \vec G_{1}=a_{11}\vec b_{1}+a_{21}\vec b_{2}.
    \label{eqn:g1}
\end{equation}
Similarly for $m_{x}=0$, $m_{y}=1$ we obtain
\begin{equation}
    \vec G_{2}=a_{12}\vec b_{1}+a_{22}\vec b_{2}
    \label{eqn:g2}
\end{equation}
for the components of the second reciprocal lattice vector.

\begin{figure}
    \centering
    \hfill
    \includegraphics[width=.25\textwidth]{recip28}
    \hfill
    \includegraphics[width=.3\textwidth]{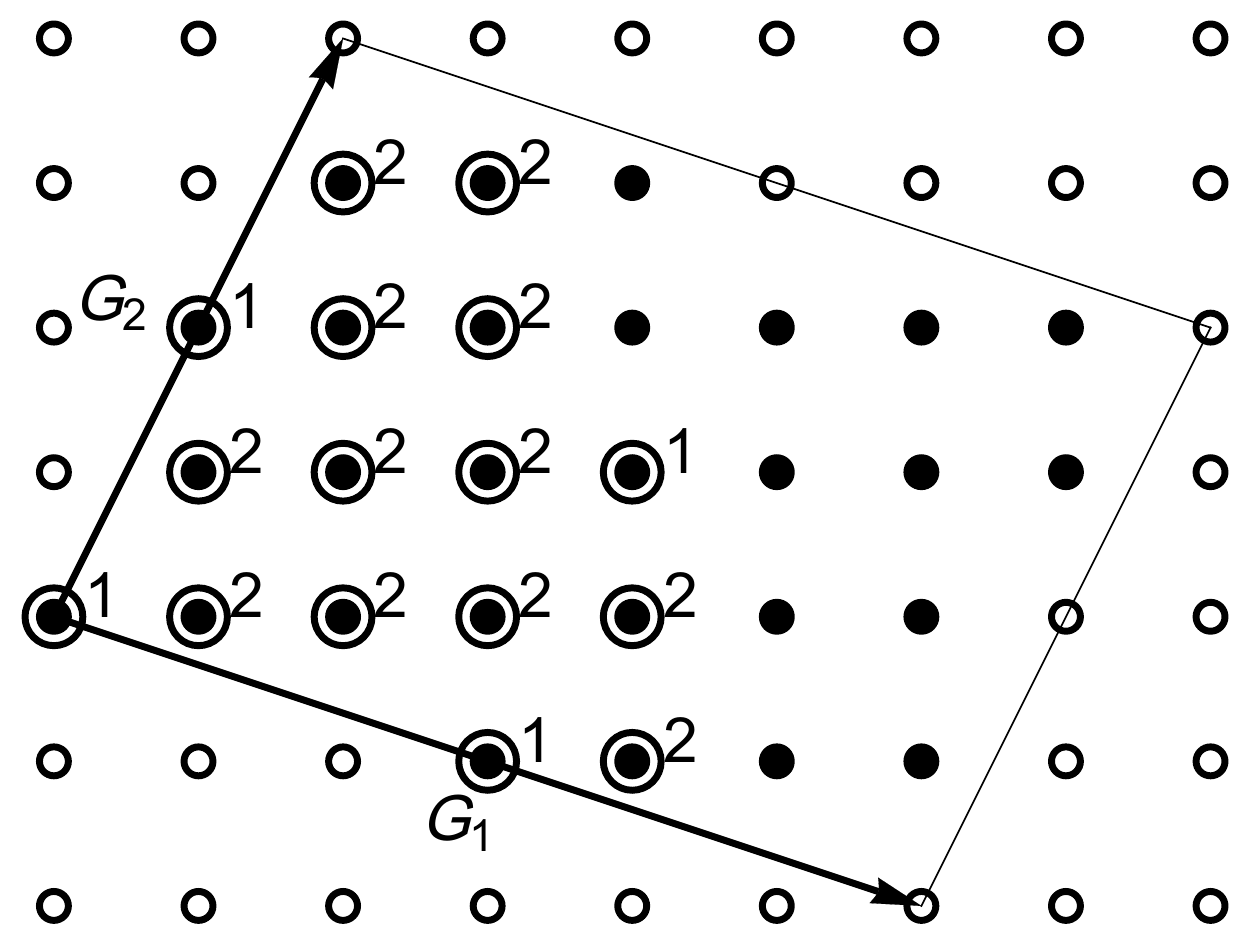}
    \hfill
    \null
    \caption{Reciprocal lattice for tile 28:1-10. Left: cartesian coordinates. Shown are the two reciprocal lattice vectors $\vec G_i$ (Eqs.~(\ref{eqn:g1}) and~(\ref{eqn:g2})) spanning the first Brillouin zone and the basis vectors $\vec b_i$ of the reciprocal lattice (Eqs.~(\ref{eqn:b1}) and~(\ref{eqn:b2})) together with reciprocal lattice points. Right: coordinate system spanned by the basis vectors $\vec b_{1}$ and $\vec b_{2}$. Black dots denote all points in the first Brillouin zone, the encircled dots denote the irreducible wedge (tile 28:2-10 has ${\cal C}_{2}$ symmetry only), numbers give the size of the stars of the corresponding wave vectors.}
    \label{fig:rlocal}
\end{figure}
With respect to the coordinate system of the reciprocal basis vectors,
the reciprocal lattice vectors $\vec G_{1}$ and $\vec G_{2}$ span the
parallelogram making up the first Brillouin zone, which contains
exactly $N$ wave vectors $\vec k=k_{1}\vec b_{1}+k_{2}\vec b_{2}$.
The coefficients $(k_{1},k_{2})$ can be found with the
following criterion: The projections of $\vec k$ onto the reciprocal
lattice vectors $\vec G_{i}$ must be positive semidefinite and less
than the length of the latter, $|\vec G_{i}|=2\pi$. Using Eqs.~(\ref{eqn:g1}) and~(\ref{eqn:g2}), and Eq.~(\ref{eqn:n})
for the area of the Brillouin zone, we get the conditions $0\le\vec k\cdot\vec G_{i}<4\pi^{2}$ ($i=1,2$)
which simplify to
\begin{equation}
    \begin{array}{ccccc}
  0&\le&k_{1}a_{22}-k_{2}a_{12}&<&N,
  \\
  0&\le&-k_{1}a_{21}+k_{2}a_{11}&<&N.
  \end{array}
\end{equation}
Inserting back into the coefficients $g_{i}$ yields 
\begin{equation}
    0\le\mathop{\rm Det}(\vec k,\vec G_{2})<N
    \ \wedge\ 
    0\le-\mathop{\rm Det}(\vec k,\vec G_{1})<N,
    \label{eqn:cond}
\end{equation}
writing the column vectors $\vec G_{i}$ and $\vec k$ in the coordinate
system spanned by the basis vectors $\vec b_{i}$.
Expression~(\ref{eqn:cond}) specifies exactly $N$ wave vectors $\vec
k=k_{1}\vec b_{1}+k_{2}\vec b_{2}$. An example for $N=28$ is given in Fig.~\ref{fig:rlocal}.

\subsubsection*{Space group symmetry}
To avoid unnecessary computations, we calculate Lanczos eigensystems
only at those wave vectors $\vec k$ which are not related by a space
group operation.  Therefore we have to determine the irreducible wedge
of the Brillouin zone.  Figure~\ref{fig:rlocal} illustrates this for
tile 28:1-10, which has ${\cal C}_2$ point group symmetry: The number
of wave vectors is reduced from a total of 28 to 16 inequivalent
ones.  Applying the point group operations to the wave vectors of a
tile (${\cal C}_2$ has only two elements), together with a
translation mapping the tile back to the origin, leaves some $\vec k$
points invariant, while others are mapped onto each other.  The small
numbers in the figure denotes the size of the star of the
corresponding wave vector.

Since we are working with mappings of the full tile, it is useful to
work in a coordinate system of the reciprocal lattice vectors $\vec
G_{1}$ and $\vec G_{2}$.  For the projection of an arbitrary wave
vector $\vec k=\hat k_{1}\vec G_{1}+\hat k_{2}\vec G_{2}$ (not
necessarily located in the first Brillouin zone) onto the reciprocal
lattice vectors, we can write
\begin{equation}
    \hat k_{i}=\frac{n_{i}}{N}+m_{i},\quad n_{i}<N,
    \quad n_{i}\in{\mathbb N}_{0},\ m_{i}\in{\mathbb Z}.
\end{equation}
Mapping $\vec k$ back into the
first Brillouin zone amounts to setting $m_{i}=0$ in the equation
above, equivalent to taking the remainder of $N\hat k_{i}$ when
divided by $N$.  This gives us a convenient and numerically well
defined procedure for mapping the Brillouin zone onto its irreducible
part.

\begin{table*}
    \small
    \[
\begin{array}{cccccccccc}
	    \text{Tile} & \text{NAF} & \text{CAFa} & \text{CAFb} & 
            \Box &
            {\cal C}_{2} & {\cal C}_{2V\,\text{rect}} &
	    {\cal C}_{2V\,\text{dia}} &
	    {\cal C}_{4} & {\cal C}_{4V} \\
	    \hline
 \text{\bf\underline{8:2-2}} & \bullet  & \bullet  & \bullet  &
\text{ 1.000} & \bullet  & \bullet  & \bullet  & \bullet  & \bullet
\\
 \hline
 \text{10:1-3} & \bullet  & - & - & \text{ 1.000} & \bullet  & - & -
& \bullet  & - \\
 \text{10:1-4} & - & - & \bullet  & \text{ 0.966} & \bullet  & - & -
& - & - \\
 \text{10:2-2} & - & \bullet  & - & \text{ 0.966} & \bullet  & - & -
& - & - \\
 \hline
 \text{12:3-0} & - & - & \bullet  & \text{ 0.980} & \bullet  &
\bullet  & - & - & - \\
 \text{12:4-0} & - & \bullet  & - & \text{ 0.980} & \bullet  &
\bullet  & - & - & - \\
 \text{12:1-5} & \bullet  & - & - & \text{ 0.960} & \bullet  & - &
\bullet  & - & - \\
 \text{\bf\underline{12:2-2}} & \bullet  & \bullet  & \bullet  &
\text{ 0.901} & \bullet  & - & - & - & - \\
 \hline
 \text{14:1-3} & \bullet  & - & - & \text{ 0.961} & \bullet  & - & -
& - & - \\
 \text{14:1-4} & - & - & \bullet  & \text{ 0.938} & \bullet  & - & -
& - & - \\
 \text{14:2-3} & - & \bullet  & - & \text{ 0.938} & \bullet  & - & -
& - & - \\
 \hline
 \text{\bf\underline{16:4-0}} & \bullet  & \bullet  & \bullet  &
\text{ 1.000} & \bullet  & \bullet  & \bullet  & \bullet  & \bullet
\\
 \hline
 \text{18:3-3} & \bullet  & - & - & \text{ 1.000} & \bullet  &
\bullet  & \bullet  & \bullet  & \bullet  \\
 \text{18:1-4} & - & - & \bullet  & \text{ 0.975} & \bullet  & - & -
& - & - \\
 \text{18:2-4} & - & \bullet  & - & \text{ 0.975} & \bullet  & - & -
& - & - \\
 \hline
 \text{\bf\underline{20:2-4}} & \bullet  & \bullet  & \bullet  &
\text{ 1.000} & \bullet  & - & - & \bullet  & - \\
 \hline
 \text{22:1-6} & - & - & \bullet  & \text{ 0.981} & \bullet  & - & -
& - & - \\
 \text{22:2-4} & - & \bullet  & - & \text{ 0.981} & \bullet  & - & -
& - & - \\
 \text{22:1-5} & \bullet  & - & - & \text{ 0.961} & \bullet  & - & -
& - & - \\
 \hline
 \text{24:1-10} & - & - & \bullet  & \text{ 0.988} & \bullet  & - & -
& - & - \\
 \text{24:2-5} & - & \bullet  & - & \text{ 0.988} & \bullet  & - & -
& - & - \\
 \text{24:1-7} & \bullet  & - & - & \text{ 0.980} & \bullet  & - &
\bullet  & - & - \\
 \text{\bf\underline{24:4-0}} & \bullet  & \bullet  & \bullet  &
\text{ 0.960} & \bullet  & \bullet  & - & - & - \\
 \hline
 \text{26:1-5} & \bullet  & - & - & \text{ 1.000} & \bullet  & - & -
& \bullet  & - \\
 \text{26:1-10} & - & - & \bullet  & \text{ 0.964} & \bullet  & - & -
& - & - \\
 \text{26:2-5} & - & \bullet  & - & \text{ 0.964} & \bullet  & - & -
& - & - \\
 \hline
 \text{28:1-8} & - & - & \bullet  & \text{ 0.986} & \bullet  & - & -
& - & - \\
 \text{28:4-3} & - & \bullet  & - & \text{ 0.986} & \bullet  & - & -
& - & - \\
 \text{\bf\underline{28:2-4}} & \bullet  & \bullet  & \bullet  &
\text{ 0.961} & \bullet  & - & - & - & - \\
 \hline
 \text{30:5-0} & - & - & \bullet  & \text{ 0.992} & \bullet  &
\bullet  & - & - & - \\
 \text{30:6-0} & - & \bullet  & - & \text{ 0.992} & \bullet  &
\bullet  & - & - & - \\
 \text{30:1-5} & \bullet  & - & - & \text{ 0.974} & \bullet  & - & -
& - & - \\
 \hline
 \text{\bf\underline{32:4-4}} & \bullet  & \bullet  & \bullet  &
\text{ 1.000} & \bullet  & \bullet  & \bullet  & \bullet  & \bullet 
\end{array}
    \]
    \caption{Label, classical phase compatibility, squareness, and
    point groups for selected lattice tilings between 8 and 32 sites.
    For each even area $N$ and for each classical phase, the list
    contains the compatible tile with maximum squareness as defined in
    Eqs.~(\ref{eqn:squareness}) or~(\ref{app:eqn:squareness}).  For $N=12$ and~$24$, the tiles
    compatible with all classical phases are included, too, although
    they have a comparatively small squareness.  The tile labels have
    the form $N$:$h_{11}$-$h_{12}$, where $h_{1j}$ are the components
    of the first edge vector of a tile in HNF representation.  Those
    tiles compatible with all four classical phases, required for the
    discussion of the spatially isotropic model with columnar and Néel order,
    are underlined and typeset in bold.}
    \label{tbl:pg}
\end{table*}
After applying all symmetry operations of a particular point group
$\cal G$ to all wave vectors $\vec k$ of a tile $\cal T$, we have a
list of lists of equivalent wave vectors, which we call equivalence
classes.  Each wave vector belongs to exactly one equivalence class.
If one or more $\vec k$ occur in more than one class,
$\cal T$ is not invariant under $\cal G$, providing us with an
algorithmical criterion for selecting the appropriate point groups for
$\cal T$.  The resulting list is displayed in Table~\ref{tbl:pg}. Each line contains the tile label, its compatibility with classical phases (see below), its squareness according to Eqs.~(\ref{eqn:squareness}) or~(\ref{app:eqn:squareness}), and its point group symmetry.
For
the ${\cal C}_{2V}$ group, two sets of mirror ``planes'' exists: The
point group ${\cal C}_{4V}$ contains two isomorphic subgroups ${\cal
C}_{2V\,\text{rect}}$ and ${\cal C}_{2V\,\text{dia}}$, corresponding
to a tile with either a rectangular shape (mirrors parallel to the
edges) or a diamond-like shape (mirrors along the diagonals).  For
simplicity, we also label those tiles having ${\cal C}_{2V}$ but not
${\cal C}_{4V}$ symmetry accordingly in Table~\ref{tbl:pg}.
    
\subsection{Ordered ground states}
The classical $J_{1}$-$J_{2}$ model on the square lattice has three
well-known ordered ground states (phases): Néel type with ordering
wave vector $\vec Q=(\pi,\pi)$, and two columnar ordered phases CAFa,b
with wave vectors $\vec Q=(\pi,0)$ or $(0,\pi)$.  Although the
corresponding wave functions are not eigenstates of the quantum
Hamiltonian, it is important that the tilings of the infinite lattice
are chosen such that these states corresponding to the classically
ordered phases are not suppressed when applying periodic boundary 
conditions.

We can determine those tiles compatible with a classical ground state 
by applying a test for the existence of the corresponding classical 
ordering vector $\vec Q$ in the list of wave vectors for a given tile,
\begin{equation}
    n_{1}\vec b_{1}+n_{2}\vec b_{2}=\vec Q,\quad n_{i}\in{\mathbb Z}.
\end{equation}
From Eqs.~(\ref{eqn:b1}) and~(\ref{eqn:b2}), we get
\begin{equation}
    \frac{2\pi}{N}\left[
    n_{1}\left(a_{22}\atop{-a_{21}}\right)+
    n_{2}\left({a_{12}}\atop a_{11}\right)
    \right]=
    \left(Q_{1}\atop Q_{2}\right),
    \label{eqn:qtest}
\end{equation}
which has to be fulfilled for integer coefficients $n_{i}$.
    
\begin{itemize}
    \item CAFa phase: $\vec Q=(\pi,0)$, and we obtain
    \begin{equation}
        n_{1}=\frac{a_{11}}{2},\quad
        n_{2}=\frac{a_{21}}{2},
    \end{equation}
    thus the coefficients $a_{11}$ and $a_{21}$ must be even numbers.
    All tiles having edge vectors $\vec a_{1}$ and $\vec a_{2}$
    compatible with this condition contain the columnar ordered state
    along the $x$ direction.  Stated physically, the Mannheim distance $d_\text M(\vec p_i,\vec p_j)$
    of the components of any two lattice points $\vec p_{i}$ and $\vec
    p_{j}$ with $S_{i}^{z}=S_{j}^{z}$ projected onto the $x$ direction
    of the lattice must be even,
    \begin{equation}
  d_{\text M}(p_{ix},p_{jx})=|p_{jx}-p_{ix}|
  \equiv 2n,\quad n\in{\mathbb N}_{0}.
    \end{equation}
    
    \item CAFb phase: $\vec Q=(0,\pi)$. We get from 
    condition~(\ref{eqn:qtest})
        \begin{equation}
        n_{1}=\frac{a_{12}}{2},\quad
        n_{2}=\frac{a_{22}}{2},
    \end{equation}
    so the components $a_{12}$ and $a_{22}$ parallel to $\vec e_{y}$
    must be even numbers.  Just for completeness: this requires
    \begin{equation}
  d_{\text M}(p_{iy},p_{jy})=|p_{jy}-p_{iy}|
  \equiv 2n,\quad n\in{\mathbb N}_{0}.
    \end{equation}
        
    \item Néel phase: $\vec Q=(\pi,\pi)$. We obtain in this case
    \begin{equation}
        n_{1}=\frac{a_{11}+a_{22}}{2},\quad
        n_{2}=\frac{a_{21}+a_{22}}{2},
    \end{equation}
    which is equivalent to 
    \begin{equation}
  d_{\text M}(\vec p_{i},\vec p_{j})=|p_{jx}-p_{ix}|+|p_{jy}-p_{iy}|
  \equiv 2n,\quad n\in{\mathbb N}_{0}.
    \end{equation}
    
    \item All three phases: All components of the edge vectors $\vec
    a_{1}$ and $\vec a_{2}$ must be even individually to be compatible
    with the full classical phase diagram.
\end{itemize}

\end{appendices}

\section*{References}
\bibliography{review}

\begin{thebibliography}{100}
\expandafter\ifx\csname url\endcsname\relax
  \def\url#1{\texttt{#1}}\fi
\expandafter\ifx\csname urlprefix\endcsname\relax\def\urlprefix{URL }\fi
\expandafter\ifx\csname href\endcsname\relax
  \def\href#1#2{#2} \def\path#1{#1}\fi

\bibitem{diep:94}
H.~T. Diep, Magnetic {Systems} with {Competing} {Interactions}, World
  Scientific, 1994.

\bibitem{diep:13}
H.~T. Diep (Ed.), Frustrated {Spin} {Systems}, World Scientific, 2013.

\bibitem{obradors:88}
X.~Obradors, A.~Labarta, A.~Isalgu{\'e}, J.~Tejada, J.~Rodriguez, M.~Pernet,
  Magnetic frustration and lattice dimensionality in
  {$\text{SrCr}_8\text{Ga}_4\text{O}_{19}$}, Solid State Commun. 65~(3) (1988)
  189--192.
\newblock \href {http://dx.doi.org/10.1016/0038-1098(88)90885-X}
  {\path{doi:10.1016/0038-1098(88)90885-X}}.

\bibitem{ramirez:94}
A.~P. Ramirez, Strongly {Geometrically} {Frustrated} {Magnets}, Annu. Rev.
  Mater. Sci. 24~(1) (1994) 453--480.
\newblock \href {http://dx.doi.org/10.1146/annurev.ms.24.080194.002321}
  {\path{doi:10.1146/annurev.ms.24.080194.002321}}.

\bibitem{kaul:05}
E.~E. Kaul, Experimental {Investigation} of {New} {Low}-{Dimensional} {Spin}
  {Systems} in {Vanadium} {Oxides}, Ph.D. thesis, Technische Universit{\"a}t
  Dresden (Jan. 2005).

\bibitem{norman:16}
M.~R. Norman, Herbertsmithite and the search for the quantum spin liquid, Rev.
  Mod. Phys. 88~(4) (2016) 041002.
\newblock \href {http://dx.doi.org/10.1103/RevModPhys.88.041002}
  {\path{doi:10.1103/RevModPhys.88.041002}}.

\bibitem{chernyshev:15}
A.~L. Chernyshev, M.~E. Zhitomirsky, Order and excitations in
  $\text{large}\ensuremath{-}{S}$ kagome-lattice antiferromagnets, Phys. Rev. B
  92~(14) (2015) 144415.
\newblock \href {http://dx.doi.org/10.1103/PhysRevB.92.144415}
  {\path{doi:10.1103/PhysRevB.92.144415}}.

\bibitem{nishimoto:13}
S.~Nishimoto, N.~Shibata, C.~Hotta, Controlling frustrated liquids and solids
  with an applied field in a kagome {Heisenberg} antiferromagnet, Nature
  Communications 4 (2013) ncomms3287.
\newblock \href {http://dx.doi.org/10.1038/ncomms3287}
  {\path{doi:10.1038/ncomms3287}}.

\bibitem{kalz:12}
A.~Kalz, M.~Arlego, D.~Cabra, A.~Honecker, G.~Rossini, Anisotropic frustrated
  {Heisenberg} model on the honeycomb lattice, Phys. Rev. B 85~(10) (2012)
  104505.
\newblock \href {http://dx.doi.org/10.1103/PhysRevB.85.104505}
  {\path{doi:10.1103/PhysRevB.85.104505}}.

\bibitem{zhang:13}
H.~Zhang, C.~A. Lamas, Exotic disordered phases in the quantum
  ${{J}}_{1}$-${{J}}_{2}$ model on the honeycomb lattice, Phys. Rev. B 87~(2)
  (2013) 024415.
\newblock \href {http://dx.doi.org/10.1103/PhysRevB.87.024415}
  {\path{doi:10.1103/PhysRevB.87.024415}}.

\bibitem{li:14}
P.~H.~Y. Li, R.~F. Bishop, C.~E. Campbell, Phase diagram of a frustrated
  spin-$\frac{1}{2}$ ${{J}}_{1}$-${{J}}_{2}$ $\mathit{{X}}{XZ}$ model on the
  honeycomb lattice, Phys. Rev. B 89~(22) (2014) 220408.
\newblock \href {http://dx.doi.org/10.1103/PhysRevB.89.220408}
  {\path{doi:10.1103/PhysRevB.89.220408}}.

\bibitem{rau:16}
J.~G. Rau, E.~K.-H. Lee, H.-Y. Kee, Spin-{Orbit} {Physics} {Giving} {Rise} to
  {Novel} {Phases} in {Correlated} {Systems}: {Iridates} and {Related}
  {Materials}, Annual Review of Condensed Matter Physics 7~(1) (2016) 195--221.
\newblock \href {http://dx.doi.org/10.1146/annurev-conmatphys-031115-011319}
  {\path{doi:10.1146/annurev-conmatphys-031115-011319}}.

\bibitem{koga:00}
A.~Koga, N.~Kawakami, Quantum {Phase} {Transitions} in the
  {Shastry}-{Sutherland} {Model} for
  {${\text{{SrCu}}}_{2}({\text{{BO}}}_{3}{)}_{2}$}, Phys. Rev. Lett. 84~(19)
  (2000) 4461--4464.
\newblock \href {http://dx.doi.org/10.1103/PhysRevLett.84.4461}
  {\path{doi:10.1103/PhysRevLett.84.4461}}.

\bibitem{matsuda:13}
Y.~H. Matsuda, N.~Abe, S.~Takeyama, H.~Kageyama, P.~Corboz, A.~Honecker, S.~R.
  Manmana, G.~R. Foltin, K.~P. Schmidt, F.~Mila, Magnetization of
  ${\text{{srcu}}}_{2}({\text{{bo}}}_{3}{)}_{2}$ in {Ultrahigh} {Magnetic}
  {Fields} up to 118 {T}, Phys. Rev. Lett. 111~(13) (2013) 137204.
\newblock \href {http://dx.doi.org/10.1103/PhysRevLett.111.137204}
  {\path{doi:10.1103/PhysRevLett.111.137204}}.

\bibitem{corboz:14}
P.~Corboz, F.~Mila, Crystals of {Bound} {States} in the {Magnetization}
  {Plateaus} of the {Shastry}-{Sutherland} {Model}, Phys. Rev. Lett. 112~(14)
  (2014) 147203.
\newblock \href {http://dx.doi.org/10.1103/PhysRevLett.112.147203}
  {\path{doi:10.1103/PhysRevLett.112.147203}}.

\bibitem{bishop:12}
R.~F. Bishop, P.~H.~Y. Li, D.~J.~J. Farnell, J.~Richter, C.~E. Campbell,
  Frustrated {Heisenberg} antiferromagnet on the checkerboard lattice:
  ${{J}}_{1}$-${{J}}_{2}$ model, Phys. Rev. B 85~(20) (2012) 205122.
\newblock \href {http://dx.doi.org/10.1103/PhysRevB.85.205122}
  {\path{doi:10.1103/PhysRevB.85.205122}}.

\bibitem{richter:04}
J.~Richter, J.~Schulenburg, A.~Honecker, Quantum magnetism in two dimensions:
  {From} semi-classical {N{\'e}el} order to magnetic disorder, in: Quantum
  {Magnetism}, Lecture {Notes} in {Physics}, Springer, Berlin, Heidelberg,
  2004, pp. 85--153.
\newblock \href {http://dx.doi.org/10.1007/BFb0119592}
  {\path{doi:10.1007/BFb0119592}}.

\bibitem{farnell:14}
D.~J.~J. Farnell, O.~G{\"o}tze, J.~Richter, R.~F. Bishop, P.~H.~Y. Li, Quantum
  ${S}=\frac{1}{2}$ antiferromagnets on {Archimedean} lattices: {The} route
  from semiclassical magnetic order to nonmagnetic quantum states, Phys. Rev. B
  89~(18) (2014) 184407.
\newblock \href {http://dx.doi.org/10.1103/PhysRevB.89.184407}
  {\path{doi:10.1103/PhysRevB.89.184407}}.

\bibitem{kudasov:06}
Y.~B. Kudasov, Steplike {Magnetization} in a {Spin}-{Chain} {System}:
  {${\text{{Ca}}}_{3}{\text{{Co}}}_{2}{\text{{O}}}_{6}$}, Phys. Rev. Lett.
  96~(2) (2006) 027212.
\newblock \href {http://dx.doi.org/10.1103/PhysRevLett.96.027212}
  {\path{doi:10.1103/PhysRevLett.96.027212}}.

\bibitem{bak:82}
P.~Bak, Commensurate phases, incommensurate phases and the devil's staircase,
  Rep. Prog. Phys. 45~(6) (1982) 587.
\newblock \href {http://dx.doi.org/10.1088/0034-4885/45/6/001}
  {\path{doi:10.1088/0034-4885/45/6/001}}.

\bibitem{selke:88}
W.~Selke, The {ANNNI} model {\textemdash} {Theoretical} analysis and
  experimental application, Physics Reports 170~(4) (1988) 213--264.
\newblock \href {http://dx.doi.org/10.1016/0370-1573(88)90140-8}
  {\path{doi:10.1016/0370-1573(88)90140-8}}.

\bibitem{wannier:50}
G.~H. Wannier, Antiferromagnetism. {The} {Triangular} {Ising} {Net}, Phys. Rev.
  79~(2) (1950) 357--364.
\newblock \href {http://dx.doi.org/10.1103/PhysRev.79.357}
  {\path{doi:10.1103/PhysRev.79.357}}.

\bibitem{shirahata:01}
T.~Shirahata, T.~Nakamura, Infinitesimal incommensurate stripe phase in an
  axial next-nearest-neighbor {Ising} model in two dimensions, Phys. Rev. B
  65~(2) (2001) 024402.
\newblock \href {http://dx.doi.org/10.1103/PhysRevB.65.024402}
  {\path{doi:10.1103/PhysRevB.65.024402}}.

\bibitem{sato:13}
M.~Sato, N.~Watanabe, N.~Furukawa, Quasi {Long} {Range} {Order} of {Defects} in
  {Frustrated} {Antiferromagnetic} {Ising} {Models} on {Spatially}
  {Anisotropic} {Triangular} {Lattices}, J. Phys. Soc. Jpn. 82~(7) (2013)
  073002.
\newblock \href {http://dx.doi.org/10.7566/JPSJ.82.073002}
  {\path{doi:10.7566/JPSJ.82.073002}}.

\bibitem{derian:06}
R.~Derian, A.~Gendiar, T.~Nishino, Modulation of {Local} {Magnetization} in
  {Two}-{Dimensional} {Axial}-{Next}-{Nearest}-{Neighbor} {Ising} {Model}, J.
  Phys. Soc. Jpn. 75~(11) (2006) 114001.
\newblock \href {http://dx.doi.org/10.1143/JPSJ.75.114001}
  {\path{doi:10.1143/JPSJ.75.114001}}.

\bibitem{bak:80}
P.~Bak, J.~von Boehm, Ising model with solitons, phasons, and "the devil's
  staircase", Phys. Rev. B 21~(11) (1980) 5297--5308.
\newblock \href {http://dx.doi.org/10.1103/PhysRevB.21.5297}
  {\path{doi:10.1103/PhysRevB.21.5297}}.

\bibitem{selke:84}
W.~Selke, P.~M. Duxbury, The mean field theory of the three-dimensional {ANNNI}
  model, Z. Physik B - Condensed Matter 57~(1) (1984) 49--58.
\newblock \href {http://dx.doi.org/10.1007/BF01679925}
  {\path{doi:10.1007/BF01679925}}.

\bibitem{gendiar:05}
A.~Gendiar, T.~Nishino, Phase diagram of the three-dimensional axial
  next-nearest-neighbor {Ising} model, Phys. Rev. B 71~(2) (2005) 024404.
\newblock \href {http://dx.doi.org/10.1103/PhysRevB.71.024404}
  {\path{doi:10.1103/PhysRevB.71.024404}}.

\bibitem{ohwada:01}
K.~Ohwada, Y.~Fujii, N.~Takesue, M.~Isobe, Y.~Ueda, H.~Nakao, Y.~Wakabayashi,
  Y.~Murakami, K.~Ito, Y.~Amemiya, H.~Fujihisa, K.~Aoki, T.~Shobu, Y.~Noda,
  N.~Ikeda, ``{Devil}'s {Staircase}''-{Type} {Phase} {Transition} in
  {${\text{{NaV}}}_{2}\text{{O}}_{5}$} under {High} {Pressure}, Phys. Rev.
  Lett. 87~(8) (2001) 086402.
\newblock \href {http://dx.doi.org/10.1103/PhysRevLett.87.086402}
  {\path{doi:10.1103/PhysRevLett.87.086402}}.

\bibitem{janssen:87}
T.~Janssen, A.~Janner, Incommensurability in crystals, Advances in Physics
  36~(5) (1987) 519--624.
\newblock \href {http://dx.doi.org/10.1080/00018738700101052}
  {\path{doi:10.1080/00018738700101052}}.

\bibitem{kohgi:00}
M.~Kohgi, K.~Iwasa, T.~Osakabe, Physics of low-carrier system detected by
  neutron and {X}-ray scattering: {Ce}-monopnictides case, Physica B: Condensed
  Matter 281~(Supplement C) (2000) 417--422.
\newblock \href {http://dx.doi.org/10.1016/S0921-4526(99)01051-0}
  {\path{doi:10.1016/S0921-4526(99)01051-0}}.

\bibitem{chattopadhyay:86}
T.~Chattopadhyay, P.~J. Brown, P.~Thalmeier, H.~G.~v. Schnering, Incommensurate
  {Magnetic} {Phase} in {Eu}${\text{{as}}}_{3}$ with {Zone}-{Boundary}
  {Lock}-in, Phys. Rev. Lett. 57~(3) (1986) 372--375.
\newblock \href {http://dx.doi.org/10.1103/PhysRevLett.57.372}
  {\path{doi:10.1103/PhysRevLett.57.372}}.

\bibitem{chen:17}
K.-W. Chen, Y.~Lai, Y.-C. Chiu, S.~Steven, T.~Besara, D.~Graf, T.~Siegrist,
  T.~E. Albrecht-Schmitt, L.~Balicas, R.~E. Baumbach, Possible devil's
  staircase in the {Kondo} lattice {CeSbSe}, Phys. Rev. B 96~(1) (2017) 014421.
\newblock \href {http://dx.doi.org/10.1103/PhysRevB.96.014421}
  {\path{doi:10.1103/PhysRevB.96.014421}}.

\bibitem{nagai:93}
O.~Nagai, S.~Miyashita, T.~Horiguchi, Ground state of the antiferromagnetic
  {Ising} model of general spin {S} on a triangular lattice, Phys. Rev. B
  47~(1) (1993) 202--205.
\newblock \href {http://dx.doi.org/10.1103/PhysRevB.47.202}
  {\path{doi:10.1103/PhysRevB.47.202}}.

\bibitem{zukovic:13}
M.~{\v Z}ukovi{\v c}, Residual entropy of spin-s triangular {Ising}
  antiferromagnet, Eur. Phys. J. B 86~(6) (2013) 283.
\newblock \href {http://dx.doi.org/10.1140/epjb/e2013-40439-x}
  {\path{doi:10.1140/epjb/e2013-40439-x}}.

\bibitem{white:07}
S.~R. White, A.~L. Chernyshev, Ne\'el {Order} in {Square} and {Triangular}
  {Lattice} {Heisenberg} {Models}, Phys. Rev. Lett. 99~(12) (2007) 127004.
\newblock \href {http://dx.doi.org/10.1103/PhysRevLett.99.127004}
  {\path{doi:10.1103/PhysRevLett.99.127004}}.

\bibitem{bramwell:01}
S.~T. Bramwell, M.~J.~P. Gingras, Spin {Ice} {State} in {Frustrated} {Magnetic}
  {Pyrochlore} {Materials}, Science 294~(5546) (2001) 1495--1501.
\newblock \href {http://dx.doi.org/10.1126/science.1064761}
  {\path{doi:10.1126/science.1064761}}.

\bibitem{gingras:14}
M.~J.~P. Gingras, P.~A. McClarty, Quantum spin ice: a search for gapless
  quantum spin liquids in pyrochlore magnets, Rep. Prog. Phys. 77~(5) (2014)
  056501.
\newblock \href {http://dx.doi.org/10.1088/0034-4885/77/5/056501}
  {\path{doi:10.1088/0034-4885/77/5/056501}}.

\bibitem{hiroi:03}
Z.~Hiroi, K.~Matsuhira, S.~Takagi, T.~Tayama, T.~Sakakibara, Specific {Heat} of
  {Kagom{\'e}} {Ice} in the {Pyrochlore} {Oxide} {Dy}2ti2o7, J. Phys. Soc. Jpn.
  72~(2) (2003) 411--418.
\newblock \href {http://dx.doi.org/10.1143/JPSJ.72.411}
  {\path{doi:10.1143/JPSJ.72.411}}.

\bibitem{anderson:56}
P.~W. Anderson, Ordering and {Antiferromagnetism} in {Ferrites}, Phys. Rev.
  102~(4) (1956) 1008--1013.
\newblock \href {http://dx.doi.org/10.1103/PhysRev.102.1008}
  {\path{doi:10.1103/PhysRev.102.1008}}.

\bibitem{pauling:35}
L.~Pauling, The {Structure} and {Entropy} of {Ice} and of {Other} {Crystals}
  with {Some} {Randomness} of {Atomic} {Arrangement}, J. Am. Chem. Soc. 57~(12)
  (1935) 2680--2684.
\newblock \href {http://dx.doi.org/10.1021/ja01315a102}
  {\path{doi:10.1021/ja01315a102}}.

\bibitem{ramirez:99}
A.~P. Ramirez, A.~Hayashi, R.~J. Cava, R.~Siddharthan, B.~S. Shastry,
  Zero-point entropy in {\textquoteleft}spin ice{\textquoteright}, Nature
  399~(6734) (1999) 333--335.
\newblock \href {http://dx.doi.org/10.1038/20619} {\path{doi:10.1038/20619}}.

\bibitem{castelnovo:08}
C.~Castelnovo, R.~Moessner, S.~L. Sondhi, Magnetic monopoles in spin ice,
  Nature 451~(7174) (2008) 42--45.
\newblock \href {http://dx.doi.org/10.1038/nature06433}
  {\path{doi:10.1038/nature06433}}.

\bibitem{eyvazov:17}
A.~B. Eyvazov, R.~Dusad, T.~J.~S. Munsie, H.~A. Dabkowska, G.~M. Luke, E.~R.
  Kassner, J.~C.~S. Davis, A.~Eyal, Common {Glass}-{Forming} {Spin}-{Liquid}
  {State} in the {Pyrochlore} {Magnets} {Dy}$_2${Ti}$_2${O}$_7$ and
  {Ho}$_2${Ti}$_2${O}$_7$, arXiv:1707.09014 [cond-mat].

\bibitem{villain:80}
J.~Villain, R.~Bidaux, J.-P. Carton, R.~Conte, Order as an effect of disorder,
  J. Phys. France 41~(11) (1980) 1263--1272.
\newblock \href {http://dx.doi.org/10.1051/jphys:0198000410110126300}
  {\path{doi:10.1051/jphys:0198000410110126300}}.

\bibitem{shender:82}
E.~F. Shender, Antiferromagnetic garnets with fluctuationally interacting
  sublattices, Sov. Phys. JETP 56~(1) (1982) 178.

\bibitem{hermele:04}
M.~Hermele, M.~P.~A. Fisher, L.~Balents, Pyrochlore photons: {The} {${U}(1)$}
  spin liquid in a ${S}=\frac{1}{2}$ three-dimensional frustrated magnet, Phys.
  Rev. B 69~(6) (2004) 064404.
\newblock \href {http://dx.doi.org/10.1103/PhysRevB.69.064404}
  {\path{doi:10.1103/PhysRevB.69.064404}}.

\bibitem{misguich:12}
G.~Misguich, C.~Lhuillier, Two-dimensional quantum antiferromagnets, in:
  Frustrated {Spin} {Systems}, World Scientific, 2012, pp. 235--319.
\newblock \href {http://dx.doi.org/10.1142/9789814440745_0005}
  {\path{doi:10.1142/9789814440745_0005}}.

\bibitem{starykh:15}
O.~A. Starykh, Unusual ordered phases of highly frustrated magnets: a review,
  Rep. Prog. Phys. 78~(5) (2015) 052502.
\newblock \href {http://dx.doi.org/10.1088/0034-4885/78/5/052502}
  {\path{doi:10.1088/0034-4885/78/5/052502}}.

\bibitem{savary:17}
L.~Savary, L.~Balents, Quantum spin liquids: a review, Rep. Prog. Phys. 80~(1)
  (2017) 016502.
\newblock \href {http://dx.doi.org/10.1088/0034-4885/80/1/016502}
  {\path{doi:10.1088/0034-4885/80/1/016502}}.

\bibitem{yasuda:05}
C.~Yasuda, S.~Todo, K.~Hukushima, F.~Alet, M.~Keller, M.~Troyer, H.~Takayama,
  N\'eel {Temperature} of {Quasi}-{Low}-{Dimensional} {Heisenberg}
  {Antiferromagnets}, Phys. Rev. Lett. 94~(21) (2005) 217201.
\newblock \href {http://dx.doi.org/10.1103/PhysRevLett.94.217201}
  {\path{doi:10.1103/PhysRevLett.94.217201}}.

\bibitem{kanoda:11}
K.~Kanoda, R.~Kato, Mott {Physics} in {Organic} {Conductors} with {Triangular}
  {Lattices}, Annual Review of Condensed Matter Physics 2~(1) (2011) 167--188.
\newblock \href {http://dx.doi.org/10.1146/annurev-conmatphys-062910-140521}
  {\path{doi:10.1146/annurev-conmatphys-062910-140521}}.

\bibitem{dionne:92}
B.~Dionne, M.~Golubitsky, Planforms in two and three dimensions, Z. angew.
  Math. Phys. 43~(1) (1992) 36--62.
\newblock \href {http://dx.doi.org/10.1007/BF00944740}
  {\path{doi:10.1007/BF00944740}}.

\bibitem{bishop:08}
R.~F. Bishop, P.~H.~Y. Li, R.~Darradi, J.~Richter, The quantum
  ${J}_1$-${J}_1'$-${J}_2$ spin-1/2 {Heisenberg} model: influence of the
  interchain coupling on the ground-state magnetic ordering in two dimensions,
  J. Phys.: Condens. Matter 20~(25) (2008) 255251.
\newblock \href {http://dx.doi.org/10.1088/0953-8984/20/25/255251}
  {\path{doi:10.1088/0953-8984/20/25/255251}}.

\bibitem{schmidt:11}
B.~Schmidt, M.~Siahatgar, P.~Thalmeier, Ordered moment in the anisotropic and
  frustrated square lattice {Heisenberg} model, Phys. Rev. B 83~(7) (2011)
  075123.
\newblock \href {http://dx.doi.org/10.1103/PhysRevB.83.075123}
  {\path{doi:10.1103/PhysRevB.83.075123}}.

\bibitem{chandra:90}
P.~Chandra, P.~Coleman, A.~I. Larkin, Ising transition in frustrated
  {Heisenberg} models, Phys. Rev. Lett. 64~(1) (1990) 88--91.
\newblock \href {http://dx.doi.org/10.1103/PhysRevLett.64.88}
  {\path{doi:10.1103/PhysRevLett.64.88}}.

\bibitem{veillette:05}
M.~Y. Veillette, J.~T. Chalker, R.~Coldea, Ground states of a frustrated
  spin-$\frac{1}{2}$ antiferromagnet:
  {${\text{{Cs}}}_{2}\text{{Cu}}{\text{{Cl}}}_{4}$} in a magnetic field, Phys.
  Rev. B 71~(21) (2005) 214426.
\newblock \href {http://dx.doi.org/10.1103/PhysRevB.71.214426}
  {\path{doi:10.1103/PhysRevB.71.214426}}.

\bibitem{chubukov:92}
A.~Chubukov, E.~Gagliano, C.~Balseiro, Phase diagram of the frustrated spin-1/2
  {Heisenberg} antiferromagnet with cyclic-exchange interaction, Phys. Rev. B
  45~(14) (1992) 7889--7898.
\newblock \href {http://dx.doi.org/10.1103/PhysRevB.45.7889}
  {\path{doi:10.1103/PhysRevB.45.7889}}.

\bibitem{lauchli:05}
A.~L{\"a}uchli, J.~C. Domenge, C.~Lhuillier, P.~Sindzingre, M.~Troyer,
  Two-{Step} {Restoration} of {SU}(2) {Symmetry} in a {Frustrated}
  {Ring}-{Exchange} {Magnet}, Phys. Rev. Lett. 95~(13) (2005) 137206.
\newblock \href {http://dx.doi.org/10.1103/PhysRevLett.95.137206}
  {\path{doi:10.1103/PhysRevLett.95.137206}}.

\bibitem{keffer:66}
F.~Keffer, Spin {Waves}, in: Ferromagnetism / {Ferromagnetismus}, Encyclopedia
  of {Physics} / {Handbuch} der {Physik}, Springer, Berlin, Heidelberg, 1966,
  pp. 1--273.
\newblock \href {http://dx.doi.org/10.1007/978-3-642-46035-7_1}
  {\path{doi:10.1007/978-3-642-46035-7_1}}.

\bibitem{manousakis:91}
E.~Manousakis, The spin-1/2{} {Heisenberg} antiferromagnet on a square lattice
  and its application to the cuprous oxides, Rev. Mod. Phys. 63~(1) (1991)
  1--62.
\newblock \href {http://dx.doi.org/10.1103/RevModPhys.63.1}
  {\path{doi:10.1103/RevModPhys.63.1}}.

\bibitem{majlis:07}
N.~Majlis, The {Quantum} {Theory} of {Magnetism}, World Scientific Publishing
  Company, Singapore, 2007.
\newblock \href {http://dx.doi.org/10.1142/6094} {\path{doi:10.1142/6094}}.

\bibitem{schmidt:10}
B.~Schmidt, M.~Siahatgar, P.~Thalmeier, Frustrated local-moment models for iron
  pnictide magnetism, Phys. Rev. B 81~(16) (2010) 165101.
\newblock \href {http://dx.doi.org/10.1103/PhysRevB.81.165101}
  {\path{doi:10.1103/PhysRevB.81.165101}}.

\bibitem{siahatgar:12}
M.~Siahatgar, Frustrated {Quantum} {Magnets} and {Correlated} {Kondo}
  {Systems}, {Frustrierter} {Quantenmagnetismus} und korrelierte {Kondo}
  {Systeme}, Ph.D. thesis, Technische Universit{\"a}t Braunschweig (Jun. 2012).

\bibitem{thalmeier:08}
P.~Thalmeier, M.~E. Zhitomirsky, B.~Schmidt, N.~Shannon, Quantum effects in
  magnetization of {${{J}}_{1}\text{\ensuremath{-}}{{J}}_{2}$} square lattice
  antiferromagnet, Phys. Rev. B 77~(10) (2008) 104441.
\newblock \href {http://dx.doi.org/10.1103/PhysRevB.77.104441}
  {\path{doi:10.1103/PhysRevB.77.104441}}.

\bibitem{schmidt:14}
B.~Schmidt, P.~Thalmeier, Quantum fluctuations in anisotropic triangular
  lattices with ferromagnetic and antiferromagnetic exchange, Phys. Rev. B
  89~(18) (2014) 184402.
\newblock \href {http://dx.doi.org/10.1103/PhysRevB.89.184402}
  {\path{doi:10.1103/PhysRevB.89.184402}}.

\bibitem{shannon:04}
N.~Shannon, B.~Schmidt, K.~Penc, P.~Thalmeier, Finite temperature properties
  and frustrated ferromagnetism in a square lattice {Heisenberg} model, Eur.
  Phys. J. B 38~(4) (2004) 599--616.
\newblock \href {http://dx.doi.org/10.1140/epjb/e2004-00156-3}
  {\path{doi:10.1140/epjb/e2004-00156-3}}.

\bibitem{igarashi:05}
J.-i. Igarashi, T.~Nagao, $1/{S}$-expansion study of spin waves in a
  two-dimensional {Heisenberg} antiferromagnet, Phys. Rev. B 72~(1) (2005)
  014403.
\newblock \href {http://dx.doi.org/10.1103/PhysRevB.72.014403}
  {\path{doi:10.1103/PhysRevB.72.014403}}.

\bibitem{chernyshev:09}
A.~L. Chernyshev, M.~E. Zhitomirsky, Spin waves in a triangular lattice
  antiferromagnet: {Decays}, spectrum renormalization, and singularities, Phys.
  Rev. B 79~(14) (2009) 144416.
\newblock \href {http://dx.doi.org/10.1103/PhysRevB.79.144416}
  {\path{doi:10.1103/PhysRevB.79.144416}}.

\bibitem{auerbach:94}
A.~Auerbach, Interacting {Electrons} and {Quantum} {Magnetism}, Springer, New
  York, 1994.
\newblock \href {http://dx.doi.org/10.1007/978-1-4612-0869-3}
  {\path{doi:10.1007/978-1-4612-0869-3}}.

\bibitem{takahashi:89}
M.~Takahashi, Modified spin-wave theory of a square-lattice antiferromagnet,
  Phys. Rev. B 40~(4) (1989) 2494--2501.
\newblock \href {http://dx.doi.org/10.1103/PhysRevB.40.2494}
  {\path{doi:10.1103/PhysRevB.40.2494}}.

\bibitem{hauke:11}
P.~Hauke, T.~Roscilde, V.~Murg, J.~I. Cirac, R.~Schmied, Modified spin-wave
  theory with ordering vector optimization: spatially anisotropic triangular
  lattice and {J} 1 {J} 2 {J} 3 model with {Heisenberg} interactions, New J.
  Phys. 13~(7) (2011) 075017.
\newblock \href {http://dx.doi.org/10.1088/1367-2630/13/7/075017}
  {\path{doi:10.1088/1367-2630/13/7/075017}}.

\bibitem{zheng:06}
W.~Zheng, J.~O. Fj{\ae}restad, R.~R.~P. Singh, R.~H. McKenzie, R.~Coldea,
  Excitation spectra of the spin-$\frac{1}{2}$ triangular-lattice {Heisenberg}
  antiferromagnet, Phys. Rev. B 74~(22) (2006) 224420.
\newblock \href {http://dx.doi.org/10.1103/PhysRevB.74.224420}
  {\path{doi:10.1103/PhysRevB.74.224420}}.

\bibitem{shannon:06}
N.~Shannon, T.~Momoi, P.~Sindzingre, Nematic {Order} in {Square} {Lattice}
  {Frustrated} {Ferromagnets}, Phys. Rev. Lett. 96~(2) (2006) 027213.
\newblock \href {http://dx.doi.org/10.1103/PhysRevLett.96.027213}
  {\path{doi:10.1103/PhysRevLett.96.027213}}.

\bibitem{schmidt:15}
B.~Schmidt, P.~Thalmeier, Thermodynamics of anisotropic triangular magnets with
  ferro- and antiferromagnetic exchange, New J. Phys. 17~(7) (2015) 073025.
\newblock \href {http://dx.doi.org/10.1088/1367-2630/17/7/073025}
  {\path{doi:10.1088/1367-2630/17/7/073025}}.

\bibitem{thesberg:14}
M.~Thesberg, E.~S. S{\o}rensen, Exact diagonalization study of the anisotropic
  triangular lattice {Heisenberg} model using twisted boundary conditions,
  Phys. Rev. B 90~(11) (2014) 115117.
\newblock \href {http://dx.doi.org/10.1103/PhysRevB.90.115117}
  {\path{doi:10.1103/PhysRevB.90.115117}}.

\bibitem{jaklic:00}
J.~Jakli{\v c}, P.~Prelov{\v s}ek, Finite-temperature properties of doped
  antiferromagnets, Advances in Physics 49~(1) (2000) 1--92.
\newblock \href {http://dx.doi.org/10.1080/000187300243381}
  {\path{doi:10.1080/000187300243381}}.

\bibitem{betts:99}
D.~D. Betts, H.~Q. Lin, J.~S. Flynn, Improved finite-lattice estimates of the
  properties of two quantum spin models on the infinite square lattice, Can. J.
  Phys. 77~(5) (1999) 353--369.
\newblock \href {http://dx.doi.org/10.1139/p99-041}
  {\path{doi:10.1139/p99-041}}.

\bibitem{hasenfratz:93}
P.~Hasenfratz, F.~Niedermayer, Finite size and temperature effects in the {AF}
  {Heisenberg} model, Z. Physik B - Condensed Matter 92~(1) (1993) 91--112.
\newblock \href {http://dx.doi.org/10.1007/BF01309171}
  {\path{doi:10.1007/BF01309171}}.

\bibitem{sandvik:97}
A.~W. Sandvik, Finite-size scaling of the ground-state parameters of the
  two-dimensional {Heisenberg} model, Phys. Rev. B 56~(18) (1997) 11678--11690.
\newblock \href {http://dx.doi.org/10.1103/PhysRevB.56.11678}
  {\path{doi:10.1103/PhysRevB.56.11678}}.

\bibitem{schulz:96}
H.~J. Schulz, T.~a.~L. Ziman, D.~Poilblanc, Magnetic {Order} and {Disorder} in
  the {Frustrated} {Quantum} {Heisenberg} {Antiferromagnet} in {Two}
  {Dimensions}, J. Phys. I France 6~(5) (1996) 675--703.
\newblock \href {http://dx.doi.org/10.1051/jp1:1996236}
  {\path{doi:10.1051/jp1:1996236}}.

\bibitem{siahatgar:11}
M.~Siahatgar, B.~Schmidt, P.~Thalmeier, Staggered-moment dependence on
  field-tuned quantum fluctuations in two-dimensional frustrated
  antiferromagnets, Phys. Rev. B 84~(6) (2011) 064431.
\newblock \href {http://dx.doi.org/10.1103/PhysRevB.84.064431}
  {\path{doi:10.1103/PhysRevB.84.064431}}.

\bibitem{schmidt:07}
B.~Schmidt, P.~Thalmeier, N.~Shannon, Magnetocaloric effect in the frustrated
  square lattice {${{J}}_{1}\text{\ensuremath{-}}{{J}}_{2}$} model, Phys. Rev.
  B 76~(12) (2007) 125113.
\newblock \href {http://dx.doi.org/10.1103/PhysRevB.76.125113}
  {\path{doi:10.1103/PhysRevB.76.125113}}.

\bibitem{bonner:64}
J.~C. Bonner, M.~E. Fisher, Linear {Magnetic} {Chains} with {Anisotropic}
  {Coupling}, Phys. Rev. 135~(3A) (1964) A640--A658.
\newblock \href {http://dx.doi.org/10.1103/PhysRev.135.A640}
  {\path{doi:10.1103/PhysRev.135.A640}}.

\bibitem{honecker:99}
A.~Honecker, A comparative study of the magnetization process of
  two-dimensional antiferromagnets, J. Phys.: Condens. Matter 11~(24) (1999)
  4697.
\newblock \href {http://dx.doi.org/10.1088/0953-8984/11/24/311}
  {\path{doi:10.1088/0953-8984/11/24/311}}.

\bibitem{coletta:13}
T.~Coletta, M.~E. Zhitomirsky, F.~Mila, Quantum stabilization of classically
  unstable plateau structures, Phys. Rev. B 87~(6) (2013) 060407.
\newblock \href {http://dx.doi.org/10.1103/PhysRevB.87.060407}
  {\path{doi:10.1103/PhysRevB.87.060407}}.

\bibitem{oitmaa:06}
J.~Oitmaa, C.~Hamer, W.~Zheng, Series {Expansion} {Methods} for {Strongly}
  {Interacting} {Lattice} {Models} by {Jaan} {Oitmaa}, Cambridge University
  Press, 2006.
\newblock \href {http://dx.doi.org/10.1017/CBO9780511584398}
  {\path{doi:10.1017/CBO9780511584398}}.

\bibitem{hehn:17}
A.~Hehn, N.~van Well, M.~Troyer, High-temperature series expansion for
  spin-$1/2$ {Heisenberg} models, Computer Physics Communications
  212~(Supplement C) (2017) 180--188.
\newblock \href {http://dx.doi.org/10.1016/j.cpc.2016.09.003}
  {\path{doi:10.1016/j.cpc.2016.09.003}}.

\bibitem{misguich:03}
G.~Misguich, B.~Bernu, L.~Pierre, Determination of the exchange energies in
  {${\text{{Li}}}_{2}{\text{{VOSiO}}}_{4}$} from a high-temperature series
  analysis of the square-lattice {${{J}}_{1}{\ensuremath{-}{J}}_{2}$}
  {Heisenberg} model, Phys. Rev. B 68~(11) (2003) 113409.
\newblock \href {http://dx.doi.org/10.1103/PhysRevB.68.113409}
  {\path{doi:10.1103/PhysRevB.68.113409}}.

\bibitem{rosner:03}
H.~Rosner, R.~R.~P. Singh, W.~H. Zheng, J.~Oitmaa, W.~E. Pickett,
  High-temperature expansions for the {${{J}}_{1}\ensuremath{-}{{J}}_{2}$}
  {Heisenberg} models: {Applications} to ab initio calculated models for
  {${\text{{Li}}}_{2}{\text{{VOSiO}}}_{4}$} and
  {${\text{{Li}}}_{2}{\text{{VOGeO}}}_{4}$}, Phys. Rev. B 67~(1) (2003) 014416.
\newblock \href {http://dx.doi.org/10.1103/PhysRevB.67.014416}
  {\path{doi:10.1103/PhysRevB.67.014416}}.

\bibitem{schmidt:11-1}
H.-J. Schmidt, A.~Lohmann, J.~Richter, Eighth-order high-temperature expansion
  for general {Heisenberg} {Hamiltonians}, Phys. Rev. B 84~(10) (2011) 104443.
\newblock \href {http://dx.doi.org/10.1103/PhysRevB.84.104443}
  {\path{doi:10.1103/PhysRevB.84.104443}}.

\bibitem{schmidt:09}
B.~Schmidt, P.~Thalmeier, N.~Shannon, High field properties of ${S}=1/2$ square
  lattice ${J}_1$-${J}_2$ magnetic compounds, J. Phys.: Conf. Ser. 145~(1)
  (2009) 012054.
\newblock \href {http://dx.doi.org/10.1088/1742-6596/145/1/012054}
  {\path{doi:10.1088/1742-6596/145/1/012054}}.

\bibitem{hanebaum:14}
O.~Hanebaum, J.~Schnack, Advanced finite-temperature {Lanczos} method for
  anisotropic spin systems, Eur. Phys. J. B 87~(9) (2014) 194.
\newblock \href {http://dx.doi.org/10.1140/epjb/e2014-50360-5}
  {\path{doi:10.1140/epjb/e2014-50360-5}}.

\bibitem{pecharsky:99}
V.~K. Pecharsky, K.~A. Gschneidner~Jr, Magnetocaloric effect and magnetic
  refrigeration, Journal of Magnetism and Magnetic Materials 200~(1) (1999)
  44--56.
\newblock \href {http://dx.doi.org/10.1016/S0304-8853(99)00397-2}
  {\path{doi:10.1016/S0304-8853(99)00397-2}}.

\bibitem{tishin:99}
A.~M. Tishin, K.~A. Gschneidner, V.~K. Pecharsky, Magnetocaloric effect and
  heat capacity in the phase-transition region, Phys. Rev. B 59~(1) (1999)
  503--511.
\newblock \href {http://dx.doi.org/10.1103/PhysRevB.59.503}
  {\path{doi:10.1103/PhysRevB.59.503}}.

\bibitem{honecker:06}
A.~Honecker, S.~Wessel, Magnetocaloric effect in two-dimensional spin-1/2
  antiferromagnets, Physica B 378~(Supplement C) (2006) 1098--1099.
\newblock \href {http://dx.doi.org/10.1016/j.physb.2006.01.436}
  {\path{doi:10.1016/j.physb.2006.01.436}}.

\bibitem{zhitomirsky:03}
M.~E. Zhitomirsky, Enhanced magnetocaloric effect in frustrated magnets, Phys.
  Rev. B 67~(10) (2003) 104421.
\newblock \href {http://dx.doi.org/10.1103/PhysRevB.67.104421}
  {\path{doi:10.1103/PhysRevB.67.104421}}.

\bibitem{skoulatos:08}
M.~Skoulatos, Spin correlations and orbital physics in vanadates, Ph.D. thesis,
  University of Liverpool (Mar. 2008).

\bibitem{karbach:97}
M.~Karbach, G.~M{\"u}ller, A.~H. Bougourzi, A.~Fledderjohann, K.-H. M{\"u}tter,
  Two-spinon dynamic structure factor of the one-dimensional ${S}=1/2$
  {Heisenberg} antiferromagnet, Phys. Rev. B 55~(18) (1997) 12510--12517.
\newblock \href {http://dx.doi.org/10.1103/PhysRevB.55.12510}
  {\path{doi:10.1103/PhysRevB.55.12510}}.

\bibitem{kaul:04}
E.~E. Kaul, H.~Rosner, N.~Shannon, R.~V. Shpanchenko, C.~Geibel, Evidence for a
  frustrated square lattice with ferromagnetic nearest-neighbor interaction in
  the new compound {$\text{Pb}_2\text{VO}(\text{PO}_4)_2$}, Journal of
  Magnetism and Magnetic Materials 272~(Part 2) (2004) 922--923.
\newblock \href {http://dx.doi.org/10.1016/j.jmmm.2003.12.002}
  {\path{doi:10.1016/j.jmmm.2003.12.002}}.

\bibitem{kini:06}
N.~S. Kini, E.~E. Kaul, C.~Geibel, Zn 2 {VO}({PO} 4 ) 2 : an {S} = 1/2
  {Heisenberg} antiferromagnetic square lattice system, J. Phys.: Condens.
  Matter 18~(4) (2006) 1303.
\newblock \href {http://dx.doi.org/10.1088/0953-8984/18/4/015}
  {\path{doi:10.1088/0953-8984/18/4/015}}.

\bibitem{yusuf:10}
S.~M. Yusuf, A.~K. Bera, N.~S. Kini, I.~Mirebeau, S.~Petit, Two- and
  three-dimensional magnetic correlations in the spin-$\frac{1}{2}$
  square-lattice system
  {${\text{{Zn}}}_{2}\text{{VO}}{({\text{{PO}}}_{4})}_{2}$}, Phys. Rev. B
  82~(9) (2010) 094412.
\newblock \href {http://dx.doi.org/10.1103/PhysRevB.82.094412}
  {\path{doi:10.1103/PhysRevB.82.094412}}.

\bibitem{sushkov:01}
O.~P. Sushkov, J.~Oitmaa, Z.~Weihong, Quantum phase transitions in the
  two-dimensional {${{J}}_{1}{\ensuremath{-}{J}}_{2}$} model, Phys. Rev. B
  63~(10) (2001) 104420.
\newblock \href {http://dx.doi.org/10.1103/PhysRevB.63.104420}
  {\path{doi:10.1103/PhysRevB.63.104420}}.

\bibitem{anderson:73}
P.~W. Anderson, Resonating valence bonds: {A} new kind of insulator?, Mater.
  Res. Bull. 8~(2) (1973) 153--160.
\newblock \href {http://dx.doi.org/10.1016/0025-5408(73)90167-0}
  {\path{doi:10.1016/0025-5408(73)90167-0}}.

\bibitem{kotov:99}
V.~N. Kotov, J.~Oitmaa, O.~P. Sushkov, Z.~Weihong, Low-energy singlet and
  triplet excitations in the spin-liquid phase of the two-dimensional
  {${{J}}_{1}{\ensuremath{-}{J}}_{2}$} model, Phys. Rev. B 60~(21) (1999)
  14613--14616.
\newblock \href {http://dx.doi.org/10.1103/PhysRevB.60.14613}
  {\path{doi:10.1103/PhysRevB.60.14613}}.

\bibitem{singh:99}
R.~R.~P. Singh, Z.~Weihong, C.~J. Hamer, J.~Oitmaa, Dimer order with striped
  correlations in the {${{J}}_{1}{\ensuremath{-}{J}}_{2}$} {Heisenberg} model,
  Phys. Rev. B 60~(10) (1999) 7278--7283.
\newblock \href {http://dx.doi.org/10.1103/PhysRevB.60.7278}
  {\path{doi:10.1103/PhysRevB.60.7278}}.

\bibitem{gelfand:90}
M.~P. Gelfand, Series investigations of magnetically disordered ground states
  in two-dimensional frustrated quantum antiferromagnets, Phys. Rev. B 42~(13)
  (1990) 8206--8213.
\newblock \href {http://dx.doi.org/10.1103/PhysRevB.42.8206}
  {\path{doi:10.1103/PhysRevB.42.8206}}.

\bibitem{jiang:12}
H.-C. Jiang, H.~Yao, L.~Balents, Spin liquid ground state of the
  spin-$\frac{1}{2}$ square {${{J}}_{1}$}-{${{J}}_{2}$} {Heisenberg} model,
  Phys. Rev. B 86~(2) (2012) 024424.
\newblock \href {http://dx.doi.org/10.1103/PhysRevB.86.024424}
  {\path{doi:10.1103/PhysRevB.86.024424}}.

\bibitem{smerald:15}
A.~Smerald, H.~T. Ueda, N.~Shannon, Theory of inelastic neutron scattering in a
  field-induced spin-nematic state, Phys. Rev. B 91~(17) (2015) 174402.
\newblock \href {http://dx.doi.org/10.1103/PhysRevB.91.174402}
  {\path{doi:10.1103/PhysRevB.91.174402}}.

\bibitem{iqbal:16}
Y.~Iqbal, P.~Ghosh, R.~Narayanan, B.~Kumar, J.~Reuther, R.~Thomale, Intertwined
  nematic orders in a frustrated ferromagnet, Phys. Rev. B 94~(22) (2016)
  224403.
\newblock \href {http://dx.doi.org/10.1103/PhysRevB.94.224403}
  {\path{doi:10.1103/PhysRevB.94.224403}}.

\bibitem{trumper:99}
A.~E. Trumper, Spin-wave analysis to the spatially anisotropic {Heisenberg}
  antiferromagnet on a triangular lattice, Phys. Rev. B 60~(5) (1999)
  2987--2989.
\newblock \href {http://dx.doi.org/10.1103/PhysRevB.60.2987}
  {\path{doi:10.1103/PhysRevB.60.2987}}.

\bibitem{hauke:13}
P.~Hauke, Quantum disorder in the spatially completely anisotropic triangular
  lattice, Phys. Rev. B 87~(1) (2013) 014415.
\newblock \href {http://dx.doi.org/10.1103/PhysRevB.87.014415}
  {\path{doi:10.1103/PhysRevB.87.014415}}.

\bibitem{zheng:99}
W.~Zheng, R.~H. McKenzie, R.~R.~P. Singh, Phase diagram for a class of
  spin-$\frac{1}{2}$ {Heisenberg} models interpolating between the
  square-lattice, the triangular-lattice, and the linear-chain limits, Phys.
  Rev. B 59~(22) (1999) 14367--14375.
\newblock \href {http://dx.doi.org/10.1103/PhysRevB.59.14367}
  {\path{doi:10.1103/PhysRevB.59.14367}}.

\bibitem{ghorbani:16}
E.~Ghorbani, L.~F. Tocchio, F.~Becca, Variational wave functions for the
  ${S}=\frac{1}{2}$ {Heisenberg} model on the anisotropic triangular lattice:
  {Spin} liquids and spiral orders, Phys. Rev. B 93~(8) (2016) 085111.
\newblock \href {http://dx.doi.org/10.1103/PhysRevB.93.085111}
  {\path{doi:10.1103/PhysRevB.93.085111}}.

\bibitem{sachdev:90}
S.~Sachdev, R.~N. Bhatt, Bond-operator representation of quantum spins:
  {Mean}-field theory of frustrated quantum {Heisenberg} antiferromagnets,
  Phys. Rev. B 41~(13) (1990) 9323--9329.
\newblock \href {http://dx.doi.org/10.1103/PhysRevB.41.9323}
  {\path{doi:10.1103/PhysRevB.41.9323}}.

\bibitem{shindou:11}
R.~Shindou, S.~Yunoki, T.~Momoi, Projective studies of spin nematics in a
  quantum frustrated ferromagnet, Phys. Rev. B 84~(13) (2011) 134414.
\newblock \href {http://dx.doi.org/10.1103/PhysRevB.84.134414}
  {\path{doi:10.1103/PhysRevB.84.134414}}.

\bibitem{zhitomirsky:10}
M.~E. Zhitomirsky, H.~Tsunetsugu, Magnon pairing in quantum spin nematic, EPL
  92~(3) (2010) 37001.
\newblock \href {http://dx.doi.org/10.1209/0295-5075/92/37001}
  {\path{doi:10.1209/0295-5075/92/37001}}.

\bibitem{andreev:84}
A.~F. Andreev, I.~A. Grishchuk, Spin nematics, Sov. Phys. JETP (1984) 267.

\bibitem{singh:03}
R.~R.~P. Singh, W.~Zheng, J.~Oitmaa, O.~P. Sushkov, C.~J. Hamer, Symmetry
  {Breaking} in the {Collinear} {Phase} of the
  {${{J}}_{1}\ensuremath{-}{{J}}_{2}$} {Heisenberg} {Model}, Phys. Rev. Lett.
  91~(1) (2003) 017201.
\newblock \href {http://dx.doi.org/10.1103/PhysRevLett.91.017201}
  {\path{doi:10.1103/PhysRevLett.91.017201}}.

\bibitem{weber:03}
C.~Weber, L.~Capriotti, G.~Misguich, F.~Becca, M.~Elhajal, F.~Mila, Ising
  {Transition} {Driven} by {Frustration} in a 2d {Classical} {Model} with
  {Continuous} {Symmetry}, Phys. Rev. Lett. 91~(17) (2003) 177202.
\newblock \href {http://dx.doi.org/10.1103/PhysRevLett.91.177202}
  {\path{doi:10.1103/PhysRevLett.91.177202}}.

\bibitem{chubokov:91}
A.~V. Chubokov, D.~I. Golosov, Quantum theory of an antiferromagnet on a
  triangular lattice in a magnetic field, J. Phys.: Condens. Matter 3~(1)
  (1991) 69.
\newblock \href {http://dx.doi.org/10.1088/0953-8984/3/1/005}
  {\path{doi:10.1088/0953-8984/3/1/005}}.

\bibitem{coletta:16}
T.~Coletta, T.~A. T{\'o}th, K.~Penc, F.~Mila, Semiclassical theory of the
  magnetization process of the triangular lattice {Heisenberg} model, Phys.
  Rev. B 94~(7) (2016) 075136.
\newblock \href {http://dx.doi.org/10.1103/PhysRevB.94.075136}
  {\path{doi:10.1103/PhysRevB.94.075136}}.

\bibitem{hida:05}
K.~Hida, I.~Affleck, Quantum vs {Classical} {Magnetization} {Plateaus} of
  ${S}=1/2$ {Frustrated} {Heisenberg} {Chains}, J. Phys. Soc. Jpn. 74~(6)
  (2005) 1849--1857.
\newblock \href {http://dx.doi.org/10.1143/JPSJ.74.1849}
  {\path{doi:10.1143/JPSJ.74.1849}}.

\bibitem{alicea:09}
J.~Alicea, A.~V. Chubukov, O.~A. Starykh, Quantum {Stabilization} of the
  $1/3$-{Magnetization} {Plateau} in {${\text{{Cs}}}_{2}{\text{{CuBr}}}_{4}$},
  Phys. Rev. Lett. 102~(13) (2009) 137201.
\newblock \href {http://dx.doi.org/10.1103/PhysRevLett.102.137201}
  {\path{doi:10.1103/PhysRevLett.102.137201}}.

\bibitem{chubukov:13}
A.~V. Chubukov, O.~A. Starykh, Spin-{Current} {Order} in {Anisotropic}
  {Triangular} {Antiferromagnets}, Phys. Rev. Lett. 110~(21) (2013) 217210.
\newblock \href {http://dx.doi.org/10.1103/PhysRevLett.110.217210}
  {\path{doi:10.1103/PhysRevLett.110.217210}}.

\bibitem{parker:17}
E.~Parker, L.~Balents, Semiclassical analysis of a magnetization plateau in a
  two-dimensional frustrated ferrimagnet, Phys. Rev. B 95~(10) (2017) 104411.
\newblock \href {http://dx.doi.org/10.1103/PhysRevB.95.104411}
  {\path{doi:10.1103/PhysRevB.95.104411}}.

\bibitem{chen:13}
R.~Chen, H.~Ju, H.-C. Jiang, O.~A. Starykh, L.~Balents, Ground states of
  spin-$\frac{1}{2}$ triangular antiferromagnets in a magnetic field, Phys.
  Rev. B 87~(16) (2013) 165123.
\newblock \href {http://dx.doi.org/10.1103/PhysRevB.87.165123}
  {\path{doi:10.1103/PhysRevB.87.165123}}.

\bibitem{seabra:16}
L.~Seabra, P.~Sindzingre, T.~Momoi, N.~Shannon, Novel phases in a
  square-lattice frustrated ferromagnet : $\frac{1}{3}$-magnetization plateau,
  helicoidal spin liquid, and vortex crystal, Phys. Rev. B 93~(8) (2016)
  085132.
\newblock \href {http://dx.doi.org/10.1103/PhysRevB.93.085132}
  {\path{doi:10.1103/PhysRevB.93.085132}}.

\bibitem{sakai:11}
T.~Sakai, H.~Nakano, Critical magnetization behavior of the triangular- and
  kagome-lattice quantum antiferromagnets, Phys. Rev. B 83~(10) (2011) 100405.
\newblock \href {http://dx.doi.org/10.1103/PhysRevB.83.100405}
  {\path{doi:10.1103/PhysRevB.83.100405}}.

\bibitem{honecker:01}
A.~Honecker, Lanczos study of the {S} = 1/2 frustrated square-lattice
  anti-ferromagnet in a magnetic field, Can. J. Phys. 79~(11-12) (2001)
  1557--1563.
\newblock \href {http://dx.doi.org/10.1139/p01-095}
  {\path{doi:10.1139/p01-095}}.

\bibitem{morita:16}
K.~Morita, N.~Shibata, Field-{Induced} {Quantum} {Phase} {Transitions} in {S} =
  1/2 {J}1{\textendash}{J}2 {Heisenberg} {Model} on {Square} {Lattice}, J.
  Phys. Soc. Jpn. 85~(9) (2016) 094708.
\newblock \href {http://dx.doi.org/10.7566/JPSJ.85.094708}
  {\path{doi:10.7566/JPSJ.85.094708}}.

\bibitem{ono:03}
T.~Ono, H.~Tanaka, H.~Aruga~Katori, F.~Ishikawa, H.~Mitamura, T.~Goto,
  Magnetization plateau in the frustrated quantum spin system
  {${\text{{Cs}}}_{2}{\text{{CuBr}}}_{4}$}, Phys. Rev. B 67~(10) (2003) 104431.
\newblock \href {http://dx.doi.org/10.1103/PhysRevB.67.104431}
  {\path{doi:10.1103/PhysRevB.67.104431}}.

\bibitem{jacobs:93}
A.~E. Jacobs, T.~Nikuni, H.~Shiba, Theory of {Magnetic} {Structures} of
  {$\text{CsCuCl}_3$} in {Transverse} {Magnetic} {Field}, J. Phys. Soc. Jpn.
  62~(11) (1993) 4066--4080.
\newblock \href {http://dx.doi.org/10.1143/JPSJ.62.4066}
  {\path{doi:10.1143/JPSJ.62.4066}}.

\bibitem{koutroulakis:15}
G.~Koutroulakis, T.~Zhou, Y.~Kamiya, J.~D. Thompson, H.~D. Zhou, C.~D. Batista,
  S.~E. Brown, Quantum phase diagram of the ${S}=\frac{1}{2}$
  triangular-lattice antiferromagnet
  {${\text{{Ba}}}_{3}{\text{{CoSb}}}_{2}{\text{{O}}}_{9}$}, Phys. Rev. B 91~(2)
  (2015) 024410.
\newblock \href {http://dx.doi.org/10.1103/PhysRevB.91.024410}
  {\path{doi:10.1103/PhysRevB.91.024410}}.

\bibitem{svistov:06}
L.~E. Svistov, A.~I. Smirnov, L.~A. Prozorova, O.~A. Petrenko, A.~Micheler,
  N.~B{\"u}ttgen, A.~Y. Shapiro, L.~N. Demianets, Magnetic phase diagram,
  critical behavior, and two-dimensional to three-dimensional crossover in the
  triangular lattice antiferromagnet
  {$\text{{Rb}}\text{{Fe}}{(\text{{Mo}}{\text{{O}}}_{4})}_{2}$}, Phys. Rev. B
  74~(2) (2006) 024412.
\newblock \href {http://dx.doi.org/10.1103/PhysRevB.74.024412}
  {\path{doi:10.1103/PhysRevB.74.024412}}.

\bibitem{ye:17}
M.~Ye, A.~V. Chubukov, Half-magnetization plateau in a {Heisenberg}
  antiferromagnet on a triangular lattice, arXiv:1705.08515 [cond-mat].

\bibitem{millet:98}
P.~Millet, C.~Satto, Synthesis and structures of the layered vanadyl({IV})
  silico-germanates {Li}2vo({Si}1-{xGex}){O}4 (x = 0, 0.5, 1), Mater. Res.
  Bull. 33~(9) (1998) 1339--1345.
\newblock \href {http://dx.doi.org/10.1016/S0025-5408(98)00122-6}
  {\path{doi:10.1016/S0025-5408(98)00122-6}}.

\bibitem{melzi:00}
R.~Melzi, P.~Carretta, A.~Lascialfari, M.~Mambrini, M.~Troyer, P.~Millet,
  F.~Mila, {${\text{{Li}}}_{2}\text{{VO}}(\text{Si},\text{Ge}){\text{O}}_{4}$},
  Phys. Rev. Lett. 85~(6) (2000) 1318--1321.
\newblock \href {http://dx.doi.org/10.1103/PhysRevLett.85.1318}
  {\path{doi:10.1103/PhysRevLett.85.1318}}.

\bibitem{melzi:01}
R.~Melzi, S.~Aldrovandi, F.~Tedoldi, P.~Carretta, P.~Millet, F.~Mila, Magnetic
  and thermodynamic properties of {${\text{{Li}}}_{2}{\text{{VOSiO}}}_{4}:$}
  {A} two-dimensional ${S}=1/2$ frustrated antiferromagnet on a square lattice,
  Phys. Rev. B 64~(2) (2001) 024409.
\newblock \href {http://dx.doi.org/10.1103/PhysRevB.64.024409}
  {\path{doi:10.1103/PhysRevB.64.024409}}.

\bibitem{carretta:04}
P.~Carretta, N.~Papinutto, R.~Melzi, P.~Millet, S.~Gonthier, P.~Mendels, {Pawel
  Wzietek}, Magnetic properties of frustrated two-dimensional {S} = 1/2
  antiferromagnets on a square lattice, J. Phys.: Condens. Matter 16~(11)
  (2004) S849.
\newblock \href {http://dx.doi.org/10.1088/0953-8984/16/11/040}
  {\path{doi:10.1088/0953-8984/16/11/040}}.

\bibitem{nath:08}
R.~Nath, A.~A. Tsirlin, H.~Rosner, C.~Geibel, Magnetic properties of
  {$\text{{BaCdVO}}{({\text{{PO}}}_{4})}_{2}$}: {A} strongly frustrated
  spin-$\frac{1}{2}$ square lattice close to the quantum critical regime, Phys.
  Rev. B 78~(6) (2008) 064422.
\newblock \href {http://dx.doi.org/10.1103/PhysRevB.78.064422}
  {\path{doi:10.1103/PhysRevB.78.064422}}.

\bibitem{nath:09}
R.~Nath, Y.~Furukawa, F.~Borsa, E.~E. Kaul, M.~Baenitz, C.~Geibel, D.~C.
  Johnston, Single-crystal {$^{31}\text{{P}}$} {NMR} studies of the frustrated
  square-lattice compound
  {${\text{{Pb}}}_{2}(\text{{VO}}){({\text{{PO}}}_{4})}_{2}$}, Phys. Rev. B
  80~(21) (2009) 214430.
\newblock \href {http://dx.doi.org/10.1103/PhysRevB.80.214430}
  {\path{doi:10.1103/PhysRevB.80.214430}}.

\bibitem{skoulatos:07}
M.~Skoulatos, J.~P. Goff, N.~Shannon, E.~E. Kaul, C.~Geibel, A.~P. Murani,
  M.~Enderle, A.~R. Wildes, Spin correlations in the frustrated square lattice
  {Pb}2vo({PO}4)2, J. Magn. Magn. Mater. 310~(2, Part 2) (2007) 1257--1259.
\newblock \href {http://dx.doi.org/10.1016/j.jmmm.2006.10.379}
  {\path{doi:10.1016/j.jmmm.2006.10.379}}.

\bibitem{skoulatos:09}
M.~Skoulatos, J.~P. Goff, C.~Geibel, E.~E. Kaul, R.~Nath, N.~Shannon,
  B.~Schmidt, A.~P. Murani, P.~P. Deen, M.~Enderle, A.~R. Wildes, Spin
  correlations and exchange in square-lattice frustrated ferromagnets, EPL
  88~(5) (2009) 57005.
\newblock \href {http://dx.doi.org/10.1209/0295-5075/88/57005}
  {\path{doi:10.1209/0295-5075/88/57005}}.

\bibitem{tsirlin:10}
A.~A. Tsirlin, R.~Nath, A.~M. Abakumov, R.~V. Shpanchenko, C.~Geibel,
  H.~Rosner, Frustrated square lattice with spatial anisotropy: {Crystal}
  structure and magnetic properties of
  {$\text{{PbZnVO}}{({\text{{PO}}}_{4})}_{2}$}, Phys. Rev. B 81~(17) (2010)
  174424.
\newblock \href {http://dx.doi.org/10.1103/PhysRevB.81.174424}
  {\path{doi:10.1103/PhysRevB.81.174424}}.

\bibitem{tsirlin:09}
A.~A. Tsirlin, B.~Schmidt, Y.~Skourski, R.~Nath, C.~Geibel, H.~Rosner,
  Exploring the spin-{$\frac{1}{2}$} frustrated square lattice model with
  high-field magnetization studies, Phys. Rev. B 80~(13) (2009) 132407.
\newblock \href {http://dx.doi.org/10.1103/PhysRevB.80.132407}
  {\path{doi:10.1103/PhysRevB.80.132407}}.

\bibitem{xiao:09}
F.~Xiao, F.~M. Woodward, C.~P. Landee, M.~M. Turnbull, C.~Mielke, N.~Harrison,
  T.~Lancaster, S.~J. Blundell, P.~J. Baker, P.~Babkevich, F.~L. Pratt,
  Two-dimensional ${XY}$ behavior observed in quasi-two-dimensional quantum
  {Heisenberg} antiferromagnets, Phys. Rev. B 79~(13) (2009) 134412.
\newblock \href {http://dx.doi.org/10.1103/PhysRevB.79.134412}
  {\path{doi:10.1103/PhysRevB.79.134412}}.

\bibitem{tsyrulin:10}
N.~Tsyrulin, F.~Xiao, A.~Schneidewind, P.~Link, H.~M. R{\o}nnow, J.~Gavilano,
  C.~P. Landee, M.~M. Turnbull, M.~Kenzelmann, Two-dimensional square-lattice
  ${S}=\frac{1}{2}$ antiferromagnet
  $\text{{Cu}}{(\text{pz})}_{2}{({\text{{ClO}}}_{4})}_{2}$, Phys. Rev. B
  81~(13) (2010) 134409.
\newblock \href {http://dx.doi.org/10.1103/PhysRevB.81.134409}
  {\path{doi:10.1103/PhysRevB.81.134409}}.

\bibitem{tsyrulin:09}
N.~Tsyrulin, T.~Pardini, R.~R.~P. Singh, F.~Xiao, P.~Link, A.~Schneidewind,
  A.~Hiess, C.~P. Landee, M.~M. Turnbull, M.~Kenzelmann, Quantum {Effects} in a
  {Weakly} {Frustrated} ${S}=1/2$ {Two}-{Dimensional} {Heisenberg}
  {Antiferromagnet} in an {Applied} {Magnetic} {Field}, Phys. Rev. Lett.
  102~(19) (2009) 197201.
\newblock \href {http://dx.doi.org/10.1103/PhysRevLett.102.197201}
  {\path{doi:10.1103/PhysRevLett.102.197201}}.

\bibitem{majlis:92}
N.~Majlis, S.~Selzer, G.~C. Strinati, Dimensional crossover in the magnetic
  properties of highly anisotropic antiferromagnets, Phys. Rev. B 45~(14)
  (1992) 7872--7881.
\newblock \href {http://dx.doi.org/10.1103/PhysRevB.45.7872}
  {\path{doi:10.1103/PhysRevB.45.7872}}.

\bibitem{schmidt:13}
B.~Schmidt, M.~Siahatgar, P.~Thalmeier, Stabilization of {N{\'e}el} order in
  frustrated magnets with increasing magnetic field, EPJ Web of Conferences 40
  (2013) 04001.
\newblock \href {http://dx.doi.org/10.1051/epjconf/20134004001}
  {\path{doi:10.1051/epjconf/20134004001}}.

\bibitem{carlin:85}
R.~L. Carlin, R.~Burriel, F.~Palacio, R.~A. Carlin, S.~F. Keij, D.~W. Carnegie,
  Linear chain antiferromagnetic interactions in {Cs}2cucl4, Journal of Applied
  Physics 57~(8) (1985) 3351--3352.
\newblock \href {http://dx.doi.org/10.1063/1.335093}
  {\path{doi:10.1063/1.335093}}.

\bibitem{cong:11}
P.~T. Cong, B.~Wolf, M.~de~Souza, N.~Kr{\"u}ger, A.~A. Haghighirad,
  S.~Gottlieb-Schoenmeyer, F.~Ritter, W.~Assmus, I.~Opahle, K.~Foyevtsova,
  H.~O. Jeschke, R.~Valent{\'\i}, L.~Wiehl, M.~Lang, Distinct magnetic regimes
  through site-selective atom substitution in the frustrated quantum
  antiferromagnet {Cs}${}_{2}${CuCl}${}_{4\ensuremath{-}x}${Br}${}_{x}$, Phys.
  Rev. B 83~(6) (2011) 064425.
\newblock \href {http://dx.doi.org/10.1103/PhysRevB.83.064425}
  {\path{doi:10.1103/PhysRevB.83.064425}}.

\bibitem{radu:05}
T.~Radu, H.~Wilhelm, V.~Yushankhai, D.~Kovrizhin, R.~Coldea, Z.~Tylczynski,
  T.~L{\"u}hmann, F.~Steglich, Bose-{Einstein} {Condensation} of {Magnons} in
  {${\text{{Cs}}}_{2}{\text{{CuCl}}}_{4}$}, Phys. Rev. Lett. 95~(12) (2005)
  127202.
\newblock \href {http://dx.doi.org/10.1103/PhysRevLett.95.127202}
  {\path{doi:10.1103/PhysRevLett.95.127202}}.

\bibitem{coldea:96}
R.~Coldea, D.~A. Tennant, R.~A. Cowley, D.~F. McMorrow, B.~Dorner,
  Z.~Tylczynski, Neutron scattering study of the magnetic structure of
  {$\text{Cs}_2\text{CuCl}_4$}, J. Phys.: Condens. Matter 8~(40) (1996) 7473.
\newblock \href {http://dx.doi.org/10.1088/0953-8984/8/40/012}
  {\path{doi:10.1088/0953-8984/8/40/012}}.

\bibitem{coldea:02}
R.~Coldea, D.~A. Tennant, K.~Habicht, P.~Smeibidl, C.~Wolters, Z.~Tylczynski,
  Direct {Measurement} of the {Spin} {Hamiltonian} and {Observation} of
  {Condensation} of {Magnons} in the 2d {Frustrated} {Quantum} {Magnet}
  {${\text{{Cs}}}_{2}{\text{{CuCl}}}_{4}$}, Phys. Rev. Lett. 88~(13) (2002)
  137203.
\newblock \href {http://dx.doi.org/10.1103/PhysRevLett.88.137203}
  {\path{doi:10.1103/PhysRevLett.88.137203}}.

\bibitem{veillette:05-1}
M.~Y. Veillette, A.~J.~A. James, F.~H.~L. Essler, Spin dynamics of the
  quasi-two-dimensional spin-$\frac{1}{2}$ quantum magnet
  {${\text{{Cs}}}_{2}{\text{{CuCl}}}_{4}$}, Phys. Rev. B 72~(13) (2005) 134429.
\newblock \href {http://dx.doi.org/10.1103/PhysRevB.72.134429}
  {\path{doi:10.1103/PhysRevB.72.134429}}.

\bibitem{zvyagin:14}
S.~A. Zvyagin, D.~Kamenskyi, M.~Ozerov, J.~Wosnitza, M.~Ikeda, T.~Fujita,
  M.~Hagiwara, A.~I. Smirnov, T.~A. Soldatov, A.~Y. Shapiro, J.~Krzystek,
  R.~Hu, H.~Ryu, C.~Petrovic, M.~E. Zhitomirsky, Direct {Determination} of
  {Exchange} {Parameters} in {${\text{{Cs}}}_{2}{\text{{CuBr}}}_{4}$} and
  {${\text{{Cs}}}_{2}{\text{{CuCl}}}_{4}$}: {High}-{Field}
  {Electron}-{Spin}-{Resonance} {Studies}, Phys. Rev. Lett. 112~(7) (2014)
  077206.
\newblock \href {http://dx.doi.org/10.1103/PhysRevLett.112.077206}
  {\path{doi:10.1103/PhysRevLett.112.077206}}.

\bibitem{coldea:03}
R.~Coldea, D.~A. Tennant, Z.~Tylczynski, Extended scattering continua
  characteristic of spin fractionalization in the two-dimensional frustrated
  quantum magnet {${\text{{Cs}}}_{2}{\text{{CuCl}}}_{4}$} observed by neutron
  scattering, Phys. Rev. B 68~(13) (2003) 134424.
\newblock \href {http://dx.doi.org/10.1103/PhysRevB.68.134424}
  {\path{doi:10.1103/PhysRevB.68.134424}}.

\bibitem{coldea:01}
R.~Coldea, S.~M. Hayden, G.~Aeppli, T.~G. Perring, C.~D. Frost, T.~E. Mason,
  S.-W. Cheong, Z.~Fisk, Spin {Waves} and {Electronic} {Interactions} in
  {${\text{{La}}}_{2}{\text{{CuO}}}_{4}$}, Phys. Rev. Lett. 86~(23) (2001)
  5377--5380.
\newblock \href {http://dx.doi.org/10.1103/PhysRevLett.86.5377}
  {\path{doi:10.1103/PhysRevLett.86.5377}}.

\bibitem{fjaerestad:07}
J.~O. Fj{\ae}restad, W.~Zheng, R.~R.~P. Singh, R.~H. McKenzie, R.~Coldea,
  Excitation spectra and ground state properties of the layered
  spin-$\frac{1}{2}$ frustrated antiferromagnets
  {${\text{{Cs}}}_{2}\text{{Cu}}{\text{{Cl}}}_{4}$} and
  {${\text{{Cs}}}_{2}\text{{Cu}}{\text{{Br}}}_{4}$}, Phys. Rev. B 75~(17)
  (2007) 174447.
\newblock \href {http://dx.doi.org/10.1103/PhysRevB.75.174447}
  {\path{doi:10.1103/PhysRevB.75.174447}}.

\bibitem{cong:14}
P.~T. Cong, B.~Wolf, N.~v. Well, A.~A. Haghighirad, F.~Ritter, W.~Assmus,
  C.~Krellner, M.~Lang, Structural {Variations} and {Magnetic} {Properties} of
  the {Quantum} {Antiferromagnets} {$\text{Cs}_2\text{CuCl}_{4-x}\text{Br}_x$},
  IEEE Transactions on Magnetics 50~(6) (2014) 1--4.
\newblock \href {http://dx.doi.org/10.1109/TMAG.2014.2298496}
  {\path{doi:10.1109/TMAG.2014.2298496}}.

\bibitem{tsujii:07}
H.~Tsujii, C.~R. Rotundu, T.~Ono, H.~Tanaka, B.~Andraka, K.~Ingersent,
  Y.~Takano, Thermodynamics of the up-up-down phase of the ${S}=\frac{1}{2}$
  triangular-lattice antiferromagnet
  {${\text{{Cs}}}_{2}\text{{Cu}}{\text{{Br}}}_{4}$}, Phys. Rev. B 76~(6) (2007)
  060406.
\newblock \href {http://dx.doi.org/10.1103/PhysRevB.76.060406}
  {\path{doi:10.1103/PhysRevB.76.060406}}.

\bibitem{fortune:09}
N.~A. Fortune, S.~T. Hannahs, Y.~Yoshida, T.~E. Sherline, T.~Ono, H.~Tanaka,
  Y.~Takano, Cascade of {Magnetic}-{Field}-{Induced} {Quantum} {Phase}
  {Transitions} in a {Spin}-$\frac{1}{2}$ {Triangular}-{Lattice}
  {Antiferromagnet}, Phys. Rev. Lett. 102~(25) (2009) 257201.
\newblock \href {http://dx.doi.org/10.1103/PhysRevLett.102.257201}
  {\path{doi:10.1103/PhysRevLett.102.257201}}.

\bibitem{radu:06}
T.~Radu, H.~Wilhelm, V.~Yushankhai, D.~Kovrizhin, R.~Coldea, Z.~Tylczynski,
  T.~L{\"u}hmann, F.~Steglich, Radu et al. {Reply}:, Phys. Rev. Lett. 96~(18)
  (2006) 189704.
\newblock \href {http://dx.doi.org/10.1103/PhysRevLett.96.189704}
  {\path{doi:10.1103/PhysRevLett.96.189704}}.

\bibitem{zapf:14}
V.~Zapf, M.~Jaime, C.~D. Batista, Bose-{Einstein} condensation in quantum
  magnets, Rev. Mod. Phys. 86~(2) (2014) 563--614.
\newblock \href {http://dx.doi.org/10.1103/RevModPhys.86.563}
  {\path{doi:10.1103/RevModPhys.86.563}}.

\bibitem{kovrizhin:06}
D.~L. Kovrizhin, V.~Yushankhai, L.~Siurakshina, Bose-{Einstein} condensation of
  magnons in {${\text{{Cs}}}_{2}{\text{{CuCl}}}_{4}$}: {A} dilute gas limit
  near the saturation magnetic field, Phys. Rev. B 74~(13) (2006) 134417.
\newblock \href {http://dx.doi.org/10.1103/PhysRevB.74.134417}
  {\path{doi:10.1103/PhysRevB.74.134417}}.

\bibitem{nikuni:00}
T.~Nikuni, M.~Oshikawa, A.~Oosawa, H.~Tanaka, Bose-{Einstein} {Condensation} of
  {Dilute} {Magnons} in {${\text{{TlCuCl}}}_{3}$}, Phys. Rev. Lett. 84~(25)
  (2000) 5868--5871.
\newblock \href {http://dx.doi.org/10.1103/PhysRevLett.84.5868}
  {\path{doi:10.1103/PhysRevLett.84.5868}}.

\bibitem{han:09}
M.~J. Han, Q.~Yin, W.~E. Pickett, S.~Y. Savrasov, Anisotropy, {Itineracy}, and
  {Magnetic} {Frustration} in {High}-{${{T}}_{\text{C}}$} {Iron} {Pnictides},
  Phys. Rev. Lett. 102~(10) (2009) 107003.
\newblock \href {http://dx.doi.org/10.1103/PhysRevLett.102.107003}
  {\path{doi:10.1103/PhysRevLett.102.107003}}.

\bibitem{diallo:09}
S.~O. Diallo, V.~P. Antropov, T.~G. Perring, C.~Broholm, J.~J. Pulikkotil,
  N.~Ni, S.~L. Bud{\textquoteright}ko, P.~C. Canfield, A.~Kreyssig, A.~I.
  Goldman, R.~J. McQueeney, Itinerant {Magnetic} {Excitations} in
  {Antiferromagnetic} {${\text{{CaFe}}}_{2}{\text{{As}}}_{2}$}, Phys. Rev.
  Lett. 102~(18) (2009) 187206.
\newblock \href {http://dx.doi.org/10.1103/PhysRevLett.102.187206}
  {\path{doi:10.1103/PhysRevLett.102.187206}}.

\bibitem{eremin:10}
I.~Eremin, A.~V. Chubukov, Magnetic degeneracy and hidden metallicity of the
  spin-density-wave state in ferropnictides, Phys. Rev. B 81~(2) (2010) 024511.
\newblock \href {http://dx.doi.org/10.1103/PhysRevB.81.024511}
  {\path{doi:10.1103/PhysRevB.81.024511}}.

\bibitem{wang:13}
C.~Wang, R.~Zhang, F.~Wang, H.~Luo, L.~P. Regnault, P.~Dai, Y.~Li, Longitudinal
  {Spin} {Excitations} and {Magnetic} {Anisotropy} in {Antiferromagnetically}
  {Ordered} {$\text{{Ba}}{\text{{Fe}}}_{2}{\text{{As}}}_{2}$}, Phys. Rev. X
  3~(4) (2013) 041036.
\newblock \href {http://dx.doi.org/10.1103/PhysRevX.3.041036}
  {\path{doi:10.1103/PhysRevX.3.041036}}.

\bibitem{yamamoto:14}
D.~Yamamoto, G.~Marmorini, I.~Danshita, Quantum {Phase} {Diagram} of the
  {Triangular}-{Lattice} ${XXZ}$ {Model} in a {Magnetic} {Field}, Phys. Rev.
  Lett. 112~(12) (2014) 127203.
\newblock \href {http://dx.doi.org/10.1103/PhysRevLett.112.127203}
  {\path{doi:10.1103/PhysRevLett.112.127203}}.

\bibitem{sellmann:15}
D.~Sellmann, X.-F. Zhang, S.~Eggert, Phase diagram of the antiferromagnetic
  {XXZ} model on the triangular lattice, Phys. Rev. B 91~(8) (2015) 081104.
\newblock \href {http://dx.doi.org/10.1103/PhysRevB.91.081104}
  {\path{doi:10.1103/PhysRevB.91.081104}}.

\bibitem{yamamoto:17}
D.~Yamamoto, H.~Ueda, I.~Danshita, G.~Marmorini, T.~Momoi, T.~Shimokawa, Exact
  diagonalization and cluster mean-field study of triangular-lattice {XXZ}
  antiferromagnets near saturation, Phys. Rev. B 96~(1) (2017) 014431.
\newblock \href {http://dx.doi.org/10.1103/PhysRevB.96.014431}
  {\path{doi:10.1103/PhysRevB.96.014431}}.

\bibitem{nikuni:95}
T.~Nikuni, H.~Shiba, Hexagonal {Antiferromagnets} in {Strong} {Magnetic}
  {Field}: {Mapping} onto {Bose} {Condensation} of {Low}-{Density} {Bose}
  {Gas}, J. Phys. Soc. Jpn. 64~(9) (1995) 3471--3483.
\newblock \href {http://dx.doi.org/10.1143/JPSJ.64.3471}
  {\path{doi:10.1143/JPSJ.64.3471}}.

\bibitem{yamamoto:14-1}
D.~Yamamoto, G.~Marmorini, I.~Danshita, Erratum: {Quantum} {Phase} {Diagram} of
  the {Triangular}-{Lattice} ${XXZ}$ {Model} in a {Magnetic} {Field} [{Phys}.
  {Rev}. {Lett}. 112, 127203 (2014)], Phys. Rev. Lett. 112~(25) (2014) 259901.
\newblock \href {http://dx.doi.org/10.1103/PhysRevLett.112.259901}
  {\path{doi:10.1103/PhysRevLett.112.259901}}.

\bibitem{li:15}
Y.~Li, G.~Chen, W.~Tong, L.~Pi, J.~Liu, Z.~Yang, X.~Wang, Q.~Zhang,
  Rare-{Earth} {Triangular} {Lattice} {Spin} {Liquid}: {A} {Single}-{Crystal}
  {Study} of {${\text{{YbMgGaO}}}_{4}$}, Phys. Rev. Lett. 115~(16) (2015)
  167203.
\newblock \href {http://dx.doi.org/10.1103/PhysRevLett.115.167203}
  {\path{doi:10.1103/PhysRevLett.115.167203}}.

\bibitem{shen:16}
Y.~Shen, Y.-D. Li, H.~Wo, Y.~Li, S.~Shen, B.~Pan, Q.~Wang, H.~C. Walker,
  P.~Steffens, M.~Boehm, Y.~Hao, D.~L. Quintero-Castro, L.~W. Harriger, M.~D.
  Frontzek, L.~Hao, S.~Meng, Q.~Zhang, G.~Chen, J.~Zhao, Evidence for a spinon
  {Fermi} surface in a triangular-lattice quantum-spin-liquid candidate, Nature
  540~(7634) (2016) 559--562.
\newblock \href {http://dx.doi.org/10.1038/nature20614}
  {\path{doi:10.1038/nature20614}}.

\bibitem{ma:17}
Z.~Ma, J.~Wang, Z.-Y. Dong, J.~Zhang, S.~Li, S.-H. Zheng, Y.~Yu, W.~Wang,
  L.~Che, K.~Ran, S.~Bao, Z.~Cai, P.~{\v C}erm{\'a}k, A.~Schneidewind, S.~Yano,
  J.~S. Gardner, X.~Lu, S.-L. Yu, J.-M. Liu, S.~Li, J.-X. Li, J.~Wen, Quantum
  spin liquid or spin glass: identifying the triangular-lattice compounds
  {YbZnGaO}$_4$ and {YbMgGaO}$_4$, arXiv:1709.00256 [cond-mat].

\bibitem{danu:16}
B.~Danu, G.~Nambiar, R.~Ganesh, Extended degeneracy and order by disorder in
  the square lattice {${{J}}_{1}$-${{J}}_{2}$-${{J}}_{3}$} model, Phys. Rev. B
  94~(9) (2016) 094438.
\newblock \href {http://dx.doi.org/10.1103/PhysRevB.94.094438}
  {\path{doi:10.1103/PhysRevB.94.094438}}.

\bibitem{kaneko:14}
R.~Kaneko, S.~Morita, M.~Imada, Gapless {Spin}-{Liquid} {Phase} in an
  {Extended} {Spin} 1/2 {Triangular} {Heisenberg} {Model}, J. Phys. Soc. Jpn.
  83~(9) (2014) 093707.
\newblock \href {http://dx.doi.org/10.7566/JPSJ.83.093707}
  {\path{doi:10.7566/JPSJ.83.093707}}.

\bibitem{iqbal:16-1}
Y.~Iqbal, W.-J. Hu, R.~Thomale, D.~Poilblanc, F.~Becca, Spin liquid nature in
  the {Heisenberg} {${{J}}_{1}\ensuremath{-}{{J}}_{2}$} triangular
  antiferromagnet, Phys. Rev. B 93~(14) (2016) 144411.
\newblock \href {http://dx.doi.org/10.1103/PhysRevB.93.144411}
  {\path{doi:10.1103/PhysRevB.93.144411}}.

\bibitem{toader:05}
A.~M. Toader, J.~P. Goff, M.~Roger, N.~Shannon, J.~R. Stewart, M.~Enderle, Spin
  {Correlations} in the {Paramagnetic} {Phase} and {Ring} {Exchange} in
  {${\text{{La}}}_{2}{\text{{CuO}}}_{4}$}, Phys. Rev. Lett. 94~(19) (2005)
  197202.
\newblock \href {http://dx.doi.org/10.1103/PhysRevLett.94.197202}
  {\path{doi:10.1103/PhysRevLett.94.197202}}.

\bibitem{wietek:17}
A.~Wietek, A.~M. L{\"a}uchli, Chiral spin liquid and quantum criticality in
  extended {${S}=\frac{1}{2}$} {Heisenberg} models on the triangular lattice,
  Phys. Rev. B 95~(3) (2017) 035141.
\newblock \href {http://dx.doi.org/10.1103/PhysRevB.95.035141}
  {\path{doi:10.1103/PhysRevB.95.035141}}.

\bibitem{lauchli:06}
A.~L{\"a}uchli, F.~Mila, K.~Penc, Quadrupolar {Phases} of the ${S}=1$
  {Bilinear}-{Biquadratic} {Heisenberg} {Model} on the {Triangular} {Lattice},
  Phys. Rev. Lett. 97~(8) (2006) 087205.
\newblock \href {http://dx.doi.org/10.1103/PhysRevLett.97.087205}
  {\path{doi:10.1103/PhysRevLett.97.087205}}.

\bibitem{tsunetsugu:06}
H.~Tsunetsugu, M.~Arikawa, Spin {Nematic} {Phase} in ${S}=1$ {Triangular}
  {Antiferromagnets}, J. Phys. Soc. Jpn. 75~(8) (2006) 083701.
\newblock \href {http://dx.doi.org/10.1143/JPSJ.75.083701}
  {\path{doi:10.1143/JPSJ.75.083701}}.

\bibitem{nakatsuji:10}
S.~Nakatsuji, Y.~Nambu, S.~Onoda, Novel {Geometrical} {Frustration} {Effects}
  in the {Two}-{Dimensional} {Triangular}-{Lattice} {Antiferromagnet}
  {$\text{NiGa}_2\text S_4$} and {Related} {Compounds}, J. Phys. Soc. Jpn.
  79~(1) (2010) 011003.
\newblock \href {http://dx.doi.org/10.1143/JPSJ.79.011003}
  {\path{doi:10.1143/JPSJ.79.011003}}.

\bibitem{ohkawa:85}
F.~J. Ohkawa, Orbital {Antiferromagnetism} in {$\text{CeB}_6$}, J. Phys. Soc.
  Jpn. 54~(10) (1985) 3909--3914.
\newblock \href {http://dx.doi.org/10.1143/JPSJ.54.3909}
  {\path{doi:10.1143/JPSJ.54.3909}}.

\bibitem{shiina:97}
R.~Shiina, H.~Shiba, P.~Thalmeier, Magnetic-{Field} {Effects} on {Quadrupolar}
  {Ordering} in a {$\Gamma_8$}-{Quartet} {System} {$\text{CeB}_6$}, J. Phys.
  Soc. Jpn. 66~(6) (1997) 1741--1755.
\newblock \href {http://dx.doi.org/10.1143/JPSJ.66.1741}
  {\path{doi:10.1143/JPSJ.66.1741}}.

\bibitem{penc:03}
K.~Penc, M.~Mambrini, P.~Fazekas, F.~Mila, Quantum phase transition in the
  {SU}(4) spin-orbital model on the triangular lattice, Phys. Rev. B 68~(1)
  (2003) 012408.
\newblock \href {http://dx.doi.org/10.1103/PhysRevB.68.012408}
  {\path{doi:10.1103/PhysRevB.68.012408}}.

\bibitem{haan:92}
O.~Haan, J.-U. Klaetke, K.-H. M{\"u}tter, Ground-state staggered magnetization
  of the antiferromagnetic {Heisenberg} model, Phys. Rev. B 46~(9) (1992)
  5723--5726.
\newblock \href {http://dx.doi.org/10.1103/PhysRevB.46.5723}
  {\path{doi:10.1103/PhysRevB.46.5723}}.

\bibitem{schulz:92}
H.~J. Schulz, T.~A.~L. Ziman, Finite-{Size} {Scaling} for the
  {Two}-{Dimensional} {Frustrated} {Quantum} {Heisenberg} {Antiferromagnet},
  EPL 18~(4) (1992) 355.
\newblock \href {http://dx.doi.org/10.1209/0295-5075/18/4/013}
  {\path{doi:10.1209/0295-5075/18/4/013}}.

\bibitem{lyness:91}
J.~N. Lyness, T.~S{\o}revik, P.~Keast, Notes on {Integration} and {Integer}
  {Sublattices}, Mathematics of Computation 56~(193) (1991) 243--255.
\newblock \href {http://dx.doi.org/10.2307/2008539}
  {\path{doi:10.2307/2008539}}.

\bibitem{domich:87}
P.~D. Domich, R.~Kannan, L.~E. Trotter, Hermite {Normal} {Form} {Computation}
  {Using} {Modulo} {Determinant} {Arithmetic}, Mathematics of Operations
  Research 12~(1) (1987) 50--59.
\newblock \href {http://dx.doi.org/10.1287/moor.12.1.50}
  {\path{doi:10.1287/moor.12.1.50}}.

\bibitem{schrijver:98}
A.~Schrijver, Theory of {Linear} and {Integer} {Programming}, Wiley, 1998.

\end{thebibliography}

\end{document}